\documentclass[sn-basic]{sn-jnl}

\usepackage{graphicx,amssymb,amsmath,times}
\usepackage{longtable,rotating}
\usepackage{bm}
\usepackage{url}
\usepackage[varg]{txfonts}
\usepackage[T1]{fontenc}
\usepackage{placeins}
\usepackage{float}
\usepackage[normalem]{ulem}

\DeclareRobustCommand{\ion}[2]{\textup{#1\,\textsc{\lowercase{#2}}}}

\begin{document}

\title[Cepheids as distance indicators and stellar tracers]{Cepheids as distance indicators and stellar tracers}

\author*[1,2]{\fnm{G.} \sur{Bono}}\email{bono@roma2.infn.it}
\author[2]{\fnm{V.~F.} \sur{Braga}}\email{vittorio.braga@inaf.it}
\author[3]{\fnm{A.} \sur{Pietrinferni}}\email{adriano.pietrinferni@inaf.it}

\affil*[1]{\orgdiv{Department of Physics}, \orgname{Universit\`a di Roma Tor Vergata}, \orgaddress{\street{via della Ricerca Scientifica 1}, \city{Roma}, \postcode{00133}, \country{Italy}}}
\affil*[2]{\orgname{INAF-Osservatorio Astronomico di Roma}, \orgaddress{\street{via Frascati 33}, \city{Monte Porzio Catone}, \postcode{00078}, \country{Italy}}}
\affil*[3]{\orgname{INAF-Osservatorio Astronomico d'Abruzzo}, \orgaddress{\street{via M. Maggini s/n}, \city{Teramo}, \postcode{64100}, \country{Italy}}}

 % \date{\centering Submitted \today\ / Received / Accepted }

\abstract{
We review the phenomenology of classical Cepheids (CCs), 
Anomalous Cepheids (ACs) and type II Cepheids (TIICs) in 
the Milky Way (MW) and in the Magellanic Clouds (MCs).  
We also examine the Hertzsprung progression in different 
stellar systems by using the shape of \textit{I}-band light curves 
(Fourier parameters) and observables based on the difference 
in magnitude and in phase between the bump and the minimum 
in luminosity.  
%
%The change in the mass-luminosity relation at the transition 
%between stellar structures which ignite helium in a partially 
%electron-degenerate core (low-mass) and those which ignite 
%helium quiescently (intermediate-mass) is also discussed. 
%
The distribution of Cepheids in optical and in optical-near 
infrared (NIR) color--magnitude diagrams is investigated to 
constrain the topology of the instability strip. The use of 
Cepheids as tracers of young (CCs), intermediate (ACs) and 
old (TIICs) stellar populations are brought forward by the 
comparison between observations (MCs) and cluster isochrones 
covering a broad range in stellar ages and in chemical compositions.  
The different diagnostics adopted to estimate individual distances 
(period--luminosity, period--Wesenheit, period--luminosity--color relations) 
are reviewed together with pros and cons in the use of fundamental and 
overtones, optical and NIR photometric bands, and reddening free pseudo 
magnitudes (Wesenheit).  We also discuss the use of CCs as stellar tracers 
and the radial gradients among the different groups of elements (iron, 
$\alpha$, neutron-capture) together with their age-dependence.  
Finally, we briefly outline the role that near-future space and ground-based 
facilities will play in the astrophysical and cosmological use of Cepheids.} 

\maketitle

%editor 3 
%%%%%%%%%%%%%%%%%%%%%%%%%%%%%%%%%%%%%%%%%%%%%%%%%%%%%%%%%%%%%%%%%%%%%%%%%%%%%%%%%%%%%%%
% Table 1 
%%%%%%%%%%%%%%%%%%%%%%%%%%%%%%%%%%%%%%%%%%%%%%%%%%%%%%%%%%%%%%%%%%%%%%%%%%%%%%%%%%%%%%%
\begin{table*}[htbp]
\scriptsize
\caption{List of acronyms}
\label{tbl:acronym}
%% \centering
\begin{tabular}{ll}
%% \hline
%% \hline
AC    & Anomalous Cepheid                                                   \\
AGB   & Asymptotic Giant Branch                                             \\
AHB   & Above Horizontal Branch                                             \\
BC    & Bolometric Correction                                               \\
BHB   & Blue Horizontal Branch                                              \\
BLH   & BL Herculis                                                         \\
CC    & Classical Cepheid                                                   \\
CMD   & Color-Magnitude Diagram                                             \\
DORB  & Classical Cepheids with a bump moving down along the rising branch  \\
EHB   & Extreme Horizontal Branch                                           \\
ELT   & Extremely Large Telescope                                           \\
FO    & First Overtone                                                      \\
FU    & Fundamental                                                         \\
HB    & Horizontal Branch                                                   \\
HRD   & Hertzsprung--Russel Diagram                                         \\
HP    & Hertzsprung Progression                                             \\
IDUP  & First Dredge-up                                                     \\
%LBV  & Luminous Blue Variable                                              \\
LMC   & Large Magellanic Cloud                                              \\
LTE   & Local Thermodynamic Equilibrium                                     \\
MCs   & Magellanic Clouds                                                   \\
MIR   & Mid Infrared                                                        \\
ML    & Mass-luminosity                                                     \\
MS    & Main Sequence                                                       \\
MW    & Milky Way                                                           \\
NIR   & Near Infrared                                                       \\
OC    & Open Cluster                                                        \\
PEAGB & Post-Early Asymptotic Giant Branch                                  \\
P(HP) & Period at the center of the Hertzsprung Progression                 \\
PL    & Period--Luminosity relation                                          \\
PLC   & Period--Luminosity-Color relation                                    \\
PW    & Period--Wesenheit relation                                           \\
pWV   & peculiar W Virginis                                                 \\
RGB   & Red Giant Branch                                                    \\
RHB   & Red Horizontal Branch                                               \\
RRL   & RR Lyrae                                                            \\  
RVT   & RV Tauri                                                            \\
SMC   & Small Magellanic Cloud                                              \\
SO    & Second Overtone                                                     \\
TIIC  & Type II Cepheid                                                     \\
TO    & Third Overtone                                                      \\
TP    & Thermal Pulse                                                       \\
TPAGB & Thermal-pulsing Asymptotic Giant Branch                             \\ 
TRGB  & Tip of the Red Giant Branch                                         \\
UPDB  & Classical Cepheid with a bump moving up along the decreasing branch \\
WD    & White Dwarf                                                         \\
WV    & W Virginis                                                          \\
ZAHB  & Zero-Age Horizontal Branch                                          \\
%% \hline
\end{tabular}
\end{table*}

\clearpage
\pagestyle{myheadings}
\markright{G. Bono} %% fix to remove CONTENTS from toc below
{\small
\setcounter{tocdepth}{3} % TOC subsubsections
\tableofcontents
}

%%FIRST
%%%%%%%%%%%%%%%%%%%%%%%%%%%%%%%%%%%%%%%%%%%%%%%%%%%%%%%%%%%%%%%%%%%%%%%%%%%%%%%%
%editor_7
\section{Introduction}

Cepheids\footnote{The first Classical Cepheid -- $\eta$ Antinoi 
(the Antinous constellation was later merged with the Aquila constellation and 
the currently adopted name is $\eta$ Aql) 
was discovered by the British amateur astronomer Edward Pigott on Septemebr 10, 1784, 
who observed it with  his own telescope.  He provided a very accurate estimate of its 
pulsation period 7.18 days compared with the current value of 7.176641 days. 
However, the official discovery was attributed to another amateur astronomer, 
John Goodricke, who, one month later, on October 20, discovered with his 
own telescope that the star $\delta$ in the constellation of Cepheus was 
a variable star. Observing the same object in subsequent nights, he 
estimated a period of 5.37 days (the current estimate is 5.366249 days). 
Edward was much more experienced with observations and 
he can be considered John's mentor, since he was 17 years old. John and 
Edward were friends and neighbours in York and they often observed 
together. John also had some health problems (he was deaf and mute).  
These are probably the reasons why the discovery was assigned to him 
and the objects are now-days called Cepheids instead of Aquilaes
\citep{hoskin1979,croswell1997}.}
% ref_1_8
are radially-pulsating variables with periodic luminosity variations 
used as standardizable candles and tracers of stellar populations. This review focusses 
on the three different flavours of Cepheids: classical Cepheids (CC, pulsation 
periods ranging from roughly one day to a few hundred days, absolute visual magnitudes 
ranging from $M_V\sim-2.5$ mag to $M_V\sim-7.5$ mag) tracers of stellar populations 
younger than 200--300 Myrs; Anomalous Cepheids (ACs, pulsation periods ranging from 
% ref_1_9 
half a day to a few days, absolute visual magnitudes ranging from $M_V\sim-1$ mag to 
$M_V\sim-2.5$ mag) tracers of intermediate-age (a few Gyrs) stellar populations 
and type II Cepheids (TIICs, pulsation periods ranging from one to a hubndred
days, absolute visual magnitudes ranging from $M_V\sim-1$ mag to $M_V\sim-4.5$ mag) 
tracers of stellar populations older than 10 Gyrs. They have been the topic 
of many detailed theoretical and empirical investigations and several reviews 
% ref_1_10  
\citep{kraft1963,sandage1972,madore1985,madore1991,gautschysaio1995,gautschysaio1996,sandage06a,tammann2008,freedman10,bono99b,bono99a,bono2010,feast13,Subramanian2017,Bhardwaj2020b}

%ref_1_11
Variable stars have played a key role in the development of modern, quantitative 
Astrophysics. Baade's discovery \citep{baade56} that CCs and TIICs trace young and old 
stellar populations and obey to two different period--luminosity (PL) relations
has had an unprecedented impact upon both size and age of the Universe.
Moreover, CCs one century ago were used to estimate 
the distance to M31 and solve the Great Debate concerning the extragalactic 
nature of the Nebulae \citep{hubble1925,hubble1929,sandage2005} and to trace, for the first time, the 
rotation of the Milky Way (MW) thin disc 
\citep[][and references therein]{oort1927,joy1939,degrijs2017}. More recently,  
they have been used as fundamental laboratories to constrain evolutionary 
\citep{anderson2016,bono2020b,desomma2021} and pulsation 
\citep{bono00a,marconi05,neilson2008,neilson2011}
properties of low and intermediate-mass stars. 

The lively debate concerning the difference between evolutionary and pulsation 
mass of classical Cepheids dates back to the 1970s/80s and required significant 
improvements in 
micro-physics \citep[radiative opacities,][]{rogers1992, seaton1994}
and in macro-physics: time-dependent convection,  \citep{kippenhanhn1980,stellingwerf82,stellingwerf82b,stellingwerf1984,stellingwerf1984b,moskalik1992b,bono94a,maeder2000}; 
extra-mixing \citep{chiosi1992} and mass-loss \citep{neilson2011}. 

%_ref_2_1
The quest for accurate empirical estimates of the current mass
of CCs reached a conclusion thanks to the discovery of CCs in double eclipsing binary
systems by large photometric surveys of variable stars in the Magellanic Clouds
(MCs) provided by microlensing experiments (MACHO, OGLE). The game changer in this context
was the early discovery by \citet{udalski99} and by \citet{welch1999}, and in particular, 
the detailed and thorough investigations by \citet{soszynski08a} and 
by \citet{udalski2015b}.

%ref_2_2
The spectroscopic follow-up of these systems provided the opportunity to
measure with a geometrical method the mass of a classical Cepheid with a
1\% accuracy \citep{pietrzynski10}.
This discovery was well complemented by the discovery of a few 
systems in the Large Magellanic Cloud \citep[LMC,][]{gieren2015}, in the    
Small Magellanic Cloud \citep[SMC,][]{Graczyk2012,pietrzynski2013} 
and with the measurement of the dynamical mass of Polaris \citep{evans2008}.
This paramount observational effort provided the opportunity to measure the
SMC distance with 2\% accuracy \citep{graczyk2020}.

% ref_1_12
These measurements are crucial in order to fix the zero-point of the
mass--luminosity relation of core helium-burning stellar structures 
\citep{neilson2011,brott2011,cassisi11,pradamoroni2012},
but the number is still too limited to constrain the slope over the entire
period range of both FO \citep{pilecki2015} and FU \citep{pilecki2021} CCs. 
%ref_2_2
These systems are also fundamental laboratories to improve the 
physics of nonlinear pulsation models \citep{szabo2011, marconi13a, paxton2019, smolec2023}.

%ref_2_3 
An accurate dynamical mass-estimate has also been recently provided for
a TIIC by \citet{pilecki2018}, confirming the evolutionary channel producing 
these variables (see section~\ref{cmd_TIIC}). We still lack firm estimates of the dynamical 
mass of an AC. In a recent very detailed investigation, \citet{pilecki2017} suggested 
that OGLE-LMC-T2CEP-098 is an AC, but its period is too long and its 
color is too red to be a canonical AC. Accurate measurements await
for more solid identifications and for detailed predictions 
concerning binary evolution (mass-transfer phase) of pulsating stars.

The universality of the diagnostics adopted to estimate individual Cepheid distances 
has also been a long-standing astrophysical problem. The period--luminosity relation 
% editor_6
(PL) was discovered more than one century ago by Henrietta Leavitt, \citep{leavitt08,leavitt1912} 
in a titanic effort to identify variable stars in the MCs. A few decades 
later, it became clear - on the basis of observations and plain physical arguments - that 
accurate CC individual distances would require the use of a period--luminosity--color 
\citep[PLC,][]{sandage1972,sandage06a} relation. 
%ref_1_12
Individual distances based on the PL relation rely on the assumption that the
width in temperature of the Cepheid instability strip can be neglected
\citep{bono99b}. This assumption is less severe in the near-infrared than in the
optical regime. Nonlinear, convective pulsation models at fixed chemical composition,
stellar mass and luminosity, indicate that Cepheids in the NIR regime become
systematically brighter when moving from the blue (hotter) to the red (cooler)
edge of the instability strip, whereas their magnitudes are almost constant in the
optical regime. The difference is caused by the stronger sensitivity to effective
temperature of NIR bolometric corrections when compared with optical bolometric
corrections. As a consequence, the spread in magnitude of the NIR PL relation,
at fixed pulsation periods, is on average a factor of two smaller than the optical
PL relations \citep[see Fig.~6 in][]{bono99a}. This improvement was brought 
forward long ago on an empirical basis by \citet{mcgonegal82} and \citet{mcalary1983} 
and it is independent of the mild dependence of NIR bands on extinction corrections.

However, the key question in using these diagnostics is: Are they Universal? 
%_ref_1_13 
Can one calibrate them 
here and now and use them across the Local Universe?  A different way to rephrase the 
same question is: Are the PL and the PLC relations dependent on the metallicity? 
This is far from being an academic issue, since the current PL/PLC relations are 
calibrated by using Galactic Cepheids to fix the zero-point of the relations and the 
MC Cepheids to fix the slope of the same relations. However, the mean observed iron 
abundance of Galactic Cepheids is around solar ([Fe/H]$\sim$0) while MC Cepheids are on average 
a factor of two (LMC, [Fe/H]$\sim$--0.5) and a factor of four (SMC, [Fe/H]$\sim$--0.7) 
more metal-poor.

%%%%%%%%%%%%%%%%%%%%%%%%%%%%%%%%%%%%%%%%%%%%%%%%%%%%%%%%%%%%%%%%%%%%%%%%%%%%%%%%%%%%%%
% ref_1_14
\section{Phenomenology of Cepheids}

Cepheids are typically divided into three different sub--groups: 
classical Cepheids, Anomalous Cepheids, and Type II Cepheids. 
%BG 
%\footnote{Bacon, who is widely regarded as the founder of modern science in England, typically 
%maintains that knowledge requires careful observation of nature to avoid distortions 
%introduced by the mind. In particular, ``all depends on keeping the eye steadily fixed 
%upon the facts of nature and so receiving their images simply as they are.'' 
%from {\em Instauratio Magna} ({\em Great Instauration})
%translated by J. Spedding, R.L. Ellis and D.D. Heath, Volume 4, Book 1,
%London: Longman \& Co., 1870.}.

%_______________________________________________________________________________
\subsection{Classical Cepheids}\label{sec:ccs_intro}
%_______________________________________________________________________________

%%%%%%%%%%%%%%%%%%%%%%%%%%%%%%%%%%%%%%%%%%%%%%%%%%%%%%%%%%%%%%%%%%%%%%%%%%%%%%%%%%%%%
% 			fig 1
%%%%%%%%%%%%%%%%%%%%%%%%%%%%%%%%%%%%%%%%%%%%%%%%%%%%%%%%%%%%%%%%%%%%%%%%%%%%%%%%%%%%%
\begin{figure}[htbp]
\begin{center}
\includegraphics[height=0.8\textheight,width=\textwidth]{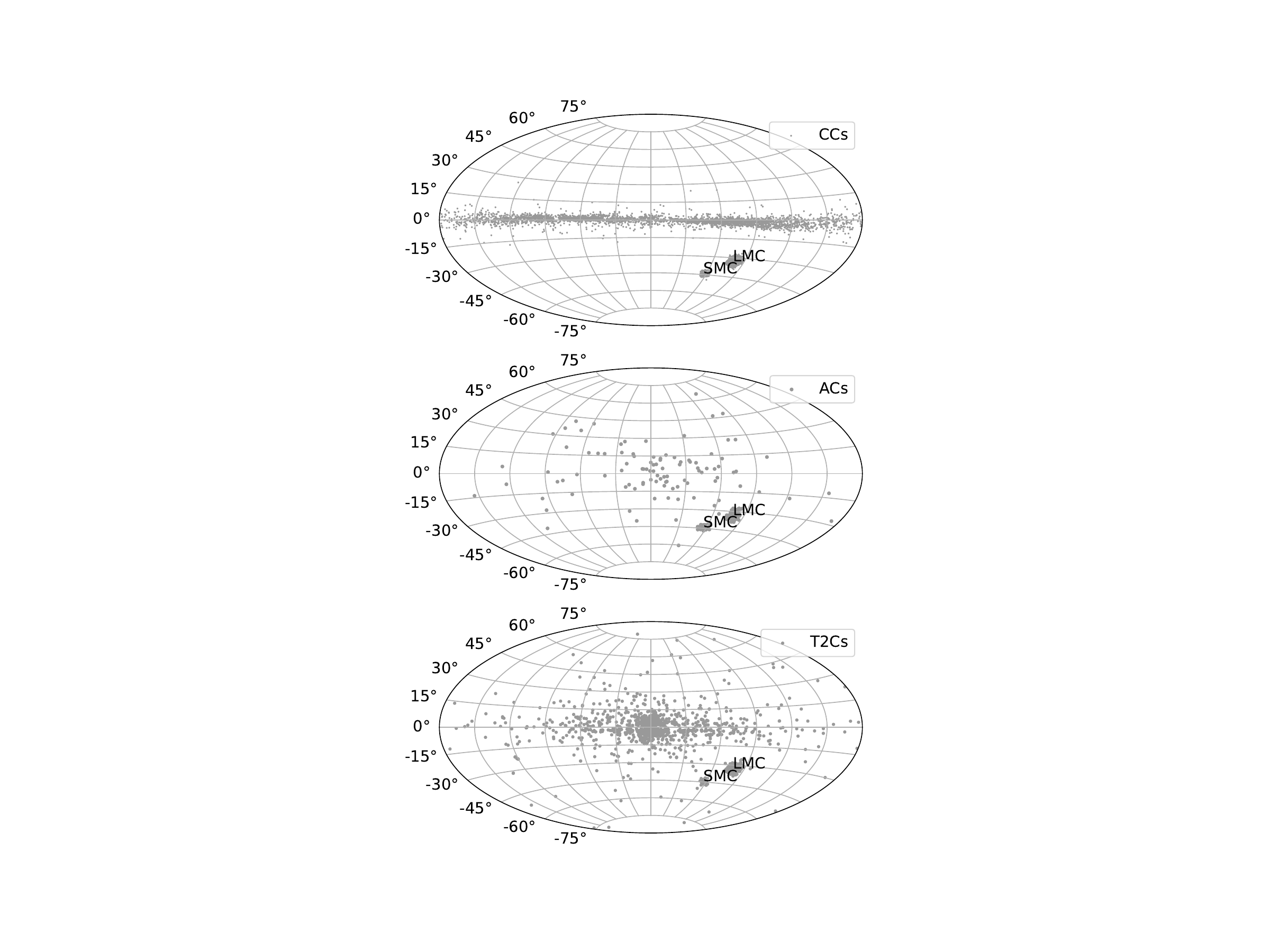}
\caption{Top -- Distribution in Galactic coordinates of currently known CCs. The current 
sample includes data from \citet[][and references therein]{pietrukowicz2021}. Galactic CCs are mainly 
distributed along the thin disk and the regions across the bulge. The overdensities associated 
with the MCs are labelled. 
Middle -- Same as the top, but for ACs. The current sample includes data available on the 
OGLE Download Site (\url{http://ogle.astrouw.edu.pl/}) and on the Gaia 
catalog \citep{ripepi2022b}.
Bottom -- Same as the middle, but for TIICs. 
}
\label{aitoff_all}
\end{center}
\end{figure}
%_______________________________________________________________________________

CCs are central helium-burning, intermediate-mass stars with ages younger than a few hundred Myrs; they are solid young stellar tracers, associated with HII regions and young open clusters typical of Galactic thin disk and of stellar systems experiencing recent star-formation episodes (spirals, dwarf irregulars; see the top panel of Fig.~\ref{aitoff_all}). CCs pulsate in a variety of radial modes, the most common are fundamental, first and second overtone. Together with the single-mode oscillation, CCs are also characterized by several mixed modes, i.e. radial oscillations in which at least two radial modes are simultaneously excited. The occurrence of overtones among classical Cepheids was still a matter of discussion till the late Eighties. In a seminal investigation \citet{bohmvitense1988} suggested 
that a good fraction of short-period Cepheids were overtone pulsators. 
In the General Catalog of Variable Stars (GCVS) they were classified 
as ``S Cepheids'' where ``S'' stands for Cepheids with sinusoidal light curves. 
% ref_1_15
The final empirical evidence was provided by the 
microlensing experiments (MACHO, EROS, OGLE) showing hundreds/thousands 
of Magellanic Cepheids pulsating in the first three radial modes and in a 
variety of mixed modes. 
%ref_1_21_22
The name adopted for mixed mode CCs was Beat Cepheids
\citep{rodgers1970}, so as to distinguish
them from the so-called bump Cepheids \citep{simon1976b},
i.e. the CCs showing the Hertzsprung progression\footnote{The most 
appropriate way to introduce and summarize the Hertzsprung progression 
is to use the sentences he wrote in his pioneering paper:
{\em \ldots At the shorter periods up to about 6 days the curves show the
characteristic regular $\delta$~Cephei form with quick rise and slow decrease,
without additional peculiarites. Above 6 days a secondary wave on the descending
branch of the light curve makes its appearance. This secondary wave is a very
characteristic feature of the following periods. For periods between 10 and 13 days,
the secondary wave is, when present, situated on top of the suppressed ordinary
maximum, the form of the light curve being nearly symmetrical. \ldots
The secondary wave has superseded the maximum shown at shorter periods. At periods
of about 14 or 15 days the new superposed maximum occurs earlier, giving the
light curve again an unsymmetrical shape with a hesitation in the increase
of the brightness about midway between minimum and maximum. This hesitation
is persistent at the periods mentioned and not found at any other period
materially different from them. \ldots At periods longer than 16 days
several curves are found rather similar to that of $\delta$~Cephei,
showing quick rise and slow decrease apparently without complications.}
%editor_8
\citep[][see also section~\ref{sec:hertzprungprog}]{hertzsprung1926}.
}
The identification of the different modes was facilitated, since MC CCs are all 
located at the same distance. Therefore, the typical diagnostics adopted to estimate 
individual distances (PL, PLC, Period-Wesenheit [PW]) can also be adopted to 
identify the pulsation mode (see section~\ref{sec:pl_vs_pw}).

%%%%%%%%%%%%%%%%%%%%%%%%%%%%%%%%%%%%%%%%%%%%%%%%%%%%%%%%%%%%%%%%%%%%%%%%%%%%%%%%%%%%%
% 			fig 29 
%%%%%%%%%%%%%%%%%%%%%%%%%%%%%%%%%%%%%%%%%%%%%%%%%%%%%%%%%%%%%%%%%%%%%%%%%%%%%%%%%%%%%
\begin{figure}[htbp]
\begin{center}
\includegraphics[width=\textwidth]{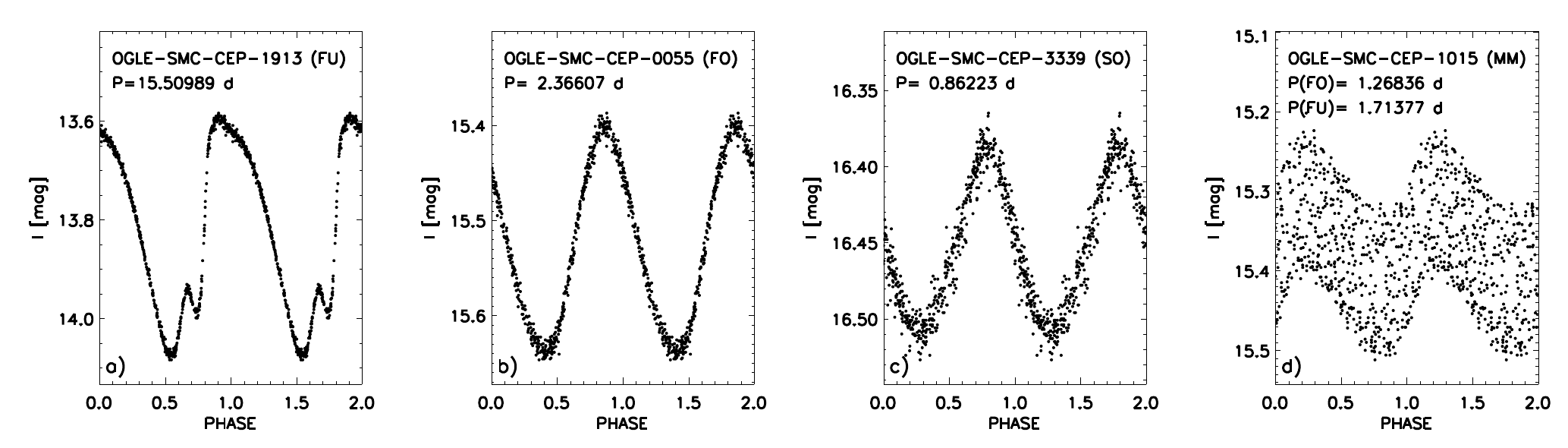}
\includegraphics[width=\textwidth]{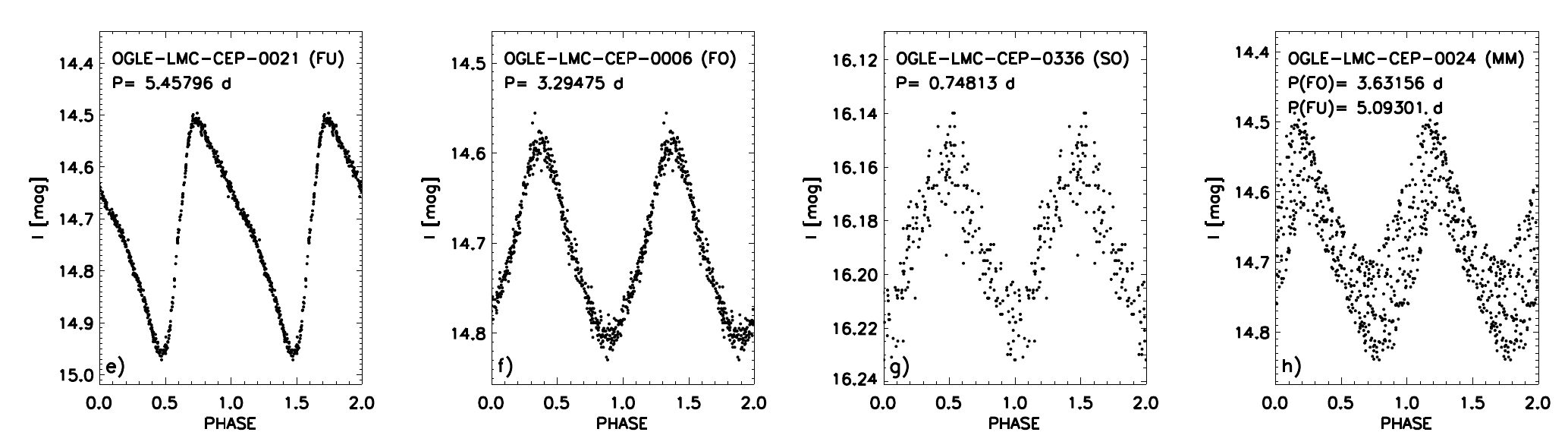}
\caption{
Top: from left to right the light curves, based on the OGLE~IV data set, for SMC CCs pulsating in the fundamental (panel a), in the first overtone (panel b), in the second overtone (panel c) and in a mixed-mode (panel d). The ID and the period (days) are labelled. For the mixed mode are labelled both primary and secondary period. 
Bottom: Same as the top, but for LMC CCs. 
\label{fig:lcvs_cc_MC}
}
\end{center}
\end{figure}
%_______________________________________________________________________________

The top panels of Fig.~\ref{fig:lcvs_cc_MC} show from left to right the light 
curves of SMC CCs pulsating in the fundamental (panel a), first overtone (panel b), 
second overtone (panel c) and in a mixed mode (first overtone/fundamental, panel d), 
while the bottom panels (e,f,g,h) display similar light curves, but for LMC CCs. 
We are showing these typical light curves for MC CCs, because the empirical scenario 
is quite rich and includes a variety of single and mixed mode variables. The reader 
interested in a more detailed discussion concerning their phenomenology is 
referred to \citet{soszynski2015b}. 
% BG_short
%Fig.~\ref{fig:lcvs_cc_GAL}  shows light curves for Galactic CCs. 
The current MW sample only includes a few first 
overtones, a single second overtone, and a very limited number of mixed-modes. 
The lack of these objects is an observational bias. 
%ref_2_4f
The current Galactic photometric surveys are far from being complete,
since the current limiting magnitude in the optical regime is
$I\approx$19 mag. The observational bias becomes more severe in the      
inner/outer disk and beyond the Galactic center, due to the high column 
% editor 10 
density of dust along the line of sight. The surface density of CCs (see 
Table~\ref{tbl:cepheid_number}) across the thin disk in the annulus in 
Galactocentric distance between 7 and 9 kpc is $\rho$=5.64 per kpc$^2$, 
while in the annulus between 5 and 7 kpc it is 20\% smaller 
% editor 10 
($\rho$=4.51 per kpc$^2$).  However, the density gradient of stars steadily 
increases when moving toward the inner disk and the surface density 
of CCs should show the same trend. This is also the reason why the 
peak in the iron distribution of CCs is at solar-iron abundance.

Pulsation models and observations indicate that the topology of the CC instability strip depends 
on the metal content. Moreover, evolutionary models also indicate that the width in temperature    
of the blue loop, and in turn, the minimum mass crossing the CC instability strip depend 
on the metal content. These circumstantial evidence takes account of the limited number 
of Galactic second overtones and mixed-mode variables currently known. This bias is going to 
be removed by the current ongoing long-term photometric surveys either of the Galactic 
plane \citep[OGLE,][]{skowron2019} or by an all-sky survey like Gaia (Gaia DR3).      

%%%%%%%%%%%%%%%%%%%%%%%%%%%%%%%%%%%%%%%%%%%%%%%%%%%%%%%%%%%%%%%%%%%%%%%%%%%%%%%%%%%%%
% 			fig 3
%%%%%%%%%%%%%%%%%%%%%%%%%%%%%%%%%%%%%%%%%%%%%%%%%%%%%%%%%%%%%%%%%%%%%%%%%%%%%%%%%%%%%
% ref_1_16 
\begin{figure}[htbp]
\begin{center}
\includegraphics[width=\textwidth]{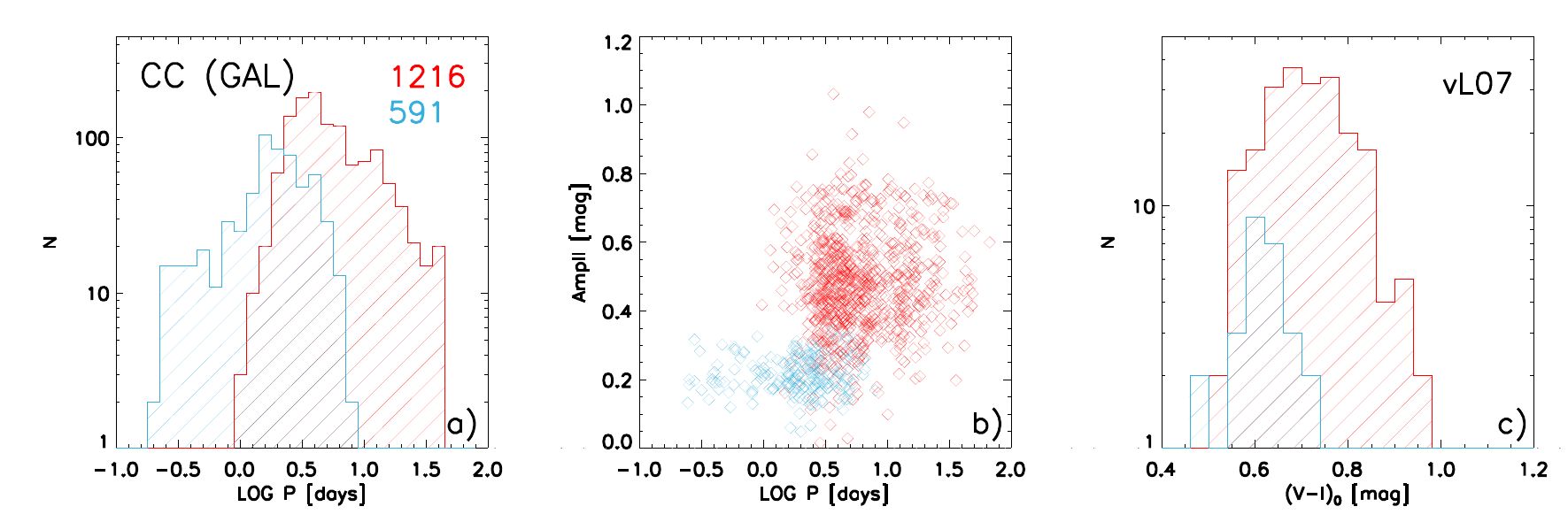}

\includegraphics[width=\textwidth]{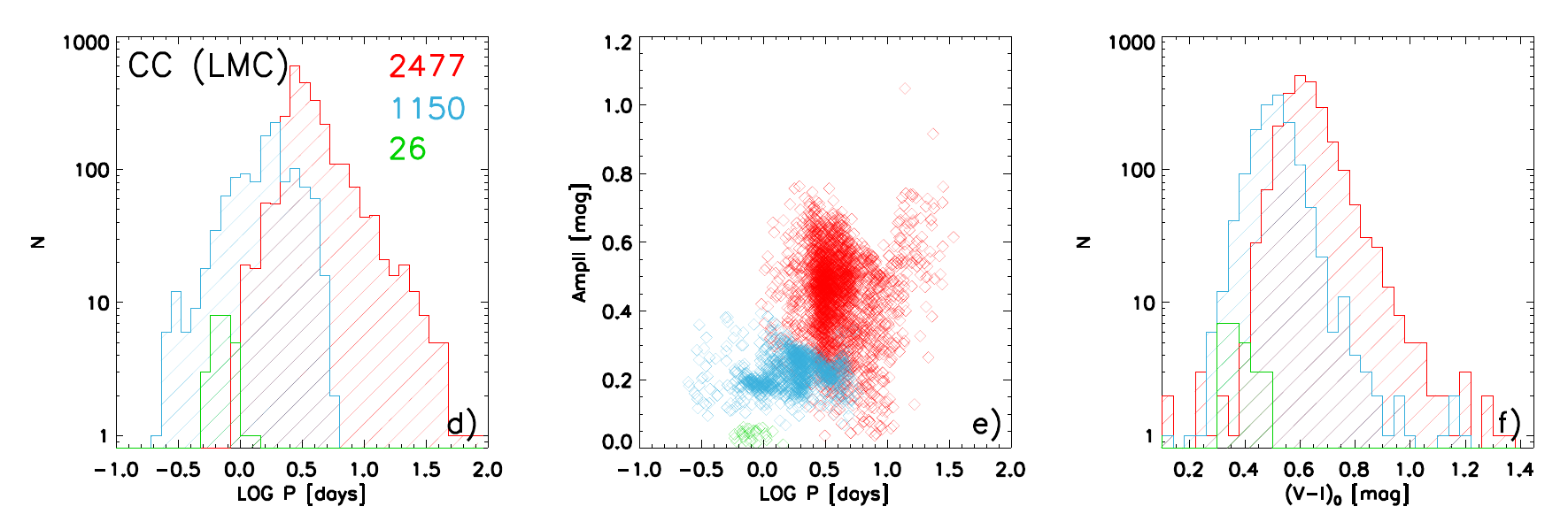}

\includegraphics[width=\textwidth]{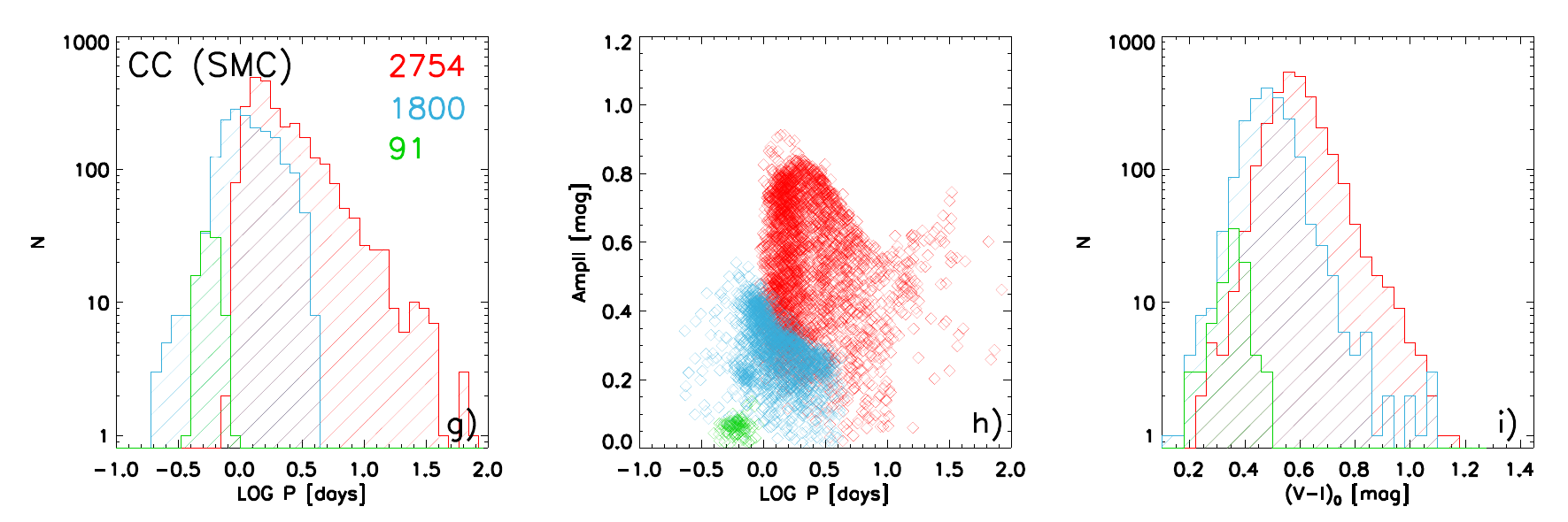}
\caption{
Top: Panel a)-- Period distribution for fundamental (red hatched area), 
and first overtone (light blue hatched area) Galactic CCs. Note that the 
Y-axis is logarithmic.    
Panel b)-- $I$-band luminosity amplitude versus logarithmic period (Bailey 
diagram) for FU (red), and FO (light blue) Galactic CCs.   
Panel c)-- Dereddened ($V-I$)$_0$ color distribution for CCs. The sample 
adopted for the color distribution is based 
on multi-band optical photometry provided by \citet[][vL07]{vanleeuwen2007} and 
on individual reddenings provided by \citet{tammann03}. 
Middle: Same as the top, but for LMC CCs (panels d,e,f). The green hatched area 
shows the period distribution of second overtone CCs. 
Bottom: Same as the middle, but for SMC CCs (panels g,h,i). 
\label{fig:bailey_cc}
}
\end{center}
\end{figure}
%_______________________________________________________________________________

Panel a) of Fig.~\ref{fig:bailey_cc} shows the period distribution of both fundamental 
(red hatched area) and first overtone CCs (light blue hatched area). This 
sample includes CCs collected by OGLE IV \citep{skowron2019,pietrukowicz2021}.
We adopted this sample because accurate and homogeneous photometry and pulsation 
parameters are available for both MC and MW variables.
The mixed mode Cepheids pulsating either in FO(primary)/FU(secondary) or  in 
FU/FO were included by using their dominant periods. 
The histograms plotted in this panel display that FUs peak at $P\sim4$ days ($\log P\sim0.60$)
and the distribution is skewed towards longer periods, with a secondary maximum 
(shoulder) located at $P\sim12.5$ days ($\log P\sim1.1$). As a whole, the periods 
range from roughly one day to 50 days. The FOs peak at $P\sim1.5$ days 
($\log P\sim0.20$), their distribution is skewed towards shorter periods and 
their periods range from $\sim$0.18 days to $\sim$5.6 days.

%ref_2_5
Note that no single-mode second overtone CC is currently known in the MW.
The variable V473 Lyr is a second overtone CC, but shows strong amplitude and 
phase modulations \citep{burki1980,breger1981,molnar2014,molnar2017}. 
Galactic double-mode and mixed-mode CCs, including a second overtone mode, 
have also been identified \citep{pardo1997,beltrame2002,sziladi2007,poretti2014}, 
but they are few and were not included.

Panel b) of Fig.~\ref{fig:bailey_cc} shows the Bailey diagram (\textit{I}-band amplitude 
versus logarithmic period) for Galactic CCs. 
% ref_1_17
This panel only includes CCs with accurate and homogeneous \textit{I}-band 
light curves collected by OGLE~IV along the Galactic plane.

The FO amplitudes are on average a factor of two smaller than FU amplitudes.
The FU amplitudes display a broad distribution with amplitudes ranging from 
$\Delta I\sim0.2$ mag to $\Delta I\sim0.8$ mag. There is a mild evidence for 
a minimum in amplitudes for periods ranging from $P\sim6.5$ to 10 days 
($\log P\sim0.8$ to $\log P\sim1$). The reason for this secondary minimum 
will become clearer in the following.

%ref_2_6
Panel c) shows the dereddened color distribution of 
both FOs and FUs. This panel only includes CCs for 
which are available accurate and homogeneous estimates of $V$,$I$ 
photometry and individual reddenings. The main source for the 
photometry is \citet{vanleeuwen2007}, while for the reddening is 
\citet{tammann03}. The latter authors performed a 
critical analysis not only of the reddening estimates available in the literature, 
but also on the absorption coefficients and on their impact on the 
Period-Color relations \citep[see also][]{laney2007}. 
As expected, the FOs are systematically bluer than FUs 
and peak at ($V-I$)$_0\sim0.63$ mag, while the FU peak at
($V-I$)$_0\sim0.73$ mag (see Table~\ref{tbl:cepheid_color_dis}).

% editor 14 
The parameters adopted to discuss the pulsation properties of CCs are
independent of uncertainties affecting either distance and reddening 
estimates (periods, luminosity amplitudes) or distance estimates (colors). 
The dispersion in period and amplitude distribution is driven by 
differences in evolutionary and/or in pulsation properties \citep{bono2020}.

The middle panels of the same figure show the same parameters, but for LMC CCs. 
Data plotted in these panels display several interesting features worth being 
discussed in detail. 

%ref_1_18
{\em i)}--FO period distribution-- The number of FOs 
(light blue diamonds) is significantly larger than for Galactic Cepheids,
indeed the population ratio between FO/FU increases by 
$\sim$50\% (0.49 versus 0.72).
% ref_1_19a 
Moreover, their periods range from 0.18 days to 6.3 days and their period 
distribution shows two well-defined shoulders for periods of $\sim$0.3~days and  
$\sim$1~day ($\log P\sim-0.5$ and 0.0).

{\em ii)}--FU period distribution-- The period distribution of FUs (red diamonds) 
is broader, indeed their periods range from about a half day to more than 
one hundred days. Moreover, the shoulder (secondary maximum) identified in  
Galactic CCs is now located at longer periods, around 20 days ($\log P\sim1.3$)
but is less evident.

{\em iii)}--SO period distribution--The LMC include a sizable sample of 
SO CCs (green diamonds) that are only minimally present among Galactic CCs.  

Panel e) shows the Bailey diagram.  A glance 
at the data plotted in this panel shows that FUs display the typical ``V''-shape 
with a well-defined maximum for periods around 2.5 days ($\log P\sim$0.4) and 
a secondary minimum for periods around ten days \citep{bono00a}. 
The main peak is connected with the peak in the period distribution, while 
the secondary minimum is associated with the Hertzsprung progression and 
to the so-called bump Cepheids (see section~\ref{sec:hertzprungprog}). 
%ref_1_19b 
For periods around ten days, the phase of the bump along the pulsation cycle
approaches the maximum in surface brightness, and the luminosity amplitude
attains a minimum (associated with a minimum in radius) when compared with 
shorter- and longer-period Cepheids.
The amplitude distribution of FO Cepheids is more complex, its 
maximum is located at periods around 1.26 days ($\log P\sim0.10$) and shows secondary 
maxima around 1 and 0.3 days ($\log P\sim 0.0$ and $-0.5$). The SOs are characterized 
by small amplitudes and also cover, as expected, a narrow range in periods.     

Panel f) shows the dereddened ($V-I$)$_0$ color distribution. The difference 
in color between Galactic and LMC CCs, the latter being systematically bluer, 
was noted more than a half century ago by \citet{gascoigne1965} and more recently by 
\citet{laney1994}. Data plotted 
in this figure show that FU, FO and SO color distributions are quite 
symmetrical and the median ($V-I$)$_0$ colors peak at 
($V-I$)$_0=0.651\pm0.107$ mag (FU), 
($V-I$)$_0\sim0.530\pm0.091$ mag (FO) and 
($V-I$)$_0\sim0.368\pm0.129$ (SO) mag, 
where the errors are the standard deviations (see Table~\ref{tbl:cepheid_color_dis}).

%_____________________________________________________________________________________________
\begin{table}[htbp]
\caption{Median \textit{V-I} colors (mag) of Magellanic Cloud and Milky Way Cepheids.}
\label{tbl:cepheid_color_dis}
\begin{center}
\begin{tabular}{lccc}
\hline
\hline
Mode & \multicolumn{3}{c}{\textit{V-I} [mag]}\\
        & SMC & LMC & MW \\ 
\hline
\multicolumn{4}{c}{---CCs---}\\
FU         & 0.613$\pm$ 0.102 & 0.651$\pm$ 0.107 & 0.726$\pm$ 0.094 \\
FO         & 0.509$\pm$ 0.095 & 0.530$\pm$ 0.091 & 0.629$\pm$ 0.059 \\
SO         & 0.373$\pm$ 0.056 & 0.368$\pm$ 0.129 &   \ldots         \\
\multicolumn{4}{c}{---ACs---}\\
FU         & 0.530$\pm$ 0.066 & 0.485$\pm$ 0.107 & 0.556$\pm$ 0.223 \\
FO         & 0.386$\pm$ 0.092 & 0.399$\pm$ 0.096 & 0.405$\pm$ 0.266 \\ 
% \multicolumn{4}{c}{---TIICs---}\\
% FU         & 0.613$\pm$ 0.102 & 0.651$\pm$ 0.107 & 0.399$\pm$ 0.820 \\
\hline
\end{tabular}
\end{center}
\footnotetext{The errors associated with the median colors are the standard deviations.}
\end{table}
%_____________________________________________________________________________________________

The bottom panels show the same parameters of middle and top panels, but 
for SMC CCs. The period distributions plotted in panel g) show that 
SOs are more representative compared with the LMC, and indeed the population ratio 
SO/FO increases by a factor of three (0.01 versus 0.03). They also cover a 
% editor 11 
narrower range in period ($0.40\le P(\rm SO)\le0.92$ vs $0.58\le P(\rm SO)\le1.46$ days) 
and a broader range in luminosity amplitudes (panel h) when compared with 
LMC SOs. 

The peak in the period distribution of SMC FUs and FOs shifts towards 
shorter periods ($\sim$0.9 days (FO), $\sim$1.5 days (FU) vs $\sim$1.8 (FO) days, 
$\sim$2.8 days (FU)). The range in periods shows the same trend: 
$0.25\le P(\rm FO)\le4.5$ vs $0.25\le P(\rm FO)\le6.0$ days and 
$0.84\le P(\rm FU)\le129$\footnote{The SMC sample includes a long-period 
CC with $P\sim208$ days.} vs $0.97\le P(\rm FU)\le134$ days, respectively.         
Note that the shoulder located in the long period tail of FU CCs 
is now placed at even longer periods when compared with Galactic and LMC CCs, 
i.e. around 25 days ($\log P\sim1.4$). 

The $I$-band luminosity amplitudes plotted in 
panel h) display quite clearly that the secondary minimum linked with 
the Hertzsprung progression moves SMC bump Cepheids towards longer 
periods, $P\sim9.8\pm0.1$ days. Moreover, the peak in the $I$-band luminosity 
amplitudes moves towards shorter periods, namely
1.58 days (FU, $\log P\sim0.2$), 0.79 days (FO, $\log P\sim-0.1$) 
and 0.63 days (SO, $\log P\sim-0.2$). 
Finally, the dereddened ($V-I$)$_0$ colors of 
SMC CCs plotted in panel f)  and listed in Table~\ref{tbl:cepheid_color_dis} 
are either similar (SO) or systematically bluer than LMC and Galactic CCs. 

The circumstantial evidence emerging in the comparison between Galactic 
and Magellanic CCs can be summarized as follows: 

{\em i)}-- The peak in the period distributions and the range in period covered 
by FU, FO and SO CCs steadily move towards shorter periods when moving from 
Galactic to LMC and SMC CCs. On the other hand, the shoulder located in the long 
period tale of FU CCs, moves in the opposite direction: it is placed at longer 
periods when moving from Galactic to LMC and SMC CCs.

%ref_1_20 
{\em ii)}-- The center of the Hertzsprung progression, 
i.e. the period in which bump Cepheids attain a well-defined minimum in 
luminosity amplitude, steadily moves towards longer periods when shifting 
from Galactic to LMC and SMC FU CCs.

{\em iii)}-- The peak in the Bailey diagram for FU and FO CCs 
systematically shifts towards shorter periods when moving from 
metal-rich to more metal-poor stellar systems (Galaxy--LMC--SMC). 

{\em iv)}-- The color distribution of FU and FO CCs becomes 
systematically bluer when moving from Galactic to LMC and to SMC
(see Table~\ref{tbl:cepheid_color_dis}).
This evidence cannot be extended to SO CCs, since they are vanishing 
in the MW and only two dozens are currently known in the LMC. 
The evidence that the peak in color when moving from FU to 
FO and to SO MC CCs  becomes systematically bluer, fully supports pulsation 
predictions concerning the topology of the instability strip. Indeed, the 
regions in which they show a stable limit cycle become, at fixed 
chemical composition, systematically hotter when moving from the 
fundamental mode to the overtones.

%editor 15
%%%%%%%%%%%%%%%%%%%%%%%%%%%%%%%%%%%%%%%%%%%%%%%%%%%%%%%%%%%%%%%%%%%%%%%%%%%%%%%%%%%%%
% 			fig 4
%%%%%%%%%%%%%%%%%%%%%%%%%%%%%%%%%%%%%%%%%%%%%%%%%%%%%%%%%%%%%%%%%%%%%%%%%%%%%%%%%%%%%
\begin{figure}[htbp]
\begin{center}
\includegraphics[width=0.66\textwidth]{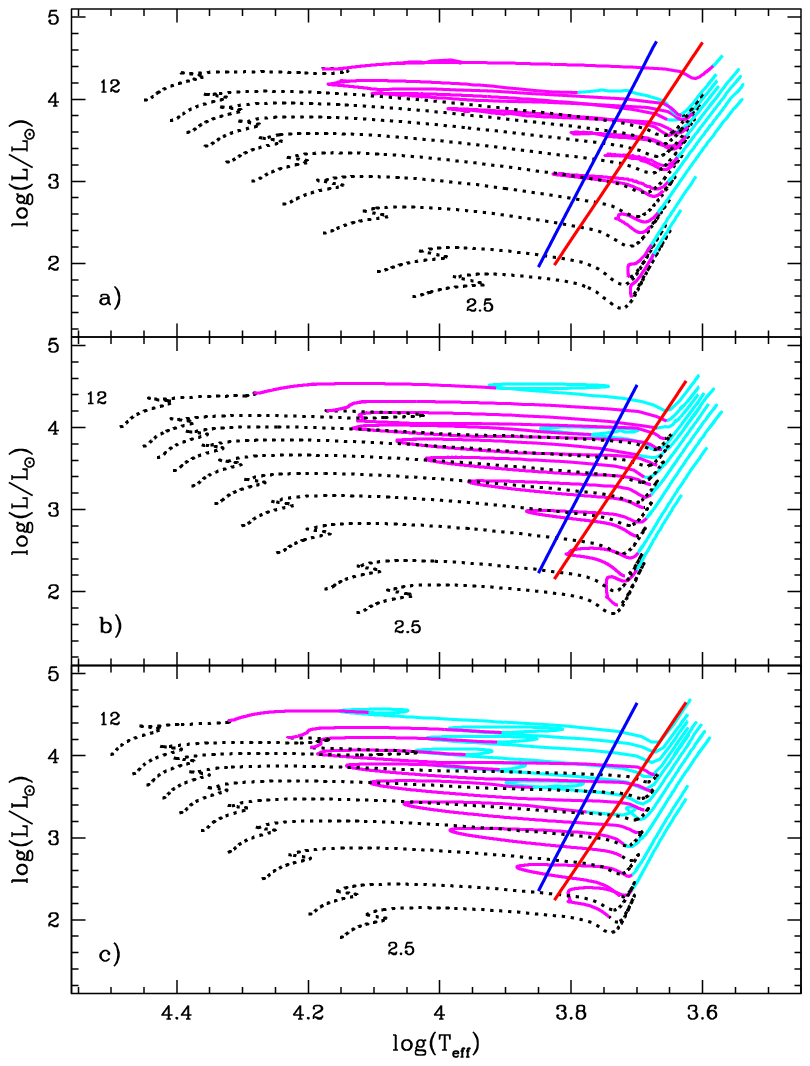}
\caption{
%\vspace{-0.5truecm}
Panel a)-- Scaled solar evolutionary tracks for young stellar structures computed by
neglecting convective overshooting during core hydrogen-burning phases 
in the Hertzsprung--Russell diagram at fixed chemical composition 
(metal mass fraction, Z=0.02)  
and stellar masses ranging from 2.5 to $12\,M/M_\odot$ (see labels). 
The dashed lines display hydrogen-burning 
phases, while the magenta color marks central helium-burning phases, the cyan color 
marks double-shell (H and He) burning phases (Asymptotic Giant Branch). The almost 
vertical solid lines display the blue (hot, first overtone) and the red 
(cool, fundamental) edge of the CC instability strip \citep{fiorentino02, desomma2021}.   
Panel b)-- Same as Panel a), but for a more metal-poor chemical composition Z=0.01. 
Panel c)-- Same as Panel b), but for a more metal-poor chemical composition Z=0.001.  
}
\label{fig:hrd_three}
\end{center}
\end{figure}
%_________________________________________________________________________

Theoretical and empirical evidence indicate that the difference in the 
evolutionary and pulsation properties of CCs when moving from the MW to 
the MCs is caused by the iron content. Indeed, current metallicity 
determinations based on high-resolution spectra indicate that the mean 
observed iron abundance of MW CCs is solar, whereas LMC CCs are on average a factor of 
two more metal-poor and SMC CCs are a factor of four more metal-poor.    
Panels a), b) and c) of Fig.~\ref{fig:hrd_three} display the comparison 
between evolutionary and pulsation predictions over a broad range of 
%chemical compositions (for more details see section~\ref{evolution}).
chemical compositions (for more details see section~C in Appendix).

In the following, we will discuss the population ratios of the different pulsation modes. 
Before discussing them, we point out that, for the MW samples, we will only derive the 
ratios based on the OGLE sample. In fact, OGLE is the only survey allowing an extremely
detailed and accurate analysis of the mixed-mode classifications. We did not include 
Gaia classification,
because the sampling of the light curves is still based on a 
modest number of phase points.

Data listed in Table~\ref{tbl:cepheid_number} show that the population ratio 
between first overtone and fundamental CCs is 
0.65$\pm$0.02 in the SMC, 
0.72$\pm$0.02 in the LMC and 
0.49$\pm$0.01 in the MW. 
The observed population ratios do not show a clear dependence on the 
observed iron abundance 
and probably hint to a possible observational bias in the Galactic sample. 
The population ratio between 
second overtone and fundamental CCs decreases, as expected, when  
moving from more metal-poor to more metal-rich stellar systems 
(0.03 [SMC] vs 0.01 [LMC]) and it is vanishing for the MW. 
On the other hand, the fraction of mixed mode variables 
does not show a clear dependence on the metallicity
(0.11$\pm$0.01, [SMC]; 0.17$\pm$0.01, [LMC]; 0.14$\pm$0.01, [MW]).

The population ratio between first overtone and fundamental ACs, 
listed in Table~\ref{tbl:cepheid_number}, is similar in the MCs 
(0.54$\pm$0.10, SMC; 0.44$\pm$0.08, LMC), while in the  MW attain 
a larger value (0.76$\pm$0.11). However, the Poissonian 
uncertainties are large and do not allow us to reach firm conclusions. 
The reader interested in a more detailed discussion concerning the 
statistics of ACs in nearby dwarf galaxies is referred to 
\citet{monelli2022}.  

The relative number of the different subgroups of TIICs changes when 
moving from the MCs to the MW. The MW has the highest fraction of BLHer 
variables (0.38$\pm$0.02) compared to the LMC (0.34$\pm$0.01)  
and to the SMC (0.28$\pm$0.01). 
The fraction of WVir, including pWVir, follows an opposite trend, 
decreasing from the SMC (0.51$\pm$0.01), to the LMC (0.46$\pm$0.01), 
and the MW (0.43$\pm$0.02). Interestingly enough, the fraction of RVTau 
variables is, within the errors, constant in the 
three different stellar systems, namely 0.19$\pm$0.01 (MW), 
0.19$\pm$0.01 (LMC) and, 0.21$\pm$0.01 (SMC).
The population 
ratio between fundamental and first overtone is meaningless for TIICs, 
due to the paucity of FOs which have been identified sofar.
%ref_2_7
Indeed, after the seminal discovery of two FO TIICs in the LMC by 
\citet{soszynski2019b} only three other variables were identified in the MW 
by \citet[][see also Table~\ref{tbl:cepheid_number}]{ripepi2022b}.

%%%%%%%%%%%%%%%%%%%%%%%%%%%%%%%%%%%%%%%%%%%%%%%%%%%%%%%%%%%%%%%%%%%%%%%%%%%%%%%%%%%%%%%
\begin{table}[htbp]
\footnotesize
% editor 9
\caption{\scriptsize Number of Magellanic Cloud and Milky Way Cepheids and their pulsation mode.}
\label{tbl:cepheid_number}
\begin{center}
\begin{tabular}{lrrrrr}
\hline
\hline
Mode        & SMC & LMC & MW (OGLE) & MW (PP) & MW (all) \\
\hline
\multicolumn{6}{c}{---Classical Cepheids---} \\
FU         & 2754 & 2477 & 1216 & 2250      & \ldots \\
FO         & 1800 & 1778 &  591 & 1092      & \ldots \\
SO         &   91 &   26 &    0 &    1      & \ldots \\
\multicolumn{6}{c}{Mixed Mode}\\ 
FU/FO       &   69 &   96 &   59 &  100     & \ldots \\
FO/SO       &  239 &  322 &  161 &  201     & \ldots \\
FO/TO       &    0 &    1 &    0 &    0     & \ldots \\
SO/TO       &    0 &    1 &    3 &    3     & \ldots \\
FU/FO/SO    &    0 &    1 &    2 &    3     & \ldots \\
FO/SO/TO    &    1 &    7 &    8 &    8     & \ldots \\
\multicolumn{6}{c}{---Anomalous Cepheids---}   \\
FU          &   79 & 102  &   38  &   \ldots & 118 \\
FO          &   43 &   45 &   81  &   \ldots &90   \\
MM          &    0 &    0 &    1  &   \ldots & 1   \\
% UNCL.  &    0 &    0  &   32  &    0  \\
\multicolumn{6}{c}{---Type II Cepheids---} \\
FU(tot)        &   53 &  291 & 1666 & \ldots &2187  \\ %2186 
FU(BLHer)      &   20 &   99 &  674 & \ldots &838  \\ %837
FU(WVir)       &   15 &  108 &  609 & \ldots &869   \\
FU(pWVir)      &    7 &   27 &   67 & \ldots &67    \\
FU(RVTau)      &   11 &   55 &  313 & \ldots &410   \\
FO             &    0 &    2 &    3 & \ldots &  3   \\
\hline
\end{tabular}
\end{center}
\footnotetext{\scriptsize
The source and the mode identification for Cepheids in 
the ``SMC'', ``LMC'' and ``MW(OGLE)'' column 
come from the OGLE~IV data set (\url{http://ogle.astrouw.edu.pl/} 
and references therein). 
The source and the mode identification for CCs listed in the 
column ``MW (PP)'' come from \citet[][]{pietrukowicz2021} and 
they are based on the cross-identification they performed with 
the Gaia catalog \citep{ripepi2022b} and with catalogs available 
in the literature.
The source and the mode identification for ACs and TIICs listed in the 
column ``MW (all)'' is based on the cross-identification (matching 
radius of 1 arcsec) that we performed between the catalogs available on the 
OGLE Download Site (\url{http://ogle.astrouw.edu.pl/}) and the Gaia 
catalog \citep{ripepi2022b}. 
}
\end{table}

%\clearpage
%_______________________________________________________________________________
\subsection{Anomalous Cepheids}\label{sec:acs_intro}
%_______________________________________________________________________________ 

Anomalous Cepheids (ACs)\footnote{The definition of Anomalous Cepheids dates back to 
\citet[][see their Fig.~1]{zinn1976} in their detailed 
analysis of the mass-distribution of Draco dSph variable stars.
The anomaly was driven by the evidence that ACs, at fixed pulsation
period, are brighter than RR Lyrae (RRLs) and TIICs, and fainter than CCs. 
% editor_13
The increase of a factor of two in the stellar mass for ACs, 
when compared with RRLs and TIICs, was 
originally suggested by \citet{christy1970}. The reader interested 
in a more detailed discussion concerning the early discoveries 
of ACs in globulars and dwarf galaxies is referred to 
\citet{monelli2022}.}
are low-mass stars with stellar masses ranging from 0.8 to 1.8 
\citep{fiorentino12c} in which central helium burning takes place 
in a partially electron-degenerate helium core. They are associated 
with intermediate-age stellar tracers. The current evidence indicates 
that they might be the aftermath of both single and binary merging 
evolutionary channels. This is the reason why they are ubiquitous over the entire Galactic spheroid and in nearby galaxies (see the middle panel of Fig.~\ref{aitoff_all}). However, theory and observations suggest that only stellar populations more metal-poor than [Fe/H]$\lesssim-1.5$ cross the Cepheid instability strip. This is the reason why they mainly show up in metal-poor/metal-intermediate stellar systems.

%%%%%%%%%%%%%%%%%%%%%%%%%%%%%%%%%%%%%%%%%%%%%%%%%%%%%%%%%%%%%%%%%%%%%%%%%%%%%%%%%%%%%
% 			fig 5
%%%%%%%%%%%%%%%%%%%%%%%%%%%%%%%%%%%%%%%%%%%%%%%%%%%%%%%%%%%%%%%%%%%%%%%%%%%%%%%%%%%%%
\begin{figure}[htbp]
\centering
\includegraphics[width=\textwidth]{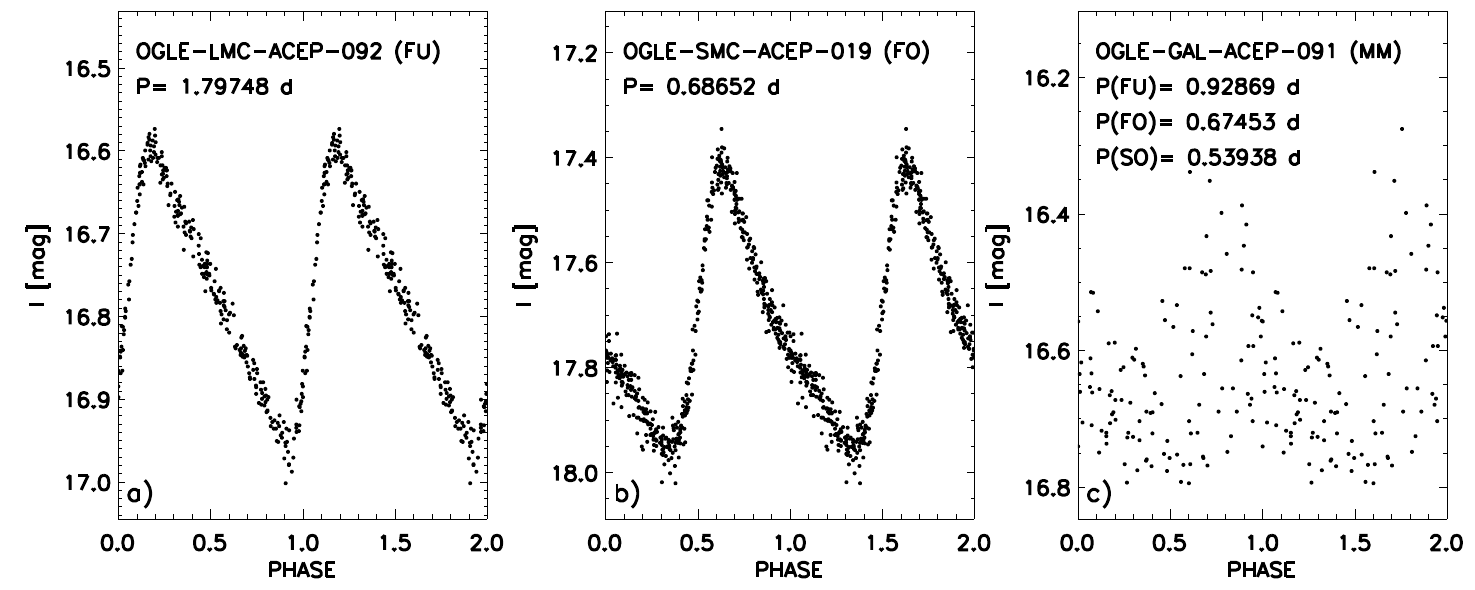}
\caption{
Panel a): Light curve for an LMC AC pulsating in the fundamental mode from the 
OGLE-IV data set. The ID and the period (days) are labelled.
Panel b): Same as the left, but for an SMC AC pulsating in the first overtone.
Panel c): Same as the left, but for the Galactic AC pulsating simultaneously 
in the first three radial modes. Primary and secondary periods (days) are labelled.}
\label{fig:lcvs_ac}
\end{figure}
%_______________________________________________________________________________

ACs display quite regular light curves for both fundamental and first overtone variables.  The majority of Magellanic ACs, the sample for which we have a more detailed knowledge, mainly includes FU variables, indeed the population ratio is $\sim$0.65--0.70. 
They were considered, until a few years ago, single mode variables, but \citet{soszynski2020}
found a Galactic triple mode AC. Figure~\ref{fig:lcvs_ac} shows representative  light curves for Galactic, 
LMC and SMC ACs in the three different pulsation modes.  

%%%%%%%%%%%%%%%%%%%%%%%%%%%%%%%%%%%%%%%%%%%%%%%%%%%%%%%%%%%%%%%%%%%%%%%%%%%%%%%%%%%%%
% 			fig 6
%%%%%%%%%%%%%%%%%%%%%%%%%%%%%%%%%%%%%%%%%%%%%%%%%%%%%%%%%%%%%%%%%%%%%%%%%%%%%%%%%%%%%
%_______________________________________________________________________________
\begin{figure}[htbp]
\centering
\includegraphics[width=\textwidth]{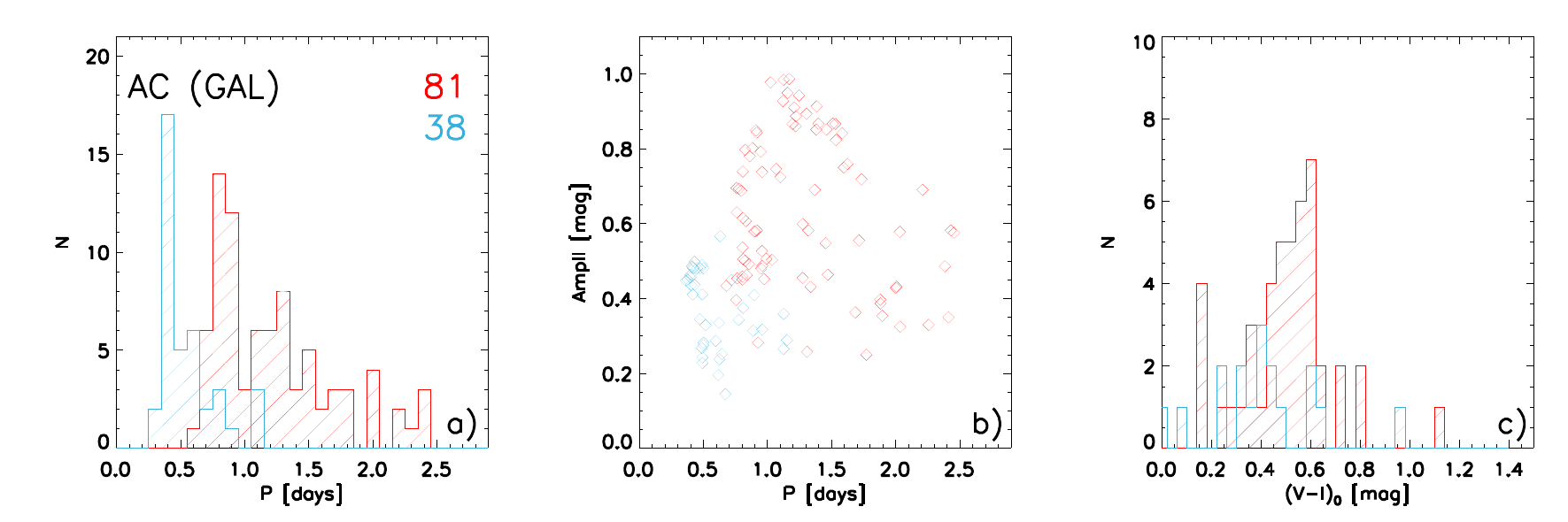}

\includegraphics[width=\textwidth]{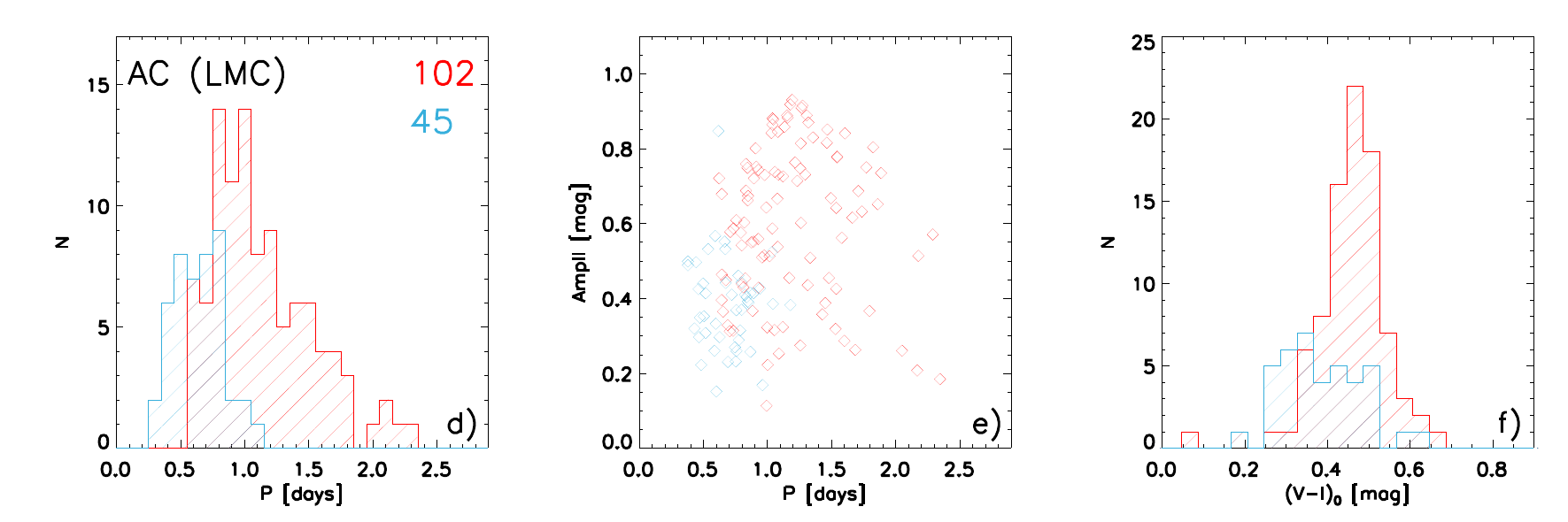}

\includegraphics[width=\textwidth]{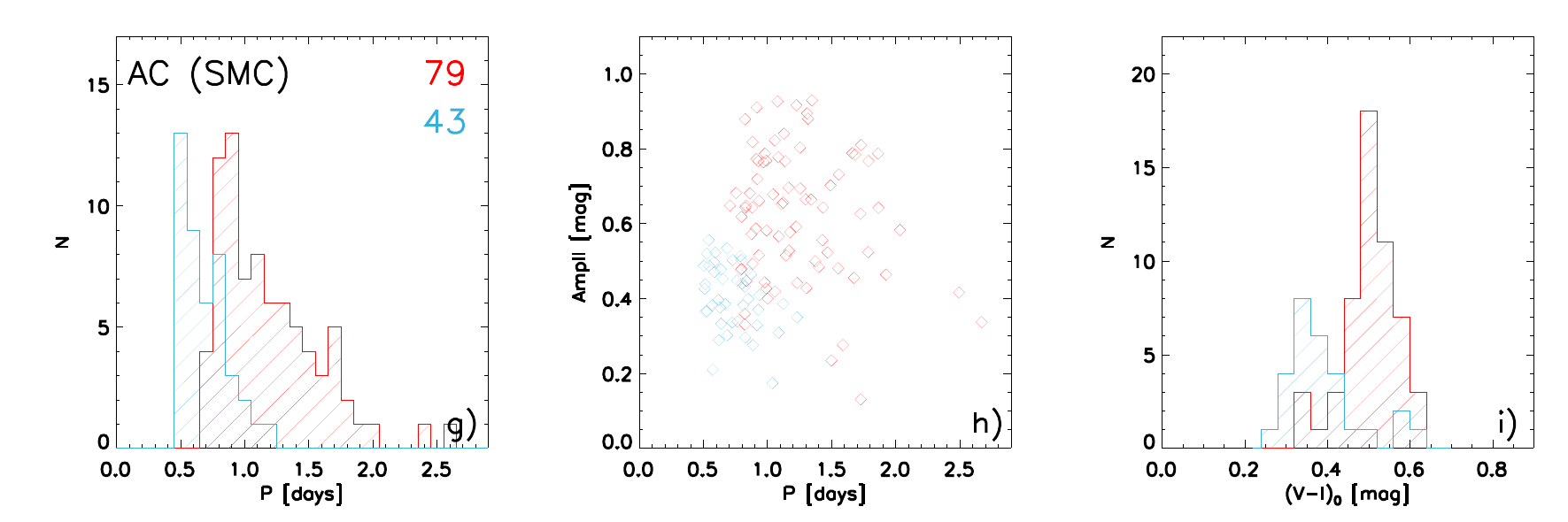}
\caption{
Top: Panel a)-- Period Distribution for fundamental (red-hatched area) and 
first overtone (light-blue hatched area) Galactic ACs.   
Panel b)-- I-band luminosity amplitude versus period (Bailey diagram) for 
both FU and FO ACs.   
Panel c)-- dereddened ($V-I$)$_0$ color distribution 
for the same ACs plotted in the left and in the middle panel.  
Middle: Same as the top, but for LMC ACs.  
Bottom: Same as the top, but for SMC ACs. 
\label{fig:bailey_ac}
}
\end{figure}
%_______________________________________________________________________________

The AC pulsating in the fundamental-mode have pulsation periods ranging from 
0.6 to 2.7 days, while the first-overtones oscillate with periods ranging from 
0.4 to 1.2 days. The period distributions are quite similar when moving from 
Galactic to Magellanic ACs  (see left panels in Fig.~\ref{fig:bailey_ac}). The 
same outcome applies to the I-band luminosity amplitudes, the FO amplitudes
are on average a factor of two smaller than fundamental ones (see middle panels 
in Fig.~\ref{fig:bailey_ac}). The color ($V-I$) distribution covered by 
Magellanic ACs are quite similar, thus suggesting a marginal dependence on the 
metallicity (see right panels in Fig.~\ref{fig:bailey_ac}).
%editor 4 and many other figures with multiple panels that were labelled
%editor 15 
%%%%%%%%%%%%%%%%%%%%%%%%%%%%%%%%%%%%%%%%%%%%%%%%%%%%%%%%%%%%%%%%%%%%%%%%%%%%%%%%%%%%%
% 			fig 7
%%%%%%%%%%%%%%%%%%%%%%%%%%%%%%%%%%%%%%%%%%%%%%%%%%%%%%%%%%%%%%%%%%%%%%%%%%%%%%%%%%%%%
\begin{figure}[htbp]
\begin{center}
\includegraphics[width=0.66\textwidth]{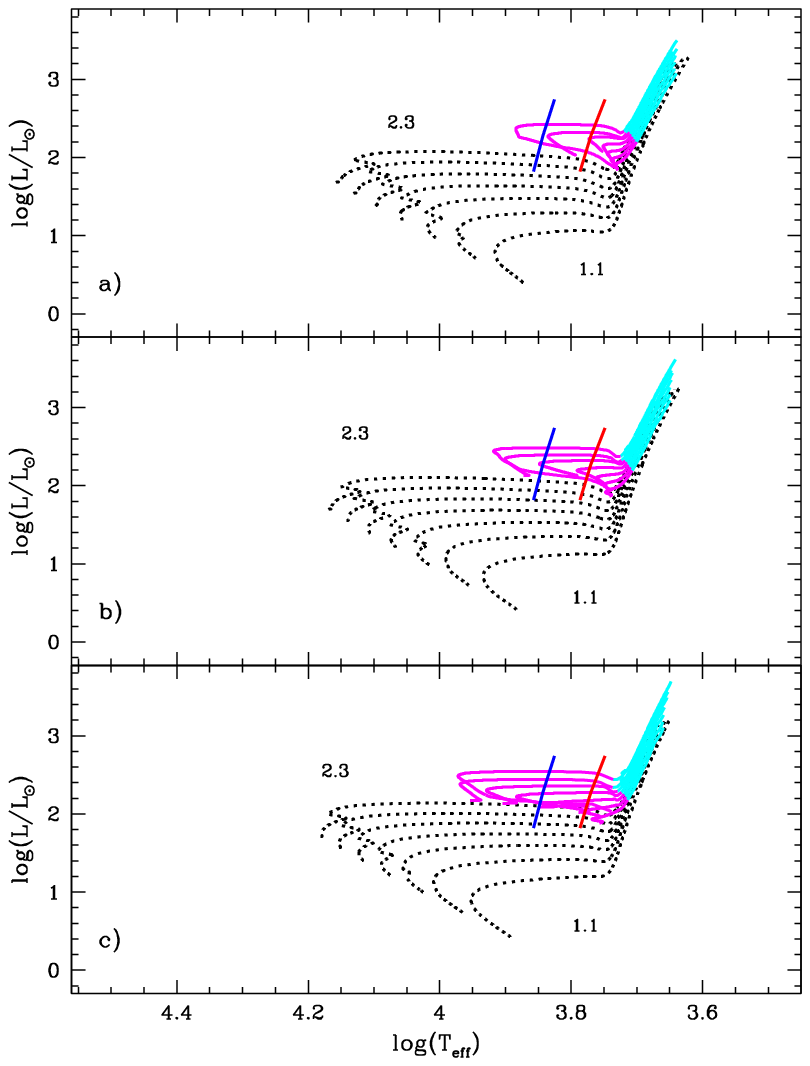}
\caption{
%\vspace{-0.5truecm}
Panel a)-- Evolutionary tracks for intermediate-age stellar structures in the 
Hertzsprung--Russell diagram at fixed metallicity (Z=0.0008) and stellar  
masses ranging from 1.1 to $2.3\,M/M_\odot$ (see labels). 
The magenta color marks central helium-burning phases, 
whereas the light blue color marks double-shell (H and He) burning phases 
(Asymptotic Giant Branch).  The almost vertical solid lines display the blue (hot) 
and the red (cool) edge of the Anomalous Cepheid instability strip 
\citep{fiorentino12c,monelli2022}.
Panel b)-- Same as panel a), but for Z=0.0004. 
Panel c)-- Same as panel a), but for Z=0.0002. 
}
\label{fig:hrd_four}
\end{center}
\end{figure}
%_________________________________________________________________________
%ref_2_8a
The comparison with Galactic ACs is hampered by the dependence of the reddening
correction in the Bulge and on the adopted reddening law in the disk.
In order to overcome possible systematics, the intrinsic \textit{(V-I)}$_0$ colors 
of ACs in the Galactic Bulge were estimated using the reddening map 
provided by \citet{nataf2013}, while for the others we adopted the 
reddening maps provided by \citet{Schlegel98} and by \citet{schlafly11} and
the \citet{cardelli89} reddening law. We found that their median 
intrinsic colors are \textit{(V-I)}$_0$=0.556$\pm$0.223 mag (FU) 
and \textit{(V-I)}$_0$=0.405$\pm$0.266 mag 
(see Table~\ref{tbl:cepheid_color_dis}) and they are quite similar, 
within the errors, to the intrinsic colors of MC ACs.     

The predicted dependence of ACs evolutionary and pulsation properties on the chemical 
composition are showed in Fig.~\ref{fig:hrd_four}). Panels a), b) and c) display 
evolutionary tracks for metal-poor (see labelled values) intermediate-age stellar 
%structures (for more details see section~\ref{evolution}).
structures (for more details see section~C in Appendix).
However, spectroscopic abundances are only available for two Galactic 
ACs (V716 Oph, BF Ser) that were originally classified as TIICs 
\citep{kovtyukh2018a}, but we still lack a detailed abundance analysis 
for both Galactic and Magellanic ACs.

%_______________________________________________________________________________
\subsection{Type II Cepheids}\label{sec:t2cs_intro}

%_______________________________________________________________________________
TIICs\footnote{
They were originally called Globular Cluster Cepheids by
\citep{baade56} thanks to the discovery of stellar populations. 
The reader interested in a more detailed discussion concerning 
early discoveries based on TIICs is referred to \citet{wallersteincox1984}.}
are low-mass stars in a double shell, helium and hydrogen, burning phase. 
According to the pulsation period, they are either Asymptotic Giant Branch (AGB) 
or post-AGB radial variables. They are solid old ($t\ge$10 Gyr) stellar tracers 
and have been identified in stellar systems hosting old stellar populations 
(Halo, Bulge, globulars, Magellanic Clouds, Andromeda group; 
see the bottom panel of Fig.~\ref{aitoff_all}). 
However, they have not been identified in nearby dwarf spheroidal galaxies. 
It is not clear yet whether this evidence is either an observational bias, 
or caused by the lack of hot and extreme horizontal branch stars in these 
stellar systems, as recently suggested by \citet{bono2020b}.    

\begin{figure}[htbp]
\begin{center}
\includegraphics[width=\textwidth]{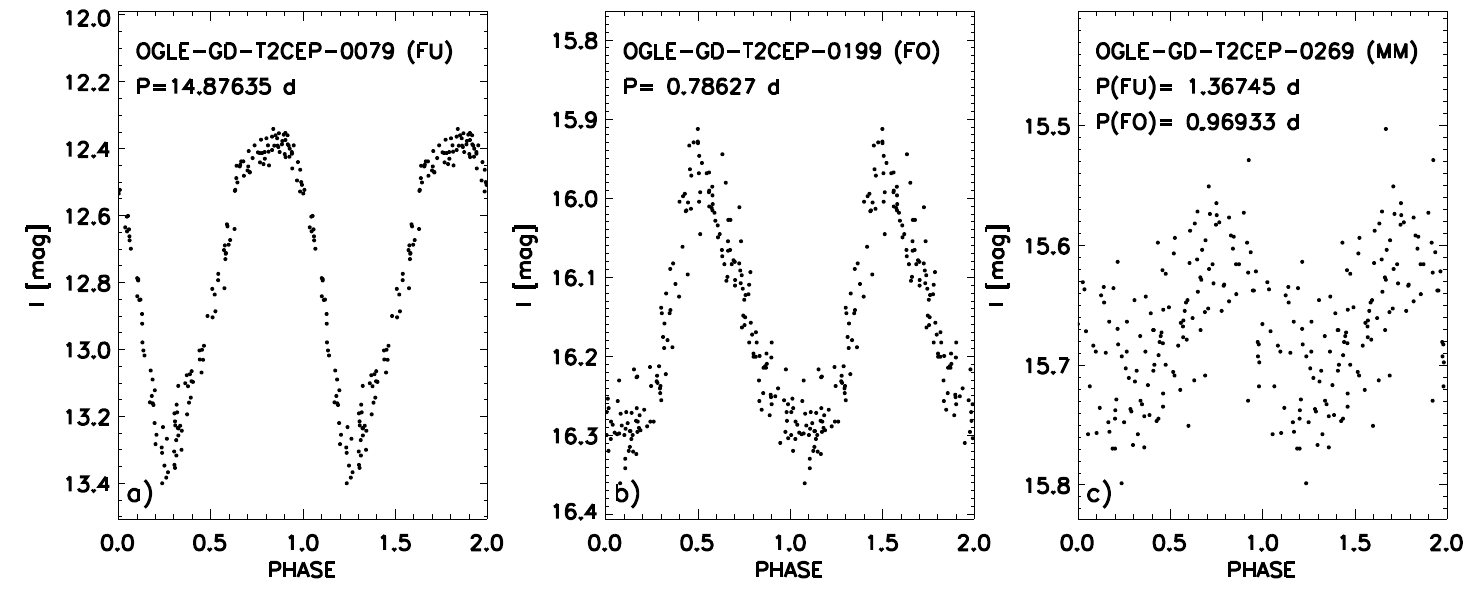}
\includegraphics[width=\textwidth]{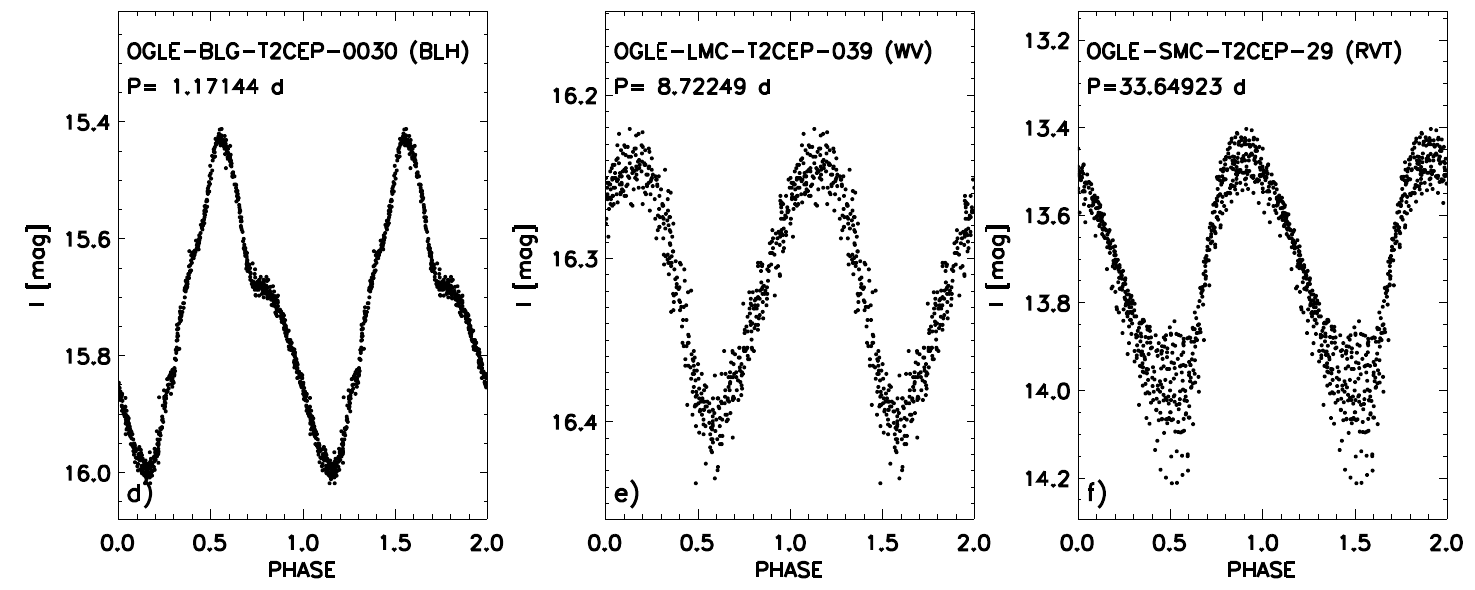}
\caption{
Panel a): Light curve for a Galactic TIIC pulsating in the fundamental mode from 
the OGLE IV dataset. The ID and the period are labelled.
Panel b): same as panel a), but for a Galactic first overtone TIIC. 
Panel c): same as panel a), but for a Galactic mixed-mode TIIC. Primary 
and secondary periods are labelled.
Panel d): same as panel a), but for a Galactic Bulge BL Herculis.
Panel e): same as panel a), but for a LMC W Virginis.
Panel f): same as panel a), but for a SMC RV Tauri.  
\label{fig:lcv_t2c}
}
\end{center}
\end{figure}

The pulsation properties of TIICs are canonical, since they pulsate as single mode variables in the fundamental and in the first overtone, but the latter group only includes a few objects both in the Galaxy and in the Magellanic Clouds \citep{soszynski2019b}. Interestingly  enough, long-term photometric surveys have also identified a few mixed-mode TIICs
%ref_2_8b
\citep{smolec2018,udalski2018}.
The top panels (a,b,c) of Fig.~\ref{fig:lcv_t2c} shows representative light curves for Galactic 
single and mixed-mode TIICs, while the bottom panels display selected light curves for TIICs 
typical of the three sub-groups: a Galactic Bulge BLH (panel d), an LMC WV (panel e) and an 
SMC RVT (panel f). 

%%%%%%%%%%%%%%%%%%%%%%%%%%%%%%%%%%%%%%%%%%%%%%%%%%%%%%%%%%%%%%%%%%%%%%%%%%%%%%%%%%%%%
% 			fig 8
%%%%%%%%%%%%%%%%%%%%%%%%%%%%%%%%%%%%%%%%%%%%%%%%%%%%%%%%%%%%%%%%%%%%%%%%%%%%%%%%%%%%%
%_______________________________________________________________________________
%_______________________________________________________________________________

The separation between RRLs and 
TIICs is a long-standing problem.  As 
a first approximation, it is possible to adopt a period 
threshold, whose exact value is still a matter of debate.
A threshold of $\sim$0.8 days was set in the review by
\citet{gautschysaio1996} where 
type 1 variables (AHB1, above the Horizontal Branch [HB]), as defined in 
\citet{strom1970} and \citet{diethelm1983,diethelm1990}, 
were considered as TIICs rather than evolved RRLs. 
This threshold is obsolete, however, because a more
extended and homogeneous investigation, based on 
period distribution and on the Fourier 
parameters of the light curve of RRLs in the Galactic
bulge, has now set the threshold at one day \citep{soszynski08c,soszynski14}.
The RRLs in the bulge have a primordial 
(or minimally enhanced) helium abundance \citep{marconiminniti2018}, 
but there is theoretical evidence
that helium enhancement increases the periods of RRLs
\citep{marconi2018}. This means that a one-day period threshold  
should be considered a particular case of a more general
chemical  and evolution-dependent threshold.
%--------------------------

\begin{figure}[htbp]
\begin{center}
\includegraphics[width=\textwidth]{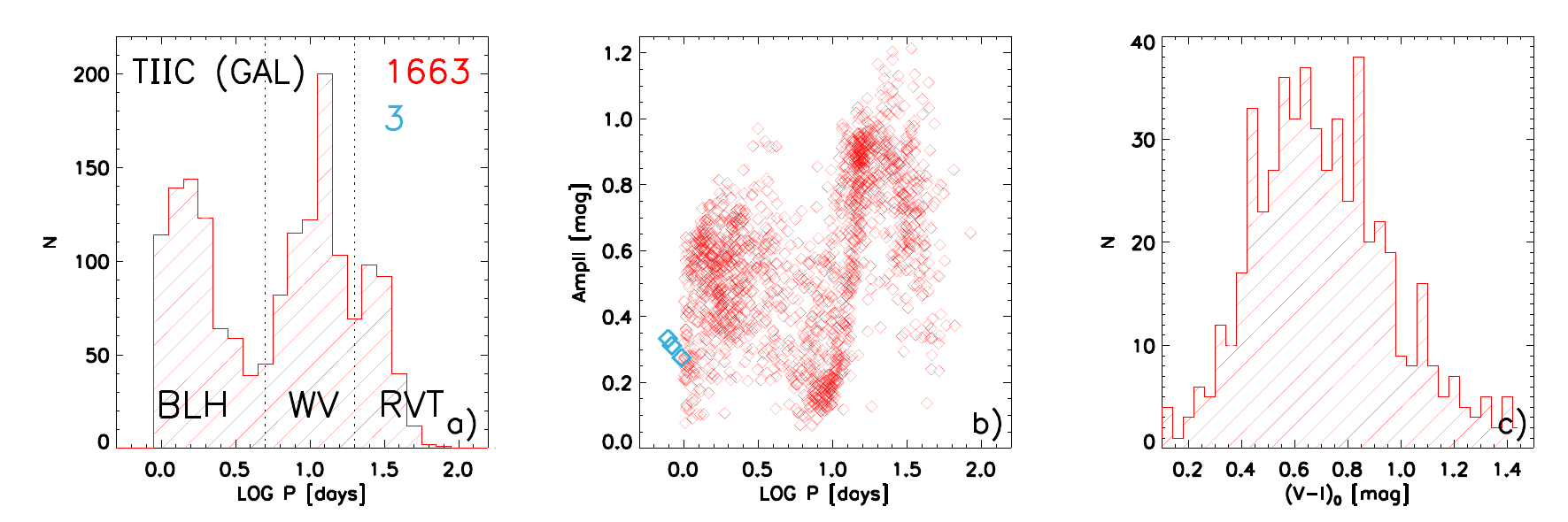}

\includegraphics[width=\textwidth]{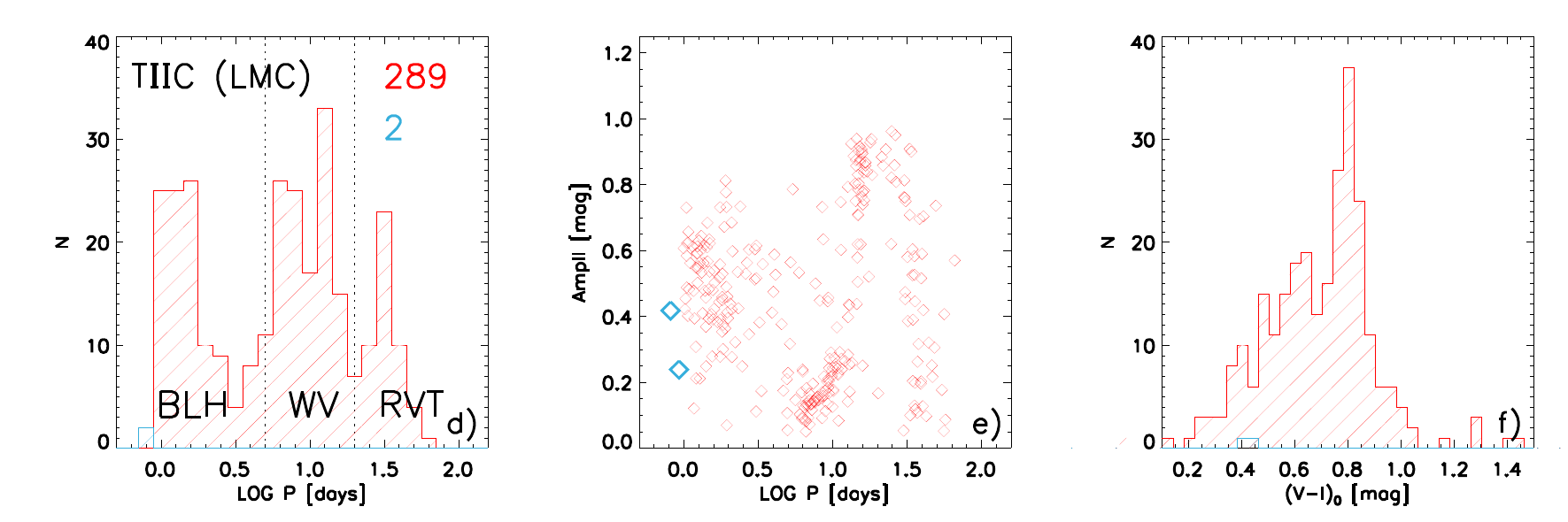}

\includegraphics[width=\textwidth]{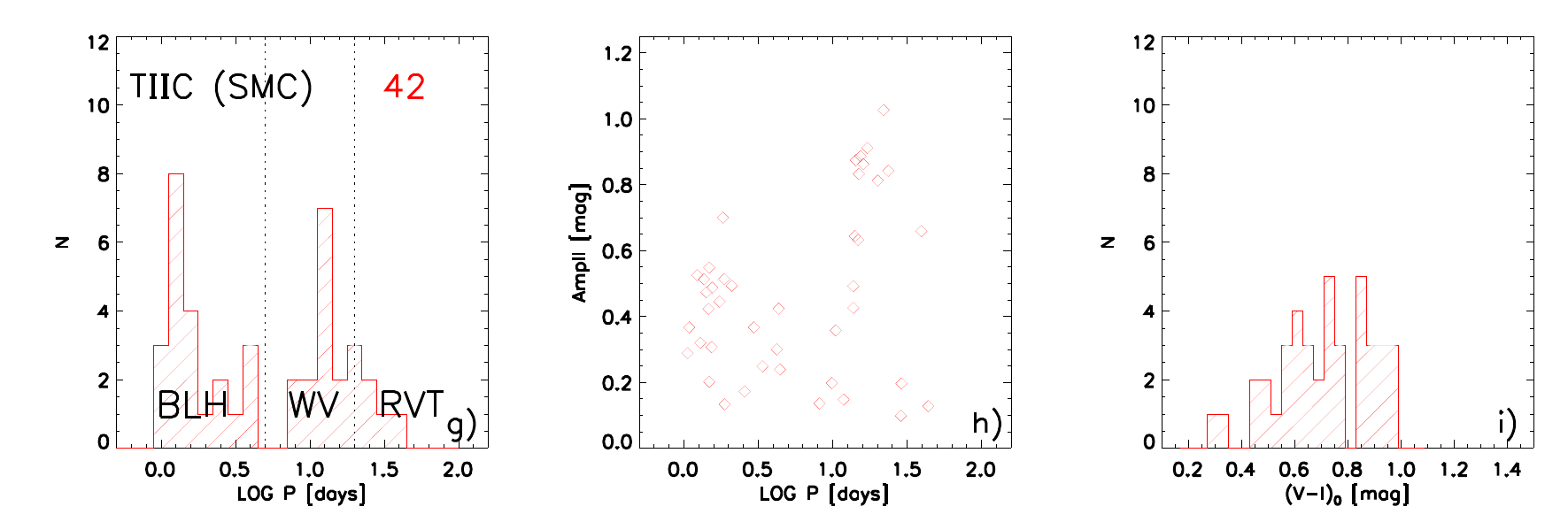}
\caption{
Top: Panel a)-- Period distribution for fundamental (red hatched area) and 
first overtone (light blue hatched area) Galactic TIICs.   
Panel b)-- I-band luminosity amplitude versus period (Bailey diagram) for 
both FU and FO TIICs.   
Panel c)-- dereddened ($V-I$)$_0$ color distribution for the same TIICs plotted 
in the left and in the middle panel.  
Middle: Same as the top, but for LMC TIICs.  
Bottom: Same as the top, but for SMC TIICs. 
\label{fig:pac_tiics}
}
\end{center}
\end{figure}

TIICs have periods ranging from 1 day to more than 100 days       
(see left panels in Fig.~\ref{fig:pac_tiics}).  
Moreover, they display two local minima in the period distribution for $P\sim5$ and      
$P\sim20$ days. The former value was adopted for separating BLHs      
from WVs, and the current data  indeed suggest that it ranges from      
about four days in the Bulge to about six days in the Galactic      
field. The latter value ($P\sim20$ days) was adopted for separating WVs from      
RVTs, and the current data  suggest that it shows up      
as a shoulder in the period distribution of Bulge and field TIICs and      
as a local minimum in GCs and in Magellanic TIICs. 

The difference in the period distribution among BLHs, WVs, and RVTs is       
fully supported by the Bailey diagrams, the $I$-band luminosity      
amplitudes versus the logarithmic period, showed in the middle panels 
of Fig.~\ref{fig:pac_tiics}. The Bailey diagram shows that   
WVs attain a well-defined minimum at $P\sim8$ days,   
with a steady increase towards longer periods. The trend      
for RVTs is far from being homogeneous, because the maximum around      
20 days is broad. Moreover, RVTs in the Bulge and in the LMC display      
a steady decrease towards longer periods and a well-defined cutoff at periods      
longer than 60 days. On the other hand, RVTs in the Galactic field      
approach 200 days and display at fixed periods a broad range in     
luminosity amplitudes.        
     
The current partition of TIICs into three sub-groups follows the classification      
suggested by \citet{soszynski08c,soszynski2011}.      
They also suggested a new group of TIICs, the      
peculiar WVs (pWVs) which have peculiar light curves. Moreover, the pWVs are,      
at fixed periods, brighter than typical TIICs.      

The possible dependence on the metallicity requires a more detailed discussion.      
We still lack spectroscopic measurements of Bulge TIICs, so we assume that their      
metallicity distribution is either similar to that of Bulge RRLs as measured by \citet{walker1991a},       
suggesting a mean [Fe/H]$=-1.0$  with a 0.16~dex standard deviation,       
or similar to Bulge red giant stars, with average [Fe/H]$=-1.5$ and 
a standard deviation equal to 0.5~dex \citep{rich2012,zoccali2017}.     
The metallicity distribution of TIICs in GCs and in the Galactic field     
ranges from [Fe/H]$\sim-2.4$ to slightly super solar [Fe/H]       
\citep[see Appendix in][]{bono2020b}.

%%%%%%%%%%%%%%%%%%%%%%%%%%%%%%%%%%%%%%%%%%%%%%%%%%%%%%%%%%%%%%%%%%%%%%%%%%%%%%%%%%%%%
% 			fig 9
%%%%%%%%%%%%%%%%%%%%%%%%%%%%%%%%%%%%%%%%%%%%%%%%%%%%%%%%%%%%%%%%%%%%%%%%%%%%%%%%%%%%%
%_______________________________________________________________________________
%______________________________________________________________________________

For LMC TIICs, we can follow two different paths.      
According to \citet{gratton04a} the iron abundance of LMC RRLs based on low-resolution      
spectra range from [Fe/H]$=-2.1$ to  [Fe/H]$=-0.3$, but only a few stars are more      
metal-rich than [Fe/H]$=-1.0$;  the mean observed iron abundance for 98 RRLs is      
[Fe/H]$=-1.48 \pm 0.03 \pm0.06$. This metallicity range is also supported by recent      
investigations of the mean metallicity of LMC globular clusters.   
Using homogeneous Str\"{o}mgren photometry, 
\citet{piatti2018} found, in agreement with      
spectroscopic measurements, that the metallicity of the ten LMC globulars ranges from          
$-2.1$ dex (NGC~1841) to $-1.1$ dex (ESO121-SC3). We still lack direct measurements of the      
metallicity distribution of truly old SMC stellar tracers.  
According to high-resolution spectroscopy \citep[][]{dalessandro2016}, 
the metallicity of NGC~121, the only SMC globular cluster, is [Fe/H]$=-1.28$.   
Metallicity estimates listed in Table~A.1 of \citet{bono2020b} indicate that TIICs
cover roughly three dex in metal abundance. However, the population ratios 
appear to be, within the errors, quite similar. This finding is also supported 
by the similarity in the color distribution between Galactic and Magellanic TIICs 
(see right panels in Fig.~\ref{fig:pac_tiics}).
%--------------------------

TIICs can be thought as  the intersection of several theoretical and empirical      
investigations; however, their evolutionary status is far from being well established.      
A first analysis of the evolutionary properties of TIICs was      
provided over 40 years ago by \cite{gingold74,gingold76,gingold85}.      
He recognized that a significant fraction of hot (blue) HB stars evolve       
off the Zero-Age Horizontal Branch (ZAHB) from the blue (hot)      
to the red (cool) region of the color--magnitude diagram (CMD).
In the approach to their AGB track, these stars are in a double shell (hydrogen and helium) burning phase      
\citep{salaris05} and cross the instability strip at luminosities systematically      
brighter than typical RRLs. The difference in luminosity and the lower mass     
  induce an increase in the pulsation period of TIICs      
when compared with RRLs. Typically, the two classes are separated by  
a period threshold at one day. This separation   is supported by a  well-defined 
minimum of the period distribution when moving from RRLs to TIICs, 
but the physical mechanism(s) causing this minimum are not yet clear, and   
the exact transition between RRLs and TIICs has not been established  
\citep{braga2020}.   
     
The quoted calculations also suggest  that blue HB stars after the first crossing of      
the instability strip experience a ``blue nose'' (then dubbed ``Gingold's nose''),     
 a blue-loop in the CMD that causes two further crossings of the instability strip before     
the tracks reach the AGB.      
These three consecutive excursions were associated with the interplay between the helium      
and hydrogen burning shells. After core-helium exhaustion, HB models with      
massive enough envelopes evolve redward in the CMD. The subsequent shell-helium ignition causes      
a further expansion of the envelope, and, in turn, a decrease in the efficiency      
of the shell-hydrogen burning, which causes a temporary contraction of the envelope,   
and a blueward evolution     
in the CMD. Once shell hydrogen burning increases  its energy production again,       
these models move back towards the AGB track.       
Finally, these models would eventually experience a fourth blueward      
crossing of the instability strip before approaching their white dwarf (WD)      
cooling sequence (see Fig.~1 in \citealt{gingold85} and Fig.~2 in \citealt{maas2007}).       
During their final crossing of the instability strip, these stellar structures  
(in the post-AGB phase) are only supported by a vanishing shell H-burning.      

Basic arguments based on their evolutionary status and on the use of the pulsation      
relation available at that time \citep{vanalbada73} allowed Gingold to      
associate the first three crossings (including  {Gingold's nose}) with      
BLHs and the fourth one with the WVs variables.       
These early analyses, however, lacked quantitative estimates of the time spent      
inside the instability strip during the different crossings, and in particular      
the period distributions associated  with  the different crossings. Moreover, 
HB evolutionary models dating back to more than 25 years ago and       
based on updated input physics      
\citep{lee90,castellani91,dorman93,brown00,pietrinferni06a,dotter08,vandenberg13}     
do not show the Gingold's nose. 
%--------------------------

The evolutionary properties of low-mass core helium burning models have been      
discussed in several recent investigations      
\citep[][and references therein]{cassisi11}. Here we summarize the main      
relevant features in order to explain the evolutionary channels producing TIICs.         

\begin{figure}[htbp]
\centering     
\includegraphics[width=0.60\textwidth]{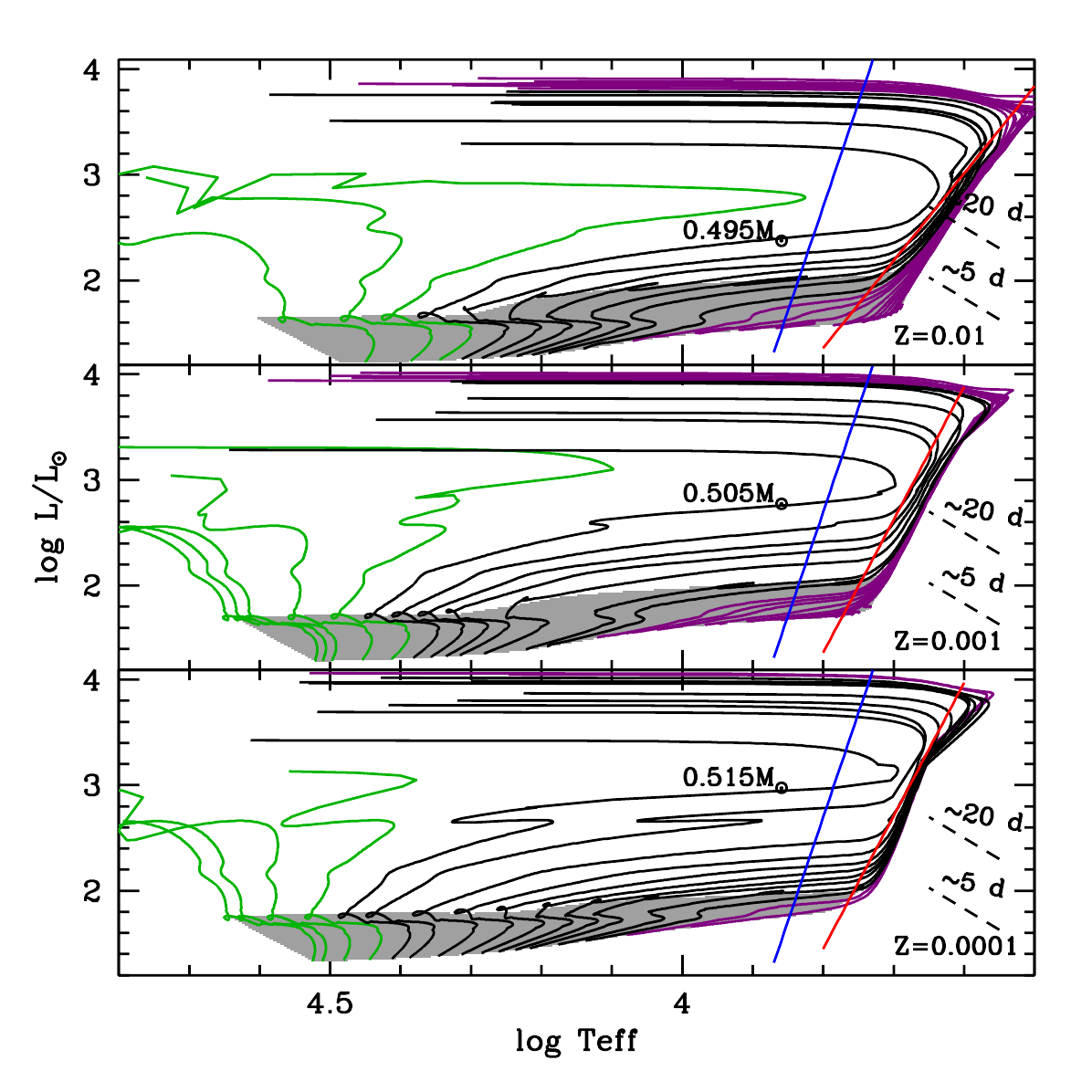}     
\caption{     
Top:  HRD of HB evolutionary models covering a broad range of stellar  masses    
($M/M_\odot$=0.48--0.90)  and the same initial chemical composition    
(Z=0.01, Y=0.259). The grey area outlines the region between ZAHB      
(faint envelope) and central-helium exhaustion (bright envelope).     
The green lines display HB models evolving as AGB-manqu\'e, black lines the post early-AGB models,      
and purple lines the thermal pulsing AGB models (see text for  details).       
The almost vertical blue and red solid lines indicate the hot and cool edge       
of the TIIC instability strip. The minimum stellar mass (in solar      
units) crossing the instability strip is labelled in black.      
The black dashed lines show two iso-periodic lines for 5 and 20 days.     
%minima 0.480 0.90    
%    
Middle: Same as the top panel, but for stellar masses ranging from    
$M/M_\odot$=0.4912 to 0.80 and for a metal-intermediate chemical composition (Z=0.001, Y=0.246).     
Bottom: Same as the top panel, but for stellar masses ranging from    
$M/M_\odot$=0.5035 to 0.70 and for a metal-poor chemical composition (Z=0.0001, Y=0.245). Image reproduced with permission from \citet[][Fig.~5]{bono2020b}, copyright by ESO.}      
\label{theo_tracks}     
\end{figure}

The grey area displayed in the top panel of Fig.~\ref{theo_tracks} outlines    
the region between the ZAHB (faint envelope) and central helium exhaustion      
(bright envelope) for a set of HB models with different masses and fixed    
chemical composition (metal--Z=0.01--and helium--Y=0.259--mass fractions).      
We have assumed an $\alpha$-enhanced chemical composition \citep{pietrinferni06b}    
and a progenitor mass according to a 13~Gyr isochrone (the mass at the main sequence    
turn off, MSTO, is equal to $0.86\,M_\odot$).

%editor 15
%%%%%%%%%%%%%%%%%%%%%%%%%%%%%%%%%%%%%%%%%%%%%%%%%%%%%%%%%%%%%%%%%%%%%%%%%%%%%%%%%%%%%
% 			fig 10
%%%%%%%%%%%%%%%%%%%%%%%%%%%%%%%%%%%%%%%%%%%%%%%%%%%%%%%%%%%%%%%%%%%%%%%%%%%%%%%%%%%%%
%^^^^^^^^^^^^^^^^^^^^^^^^^^^^^^^^^^^^^^^^^^^^^^^^^^^^^^^^^^^^^^^^^^^^^^^^^^^^^^^     
%^^^^^^^^^^^^^^^^^^^^^^^^^^^^^^^^^^^^^^^^^^^^^^^^^^^^^^^^^^^^^^^^^^^^^^^^^^^^^^^ 

The total mass along the ZAHB, as expected, decreases when moving      
from the red HB (RHB) to the blue HB (BHB) and further on to the extreme HB (EHB).      
The helium-core mass is constant along the ZAHB and is mainly fixed by the    
chemical composition  of the progenitor ($M_c^{\rm He}=0.4782\,M_\odot$) and is    
roughly independent of age for ages above a few Gyr. The mass of the envelope decreases      
from $0.4218\,M_\odot$ for RHB models to a few thousandths solar masses for EHB models.      
In an actual old stellar population with a fixed initial chemical composition, the mass    
lost along the RGB \citep[more efficient when approaching the tip of the RGB;]   
[]{origlia14} determines the final mass distribution along the ZAHB.

The bright envelope of the grey area marks the central helium exhaustion,    
corresponding formally to the beginning of the AGB phase. The small ripples    
along the helium-exhaustion sequence ($\log L/L_\odot\sim 1.8$) show that    
the lower the total mass of the HB model, the hotter the effective temperature    
at which the helium exhaustion takes place. The luminosity of the ripples      
ranges from $\log L/L_\odot\approx 2$ in the warm region to      
$\log L/L_\odot\approx 1.6$ in the hot region of the HB.      
At this point, the He burning moves smoothly to a shell around the carbon-oxygen core. The overlying H-shell    
extinguishes, due to the expansion of the structure before reigniting later with various efficiencies, depending on   
the mass thickness of the envelope around the original He core.   
   
Models with mass below $0.495\,M_{\odot}$ (corresponding to an envelope mass 
lower than $\sim0.017\,M_{\odot}$)  never reach the AGB location; they do not 
cross the instability strip and move  to their WD cooling sequence, as a carbon-oxygen      
(CO) WD \citep{castellani06,salaris13b}.     
These objects have been called AGB-manqu\'e \citep{greggio90}, 
and are shown as green tracks in the top panel of      
Fig.~\ref{theo_tracks}).   
   
More massive models  cross the instability strip while moving towards their AGB tracks.    
Models with $0.495 \leq M/M_{\odot} < 0.55 $  reach the AGB, but move back towards    
the WD sequence (hence they cross the instability strip again but at higher    
luminosities) well before reaching the thermal pulse (TP) phase. They are named    
post-early AGB (PEAGB), and are plotted as black lines in the top panel    
of Fig.~\ref{theo_tracks}. These AGB models perform several gravo-nuclear    
loops in the Hertzsprung--Russell Diagram (HRD), either during the AGB phase and/or in their approach to    
the WD cooling sequence after leaving the AGB (during this post-AGB transition    
models cross again the instability strip). Some of them may take place inside the     
instability strip. The reader interested in a more detailed discussion      
concerning their impact on the pulsation properties is referred to      
\cite{bono97d}. The evolutionary implications, and in particular the    
impact of the loops concerning the AGB lifetime have recently been discussed    
by \cite{constantino16}.

Models with $M \geq 0.54\,M_{\odot}$ (plotted as purple lines in 
Fig.~\ref{theo_tracks})  evolve along the AGB and experience the TPs. 
The number of TPs, and in turn      
the duration of their AGB phase, is once again dictated by the efficiency of     
the mass loss and by their residual envelope mass \citep{weiss09,cristallo09}.     
Calculations of TP evolution are quite demanding from the computational point    
of view, hence  we decided to use the fast and simplified synthetic AGB    
technique originally developed by \cite{iben78} and more recently by    
\cite{wagenhuber98} to compute the approach of these AGB models to the    
WD cooling sequence.    
In particular, the synthetic AGB modelling started for thermal-pulsing AGB   
(TPAGB) models just before the occurrence of the first TP, while    
for PEAGBs it was initiated at the brightest and reddest point    
along the first crossing of the HRD, towards the AGB.

The middle and the bottom panels of Fig.~\ref{theo_tracks} show the same      
predictions, but for two more metal-poor chemical compositions.      
The values of the stellar masses plotted along the ZAHBs show the      
impact of the chemical composition.   
The mass range of the tracks that cross the instability strip and produce TIICs steadily      
decreases from 0.495--0.90 $M/M_\odot$ for Z=0.01, to 0.505--0.80 $M/M_\odot$      
for Z=0.001, and to 0.515--0.70 $M/M_\odot$ for Z=0.0001. It is worth mentioning    
that the range in luminosity covered by the different sets of models is    
 similar. The mild change in stellar mass and the similarity in    
luminosities suggests a marginal dependence of the mass--luminosity (ML)    
relation of TIICs on the chemical composition.
In order to further define the theoretical framework for RVTs stars    
\citep{wallerstein2002,soszynski2011}, we suggest that they are    
the progeny of both PEAGB and TPAGB. There are two reasons supporting this hypothesis:\par       
     
a) Period range -- The theoretical periods for these models are systematically      
longer than WVs and more typical of RVTs stars. The predicted mass for these      
structures is uncertain because it depends on the efficiency of mass      
loss during the TP phase. The theoretical framework is further      
complicated by the fact that the number of TPs also depends on      
the initial mass of the progenitor and on its initial chemical composition.      
This means that  a contribution from      
intermediate-mass stars  cannot be  excluded a priori. 
However, the lack of RVTs in nearby dwarf      
spheroidal galaxies hosting a sizable fraction of intermediate-mass stars with      
ages ranging from 1~Gyr to more than 6-8~Gyr, such as      
Carina, Fornax and Sextans \citep{aparicio04,beaton2018},  suggests that this      
channel might not be very efficient. However, RVTs have been identified in      
the MCs \citep{soszynski08c,ripepi2015}.         
     
b) Alternating cycle behaviour -- There is evidence of an interaction      
between the central star and the circumstellar envelope, possibly causing the      
alternating-cycle behaviour \citep{feast08,rabidoux10}. The final crossing      
of the instability strip before approaching the WD cooling sequence either    
for PEAGB or for TPAGB models appears a very plausible explanation.    
     
The above circumstantial evidence suggests that the variable stars    
currently classified as TIICs have a range of evolutionary properties. The BLHs      
and the WVs appear to be either post-ZAHB (AGB, double shell burning) or      
post-AGB (shell hydrogen burning) stars, whereas RVTs are mainly       
post-AGB objects.

%____________________________________________________________________________________
\section{The Hertzsprung progression}\label{sec:hertzprungprog}

More than ninety years ago, \citet{hertzsprung1926} discovered that a sub-sample of Galactic classical Cepheids presents a relationship between the bump along the light curve and the pulsation period. The so-called ``Hertzsprung progression'' (HP) was subsequently discovered among Andromeda and MC Cepheids by \citet{shapley1940}, \citet{kukarkinparenago1949}, and \citet{paynegaposchkin1954}. The HP observational scenario was enriched by \citet{joy1937} and by \citet{ledoux1958} who found a similar shift in the phase of the bump in radial velocity curves.

The empirical fingerprint of the HP is the following: classical Cepheids in the period range $6<P<16$ days show a bump along both the light and the velocity curves. This secondary feature appears on the decreasing branch of the light curve for Cepheids with periods up to 9 days, while it appears close to maximum light for $9<P<11$ days and moves at earlier phases for longer periods. On the basis of this observational evidence this group of variables was christened ``Bump Cepheids'' for avoiding to be mixed-up with ``Beat Cepheids''. In fact, the latter group refer to mixed-mode variables i.e. objects in which two or more modes are simultaneously excited, and therefore both the shape of the light curves and the pulsation amplitudes change from one cycle to the next, whereas bump Cepheids are single-mode variables and their pulsation properties are characterized by a strong regularity over consecutive cycles.

%%%%%%%%%%%%%%%%%%%%%%%%%%%%%%%%%%%%%%%%%%%%%%%%%%%%%%%%%%%%%%%%%%%%%%%%%%%%%%%%%%%%%
% 			fig 11
%%%%%%%%%%%%%%%%%%%%%%%%%%%%%%%%%%%%%%%%%%%%%%%%%%%%%%%%%%%%%%%%%%%%%%%%%%%%%%%%%%%%%
%_______________________________________________________________________________
%IDLWorkspace/Default/Temp/temp_220308.pro
\begin{figure}[htbp]
\begin{center}
\includegraphics[width=0.46\textwidth]{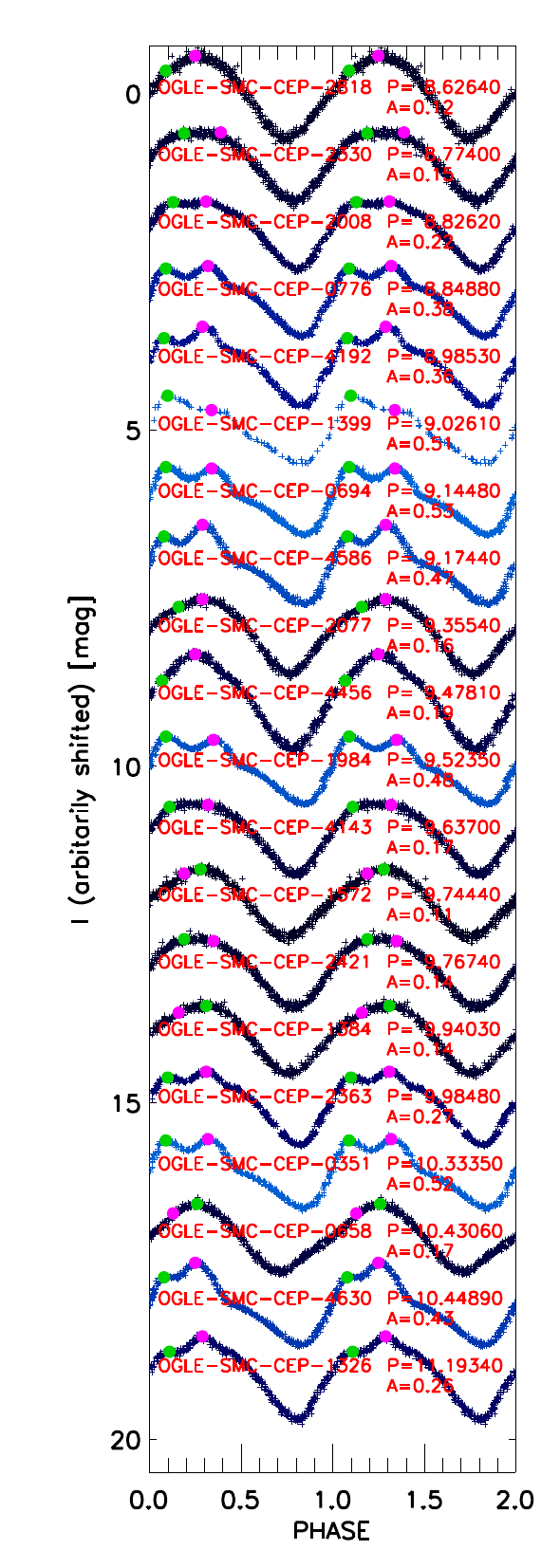}
\includegraphics[width=0.46\textwidth]{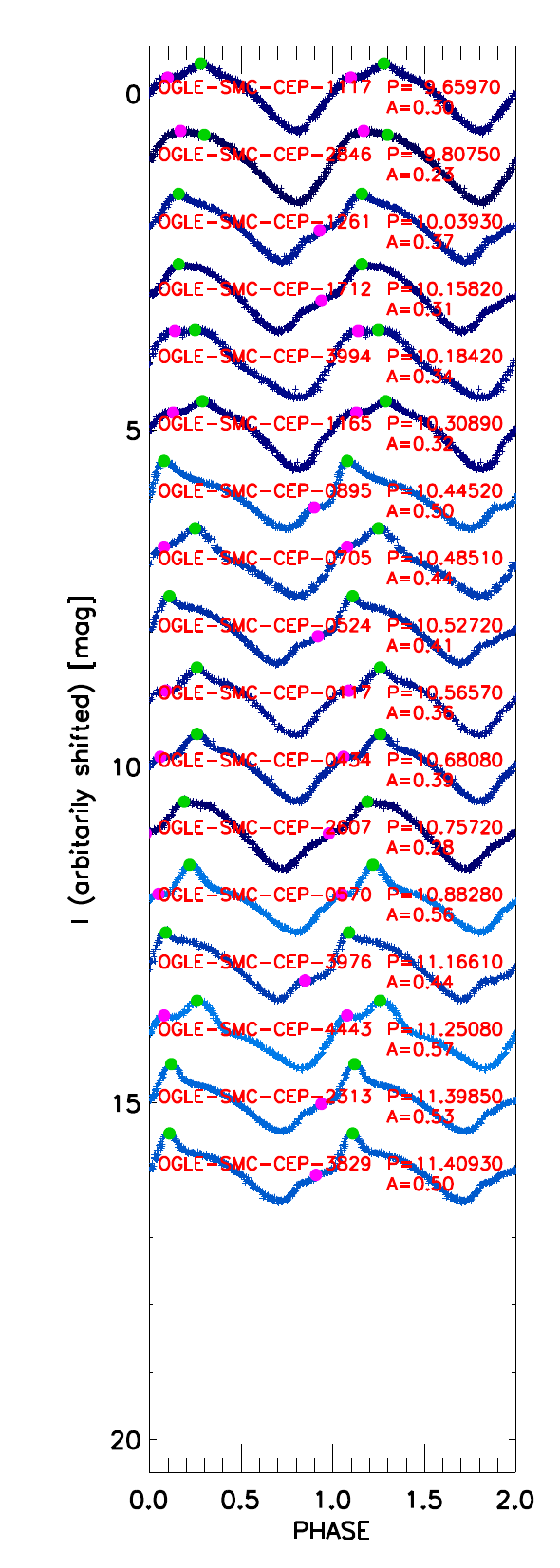}
\caption{
Left: Phased $I$-band light curves of SMC bump Cepheids with a bump moving up along the decreasing branch (UPDB). From top to bottom, the different variables are plotted with increasing pulsation period and artificially shifted in magnitude. The position of the bump is marked with a pink circle, while the pulsation maximum is marked with a green circle. The light curves are plotted twice, so as to emphasize the changes along the pulsation cycle. The light curves are color-coded according to the luminosity amplitude. They range from light blue for the largest amplitudes to dark blue for the smallest ones. The labels on the left side show the name of the variables, whereas those on the right side display the period (days) and the $I$-band amplitude (mag). 
Right: Same as the left, but for SMC bump Cepheids with a bump moving down along
the rising branch (DORB).}
\label{fig_smc_bump_light}
\end{center}
\end{figure}
%_______________________________________________________________________________

A more quantitative approach concerning bump Cepheids was originally suggested by \citet{parenago1936,paynegaposchkin1947}, and more recently by \citet{simon1976} and by \citet{simon_lee1981} who investigated the shape of the light curves by means of Fourier analysis. The latter authors found out that both the phase difference, $\phi_{21}$, and the amplitude ratio, $R_{21}$, show a sharp minimum close to the center of the HP. Following this approach, several investigations have been already devoted to Fourier parameters of Galactic and Magellanic Cepheids. In particular, \citet[][hereinafter MBM]{moskalik1992b} suggested that the minimum in the Fourier parameters for Galactic Cepheids takes place at P(HP) = 10.0 $\pm$ 0.5 days, while \citet{moskalik2000} by investigating a sample of more than 100 radial velocity curves, they found P(HP) = 9.95 $\pm$ 0.05 days \citep[see also][and references therein]{hocde2023}. At the same time, \citet{alcock1999} by investigating a large sample of LMC Cepheids estimated that the minimum in the Fourier parameters is located at P(HP) = 11.2 $\pm$ 0.8 days. Thus supporting the shift of the HP center toward longer periods originally suggested by \citet{paynegaposchkin1951} and strengthened by \citet{anderasen_petersen1987} and by \citet{andreasen1988}. More recently, \citet{beaulieu1998} suggested that the HP center in LMC and in SMC Cepheids is located at P(HP) = 10.5 $\pm$ 0.5 days and P(HP) = 11.0 $\pm$ 0.5 days, respectively. Since these three stellar systems are characterized by different mean metallicities, namely Z=0.02 (MW), Z=0.008 (LMC), and Z=0.004 (SMC), this empirical evidence suggests that a decrease in metallicity moves the HP center toward longer periods.

Up to now, two distinct models have been proposed in the literature to explain the appearance of the
HP among bump Cepheids: the echo model and the resonance model. The former was suggested by
\citet{whitney1956} and discussed by \citet{christy1968,christy1975} on the basis of Cepheid 
nonlinear, radiative models. According to Christy, during each cycle close to the phases of 
minimum radius and before the phase of maximum expansion velocity a pressure excess is 
generated in the first He ionization region. This pressure excess causes a rapid expansion 
which, in turn, generates two pressure waves moving outward and inward. The latter reaches 
the stellar core close to the phase of maximum
radius, then reflects and reaches the surface one cycle later causing the appearance of the bump.
The resonance model was suggested by \citet{simon1976} and is based on linear,
adiabatic periods. Within this theoretical framework, the bump would be caused by a resonance between the
second overtone and the fundamental mode, and takes place when the period ratio between these two
modes is close to 0.5. In particular, they suggested that the instability of the fundamental mode
drives, due to a resonance, the second overtone instability. This explanation lies on the evidence
that the nonlinear, radiative models constructed by \citet{stobie1969b} show a bump along the radial
velocity curves close to the resonance line $P_2$/$P_0 = 0.5$.

Such an extensive observational and theoretical effort devoted to bump Cepheids was not
only aimed at understanding the HP, but also at providing independent estimates of both
the mass and the radius of these variables. In fact, dating back to \citet{christy1968}, \citet{christy1975},
\citet{stobie1969b} and \citet{fricke1972} it was suggested that these two
evolutionary parameters can be constrained on the basis of period and phase of the bump.
A different method to estimate the mass, based on period ratios, was suggested by
\citep{petersen1973}. Mass determinations based on these two methods present a
compelling feature: they are based on observables, such as periods and phases of the bump,
which are not affected by systematic empirical uncertainties, since they are only
limited by photometric accuracy. However, pulsation masses based on these methods are,
with few exceptions \citep{carson_stothers1988}, systematically smaller than
the evolutionary masses. This longstanding puzzle raised the so-called bump mass discrepancy
\citep[see also ][]{cox1980} and at the same time supported the use of a ML relation based on
evolutionary models, which include either a mild or a strong convective core 
overshooting \citep{simon1995,wood1998b}. 

Even though the new radiative opacities settled down this long-standing problem 
(MBM; \citealt{kanbur_simon1994}), recent linear \citep{buchler1996,simon_young1997} 
and nonlinear \citep{wood1997} predictions for MC Cepheids present a small 
discrepancy with the ML relations predicted by current evolutionary models.

In order to provide a new quantitative spin on the HP, we decided to take 
advantage of the homogeneous and accurate data set collected by OGLE-IV 
for both Galactic and MC Cepheids to further constrain the metallicity 
dependence.

Figure~\ref{fig_smc_bump_light} shows the $I$-band light curves of SMC bump 
Cepheids. The light curves from top to bottom are plotted in order of 
increasing pulsation period and their colors are correlated with the 
luminosity amplitude. Darker colors mark low luminosity amplitude variables, 
whereas light blue colors mark large amplitudes. To help the eye in the 
identification of secondary features, the bump is marked with a pink 
circle, while the pulsation maximum is marked with a green circle. 
To properly trace the transition of the bump from the decreasing to the 
rising branch, the left panel only shows bump Cepheids with the bump 
moving UP along the decreasing branch (UPDB), whereas the right panel 
shows bump Cepheids with the bump moving DOwn along the rising branch 
(DORB). The light curves plotted in this figure display several 
distinctive features that are worth being discussed in detail. 

a) The phase of the maximum among the UPDB Cepheids (left panel) 
is quite constant and equal to $\phi\sim$0.1\footnote{The
current light curves were phased by using as reference epoch the 
phase along the rising branch in which the magnitude of the 
light curve is equal to the mean magnitude of the object 
\citep{Innoetal2013}.}, while the 
phase of the bump steadily approaches the phase of the maximum light 
when moving from shorter (top) to longer-period (bottom) bump Cepheids.

b) The bump in UPDB Cepheids with periods shorter than $\sim$9 days 
is fainter than the pulsation maximum, while for periods 
of the order of $\sim$9.3--9.5 days it attains magnitudes similar 
to the pulsation maximum and it becomes the main maximum at longer 
periods. The variation in the shape of the light curve is far from 
being smooth, indeed UPDB Cepheids with low luminosity amplitudes 
have light curves that have been defined as ``flat topped'' , i.e. 
with a luminosity maximum characterized by two secondary maxima.

c) The center of the HP is located at periods of P(HP)=9.8$\pm0.2$ days, 
in which the UPDB Cepheids attain a well-defined minimum in 
luminosity amplitude (A($I$)$\sim$0.15 mag). The light curve of these 
variables is more sinusoidal, almost featureless, with the 
bump showing up either as a small secondary maximum along the rising 
branch, or a change in the slope along the rising branch. To improve 
the identification of the bump in these variables we also took advantage 
of the OGLE-IV V-band light curves.  

d) The light curves of DORB Cepheids (right panel) show, as expected, 
an opposite trend. Close to the minimum in luminosity amplitudes, 
the light curves of UPDB and DORB Cepheids are similar. The phases 
of the maximum are once again quite constant, 
and take place around phases $\phi$=0.1--0.2. However, the phase 
of the bump steadily approaches the phase of the minimum in 
luminosity when moving from shorter to longer period DORB Cepheids. 

e) The variation of the light curve among LMC bump Cepheids 
%(see Fig.~\ref{fig_lmc_bump_light}) is similar 
(see Fig.~\ref{fig_lmc_bump_light} in Appendix) is similar 
to  SMC bump Cepheids (Fig.~\ref{fig_smc_bump_light}). The key 
difference is that the center of the HP occurs at shorter periods, 
namely P(HP)$\sim9.4\pm$0.2 days. In this period 
range, the luminosity amplitude is smaller than 0.2 magnitudes, with a 
well-defined minimum for the Cepheid 2252 with A($I$)$\sim$0.1 mag. 
There are a few exceptions concerning the shape of the light curve, 
but the global trend is well defined among both UPDB and DORB Cepheids. 

f) The similarity in the variation of the light curves across the HP also 
%ref_2_12a
%applies to Galactic bump Cepheids (see Fig.~\ref{fig_gal_bump_light}). 
applies to Galactic bump Cepheids (see Fig.~\ref{fig_lmc_bump_light} in Appendix). 
The center of the HP is located at even shorter periods 
P(HP)$\sim9.0\pm0.2$ days and $I$-band luminosity amplitudes of 
the order of 0.25 magnitudes.   

The reader interested in a more quantitative discussion concerning the 
metallicity dependence  of the HP is referred to section~A in 
Appendix.

%%%%%%%%%%%%%%%%%%%%%%%%%%%%%%%%%%%%%%%%%%%%%%%%%%%%%%%%%%%%%%%%%%%%%%%%%%%%%%%%%%%%%%
\section{Magellanic Cepheids and the color--magnitude diagram} 

\begin{figure}[htbp]
\begin{center}
\includegraphics[width=0.49\textwidth]{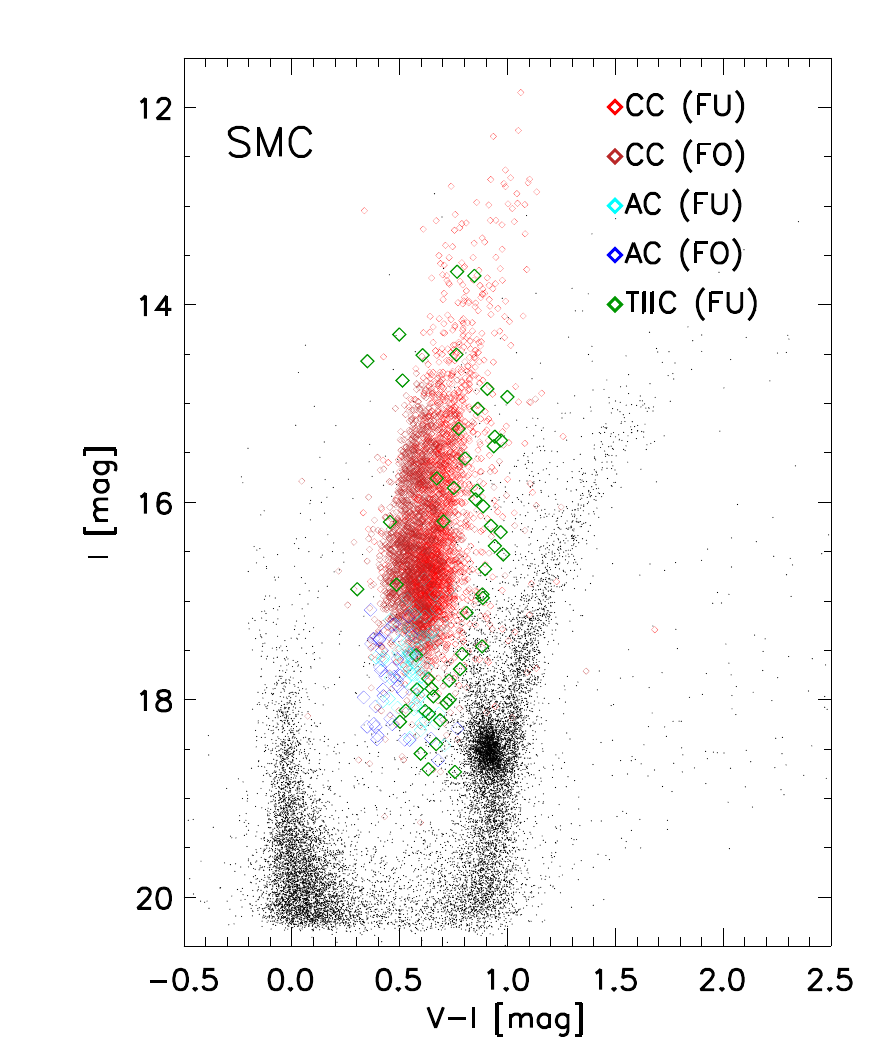}
\includegraphics[width=0.49\textwidth]{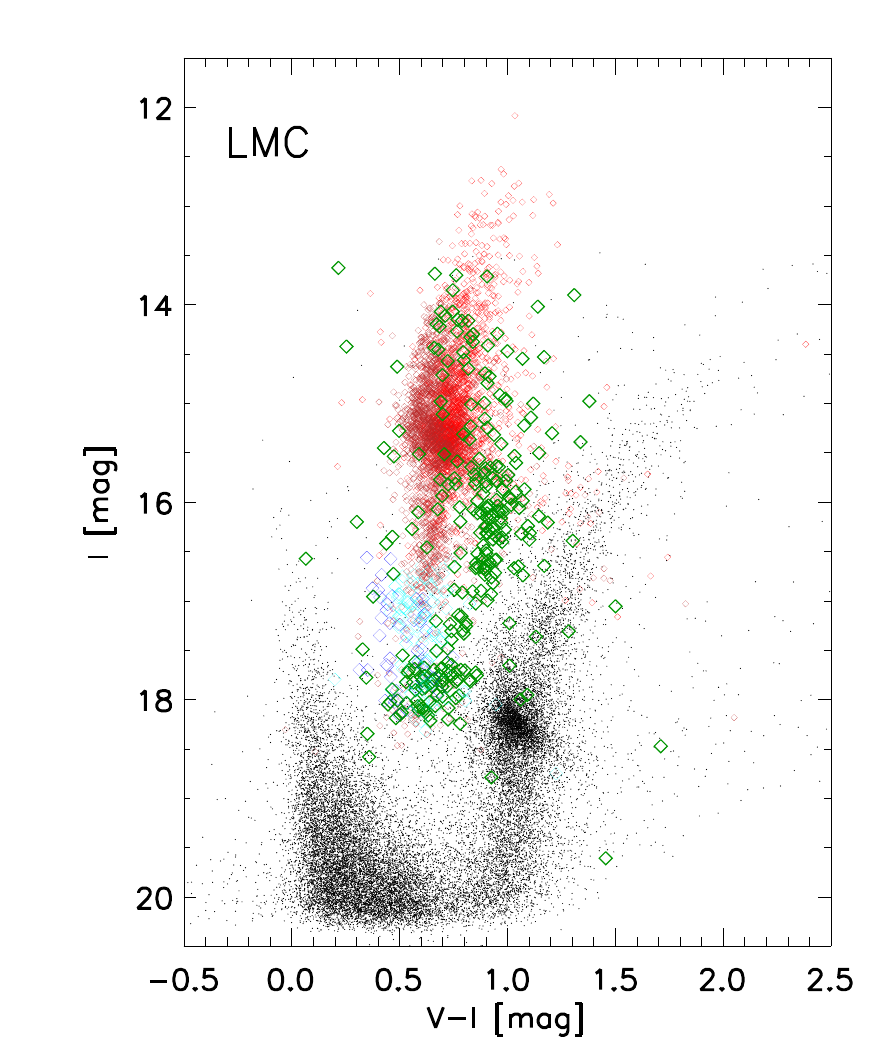}
\caption{Left: distribution of fundamental (light red) and first overtone 
(dark red) CCs, fundamental (light blue) and first overtone (dark blue) 
ACs and fundamental (green) TIICs  in the optical $I$,$V-I$ CMD of the 
SMC. Field SMC stars are plotted as grey dots. The optical CMD shows  
different triple regions, i.e. regions in which the three different groups 
of variables overlap. Data plotted in this panel come from the OGLE~IV data 
set. The mean magnitudes of variable stars were estimated with a fit of the 
light curves.
Right: Same as the left, but for LMC Cepheids.}
\label{fig_optical_cmd_MC}
\end{center}
\end{figure}

Optical, NIR and MIR photometric surveys of the MCs provide the unique opportunity to investigate the Cepheid distribution across the CMD. The main reason why the MCs play a crucial role in quantitative Astrophysics is mainly that we can, as a first approximation, assume that the MC stellar populations are all placed at the same distance, i.e., by neglecting the depth effect. Data plotted in Fig.~\ref{fig_optical_cmd_MC} display the distribution of SMC (left) and LMC (right) Cepheids in the optical $I$,$V-I$ CMD. Different groups of variable stars are plotted with different symbols, while variables pulsating in different modes are plotted with different colors. Common stars are plotted as grey dots. Data for both static and variables stars come from the OGLE~IV data set. Helium-burning variable stars are situated, as expected, between the main sequence and the red giant branch in the so-called Hertzsprung gap. 
%ref_2_9
They typically attain magnitudes either brighter (CCs, TIICs) or
similar (ACs) to red clump stars, the stellar over-density located in the SMC
at $I\sim$18.55--18.65 mag and $V-I \sim$0.90--0.95 mag and in the LMC at
$I\sim$18.25--18.35 mag and $V-I \sim$1.00--1.05 mag.
The mean magnitude of the variable stars was estimated as a mean in flux over the entire pulsation cycle, and then transformed into a mean magnitude. Variables placed outside the well-defined, almost vertical, variable sequences are either affected by differential reddening or by blending (crowded regions).  A glance at the data plotted in this figure clearly show that optical CMDs can be adopted to validate variable identification, but they are far from being an optimal diagnostic to identify different groups of variable stars, since the different Cepheid groups overlap both in magnitude and in color.  

%%%%%%%%%%%%%%%%%%%%%%%%%%%%%%%%%%%%%%%%%%%%%%%%%%%%%%%%%%%%%%%%%%%%%%%%%%%%%%%%%%%%%
% 			fig 12
%%%%%%%%%%%%%%%%%%%%%%%%%%%%%%%%%%%%%%%%%%%%%%%%%%%%%%%%%%%%%%%%%%%%%%%%%%%%%%%%%%%%%
%IDLWorkspace/Default/Temp/temp_220225b.pro
%_______________________________________________________________________
%_______________________________________________________________________________

%%%%%%%%%%%%%%%%%%%%%%%%%%%%%%%%%%%%%%%%%%%%%%%%%%%%%%%%%%%%%%%%%%%%%%%%%%%%%%%%%%%%%
% 			fig 13
%%%%%%%%%%%%%%%%%%%%%%%%%%%%%%%%%%%%%%%%%%%%%%%%%%%%%%%%%%%%%%%%%%%%%%%%%%%%%%%%%%%%%
%ref_1_23 .... improved blue loops in isochrones 

%_______________________________________________________________________________
\begin{figure}[htbp]
\begin{center}
\includegraphics[width=0.49\textwidth]{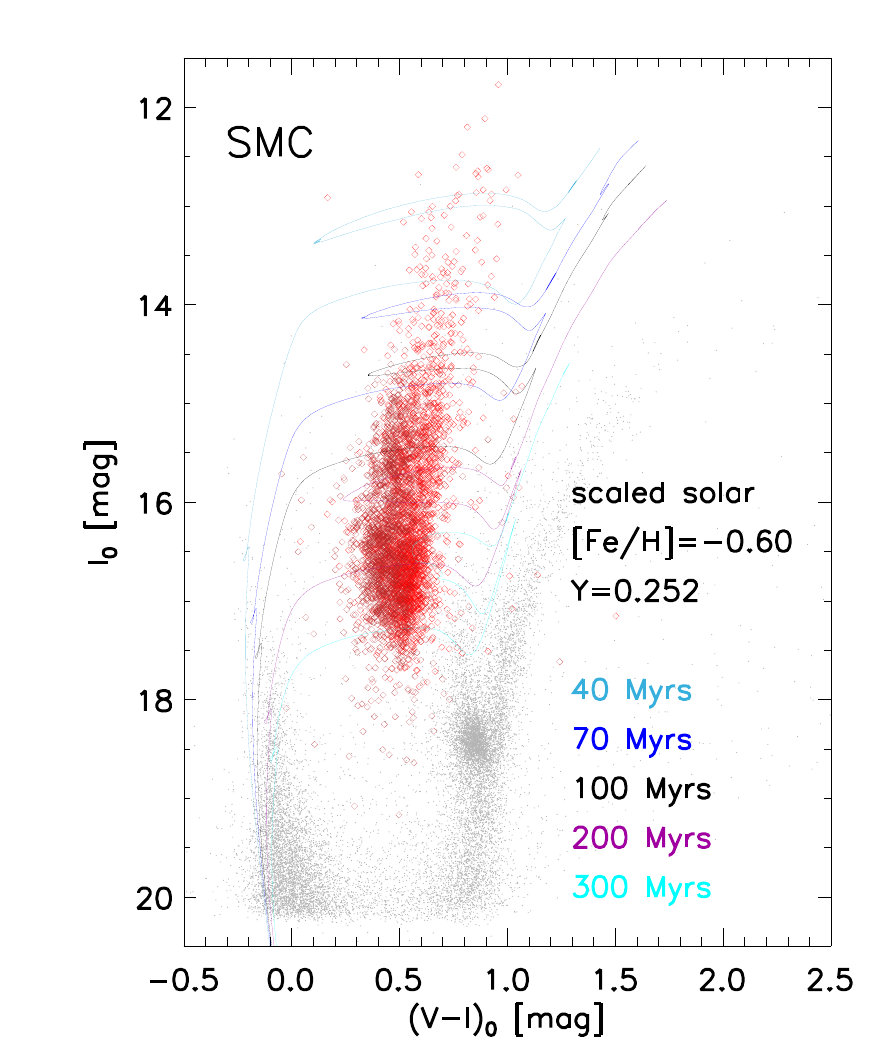}
\includegraphics[width=0.49\textwidth]{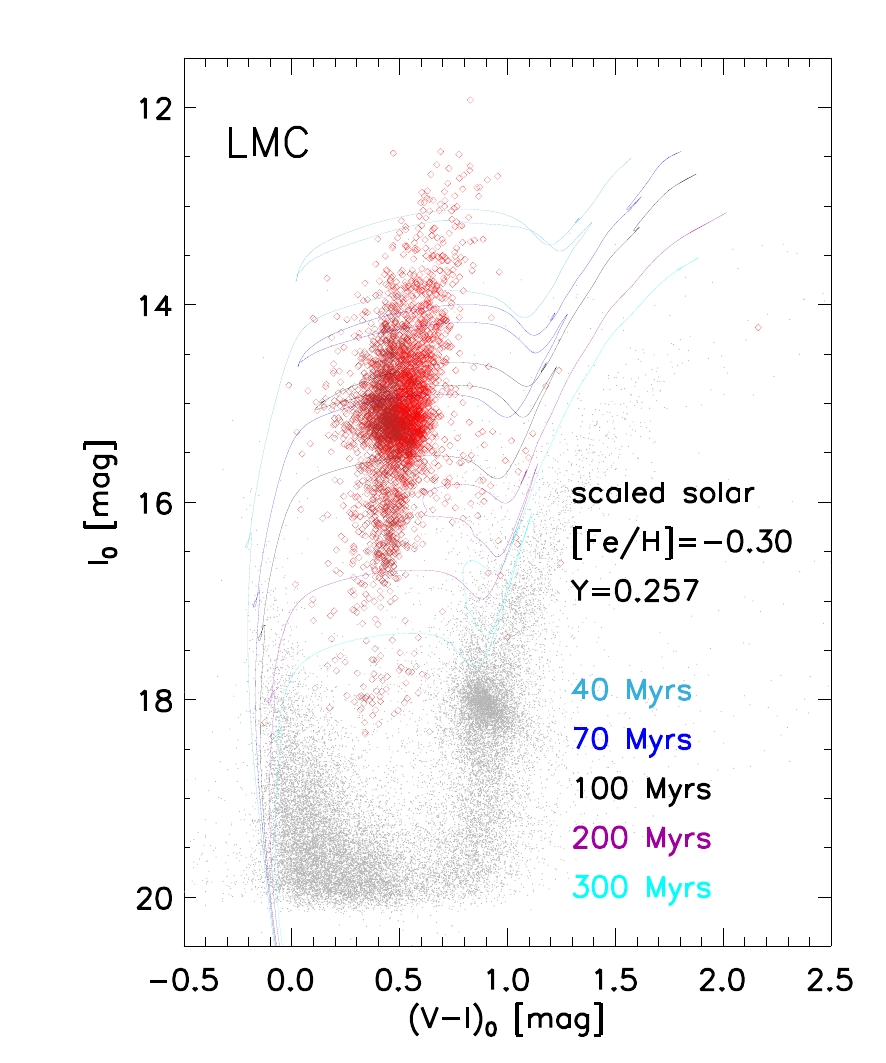}
\caption{Left: Comparison between theory and observations in the dereddened 
$I_0, (V-I)_0$ CMD for SMC CCs. CCs are marked with light (fundamental) and dark (first overtone) 
red symbols, while static stars with grey dots. Variable stars were dereddened 
by using the reddening map provided by \citet{skowron2021}, while static stars were 
dereddened through a mean reddening of E($V-I$)=0.047 mag \citep{skowron2021}.  
Solid lines display selected scaled-solar stellar isochrones from the BASTI-IAC 
database at a fixed chemical composition ([Fe/H]$=-0.60$, Y=0.252) and ages ranging 
from 30 to 300 Myrs (see labelled values). Isochrones were plotted by assuming a true 
distance modulus of $\mu=18.977 \pm 0.028$ mag \citep{graczyk2020}.
Right: Same as the left, but for LMC CCs and more metal-rich scaled-solar stellar 
isochrones ([Fe/H]$=-0.30$, Y=0.257). Isochrones were plotted by assuming a true 
distance modulus of $\mu=18.477 \pm 0.023$ mag \citep{pietrzynski2019}. The mean 
reddening adopted for static stars is E($V-I$)=0.100 mag \citep{skowron2021}.}
\label{fig_optical_isochrone_MC_CC}
\end{center}
\end{figure}
%_______________________________________________________________________________

In order to find out more about the global properties of Magellanic CCs, 
the right panel of Fig.~\ref{fig_optical_isochrone_MC_CC} shows the 
comparison between theory and 
observations in the optical dereddened $I_0,(V-I)_0$ CMD for LMC CCs. 
The symbols are the same as in Fig.~\ref{fig_optical_cmd_MC}, individual 
reddenings for CCs were estimated by using the reddening map provided by 
\citet{skowron2021}, while common stars were dereddened through a mean 
reddening of E($V-I$)=0.100 mag \citep{skowron2021}. 
The solid lines with different colors display stellar isochrones from 
the BASTI-IAC database at fixed chemical composition ([Fe/H]$=-0.30$, 
iron abundance; Y=0.252, helium mass fraction) and ages ranging from 
40 to 300 Myrs (see labelled values). Stellar isochrones are based on 
evolutionary models, constructed by assuming a scaled-solar chemical mixture 
and by neglecting convective core overshooting. They were plotted by 
assuming a true distance modulus of $\mu=18.477 \pm 0.023$ mag \citep{pietrzynski2019}.  

Detailed comparisons between theory (evolutionary and pulsation 
properties) concerning MC variable stars have been widely discussed 
in the literature \citep{cioni2014,soszynski2017b,ripepi2016}. 
Here we are mainly interested in providing a global picture of 
their properties. The agreement between theory and observations 
is quite good over the entire mass/age range. Indeed, the width 
in color of the blue loops (central helium burning phases) takes 
globally into account the observed distribution of CCs inside the 
instability strip. However, this pending issue requires more 
detailed investigations, since the extension in temperature of the 
blue loop depends on many different physical assumptions 
\citep{bono00a} and we still lack quantitative constraints on 
the impact that input physics (opacities, equation of state), 
mass loss \citep{maeder2000,marengo2010b,barmby2011}, rotation 
\citep{maeder2000,costa2019} and extra-mixing have on these 
evolutionary phases. The current theoretical framework also takes 
into consideration young MS stars (central hydrogen burning phases), 
and RGB stars (shell hydrogen burning phases). 

The left panel of Fig.~\ref{fig_optical_isochrone_MC_CC} shows the same comparison, but for SMC CCs. The solid lines display BASTI-IAC scaled-solar stellar isochrones at fixed chemical composition ([He/H]$=0.60$, Y=0.257) and different ages (see labelled values). Isochrones were plotted by assuming a true distance modulus of $\mu=18.977 \pm 0.028$ mag \citep{graczyk2020}. The mean reddening adopted for static stars is E($V-I$)=0.047 mag \citep{skowron2021}. The agreement between theory and observations is once again quite good over the entire mass/age range. 

The difference between stellar isochrones based on 
evolutionary models which either neglect or take into account 
convective core overshooting during core hydrogen-burning phases 
have been widely discussed in the literature. The main difference is 
that the isochrones taking into account overshooting are systematically 
brighter than canonical isochrones. This means that individual Cepheid ages estimated 
by using the former set are systematically younger than those based on the latter one. 
Moreover, the extent in temperature/color of the 
blue loops associated with canonical and overshooting isochrones is 
different: the former ones are on average larger. Once again, the morphology of the 
blue loops depends on a variety of micro and macro physics phenomena 
and we still lack firm theoretical predictions concerning who is 
doing what. Note that stellar isochrones were downloaded from the new 
BASTI-IAC database (\url{http://basti-iac.oa-abruzzo.inaf.it}).

However, in some specific cases, like the comparison between theory 
and observations for the MC CCs, the resolution in mass of the BASTI-IAC
evolutionary tracks was improved, and in turn, we also re-computed 
stellar isochrones.   

%%%%%%%%%%%%%%%%%%%%%%%%%%%%%%%%%%%%%%%%%%%%%%%%%%%%%%%%%%%%%%%%%%%%%%%%%%%%%%%%%%%%%
% 			fig 14
%%%%%%%%%%%%%%%%%%%%%%%%%%%%%%%%%%%%%%%%%%%%%%%%%%%%%%%%%%%%%%%%%%%%%%%%%%%%%%%%%%%%%
%_______________________________________________________________________
\subsection{Anomalous Cepheids}
\begin{figure}[htbp]
\begin{center}
\includegraphics[width=0.49\textwidth]{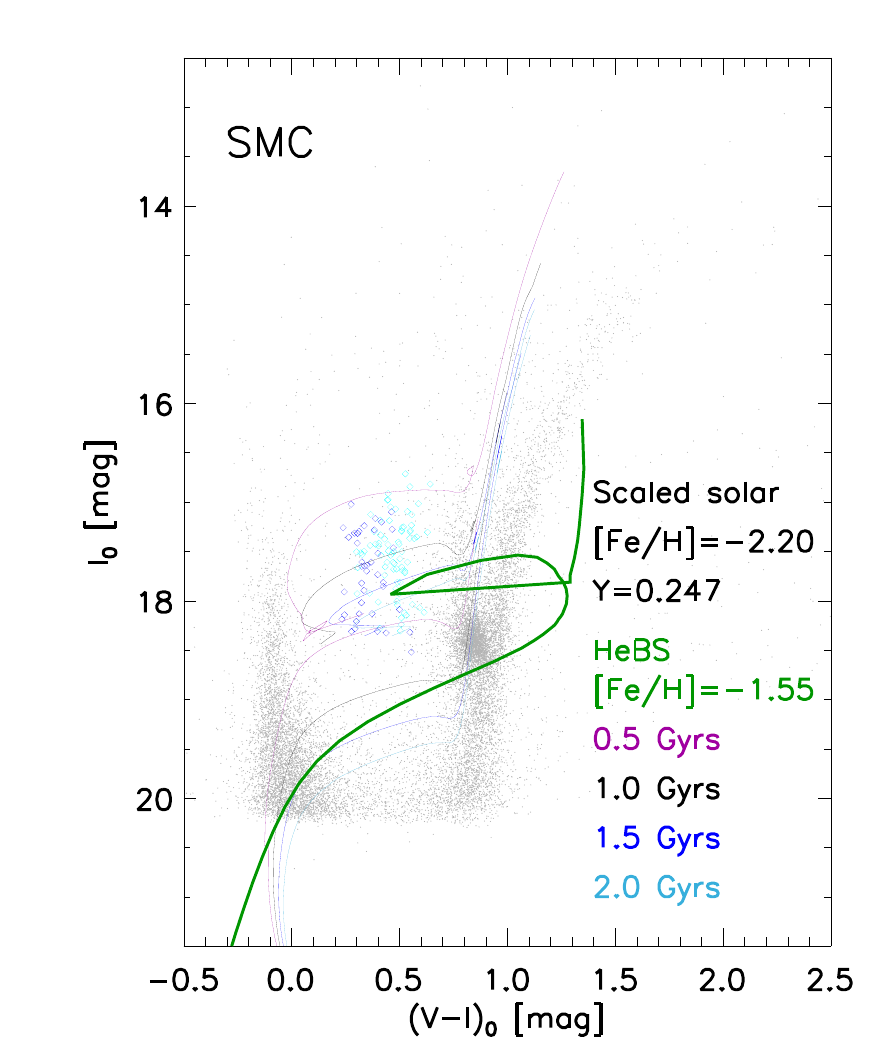}
\includegraphics[width=0.49\textwidth]{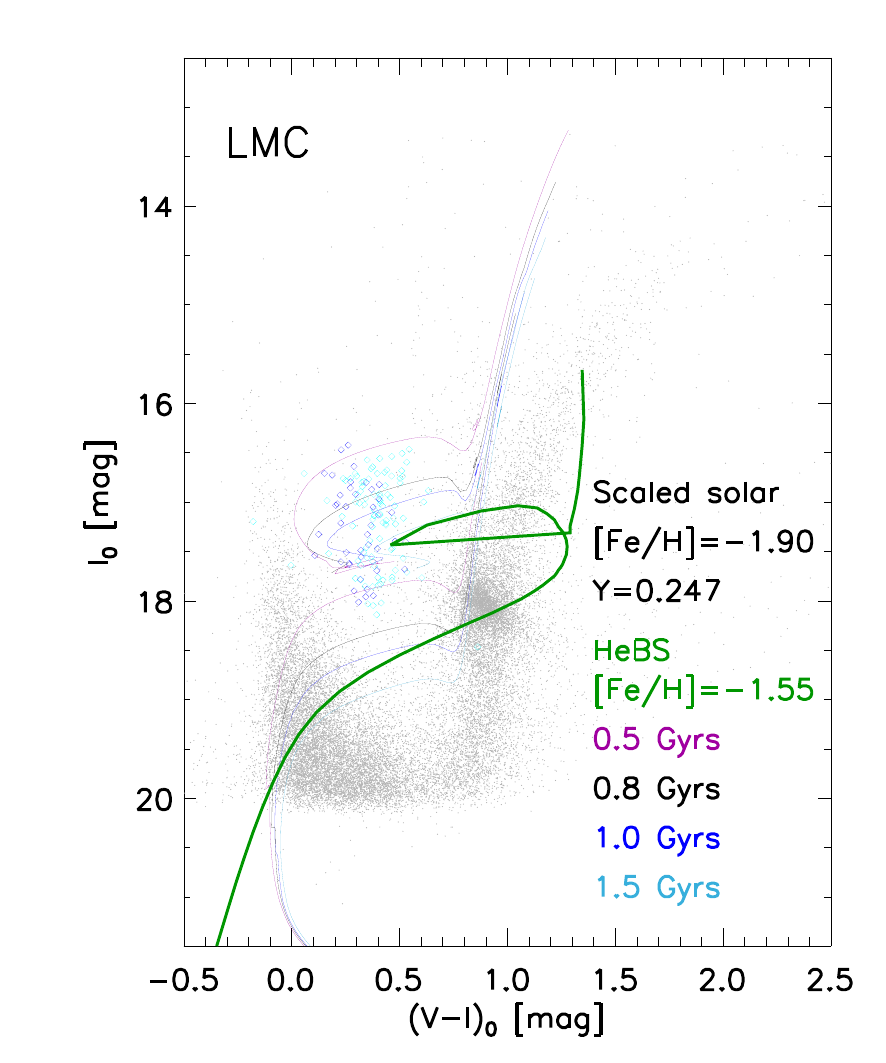}
\caption{Left: 
Comparison between theory and observations in the dereddened
$I_0, (V-I)_0$ CMD for SMC ACs. ACs are marked with light 
(fundamental) and dark (first overtone) blue symbols, while static stars with grey dots. 
Color-coded lines show stellar isochrones from the BASTI-IAC database 
at fixed chemical composition ([Fe/H]$=-2.20$, Y=0.247) and stellar ages ranging from 
0.5 Gyr to 2 Gyr (see labelled values). The green line shows the helium-burning sequence
(HeBS) for a more metal-rich ([Fe/H]$=-1.55$, Y=0.248) chemical composition.  
Right: Same as the left, but for LMC ACs and more metal-rich stellar isochrones 
([Fe/H]$=-1.90$, Y=0.247).}
\label{fig_optical_isochrone_MC_AC}
\end{center}
\end{figure}
%_______________________________________________________________________

The right panel of Fig.~\ref{fig_optical_isochrone_MC_AC} shows the  
comparison between theory and observations for LMC ACs. The LMC ACs and 
the static stars were dereddened following the same approach adopted 
for LMC CCs. The solid lines display stellar isochrones from the 
BASTI-IAC database at fixed chemical composition ([Fe/H]$=-1.90$, 
Y=0.247) and ages ranging from 0.8 to 2 Gyr (see labelled values). 
Stellar isochrones are based on evolutionary models constructed 
assuming a scaled-solar chemical mixture and by neglecting 
convective core overshooting and they were plotted by assuming 
a true distance modulus of $\mu=18.477 \pm 0.023$ mag 
\citep{pietrzynski2019}. A glance at the data plotted in this panel 
brings forward two interesting features worth looking into. 

{\em a) Metallicity distribution --}
The faint tail (m$_I\lesssim 18.7$ mag) of LMC ACs is quite metal-poor. 
The current comparison suggests a mean iron abundance of the order of 
[Fe/H]=--1.9/--2.0, while the bright tail is less metal-rich than 
[Fe/H]$=-1.55$, as suggested by the more metal-rich helium-burning 
sequence (green line). The upper limit is quite well known and fixed 
by the evidence that helium-burning loci for more metal-rich chemical 
compositions attain effective temperatures (colors) that are systematically 
hotter (bluer) than the Cepheid instability strip \citep{monelli2022}. 

{\em b) Luminosity function --} 
There is evidence that both the I- and the V-band luminosity function of LMC ACs shows a well-defined minimum, or a zone of avoidance for m$_I\approx 18.75$ mag. It is not clear whether this feature might be associated with a difference in their origin \citep[binary merging versus single star evolution,][]{fiorentino12c}. The current evidence indicates that stellar systems mainly dominated by old stellar populations (Tucana, Cetus, Sculptor, LGS3) host ACs which are on average 1.0--1.5 mag brighter than RRLs \citep{monelli2022}. Moreover, the bright tail of the ACs shows up in stellar systems showing multiple star formation episodes and a well sampled intermediate-age stellar population (MCs). This working hypothesis requires more quantitative constraints on the possible difference between faint and bright ACs. 

The left panel of Fig.~\ref{fig_optical_isochrone_MC_AC} shows the same comparison as the right panel, but for SMC ACs. The stellar isochrones were computed at fixed chemical composition ([Fe/H]$=-2.20$, Y=0.247) and the stellar ages are labelled. The global properties of the SMC ACs appear to be quite similar to LMC ACs. They only cover a narrower range in magnitudes and FO ACs also cover a narrower range in period. The comparison between the LMC and the SMC luminosity function for ACs indicates that the bright tail appears less extended. This suggests a minor contribution in the younger (more massive) range.

%_______________________________________________________________________
\subsection{Type II Cepheids}\label{cmd_TIIC}

%%%%%%%%%%%%%%%%%%%%%%%%%%%%%%%%%%%%%%%%%%%%%%%%%%%%%%%%%%%%%%%%%%%%%%%%%%%%%%%%%%%%%
% 			fig 15
%%%%%%%%%%%%%%%%%%%%%%%%%%%%%%%%%%%%%%%%%%%%%%%%%%%%%%%%%%%%%%%%%%%%%%%%%%%%%%%%%%%%%
\begin{figure}[htbp]
\begin{center}
\includegraphics[width=0.49\textwidth]{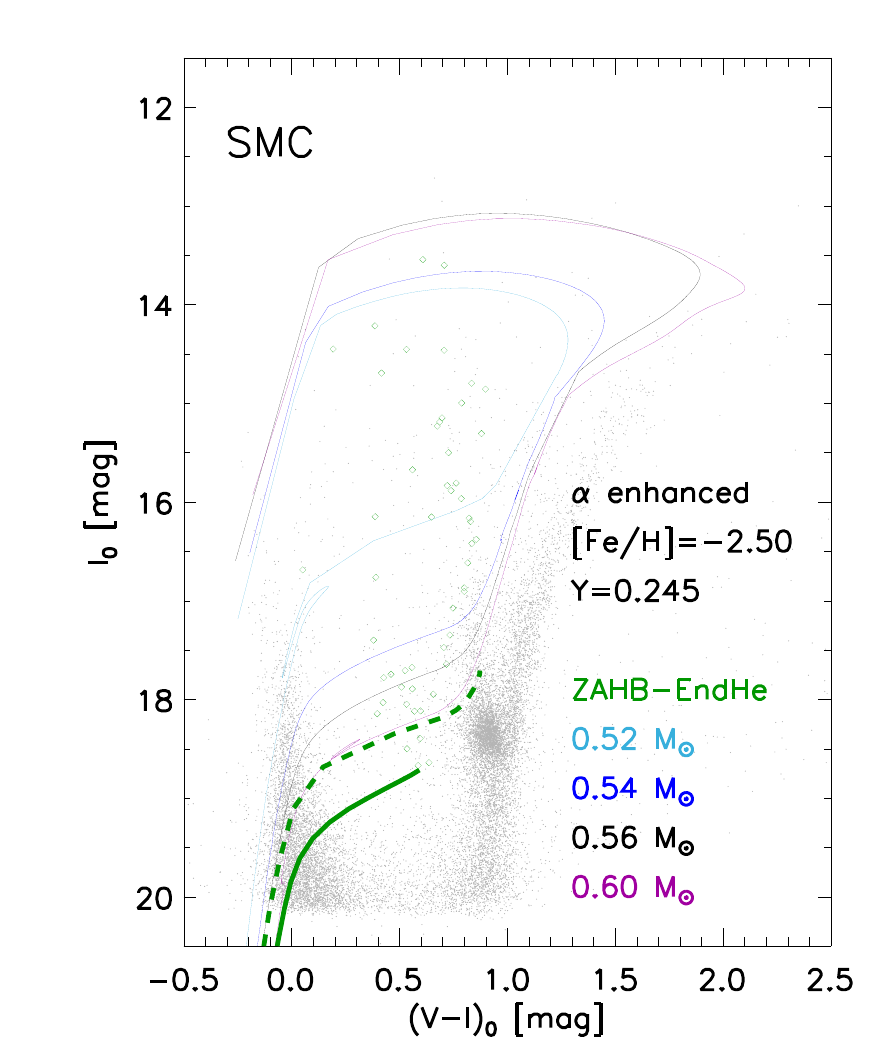}
\includegraphics[width=0.49\textwidth]{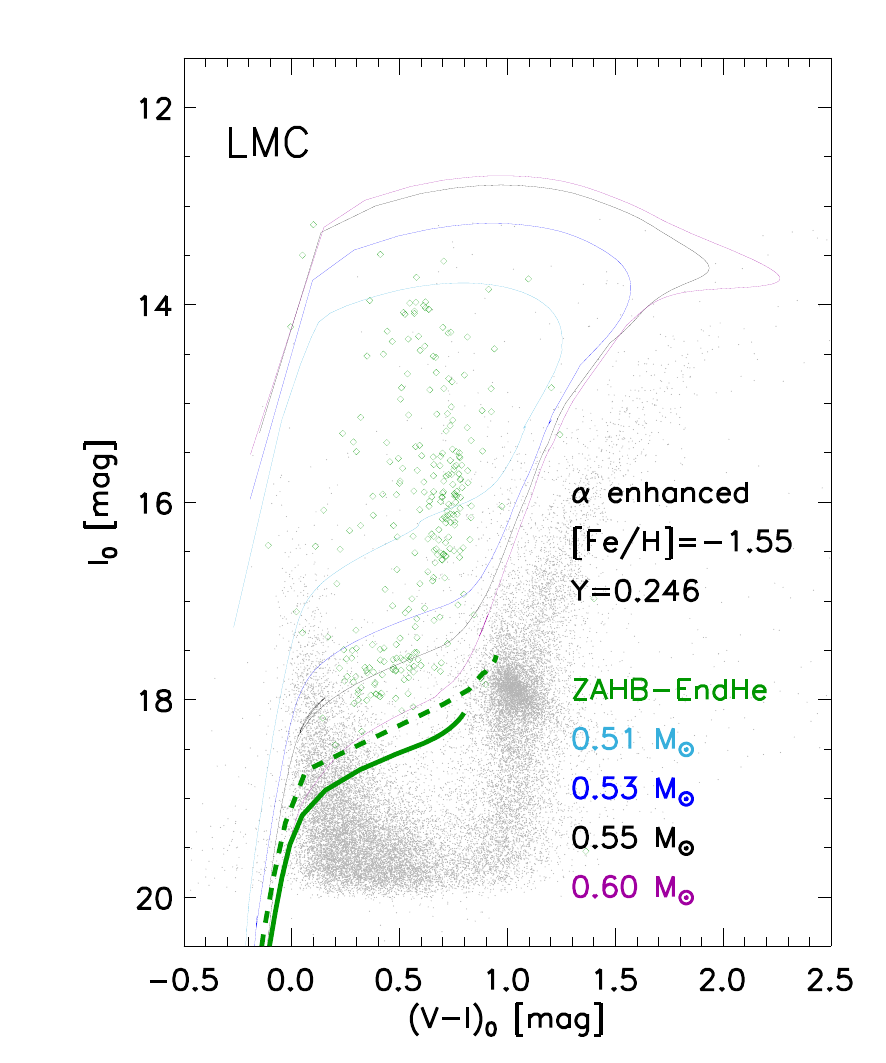}
\caption{Left: 
Comparison between theory and observations in the dereddened
$I_0, (V-I)_0$ CMD for SMC TIICs. TIICs 
%ref_2_12b
are marked with light (fundamental) green symbols, 
while static stars with grey dots. Solid lines display selected 
HB evolutionary models from the BASTI-IAC database with a $\alpha$-enhanced 
chemical composition ([Fe/H]$=-2.50$, Y=0.245) and stellar masses ranging 
from $0.51\,M/M_\odot$ to $0.60\,M/M_\odot$ (see labelled values). The thick 
green lines show the Zero Age Horizontal Branch (ZAHB) and the dashed line 
the end of core helium burning.  
Right: Same as the left, but for LMC TIICs and more metal-rich 
HB evolutionary models ([Fe/H]$=-1.55$, Y=0.246).}
\label{fig_optical_isochrone_MC_TIIC}
\end{center}
\end{figure}
%_______________________________________________________________________________

The right panel of Fig.~\ref{fig_optical_isochrone_MC_TIIC}
shows the comparison between theory and observations for the LMC TIICs following the same approach adopted for CCs and ACs. However, the solid lines display HB evolutionary models computed with an $\alpha$-enhanced chemical mixture, at fixed chemical composition  ([Fe/H]$=-1.55$, Y=0.246) and stellar masses ranging from $0.51\,M/M_\odot$ to $0.60\,M/M_\odot$ (see labelled values). 
Predictions plotted in this panel display that the bulk of TIICs are AGB stars (hydrogen and helium  shell burning). The reader interested in a more detailed discussion concerning the three different subgroups (BL~Herculis, W~Virginis, RV~Tauri) is referred to \citet{bono2020b} and to \citet{braga2020}. The solid and the dashed green lines plotted in the same panel show the ZAHB and the core helium exhaustion. They enclose the region of the CMD, in which the bulk of the RRLs are expected to be located. Clearly, this is a mere simplification, because LMC RRLs cover a broad range in metallicity \citep{gratton04a,skowron2016} and a significant overlap in the CMD is expected between these two different groups of variable stars. 

The left panel of Fig.~\ref{fig_optical_isochrone_MC_TIIC} shows the same comparison as the right panel, but for SMC TIICs. The solid lines display similar HB evolutionary models, but computed by assuming  a more metal-poor chemical composition ([Fe/H]$=-2.50$, Y=0.245). The global agreement is once again quite good over the entire magnitude and color range.

%%%FIRST
%_______________________________________________________________________
\section{Diagnostics to estimate individual Cepheid distances}\label{sec:dist_diagn}

% editor_6
More than one century ago, Henrietta Leavitt discovered that Cepheids obey to 
well-defined Period Luminosity (PL) relations. The key advantage of these 
relations is that they are linear over the entire period range. This outcome 
applies to optical, NIR and MIR regime.\footnote{RRLs obey to PL relations 
only for wavelengths longer than the $R$ bands \citep{bono01,braga15} in the 
visual band they obey to a mean absolute visual-magnitude metallicity relation.} 
There are a number of pros and cons in using similar diagnostics in different 
wavelength regimes.
%BG 
%\footnote{\em \ldots I have bored you so far with a good 
%number of details; but I trust
%it is not necessary to offer any excuse for this, since a 
%correct opinion as to whether the investigation of the parallax 
%of 61 Cygni has already led to an approximate result, or must
%still be carried further before this can be affirmed of them,
%can only be formed from the knowledge of those particulars.
%Had I merely communicated to you the result, I could not have
%expected that you would attribute to it that certainty which,
%according to my own judgment, it does have. \ldots \citep{bessel1838}.} 
% ref_1_25
The reader interested in a detailed empirical 
discussion concerning the improvements in moving from the optical to the 
NIR regime is referred to the seminal papers by 
\citet{mcgonegal82,mcgonegal83} and \citet{mcalary1983}.
In the following, we will outline the most relevant cons. 

%%%%%%%%%%%%%%%%%%%%%%%%%%%%%%%%%%%%%%%%%%%%%%%%%%%%%%%%%%%%%%%%%%%%%%%%%%%%%%%%%%%%%
% 			fig 16
%%%%%%%%%%%%%%%%%%%%%%%%%%%%%%%%%%%%%%%%%%%%%%%%%%%%%%%%%%%%%%%%%%%%%%%%%%%%%%%%%%%%%
\begin{figure}[htbp]
\begin{center}
\includegraphics[width=0.66\textwidth]{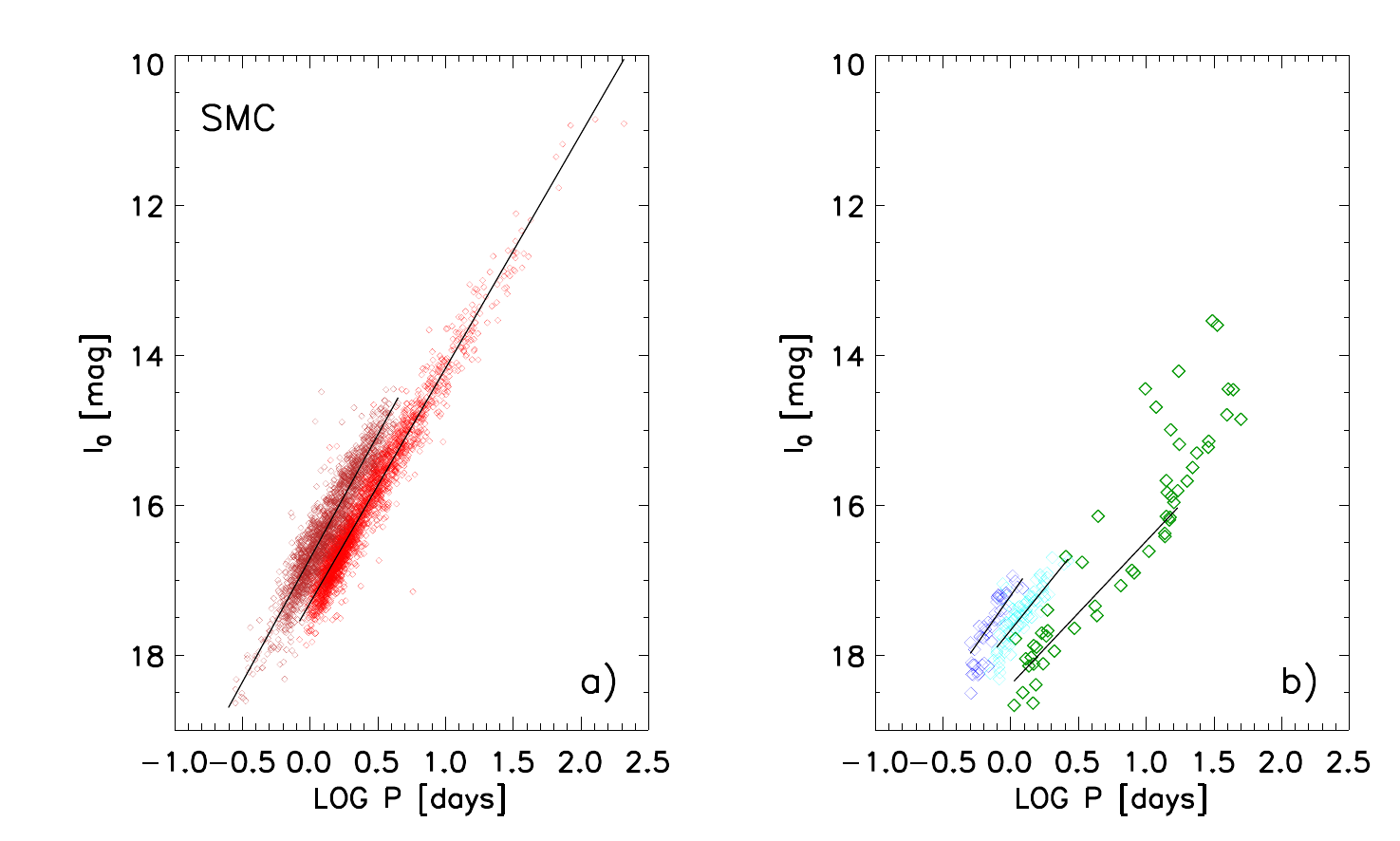}
\includegraphics[width=0.66\textwidth]{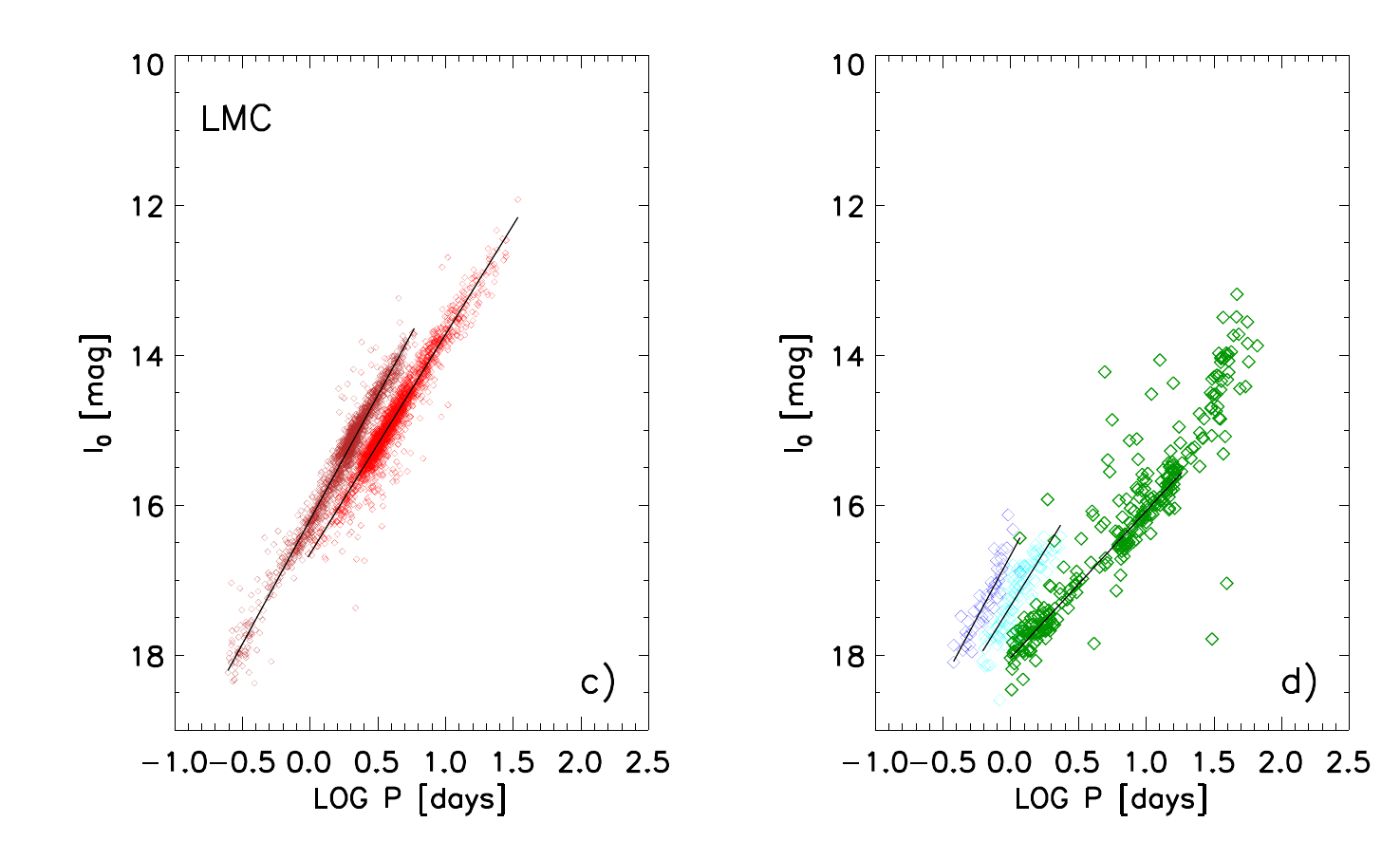}
\caption{Panel a): dereddened $I$-band PL relations for SMC CCs. From 
shorter to longer periods, the different symbols display FO (dark red) 
and fundamental (light red) CCs. The solid black lines display the linear 
fits to the PL relations (see 
% Table~\ref{tbl:cepheid_PL}). 
Table~\ref{tbl:cepheid_PL}). 
Panel b): Same as the left, but for FO (dark blue) and FU (light blue) 
ACs and FU (green) TIICs. The black lines display the linear fits to the 
PL relations  (see 
% Table~\ref{tbl:cepheid_PL}).
Table~\ref{tbl:cepheid_PL}).
%ref_2_12c
Panels c) and d): Same as the top, but for LMC Cepheids. Note the increase 
in sample size of TIICs.}
\label{fig_PLI_MC}
\end{center}
\end{figure}
%_______________________________________________________________________________

{\em i)}--Luminosity amplitude-- The identification and characterization of 
regular variables is significantly easier in the optical regime, because 
the luminosity amplitude in the $B$-band is typically a factor of three-to-five
larger than in the NIR and MIR bands. This means that period determination 
and mode identification are more straightforward. 

{\em ii)}--Time series-- Optical measurements, thanks to linearity, 
pixel scale and the size of current CCDs are less demanding 
about telescope time than NIR measurements. 

{\em iii)}--Limiting magnitude-- Current ground- and space-based observing 
facilities allow us to identify and characterize CCs and TIICs in the optical 
regime not only in Local Group (d$\lesssim$ 1~Mpc), but also in Local Volume 
(d$\sim$ 25~Mpc) galaxies \citep{freedman10,riess2021}. The limiting magnitudes 
in the NIR regime are systematically brighter, but JWST and ELTs are going 
to open new paths \citep{fiorentino2020}. There are also some 
indisputable pros.

%%%%%%%%%%%%%%%%%%%%%%%%%%%%%%%%%%%%%%%%%%%%%%%%%%%%%%%%%%%%%%%%%%%%%%%%%%%%%%%%%%%%%
% 			fig 17
%%%%%%%%%%%%%%%%%%%%%%%%%%%%%%%%%%%%%%%%%%%%%%%%%%%%%%%%%%%%%%%%%%%%%%%%%%%%%%%%%%%%%
\begin{figure}[htbp]
\begin{center}
\includegraphics[width=0.66\textwidth]{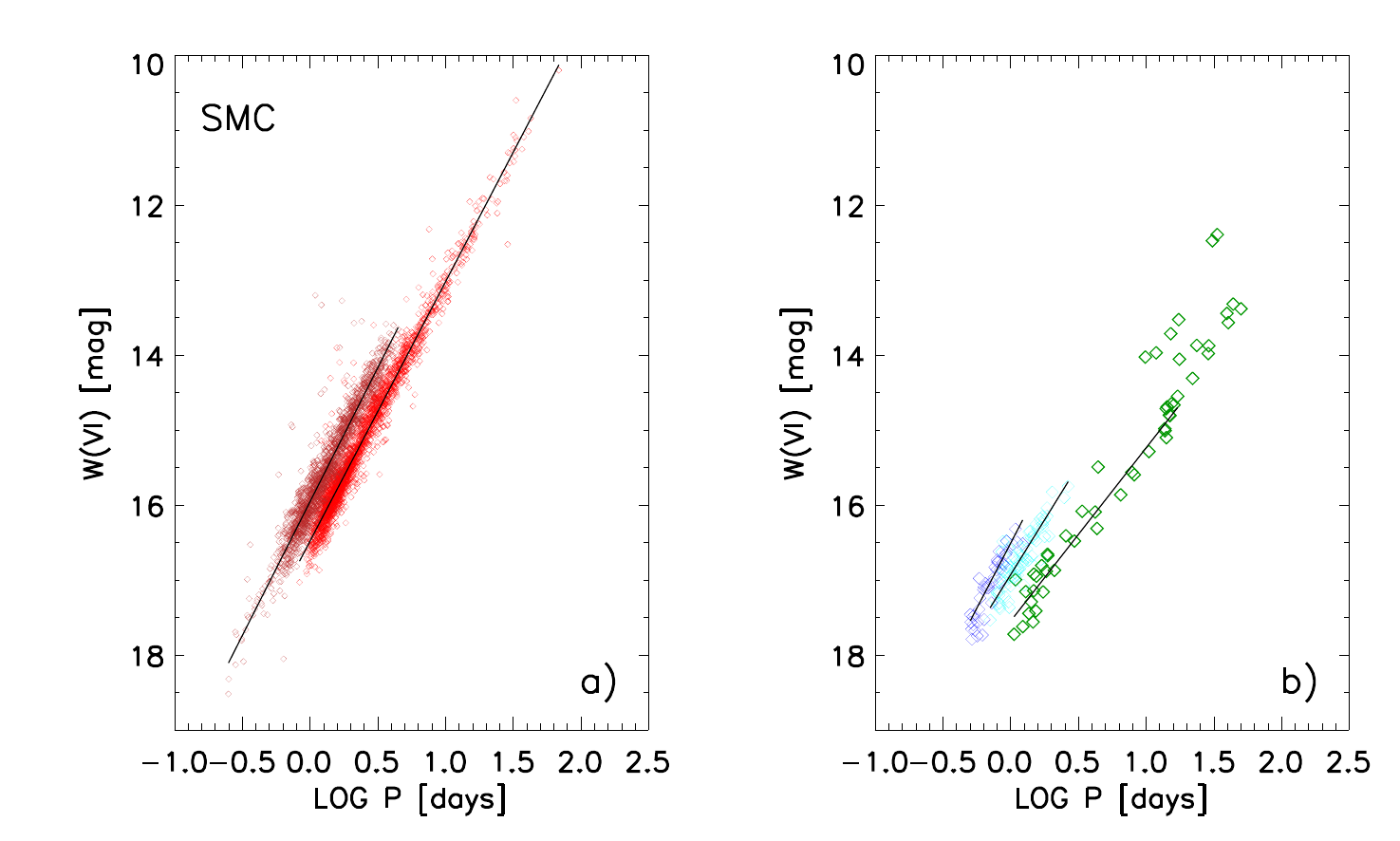}
\includegraphics[width=0.66\textwidth]{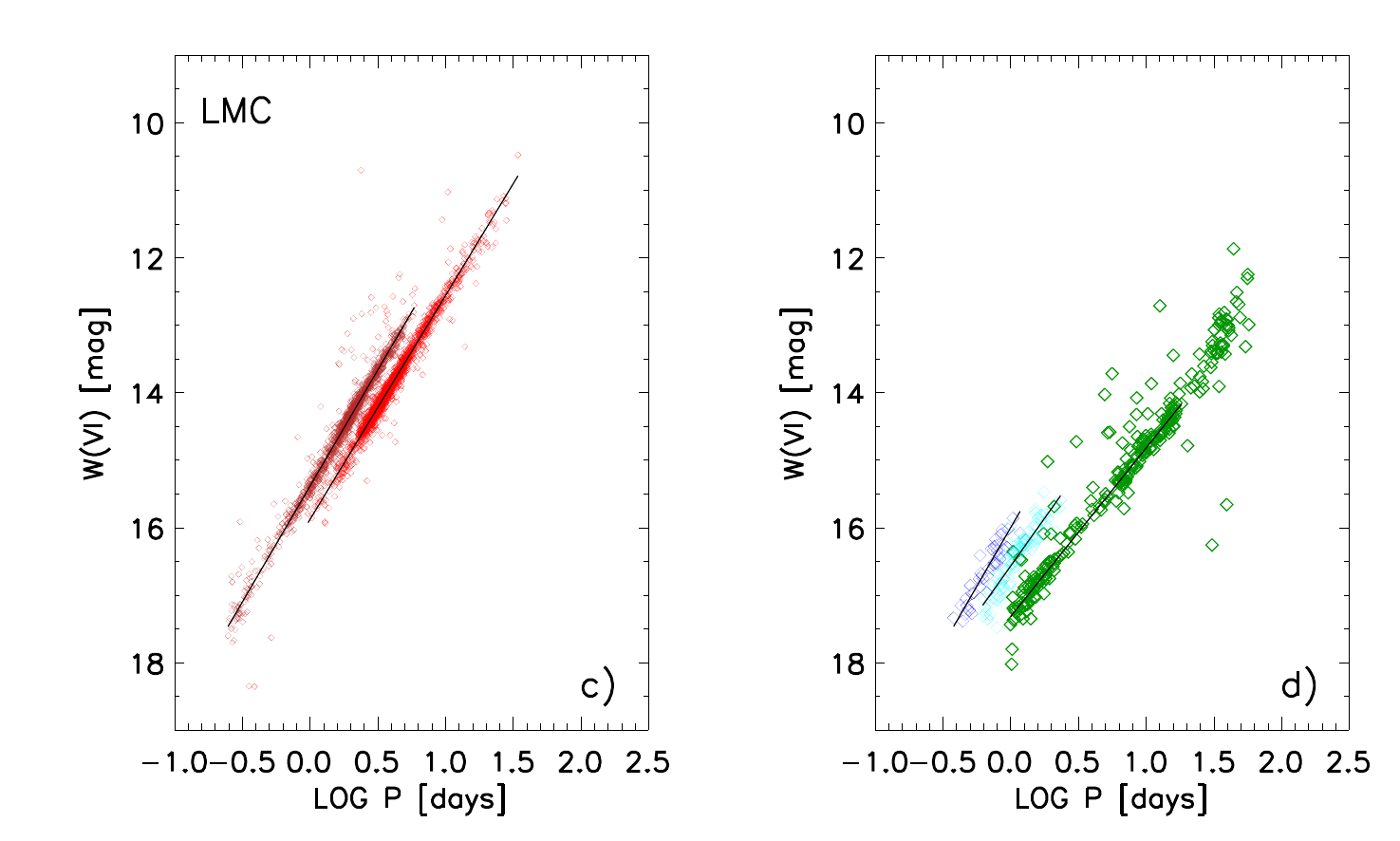}
\caption{Top: Optical $I$,$V-I$ PW relations (see 
%Table~\ref{tbl:cepheid_PW}) 
Table~\ref{tbl:cepheid_PW}) 
for SMC CCs (panel a) and for ACs plus TIICs (panel b). 
Symbols and color coding are the same as in Fig.~\ref{fig_PLI_MC}.
Bottom: Same as the top, but for LMC Cepheids 
(CCs, panel c; ACs plus TIICs, panel d).}
\label{fig_PWIVI_MC}
\end{center}
\end{figure}
%_______________________________________________________________________________

{\em i)}--Reddening correction-- Optical bands are more prone to systematics
concerning the reddening correction than NIR bands. The ratio between 
%editor 18
selective absorption 
in the $K$ and in the $V$ band is, according to 
current reddening law \citep{cardelli89}, of the order of 0.12. This means 
that measurements in the $K$-band are roughly one order of magnitude less 
affected by reddening uncertainties than the ones in the $V$-band. 

{\em ii)}--Shape of the light curves-- The shape of the light curves in the 
optical regime is affected by both temperature and radius variation. This 
means that light curves in the optical regime might display, according to 
the pulsation period and the luminosity amplitude, sharp rising branches 
and cuspy maxima/minima. The shape of the light curves in the NIR and in the 
MIR regime are mainly dominated by radius variations and they are typically 
more sinusoidal. This is the main reason why the light curve templates provide 
very accurate estimates of the mean NIR magnitudes even with a single 
measurement, once the period, the luminosity amplitude and the reference 
epoch are known \citep{inno15}. In case three independent 
measurements are available, the mean magnitude can be estimated once 
the period and the luminosity amplitude are known \citep{inno15,braga2019}.

{\em iii)}--Slope of the PL relation-- The slope of the PL relations 
becomes systematically steeper when moving from the optical to the 
NIR regime. This means that NIR PL relations are, at fixed pulsation 
period, more accurate than optical ones. In this context, it is worth 
mentioning that the luminosity amplitudes becomes almost constant at 
wavelengths longer than the $K$-band, and the same outcome applies to 
the slope of the PL relations. The approach to a constant value is due 
to the fact that luminosity variations are mainly dominated by radius 
variations. 

{\em iv)}--Width in temperature of the Instability Strip-- Radial 
variables obey to PLC relation. The use 
of the PL relation for the estimate of individual distances rely 
on the assumption that the width in temperature of the instability 
strip can be neglected. This assumption is quite severe in the 
optical regime, because optical bands are more affected by a variation 
in intrinsic parameters and in metal content when compared with 
NIR bands. The physical reason for this variation is mainly in the 
bolometric correction (BC). The BC in the optical regime is almost 
constant when moving from the blue to the red edge of the instability 
strip, while in the NIR regime it increases. This means that regular 
variables in the NIR become intrinsically brighter when moving from 
hotter (shorter periods) to cooler (longer periods) variables. The 
consequence of this difference is that the cosmic variance 
(standard deviation) of the $V$-band CC PL relation is, on average, 
a factor of two to three larger than in the $K$-band \citep{mcgonegal82,inno2016}. 
This is the reason why individual 
distances based on $V$-band PL relations need to be cautiously treated, 
since they are more affected by possible systematics affecting the 
completeness of the sample over the entire period range 
covered by the PL relation \citep{bono99b}.

%_______________________________________________________________________
\subsection{Fundamental versus overtone PL relations}

The use of overtone variables brings forward several 
key advantages when compared with fundamental variables. 

{\em i)}--Width in temperature of the Instability Strip-- Overtones 
approach a stable limit cycle in a region of the instability strip 
that is at least a factor of two narrower than fundamental variables. 
This means that the PL relations provide individual distances that are 
more accurate than distances based on fundamental PL relations, because
the dependence on the mean temperature is milder \citep{bono99b}. 

{\em ii)}--Comparison between theory and observations-- Overtone variables 
are systematically hotter than fundamental variables. This means that 
predicted pulsation properties are less prone to uncertainties affecting 
the treatment of convective transport.

However, overtones are also affected by two cons. 

{\em i)}--Identification-- Overtone variables are characterized by more 
sinusoidal light curves due to the presence of a nodal line 
\citep{Bonoetal2000a}. This means that they have on average 
smaller amplitudes, and in turn, they need very accurate time series 
to be identified. Moreover, the shape of their light curves can be 
misidentified with eclipsing binaries.

{\em ii)}--Absolute calibration-- Till a few years ago, the number of 
CCs pulsating in the first overtone known in the MW was quite 
limited. An accurate geometrical distance was only available for Polaris 
\citep{evans2002}. Fortunately, the current status concerning first 
overtone Cepheids rapidly improved \citep{ripepi2019,Breuvaletal2020}. 
It is not clear yet whether the same outcome applies to Galactic 
single-mode second overtones.

%_____________________________________________________________________________
\subsection{PL versus PW relations}\label{sec:pl_vs_pw}

%_____________________________________________________________________________

Almost half a century ago, \citet{vandenbergh75} and 
\citet{madore82} introduced the so-called 
Wesenheit\footnote{Wesenheit is a German word used by 
philosophers and its meaning is ``essence''.} magnitude, 
a pseudo magnitude that is reddening free by construction. 
Given two generic photometric bands ($\xi$, $\chi$) it can 
be defined as: 

W($\chi$,$\xi$-$\chi$)= $\chi$ - [A$_\chi$ / A$_\xi$- A$_\chi$] $\times$($\xi$-$\chi$)

\noindent 
where the coefficient in square parentheses only depends on the adopted 
reddening law. Therefore, if we assume that the reddening law is 
universal, we can easily define a pseudo magnitude that is independent 
of reddening. The use of the Wesenheit magnitudes has several 
indisputable advantages. 

{\em i)}--Individual reddenings-- CCs are young stars which trace 
the spiral arms of the thin disk \citep{medina2021}. Moreover, they are 
quite often still embedded in the relics of the giant molecular cloud 
from which they originated \citep{genovali2014}. This means that they 
are either reddened or highly reddened. This limitation applies not 
only to Galactic Cepheids, but also to CCs in nearby galaxies. 
The use of the Wesenheit magnitudes overcome the difficulty of 
individual reddening estimates/measurements.

Data plotted in  Fig.~\ref{fig_PLI_MC} and in Fig.~\ref{fig_PLK_MC} 
(Appendix) display the 
dereddened $I$- and  $K$-band PL relation. To overcome the overlap 
among different groups of variable stars the top left panel shows the $I$-band 
PL relations for FO (dark red) and FU (light red) SMC CCs, while the 
top right panel shows the PL relations for SMC ACs (FO, dark blue; 
FU, light blue) and SMC TIICs (FU, green). The bottom panels display the same data, 
but for LMC Cepheids. The coefficients of the PL relations plotted in this 
%figure are listed in Table~\ref{tbl:cepheid_PL}. A few highlights concerning 
figure are listed in Table~\ref{tbl:cepheid_PL} (Appendix). 
A few highlights concerning observed optical-NIR PL relations. 

{\em a)}-- Difference in standard deviation -- Optical and NIR PL relations 
for LMC Cepheids have, at fixed pulsation period, systematically smaller 
standard deviations. The difference is due to the fact that the LMC is 
almost face-on, while the SMC is elongated along the line of sight 
%ref_2_10
\citep[][and references therein]{inno15, JacyszynDobrzeniecka2016}.

{\em b)}-- AC number counts -- The LMC is roughly one order of magnitude 
more massive than the SMC 
($\sim$2$\times10^{11}$ \citealt{shipp2021} vs 
$\sim$3$\times10^{10}$ M$_\odot$, \citealt{besla2016}). 
However, the number of ACs differ at the 20\% level  (146 vs 122). 
The modest difference in the AC number counts is mainly due to 
evolutionary effects: the SMC is, on average, more metal poor, 
and in turn, the probability to produce ACs increases. 

{\em c)}-- TIICs number counts -- The  number of SMC TIICs is 
almost six times smaller than LMC TIICs. It is 
worth mentioning that dwarf galaxies, typically, do not host TIICs. 
The Sagittarius dSph \citep{soszynski2011} and the MCs are exceptions, 
and their occurrence seems the consequence of the total mass and/or 
of their dynamical evolution \citep{bono2020b,braga2020,neeley2021}.
The difference between the number counts of SMC and LMC TIICs seems 
to be the consequence of the difference in total mass, and in particular, 
in the mass fraction of old stellar populations. The SMC hosts a single 
GC (NGC~121), while the LMC hosts more than a dozen 
GC \citep{vanderbergh2006}.

{\em ii)}--Color information-- The Wesenheit magnitude takes into account  
the color information. Thus the PW relations are also pseudo 
PLC relations \citep{bono99a}. The main difference 
is that the color coefficient is fixed by the adopted reddening law and 
not as a least squares solution of a three-dimensional relation among 
periods, magnitudes and colors.  The reason why the intrinsic dispersion 
of the PW relations is smaller than the standard deviations of the 
associated PL relations is also due to the use of color information. 
Periods, magnitude and colors provide a more accurate location  of 
individual Cepheids inside the instability strip. 
Figures~\ref{fig_PWIVI_MC} and ~\ref{fig_PWKVK_MC} 
show the key advantage in using optical, NIR and optical--NIR  
Wesenheit magnitudes for estimating individual Cepheid distances. 
The standard deviation decreases, on average, by a factor of two 
to 20/30\% in the NIR and in the optical-NIR regime.

{\em iii)}--Triple band-- Wesenheit magnitudes can be defined on the 
basis of two magnitudes, but this means that magnitude and color are 
correlated. The use of a Wesenheit pseudo magnitude based on three 
different bands overcome this limitation. Moreover, the use of 
three bands brings forward a key advantage. Optical/NIR/MIR magnitude 
and colors have opposite trends concerning the metallicity dependence. 
Therefore, appropriate combinations of magnitude and colors might 
be, within the errors, either independent of, or minimally affected by 
the metal content. There are two key drawbacks in using the Wesenheit 
pseudo magnitudes.

{\em a)}--Universality of the reddening law-- The current empirical 
and theoretical evidence indicates that the reddening law changes 
in stellar systems that experienced different chemical enrichment 
histories \citep{calzetti2021}. However, the use of either 
optical-NIR or optical-MIR magnitudes has the key advantage that 
the coefficients of the color term are tracing the derivative 
of the reddening law rather than ``local'' value \citep{marconi15,braga15,bono2019}. 

{\em b)}--Accuracy of mean colors-- The observing strategy of 
long-term photometric surveys is mainly based on a single band 
photometric survey. This band is used to identify and characterize 
the variables. The number of measurements in the companion band(s) 
is more limited, and in some cases they do not uniformly cover the 
entire pulsation cycle. As a consequence, mean colors are affected 
by photometric errors that are slightly larger than the repeatability 
errors. 

%ref_1_24
%_____________________________________________________________________________
\subsubsection{Highlights concerning observed optical-NIR PL and PW relations}
%_____________________________________________________________________________

%%%%%%%%%%%%%%%%%%%%%%%%%%%%%%%%%%%%%%%%%%%%%%%%%%%%%%%%%%%%%%%%%%%%%%%%%%%%%%%%%%%%%
% 			fig 18
%%%%%%%%%%%%%%%%%%%%%%%%%%%%%%%%%%%%%%%%%%%%%%%%%%%%%%%%%%%%%%%%%%%%%%%%%%%%%%%%%%%%%
%_______________________________________________________________________________
\begin{figure}[htbp]
\begin{center}
\includegraphics[width=height=4truecm,width=0.60\textwidth]{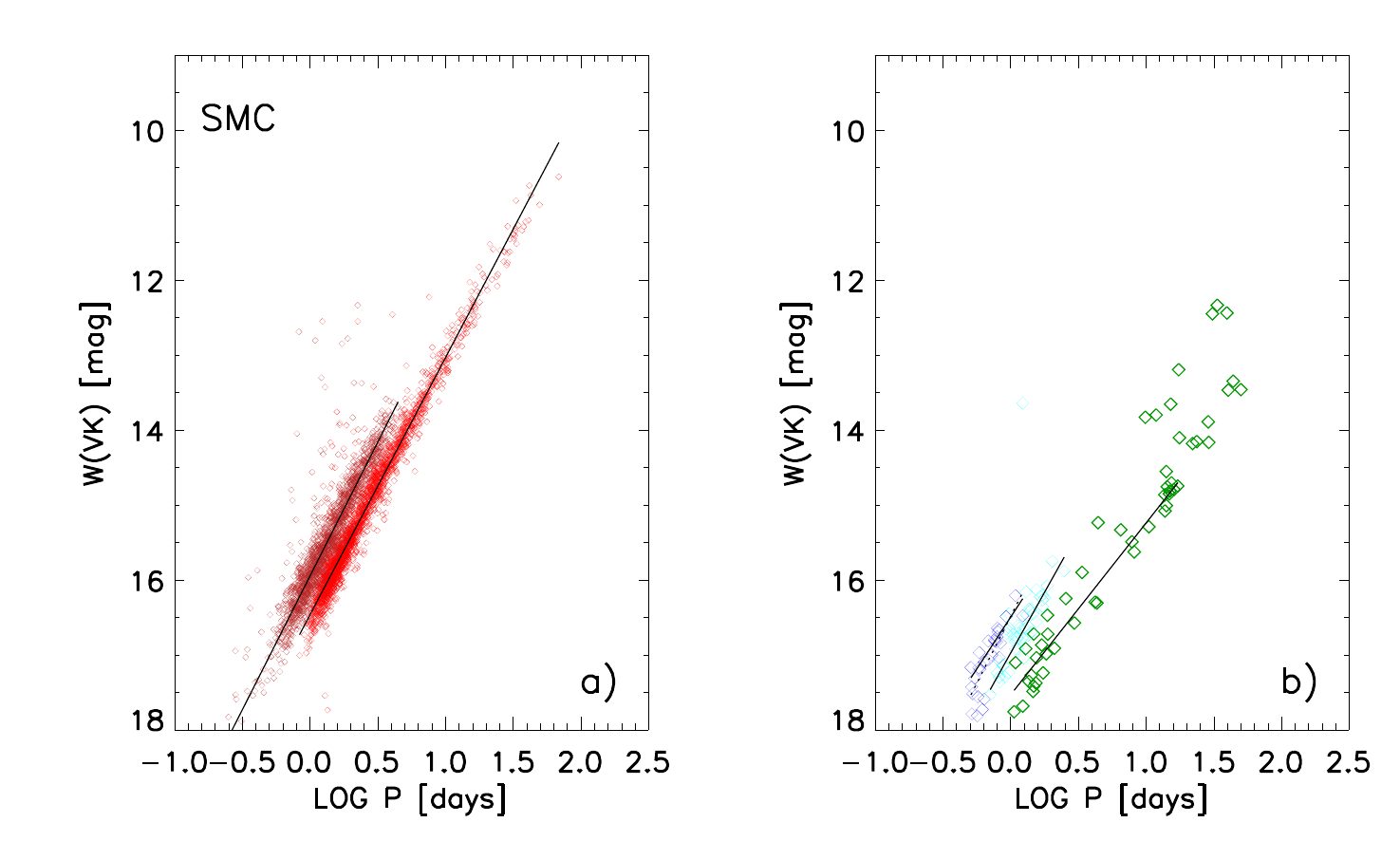}
\includegraphics[width=height=4truecm,width=0.60\textwidth]{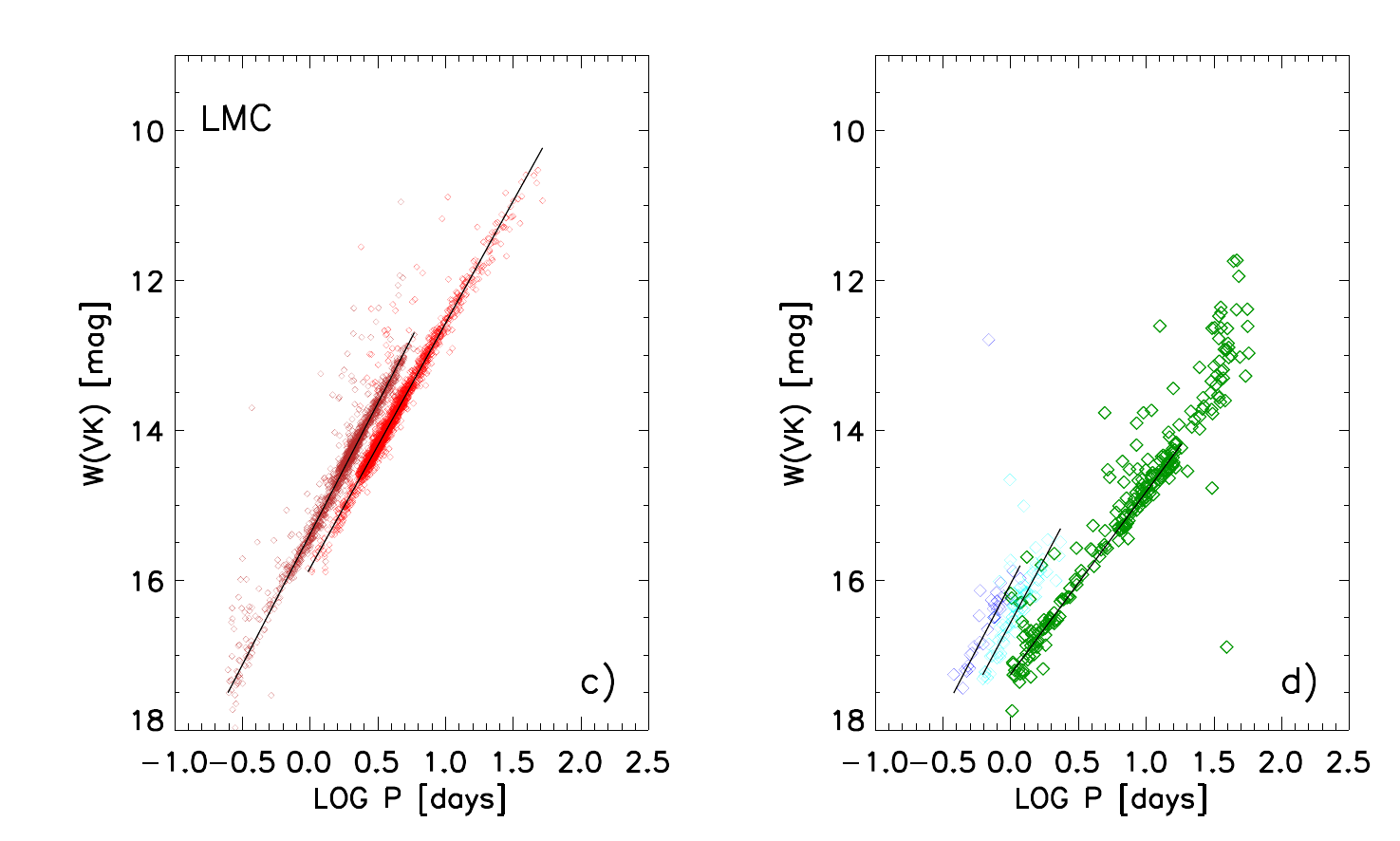}
\caption{Top: Optical--NIR $K$,$V-K$ PW relations (see Table~\ref{tbl:cepheid_PW} 
(Appendix))
for SMC CCs (panel a) and for ACs plus TIICs (panel b). 
Symbols and color coding are the same as in Fig.~\ref{fig_PLI_MC}.
Bottom: Same as the top, but for LMC Cepheids 
(CCs, panel c; ACs plus TIICs, panel d).}
\label{fig_PWKVK_MC}
\end{center}
\end{figure}
%_______________________________________________________________________________

The current empirical evidence indicates that PL and PW relations for CCs are 
linear over the entire period range. However, there is a vast literature 
concerning the possible occurrence of non-linearity in PL, 
Period-Color (PC) and Amplitude-Color relations of CCs based on 
mean magnitudes \citep{ripepi2022}. The same outcome applies to 
predicted \citep{das2020} and observed (MW, MC) diagnostics at maximum 
and minimum light \citep{kanbur2007}. More specifically, 
\citet{ngeow2005,ngeow2006} by using optical (OGLE, MACHO) and NIR (2MASS) mean 
magnitudes found that LMC CCs display either a break or a non--linearity 
for periods around 10 days. They both adopted a chi-squared test 
and an F-test and found that $V$,$R$,$J$,$H$ PL relations 
show evidence of non-linearity for P$\approx$10 days, whereas the K-band 
PL relation appear to be consistent with a single-line regression. 

A similar investigation was also performed by \citet{subramanian2015} 
by using optical ($V$,$I$) band photometry provided by OGLE to investigate
the PL relations of SMC Cepheids. They found that optical PL relations of 
both FU and FO CCs show a break for periods around 2.95 days and around 
1 day, respectively. These findings were supported and complemented by 
\citet{Bhardwaj2016a,Bhardwaj2016b} 
using optical ($V$,$I$; OGLE) and NIR \citep[$J$,$H$,$K$][]{macri2015}
photometry and several robust statistical tests. They found that optical 
PL, PW and PC relations for LMC FU CCs are non-linear at periods 
around 10 days, while the NIR PL and the optical-NIR, triple band PW  
relations are non-linear around 18 days. Furthermore, they found that 
PL, PW and PC relations for FO CCs display a significant change in the 
slope for periods around 2.5 days only at optical wavelengths.
More recently, \citet{ripepi2022} provided a variety of PL and PW 
relations for LMC CCs, by using data collected by the VISTA NIR 
survey of the MCs,  they found for the first time a break for FO CCs 
at periods of 0.58 days. The reader interested in a more detailed discussion 
addressing theoretical and empirical evidence concerning breaks in PL
relations and multi-phase relations is referred to \citet{Kurbah2023}

A few general considerations in dealing with this wealth of empirical 
evidence. Photometric accuracy and completeness are two major issues which
can affect the slope of both PL/PW relations. Fortunately, long-term 
photometric surveys (OGLE, Gaia) 
and space photometry \citep[HST][]{Riessetal2019}
are providing complete and accurate samples.
However, we still lack accurate estimates of the star formation episodes 
during the last 300 Myr in the MCs. Moreover, spectroscopic surveys are 
still lagging, and we also lack accurate estimates of the impact that 
chemical composition, and in particular iron abundance, has on the slope 
of the PL/PW relations. Moreover, \citet{soszynski2015a}  
suggested that the cleaning of the sample is a possible source of 
systematics in the identification of  break(s) in PL relations. He found 
that a solid identification of ACs among SMC Cepheids significantly 
reduces the evidence of a break for periods around 2.5 days among SMC FU CCs.  

This is the case in which the use of the Occam's razor can help in dealing 
with this heuristic hypothesis. This precaution is supported by 
evolutionary models:  there is agreement on the evidence that the slope 
of the mass--luminosity relation does not change, at fixed chemical composition, 
for stellar masses typical of CCs \citep{bono00a,anderson02,desomma2021}.

The lively discussion  concerning the linearity of PL/PW relations of CCs 
does not apply to TIICs, indeed, the RV Tauri (long-period TIICs) appear 
to be in the optical regime, at fixed pulsation period, brighter than 
expected on the basis of the global PL relation. The difference is still 
present, but reduced  in the NIR regime and in the $I$,$V-I$ PW relation. 
This issue has been widely discussed in the literature 
\citep{matsunaga06,soszynski2018,ripepi2014,bhardwaj2022}. 
In order to overcome these difficulties, in the estimate of both PL 
and the PW relations, we only included TIICs with periods 
shorter than 20 days. 

Interestingly enough, data plotted in Fig.~\ref{fig_PLI_MC} 
and in Fig.~\ref{fig_PLK_MC} (Appendix) display that SMC ACs pulsating 
in the FO appear to be for $\log P\sim-0.30$ (short period tail)  
systematically fainter than expected, according to the slope of the 
PL relation.  The same outcome applies to short period 
FU ACs, but the difference is smaller. On the other hand, 
LMC ACs show a well-defined slope over the entire period range. 

The evidence that 
the change in SMC FO ACs occurs both in optical 
($V$,$I$) and in NIR ($J$,$K$) PL/PW relations further 
supports that it is intrinsic. In order to detect the possible occurrence 
of a change in the slope, we computed the PL and the PW relations, while 
neglecting the short period tail, and we found that the slopes are 
significantly shallower (see Tables~\ref{tbl:cepheid_PL}~\ref{tbl:cepheid_PW} (Appendix) 
and Figs.~\ref{fig:slopes_pl}, \ref{fig:slopes_pw}). The current data do not allow us 
to constrain whether the difference is caused 
by a difference in chemical compositions and/or in the ML 
relation.  

The difference among optical, NIR and optical-NIR PW relations 
listed in Table~\ref{tbl:cepheid_PW} (Appendix) and plotted in 
Figs.~\ref{fig_PWIVI_MC} and \ref{fig_PWKVK_MC} 
are mainly due to the adopted color index, and to the coefficient of 
the color index. Optical-NIR colors cover a broad range in wavelength, 
this means a larger sensitivity in temperature. The range in 
$V-K$ color of LMC CCs is, on average, a factor of 2.5 larger 
than the range in $J-K$ color. The use of optical--NIR magnitudes provides 
the unique opportunity to use color coefficients in the Wesenheit 
magnitudes smaller than one, therefore limiting possible 
uncertainties in the adopted mean colors.

%%%%%%%%%%%%%%%%%%%%%%%%%%%%%%%%%%%%%%%%%%%%%%%%%%%%%%%%%%%%%%%%%%%%%%%%%%%%%%%%%%%%%
% 			fig 19
%%%%%%%%%%%%%%%%%%%%%%%%%%%%%%%%%%%%%%%%%%%%%%%%%%%%%%%%%%%%%%%%%%%%%%%%%%%%%%%%%%%%%
%_______________________________________________________________________________
\begin{figure}[htbp]
\begin{center}
\includegraphics[width=height=4truecm,width=0.60\textwidth]{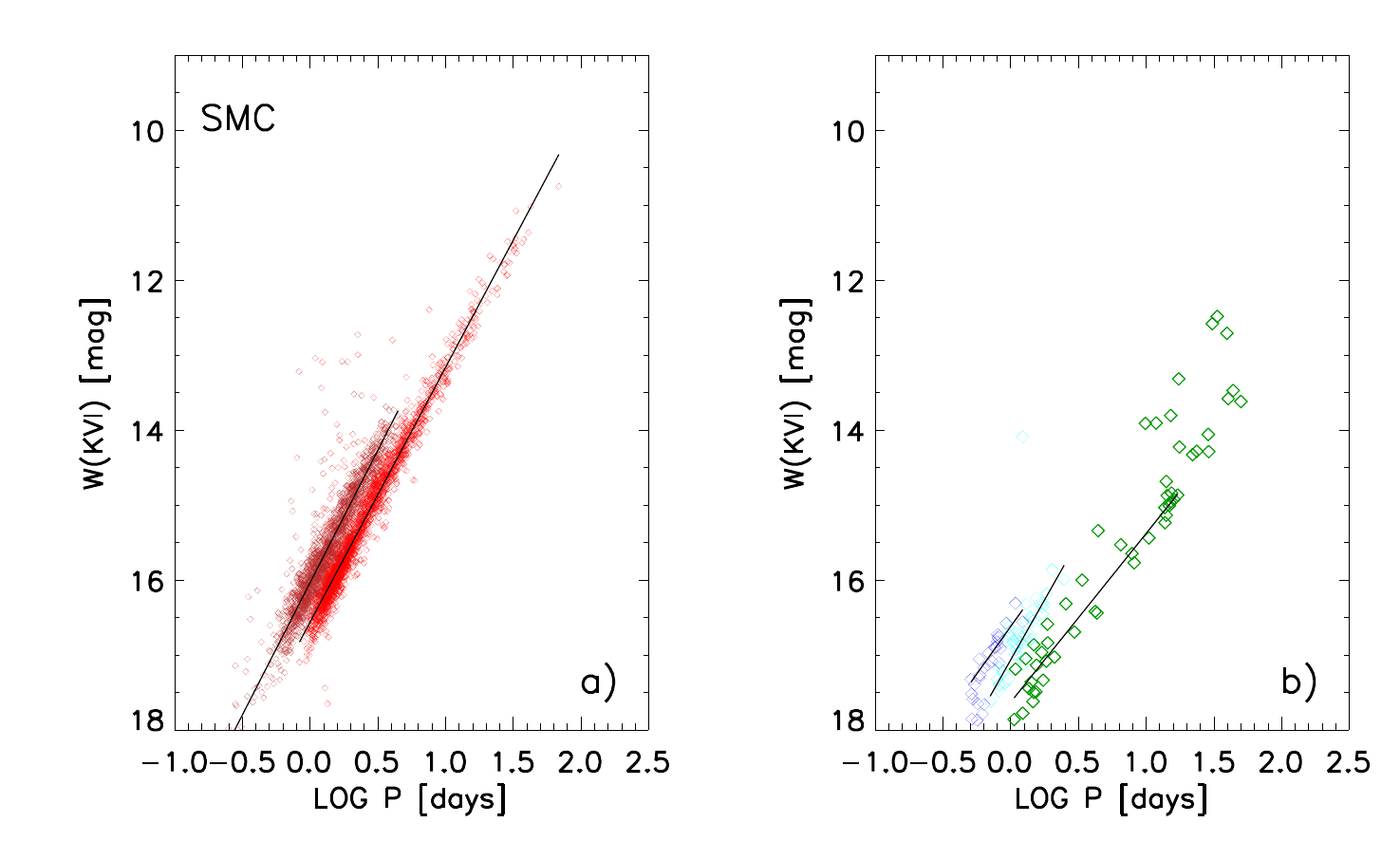}
\includegraphics[width=height=4truecm,width=0.60\textwidth]{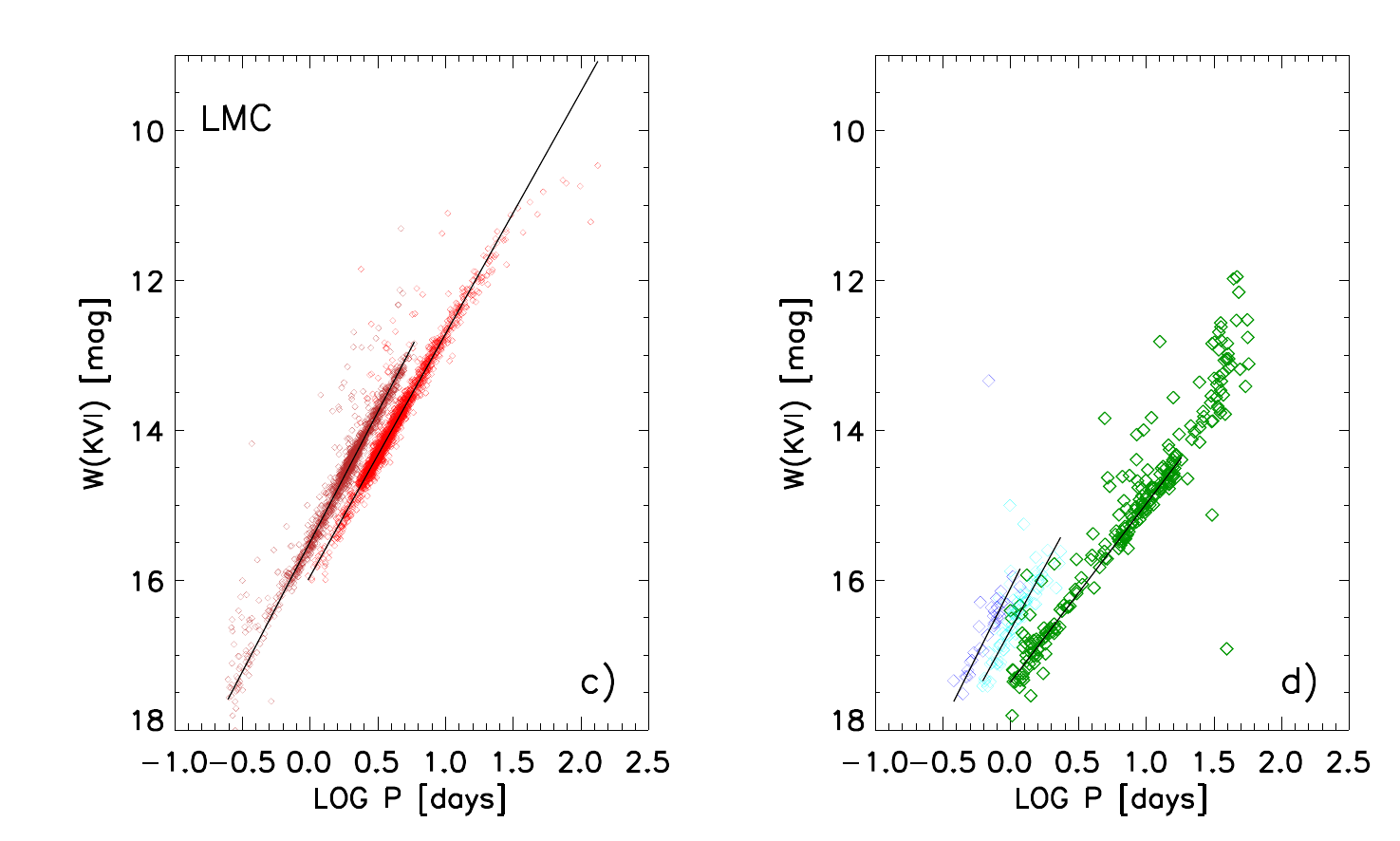}
\caption{Top: Optical--NIR, triple-band $K$,$V-I$ PW relations 
(see Table~\ref{tbl:cepheid_PW} (Appendix)) for SMC CCs (panel a) and for 
ACs plus TIICs (panel b). 
Symbols and color coding are the same as in Fig.~\ref{fig_PLI_MC}.
Bottom: Same as the top, but for LMC Cepheids
(CCs, panel c; ACs plus TIICs, panel d).}
\label{fig_PWKVI_MC}
\end{center}
\end{figure}
%_______________________________________________________________________________

\subsection{Global trend in the slopes of PL and PW relations}\label{sec:global_slopes}

In order to investigate on a more quantitative basis, the use of the triple-bands in the 
definition of the Wesenheit pseudo magnitude, Fig.~\ref{fig_PWKVI_MC} shows 
the $K$,$V-I$ PW relations for Cepheids in the Magellanic Clouds. The key 
advantage in using this optical-NIR diagnostic is three-fold. 
{\em a)}-- The mean $K$-band magnitude is minimally affected by reddening.
{\em b)}-- The $V-I$ color has a good sensitivity to the effective temperature. 
{\em c)}-- The standard deviation is slightly smaller when compared 
with similar optical-NIR PW relations (see Table~\ref{tbl:cepheid_PW} (Appendix)).

%%%%%%%%%%%%%%%%%%%%%%%%%%%%%%%%%%%%%%%%%%%%%%%%%%%%%%%%%%%%%%%%%%%%%%%%%%%%%%%%%%%%%
% 			fig 20
%%%%%%%%%%%%%%%%%%%%%%%%%%%%%%%%%%%%%%%%%%%%%%%%%%%%%%%%%%%%%%%%%%%%%%%%%%%%%%%%%%%%%
%_______________________________________________________________________________
\begin{sidewaysfigure}
\begin{center}
\includegraphics[height=0.6\textheight,width=\textwidth]{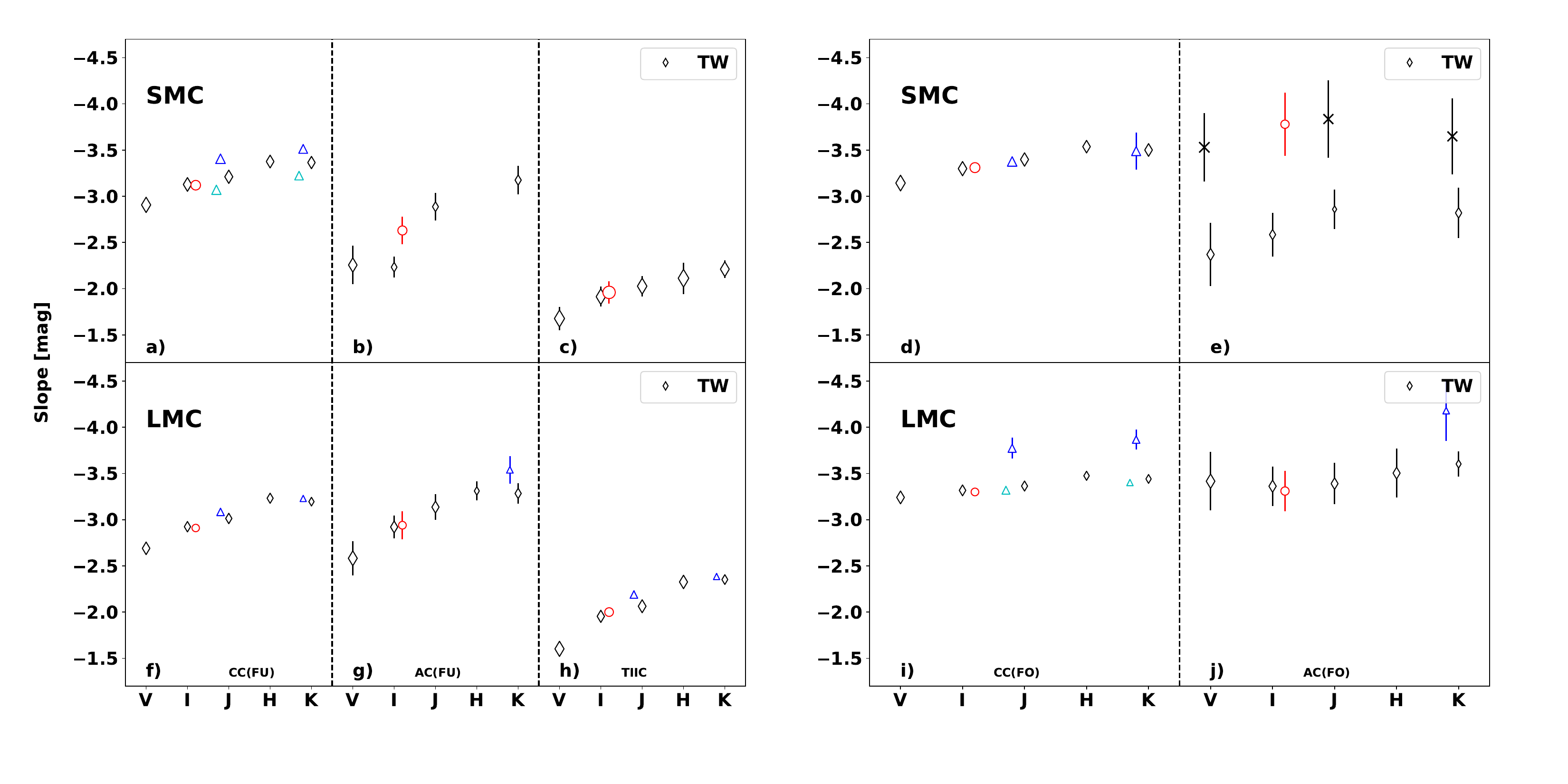}
\vspace{-1.5cm}
\caption{Top-left: from left to right the slope of the PL relations for 
SMC fundamental Cepheids: CCs (panel a), ACs (panel b), TIICs (panel c). 
The size of the symbol is proportional to the error on the slope. 
The error bar shows the standard deviation of the individual 
%PL relation (see also Table~\ref{tbl:cepheid_PL}). Black diamonds 
PL relation (see also Table~\ref{tbl:cepheid_PL} (Appendix)). Black diamonds 
display our own estimates of the PLs, while red circles show the 
estimates by \citet{soszynski2015a}, triangles (both blue and cyan) 
the estimates by \citet{ripepi2014,ripepi2015,ripepi2017,ripepi2022}. 
The blue triangles display the slope of the PL relation for CC(FU) with 
logP$<$0.47; the cyan triangles display the slope of the PL relation for 
CC(FU) with logP$>$0.47.
Top-right same as the top-left, but for SMC first-overtone CCs (panel d) and ACs (panel e).
Bottom-left: same as the top-left, but for the LMC fundamental CCs 
(CCs, panel f; ACs, panel g; TIICs, panel h).
Bottom-right: same as the top-right, but for LMC first overtone CCs 
((panel i) and ACs (panel j). According to \citet{ripepi2022} the blue triangles 
display the  slope of the PL relation for FO CCs with P$<$0.58 days while 
the cyan triangles display slope of the PL relations  for FO CCs with 
P$>$0.58 days.}
\label{fig:slopes_pl}
\end{center}
\end{sidewaysfigure}
%_______________________________________________________________________________

The current optical (OGLE-IV) and NIR (priority: VISTA \citep{cioni11}, 
secondary: IRSF \citep{ita2004}, both converted to the 2MASS system
by using the transformations provided by \citet{kato2007,gonzalezfernandez2018}) 
photometric data sets for Magellanic variables allow us to investigate 
on a more quantitative basis several interesting features of the 
PL relations. 

Data plotted in panel a) of  Fig.~\ref{fig:slopes_pl} 
show the slopes of both optical ($V,I$) and NIR ($J,H,K$) PL relations by using 
the same SMC fundamental variables plotted in  Fig.~\ref{fig_PLI_MC} and in 
%Fig.~\ref{fig_PLK_MC} and listed in Table~\ref{tbl:cepheid_PL}. 
Fig.~\ref{fig_PLK_MC} and listed in Table~\ref{tbl:cepheid_PL} (Appendix). 
The size of the symbols is correlated with the 
standard deviation of the relations, while the error bars display the uncertainties
on the slopes.  The slope, as expected, steadily increases when 
moving from optical to NIR bands. The increase is of the order of 20\% in the 
transition from optical to NIR, and it becomes negligible in the longer wavelength regime. 
Therefore the intrinsic accuracy of K-band PL relation improves because the slope 
is steeper, and because they are minimally affected by uncertainties on 
reddening \citep{bono2008b}.

The agreement with the slopes estimated by 
\citep[][red circles]{soszynski2015a} 
is quite good, since it is based on the same data, but using different selection 
criteria. The same outcome applies to the slopes (blue triangles) of the 
NIR CC PL relations provided by \citet{ripepi2017,ripepi2022}.
In this case we used the NIR mean magnitudes, based on the fit of the 
light curves provided by \citet{ripepi2017,ripepi2022}, together with the mean 
of the NIR measurements provided by IRSF \citep{ita2004,ita2018}. The red circles 
and the blue triangles have been slightly shifted along the X-axis to avoid the 
overlap with symbols used for the current estimates (diamonds). 

Panel b) of  Fig.~\ref{fig:slopes_pl} shows the optical 
NIR slopes for SMC FU ACs. The symbols are the same as in the left panel. 
The trend when moving from the optical to the NIR bands is similar to the CCs 
and the slopes also attain similar values, whereas the uncertainty on the slope 
and the standard deviations  are systematically larger (see 
%Table~\ref{tbl:cepheid_PL}). The uncertainty in the $H$-band is higher, 
Table~\ref{fig_lmc_bump_light} (Appendix)). The uncertainty in the $H$-band is higher, 
because in this band fewer measurements have been collected
compared with the $J$ and the $K$-band. 

Panel c) of  Fig.~\ref{fig:slopes_pl} show the slopes for 
fundamental SMC TIICs. The trend is, as expected, similar when moving from 
the optical to the NIR bands. However, the slope of the PL relations are, 
on average, 30--40\% shallower than the slopes of ACs and CCs, while the 
standard deviations are 1.5--2 times larger. The reasons for the difference 
have already been discussed in section~\ref{sec:pl_vs_pw}.

Bottom left panels of Fig.~\ref{fig:slopes_pl} display the same slopes 
of the top panels, but for LMC Cepheids. The trend
is similar when moving from the optical to the NIR bands 
as originally suggested by \citet{matsunaga06,matsunaga2009a}.
The blue and the cyan triangles for CCs (panel f) 
display the comparison with \citet{ripepi2017}. The two estimates
refer to the PL relations they derived for variables with periods 
longer/shorter than 4.7 days and they agree, within the errors, quite well. 
The agreement with similar estimates in the  literature is, once again, 
quite good. 
The similarity in the slope for LMC and SMC FU TIICs was also 
suggested by \citet{matsunaga11a}.
The slopes for both ACs (panel g) and TIICs (panel h) have larger errors due either to small 
statistics (ACs), or to a large dispersion at a fixed pulsation period (TIICs). 
Moreover, the slopes of both ACs and TIICs in the NIR bands appear to be 
either constant, or display a mild decrease. They need to be cautiously 
treated, since this is probably an observational bias caused by the limited 
accuracy of NIR mean magnitudes in the short period range (fainter limit) 
of ACs. 

The right panels of  Fig.~\ref{fig:slopes_pl} display the 
same optical, NIR PL relations of the left panels, but for first overtone 
CCs and ACs. The trends are similar, but there are three distinctive features 
worth being discussed. 

{\em a)}-- The slopes are on average larger, when compared with fundamental 
variables (panels d, i), since they range from $\sim$3 to $\sim$3.5. Moreover, 
the current estimates are in remarkable agreement with literature estimates.

{\em b)}-- The standard deviations of both optical and NIR PL relations are, on 
average smaller, when compared with the FU CCs. This means that at a fixed chemical 
composition and photometric band, the width in temperature of the 
instability strip in which FOs are pulsationally stable is systematically 
narrower than for FU CCs. These findings indicate that individual distances 
based on PL relations for FO CCs are, at fixed chemical composition, 
intrinsically more accurate (optical), or with an accuracy similar (NIR) to 
individual distances for FU CCs.  This evidence further supports predictions 
provided by \citet{bono01} by using pulsation models. 

{\em c)}--Panel e)  of  Fig.~\ref{fig:slopes_pl} 
displays the slopes for SMC FO ACs. The empirical scenario concerning 
the slopes of SMC FO ACs is more 
complex. Indeed, the slopes based on the entire sample (crosses) display 
values similar to the slopes of FO CCs. However, optical and NIR 
slopes become systematically shallower once the short period tails 
are neglected (diamonds). The former values agree, as expected, quite 
well with the estimates provided by \citet[][red circles]{soszynski2015a}. 
The slope of the $H$-band was not included, because the number 
of measurements in this band is either missing (VVV) or limited 
(IRSF).

%%%%%%%%%%%%%%%%%%%%%%%%%%%%%%%%%%%%%%%%%%%%%%%%%%%%%%%%%%%%%%%%%%%%%%%%%%%%%%%%%%%%%
% 			fig 21
%%%%%%%%%%%%%%%%%%%%%%%%%%%%%%%%%%%%%%%%%%%%%%%%%%%%%%%%%%%%%%%%%%%%%%%%%%%%%%%%%%%%%
%_______________________________________________________________________________
\begin{sidewaysfigure}
\begin{center}
% \makebox[\textwidth][c]{\includegraphics[height=0.7\textheight,width=1.2\textwidth]{figure/all_slopes_pw2x.pdf}}
% \includegraphics[height=0.7\textheight,width=\textwidth]{figure/all_slopes_pw2x.pdf}
\includegraphics[height=0.6\textheight,width=\textwidth]{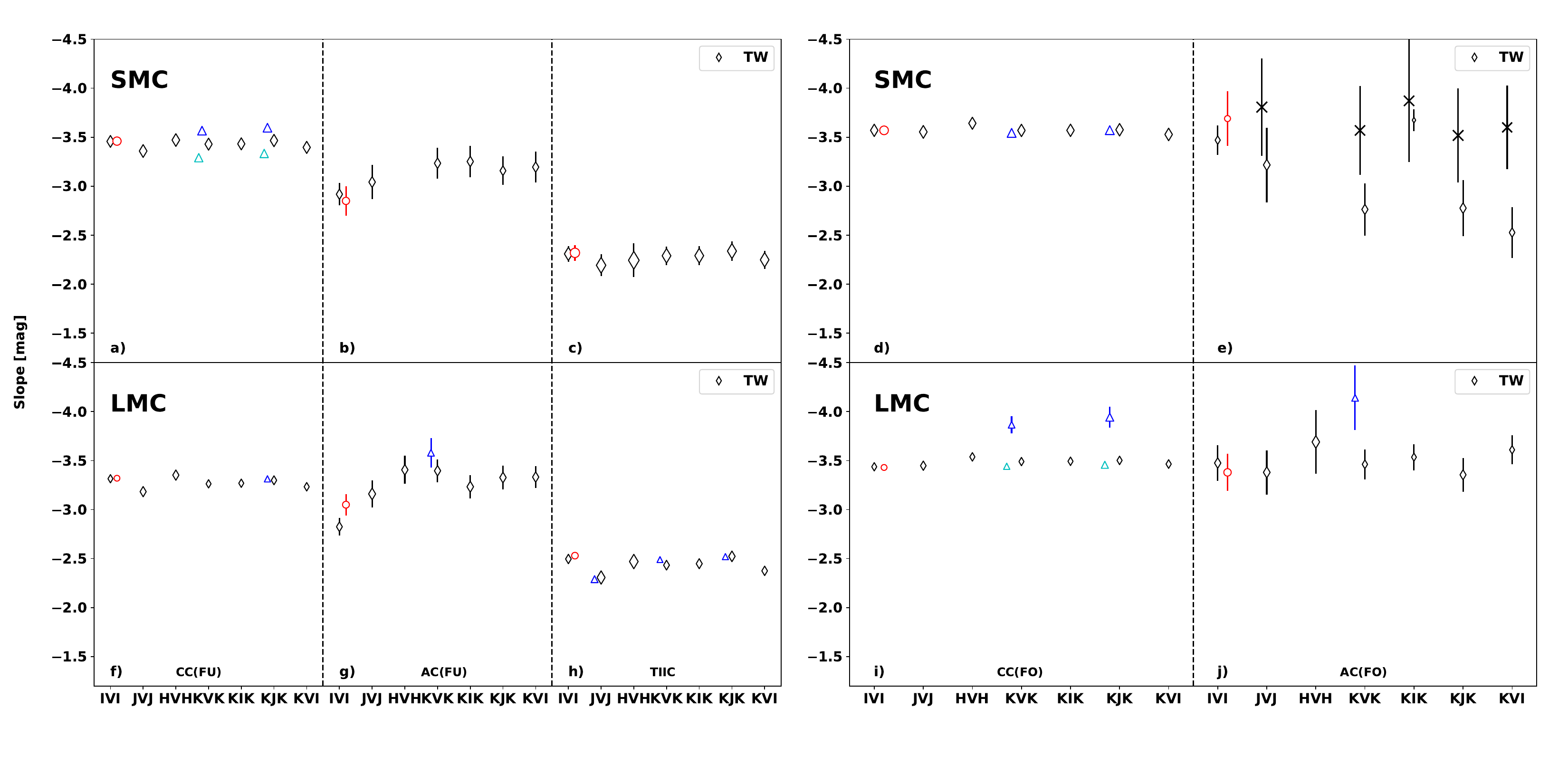}
\caption{
Top-left: from left to right the slope of optical, NIR and optical-NIR PWrelations for
SMC fundamental Cepheids: CCs (panel a), ACs (panel b), TIICs (panel c). 
The symbols are the same as in Fig.~\ref{fig:slopes_pl}. 
Top-right same as the top-left, but for SMC first-overtone CCs (panel d) and ACs (panel e).
Bottom-left: same as the top-left, but for the LMC fundamental CCs 
(CCs, panel f; ACs, panel g; TIICs, panel h).
Bottom-right: same as the top-right, but for LMC first overtone CCs 
((panel i) and ACs (panel j). 
}
\label{fig:slopes_pw}
\end{center}
\end{sidewaysfigure}
%_______________________________________________________________________________

Figure~\ref{fig:slopes_pw} shows the coefficients of the logarithmic period in both 
optical, optical-NIR and NIR PW relations. The key feature of the data plotted in 
this figure for both FU and FO Cepheids is that the slope is almost constant even 
if the coefficient of the color term decreases by one order of magnitude when  
moving from $I$,$V-I$ to $K$,$V-K$ (1.38 versus 0.13). Note once again that the 
PW relations are pseudo PLC relations, in which the coefficient of the color term 
is only fixed by the reddening law. 

The comparison with similar estimates available in the literature is, within 
the errors, quite good for the different variable groups and distance 
diagnostics. The only difference is with slopes for FO CCs provided by 
\citet{ripepi2017}, but it is due to the split of the sample in 
short and long period sub-sample. Indeed, the slopes they found using 
the entire sample (green triangles) agree quite well.

The similarity between the slopes of the PW relations for FU and FO MC CCs 
is quite remarkable and applies to optical, optical-NIR and NIR PW relations. 
This similarity fully supports the simultaneous 
use of both FU and FO CCs (fundamentalization of the periods) to estimate 
distances.

Finally, the PW relations for ACs including NIR bands need to be cautiously 
treated for the same reasons we already mentioned for the NIR PL relations. 

%_____________________________________________________________________
\subsection{PL versus PLC relations}

The pulsation relation dictates that the period of a variable depends 
on the stellar mass, luminosity and effective temperature. Stellar mass 
and luminosity are tightly correlated and provided by evolutionary 
predictions through the so-called mass--luminosity relation. 
This means that pulsation periods, if we neglect the dependence on the 
chemical composition,  mainly depend on luminosity and effective 
temperature. Therefore, an individual correlation among the 
physical parameters governing the pulsation properties of a radial 
variable requires the knowledge, together with the pulsation period 
and the pulsation mode of both mean magnitude and color. 

\begin{figure}[htbp]
\begin{center}
\includegraphics[height=0.25\textheight,width=\textwidth]{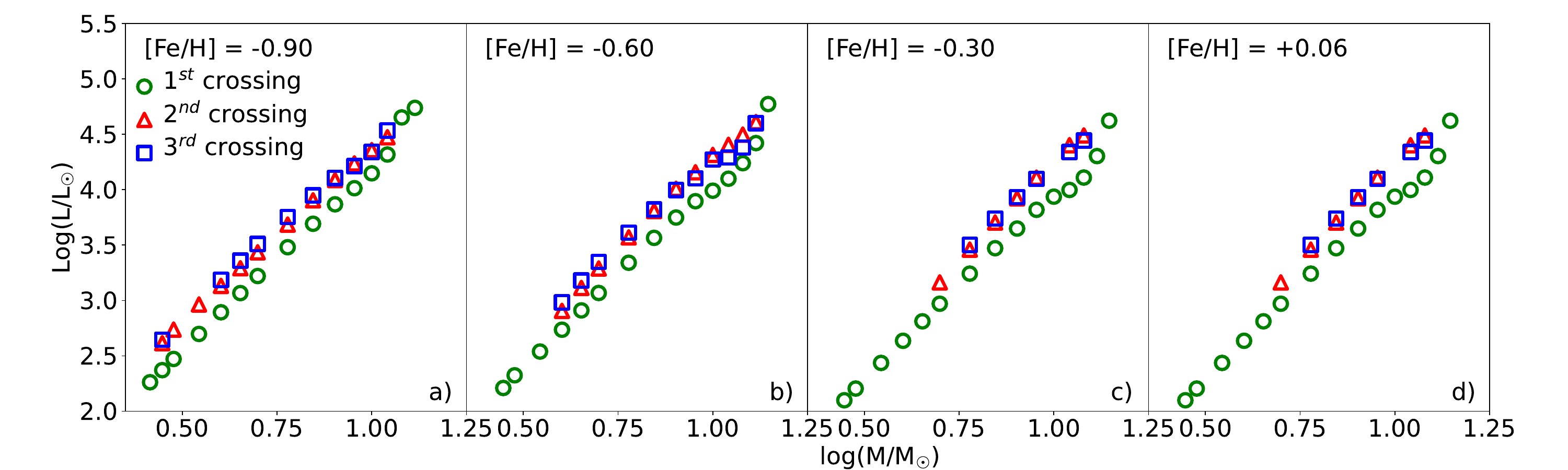}
\caption{
Panel a) -- Logarithmic stellar mass  versus logarithmic luminosity for the 
different crossings of the instability strip. Predictions plotted in this panel 
have been computed for a scaled solar mixture and at fixed chemical composition 
([Fe/H]=-0.90) representative of CCs in metal--poor disk galaxies. The three different 
crossings are plotted with different colors and symbols. In the less massive 
regime only the 1st crossing is present, because in these stellar structures the blue 
loop does not cross the instability strip. In the more massive regime, only 
the 1st crossing is present because these stellar structures  already ignited 
core helium burning and they cross once the instability strip.
Panel b) -- Same as Panel a), but for a chemical composition representative of LMC CCs.
Panel c) -- Same as Panel a), but for a chemical composition representative of SMC CCs.
Panel d) -- Same as Panel a), but for a chemical composition representative of MW CCs.}\label{cc_crossings}
\end{center}
\end{figure}

%ref_1_27
Intermediate-mass stars during core helium burning phases cross from 
one to three times the instability strip. In multiple crossings CCs can 
have the same effective temperature, 
and the same radius, and therefore the same luminosity, but different masses. 
The consequence is a difference in the mean density, and in turn in the 
pulsation period. In order to address this issue on a more quantitative basis, 
we investigated the difference in luminosity among different crossings for 
chemical compositions representative of MC and MW CCs 
(see Fig.~\ref{cc_crossings}). Predictions plotted in this figure and 
listed in Table~\ref{tbl:cc_crossing_time} (Appendix) are based on evolutionary models 
taking into account mild convective core overshooting during central hydrogen 
burning phases and at the bottom of the convective envelope along the red giant 
phases, show that the difference in the mean luminosity across the instability 
strip between 2nd and 3rd crossing are at most of the order of a few hundredths 
of a dex. The difference between 1st and 2nd/3rd is larger and of the order 
of 0.25--0.30 dex, slightly increasing from low to high metallicities. 
However, the 1st crossing plays a marginal role, since 
the evolutionary time spent along this crossing is one/two order of magnitude 
shorter than 2nd+3rd crossing. 
According to the mass--luminosity relation provided by \citet{bono00b} 
the quoted difference ($\Delta \log (L/L_\odot)\le0.05$) implies a 
difference in stellar mass of $0.08\,M_\odot$ for $M=5\,M_\odot$. 
The pulsation relation for fundamental CCs correlating 
stellar mass, luminosity and effective temperature, the so-called 
van Albada \& Baker relation \citep{vanalbada1971}, provided
by \citet[][see their Table~6]{bono00a} gives, at fixed effective 
temperature, a difference in logarithmic period of 0.06 dex. The 
current analysis is suggesting that the spread either in mass or in 
luminosity introduced by 2nd and 3rd crossing is negligible.

%%%%%%%%%%%%%%%%%%%%%%%%%%%%%%%%%%%%%%%%%%%%%%%%%%%%%%%%%%%%%%%%%%%%%%%%%%%%%%%%%%%%%
% 			fig 22
%%%%%%%%%%%%%%%%%%%%%%%%%%%%%%%%%%%%%%%%%%%%%%%%%%%%%%%%%%%%%%%%%%%%%%%%%%%%%%%%%%%%%
%_______________________________________________________________________________
%_______________________________________________________________________________

The difficulties in using mean colors have already been discussed 
in section~\ref{sec:ccs_intro}. Note that random-phase measurements  
in different bands introduce systematic offsets in the mean color. 
Moreover, PLC relations are more affected by reddening uncertainties, 
because they require dereddened magnitudes and colors. 
These are the main reasons why individual Cepheid distances are estimated 
by using either PL or PW relations. 

In passing, it is worth mentioning that, on some occasions, 
the coefficient of the color term in optical ($I$, $V$-$I$) 
and in optical-NIR \citep[$K$,$V-K$;][]{ripepi2017} 
PLC relations are,  within the errors, quite similar to the 
ratio between selective absorption coefficient and color excess. 
It is still not clear whether this similarity is fortuitous, but the
quoted PLC relations are also minimally affected by uncertainties in 
reddening uncertainties.

%%%%%%%%%%%%%%%%%%%%%%%%%%%%%%%%%%%%%%%%%%%%%%%%%%%%%%%%%%%%%%%%%%%%%%%%%%%%%%%%%%%
\subsection{Metallicity dependence}\label{sec:met_dependence}

We have already discussed in section~\ref{sec:ccs_intro}  that the pulsation properties of Cepheids depend on the metal content. As a consequence, pulsation diagnostics (PL/PW/PLC relation) adopted to estimate individual Cepheid distances are not Universal. This working hypothesis implies that not only the slope but also the zero-point of the adopted relations depend on the metallicity and they must be properly calibrated.   

There are two different paths that we can take in order to analyze the dependence of Cepheid pulsation properties on the metallicity.

The first path relies on circumstantial empirical evidence. We can use either Galactic or Magellanic Cepheids to investigate the impact that metallicity has on individual distance determinations. This approach is still affected by systematics, because quantitative constraints on the metallicity dependence require accurate and homogeneous iron abundances and distances.
Accurate iron abundances based on high-resolution spectra are available for a significant fraction of Galactic Cepheids. However, the current estimates are far from being homogeneous. Different investigations rely on different line lists and physical assumptions to estimate physical parameters (effective temperature, surface gravity, micro-turbulent velocity) and elemental abundances \citep{dasilva2022}. However, the strong limitation in the estimate of the metallicity dependence for Galactic Cepheids is the need for accurate (1--2\% level) individual geometrical distances. 

%ref_1_28
Accurate trigonometric parallaxes for a dozen Galactic CCs were
originally provided by HST, and more recently Gaia is significantly increasing
the accuracy and the sample size \citep{Breuvaletal2021}. However, the role 
that the current sample of Galactic Cepheids can play to assess the metallicity 
dependence is still lively debated. In a recent investigation, \citet{owens2022}
derived new PL relations by using optical ($B$,$V$,$I$), NIR ($J$,$H$,$K$,) and 
MIR ([3.6], [4.5]) photometric bands for three dozen Galactic CCs with periods 
ranging from 4 to 60 days and Gaia trigonometric parallaxes. The distances they 
obtain by applying the new PL relations to the MCs are discrepant when compared 
with geometrical distances based on detached eclipsing binaries. Their main conclusion 
is that the coupling between systematics affecting Gaia DR3 trigonometric parallaxes 
and uncertainties on the metallicity dependence causes the error budget to be 
at the level of 3\%. The current findings indicate that there are still some 
limitations in using bright Galactic CCs close to the saturation limit, but these 
thorny problems will be solved in a few years.   

In this context, MC Cepheids play a crucial role, because they are located, 
in first approximation, at the same distance. This means that they are a perfect 
laboratory to study the metallicity dependence. However, the number of Cepheids 
for which accurate iron abundances are available, based on high-resolution spectra, 
is only limited to less than one hundred LMC Cepheids \citep{romaniello2022}. 
The current estimates indicate a modest spread in iron abundance. This finding is also 
supported by a large sample of metallicity estimates based on the shape of both 
$V$- and $I$-band light curves (Fourier parameters) provided by  \citet{hocde2023}. 
However, the metallicity distribution of B-type stars (progenitors of Cepheids) 
indicate a spread in N abundance of the order of one dex 
\citep{trundle2007,hunter2008,hunter2009}

Accurate spectroscopic abundances for SMC Cepheids are only available 
for a few tens Cepheids and they suggest a modest chemical enrichment 
\citep{Lemasleetal2017}. The same outcome applies to metallicity estimates 
of SMC CCs based on photometric indices \citep{hocde2023} and to the 
metallicity distribution based on spectroscopy of A- and B-type stars
\citep{venn2003,trundle2005,hunter2009}

The second path relies on theoretical predictions. This approach is based on nonlinear, time-dependent convective models of CCs. The current hydro-dynamical codes allow us to construct several series of pulsation models covering a broad range in stellar masses, stellar luminosities and effective temperatures typical of CCs. Moreover, the same calculations can be performed for chemical compositions typical of both Galactic and Magellanic CCs pulsating both in the fundamental and in the first overtone. The key advantage of this approach concerning the metallicity dependence is that pulsation predictions are used to estimate the difference in metallicity and not their own metallicities. 

The current predictions indicate that the visual magnitude of CCs, at fixed pulsation period, becomes systematically fainter when moving from metal-poor to more metal-rich stellar structures.

Empirical evidence based on individual heavy element abundances and on both Galactic and Magellanic Cepheids covering a broad range in pulsation periods also indicate that metal-rich CCs are, at a fixed pulsation period, fainter than metal-poor ones. However, no general consensus has been reached on this effect. This applies not only to the derivative, the so-called gamma coefficient, but also to the sign of the coefficient \citep{storm11,Groenewegen2018, ripepi2021, breuval2022, trentin2023}. 
This is the typical long-standing open problem which requires a significant 
improvement in the accuracy and in the sample size of the 
spectroscopic measurements before we can reach a firm conclusion.

%%%%%%%%%%%%%%%%%%%%%%%%%%%%%%%%%%%%%%%%%%%%%%%%%%%%%%%%%%%%%%%%%%%%%%%%%%%%%%%%
\section{Cepheids as stellar tracers}

%_______________________________________________________________________________
\subsection{High-resolution spectroscopy}
Cepheids are typically bright stars, therefore we are dealing with very 
high S/N, broad-wavelength range spectra\footnote{
In a short communication to the National Academy of Sciences P.W. Merrill 
wrote: {\em \dots Spectrograms \ldots show several lines of neutral technetium 
in the spectra of S-type stars especially of long-period variables. \ldots 
It is surprising to find an unstable element in the stars. \ldots S-type stars 
somehow produce technetium as they go along  \dots 
\citep[][see also \citet{merrill1952b}]{merrill1952a}.}
These are probably the most dry sentences in the history of Astronomy. 
The content of these sentences not only explained the observations of Tc in 
AGB stars, but also paved the way to modern nuclear Astrophysics, to stellar 
nucleo--synthesis and to the Astrophysical origin of the Periodic Table.}. 
%The reader interested in a more detailed discussion concerning the personality 
%and the role that P.W. Merrill has had in Mount Wilson Observatory is referred to 
%\citet[][pg. XXX]{sandage2013}.
% 
The precision of the atmospheric parameters 
and of the chemical abundances, however, depend not only on the quality of the spectra, 
but also on the quality of the adopted line list. With this in mind, a large portion 
of the investigation by \citet{dasilva2022}  was dedicated to finding the best atomic 
transitions for iron and $\alpha$--elements that can be detected in Cepheids. 
This means i) collecting the most precise transition parameters, ii) removing 
absorption lines which are blended with other lines, and iii) eliminating lines 
that for any reason deviate significantly from the average value for their chemical 
species, or that display a dependence on effective temperature, surface gravity and 
micro-turbulent velocity  across the pulsation cycle.

The first point is typically addressed by using a list of hundreds of 
atomic transitions for iron and $\alpha$--elements commonly
employed in the literature. Their 
transition parameters were updated, whenever possible, with laboratory 
measurements from \citet{Ruffonietal2014}, \citet{DenHartogetal2014} and 
\citet{Belmonteetal2017} for Fe I, \citep{DenHartogetal2019} for Fe II, 
\citep{Lawleretal2013} for Ti I, and for Ti II \citep{Woodetal2013}. 
The astrophysical, but homogeneous and precise compilation of 
\citet{MelendezBarbuy2009} for Fe II lines was also adopted. For the remaining lines, the transition parameters collected and updated by the National Institute of Standards and Technology (NIST) Atomic Spectra Database \citep{Kramidaetal2020} were used. If a line is not available in any of these sources, it is eliminated from the preliminary list and not used for the computation of atmospheric parameters, nor chemical abundances.
 
The second point was addressed by removing blended lines, using 
as a reference the Solar spectrum table by \citet{Mooreetal1966} 
and synthetic spectra. Then, \citet{dasilva2022} measured their equivalent width 
(EW) using the Automatic Routine for line Equivalent widths in 
stellar Spectra  \citep[ARES,][]{Sousaetal2007,Sousaetal2015}.

Finally, the third point was taken into account by removing deviant lines. 
At first, the initial values were estimated for the atmospheric parameters and the chemical abundances. Lines are considered deviant if their abundances differ by at least 3$\sigma$ from the mean for a given element, or if their behavior across the pulsation cycle is irregular. This latter situation is investigated by plotting all line-by-line measurements for each chemical species and individual exposure of the calibrating stars versus phase, effective temperature, and equivalent width, and removing lines that show strong trends in any of those planes. The calibrating stars mostly used are those with the largest number of individual spectra covering the whole pulsation cycle, such as $\beta$ Dor, $\zeta$ Gem, and FF Aql.

The reader interested in the final clean lists adopted for the estimate of the atmospheric parameters and the chemical abundances (quality flag 1) and for comparison purposes (quality flag 0) is referred to Tables 2 and 3, and to Appendix~A by \citet{dasilva2022}.

%%%%%%%%%%%%%%%%%%%%%%%%%%%%%%%%%%%%%%%%%%%%%%%%%%%%%%%%%%%%%%%%%%%%%%%%%%%%%%%%%%%%%
\subsection{Atmospheric parameters}

\subsubsection{Effective temperature, surface gravity and micro-turbulent velocity}\label{teff_logg_vt}

The approach typically adopted to estimate the atmospheric parameters (i.e., the 
effective temperature, the surface gravity, and the micro-turbulent velocity, [$\xi$]) was 
already discussed in detail by \citet{proxauf2018}. Here we only recap the key points. 
The $T_{\rm eff}$ along the pulsation cycle was estimated by using the LDR method, which 
relies on the correlation between the line-depth ratios of pairs of absorption lines 
in the spectra of different stars and the effective temperature of the same stars. 
The surface gravity was derived through the ionization equilibrium of Fe I and Fe II lines, 
and the micro-turbulent velocity was obtained by minimizing the dependence of the abundances 
provided by single Fe I lines on their EWs. During this procedure, the effective temperature 
was kept fixed, whereas $\log g$ and $\xi$ were iteratively changed until convergence. 
The metallicity used as input by our algorithm is updated at each step, and the adopted 
value is [Fe I/H], which is the mean iron abundances provided by individual Fe I lines. 
The uncertainties on $T_{\rm eff}$ are the standard deviations calculated using the LDR method, they are typically of the order of 150 K
while the uncertainties on the individual estimates of $\log g$ and $\xi$ are assumed to be 
$\sim$0.3 dex and 0.5 km s$^{-1}$ (see \citet{genovali2014} for a detailed discussion). 

Thanks to these improvements, the errors on the derived amplitudes for the effective temperature 
curves are a factor of two smaller than our previous estimates, whereas for the surface gravity the 
differences are even more significant, with typical errors almost three times smaller. Concerning 
the micro-turbulent velocity variation, the errors on the current amplitude estimates are,
% editor 20 
on average, more than twice smaller \citep{proxauf2018}. 

The improvement in the estimate of $\log g$ and $\xi$ is mainly due to the 
improvement in the line list, which reduced the number of spurious abundance 
values provided by unreliable iron lines. Together with the removal of blended  
lines, and of lines that display a well-defined trend with effective temperature
or with equivalent width. The estimate of $T_{\rm eff}$ is based on the same line 
list that we previously used, therefore the improvement is mainly due to the 
larger number of spectra analyzed by \citet{dasilva2022}. The reader interested 
in a more detailed discussion concerning the estimate of the intrinsic parameters 
of classical Cepheids is referred to \citet{Vasilyevetal2018}. 

In this context, it is worth mentioning that the improvement in the accuracy 
and in the sampling of the pulsation cycle improved the 
empirical framework concerning the variation of atmospheric parameters along 
the pulsation cycle. Indeed, the minimum in the effective temperature is often 
approached across the same phases in which the minimum in surface gravity is 
also approached. Moreover and even more importantly, these phases across 
the minimum anticipate the increase in microturbulent velocity. This parameter, 
as expected, attains its maximum just before the maximum in effective 
temperature (see Figs.~B.2 and B.4 in \citealt{dasilva2022}).

A new and innovative approach to constrain the temperature scale of classical 
Cepheids was recently suggested by \citet{lemasle2020}. The new approach 
relies on a data-driven machine-learning technique applied to observed spectra, 
in which the flux ratios were tied to temperatures derived using the infrared 
surface-brightness method. 

%%%%%%%%%%%%%%%%%%%%%%%%%%%%%%%%%%%%%%%%%%%%%%%%%%%%%%%%%%%%%%%%%%%%%%%%%%%%%%%%%%%%%%%%%%%%%%%%%%%%%%%%%%%%%%%%%%%
\subsubsection{Effective temperature curve templates}

By taking advantage of the substantial phase coverage of the sample of 
calibrating Cepheids, the effective temperature measurements based on LDRs 
were also adopted to compute new templates for different bins 
of pulsation periods \citep{dasilva2022}. However, instead of using the $T_{\rm eff}$ curves directly, 
they preferred to provide templates for the $\theta$ parameter -- defined as  
$\theta = 5040/T_{\rm eff}$ -- given its linear dependence on the Johnson--Cousins 
$R-I$ color index \citep{Taylor1994}.

The approach is similar to the NIR light-curve templates provided by \citet{inno15}: first, 
periods and reference epochs are adopted to fold the $\theta$ curves ($\theta$Cs). 
Subsequently, the folded  $\theta$Cs were normalized by subtracting their average 
and dividing by their amplitudes. The $\theta$Cs curves are well sampled, 
meaning that the phase points are large enough to separate the Cepheids into 
different period bins and to provide analytical relations for the  $\theta$ curve 
templates in each bin. We adopted the same period thresholds introduced 
by \citet{inno15}. Therefore, based on their Table~1, \citet{dasilva2022} 
generated four cumulative and normalized theta curves for the bins 2 
(3--5 days, three Cepheids), 3 (5--7 days, three Cepheids), 
4 (7--9.5 days, two Cepheids), and 5 (9.5--10.5 days, two Cepheids). 
By adopting the same period bins, they were able to provide  
$\theta$C templates homogeneous with those adopted for 
the NIR light curve templates. The reader interested in a more detailed 
and quantitative discussion about the use of both cumulative and 
normalized curves to derive the analytical fits, together with the 
adopted thresholds for the different period bins, is referred to \citet{inno15}.

%%%%%%%%%%%%%%%%%%%%%%%%%%%%%%%%%%%%%%%%%%%%%%%%%%%%%%%%%%%%%%%%%%%%%%%%%%%%%%%%%%%%%
% 			fig 23
%%%%%%%%%%%%%%%%%%%%%%%%%%%%%%%%%%%%%%%%%%%%%%%%%%%%%%%%%%%%%%%%%%%%%%%%%%%%%%%%%%%%%
%\clearpage
\begin{figure}[ht]
\begin{center}
\includegraphics[width=0.5\columnwidth]{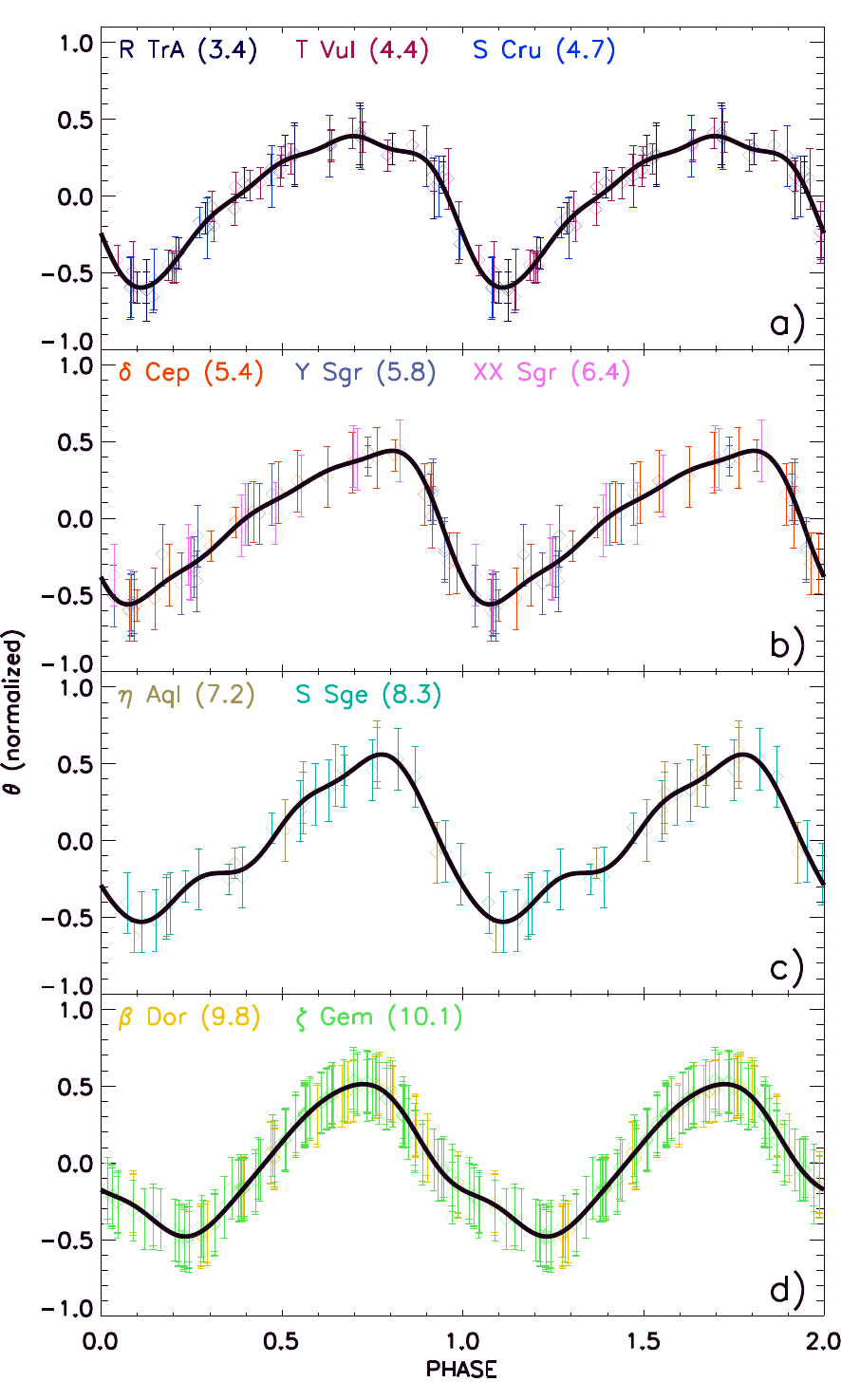}
\caption{From top to bottom: Normalized $\theta=5040/T_{\rm eff}$ curves of CCs 
in different period bins (3--5 days, panel a; 5--7 days, panel b; 7--9 days, panel d; 9--11 days, panel e). Different colors are associated to different CCs. The names of the CCs are labelled and the numbers 
in parentheses show the pulsation period in days. The black line is the 
analytical fit to the $\theta$-curve template. Adapted from \citet[][Fig.~5]{dasilva2022}. Image credit: R. da Silva.}
\label{fig:teff_templates}
\end{center}
\end{figure}
%_______________________________________________________________________________

The normalized  $\theta$ curves for the four period bins are plotted 
in Fig.~\ref{fig:teff_templates} together with the analytical fits 
(black lines), based on Fourier series, to the data 
\citep[see Fig.~7 in][]{dasilva2022}. The separation of the Cepheids 
into different period bins is a mandatory step, because not only the 
amplitude of the $\theta$ curves, but also their shape changes with 
period. Note that spectroscopic data for Cepheids with periods outside the 
selected period bins are available, but they are either not as 
well-sampled as the data for the other Cepheids, or the $\theta$ curves 
do not overlap very well because of the limited number of measurements. 
More accurate and homogeneous data for FU and FO CCs are 
required in order to overcome this limitation.

The use of the effective temperature curve templates to
derive the mean effective temperature requires the ephemerides,
the luminosity amplitude ($\Delta$V), and a single spectroscopic 
measurement of the $T_{\rm eff}$ together with a linear relation 
between  $\Delta$V and $\Delta\theta$: $\Delta\theta = (0.184 \pm 0.014) 
\Delta {\rm V} + (0.000 \pm 0.011)$; $(\sigma = 0.013)$.

The estimate of the $\Delta$ $\theta$ amplitude must be applied 
to the normalized template as a multiplier factor. Only after 
this re-scaling operation can the template be anchored to the 
empirical data and used to derive the mean $T_{\rm eff}$.

%%%%%%%%%%%%%%%%%%%%%%%%%%%%%%%%%%%%%%%%%%%%%%%%%%%%%%%%%%%%%%%%%%%%%%%%%%%%%%%%%%%%%
% 			fig 24
%%%%%%%%%%%%%%%%%%%%%%%%%%%%%%%%%%%%%%%%%%%%%%%%%%%%%%%%%%%%%%%%%%%%%%%%%%%%%%%%%%%%%
%\clearpage
\begin{figure}[!ht]
\begin{center}
\includegraphics[width=0.5\textwidth]{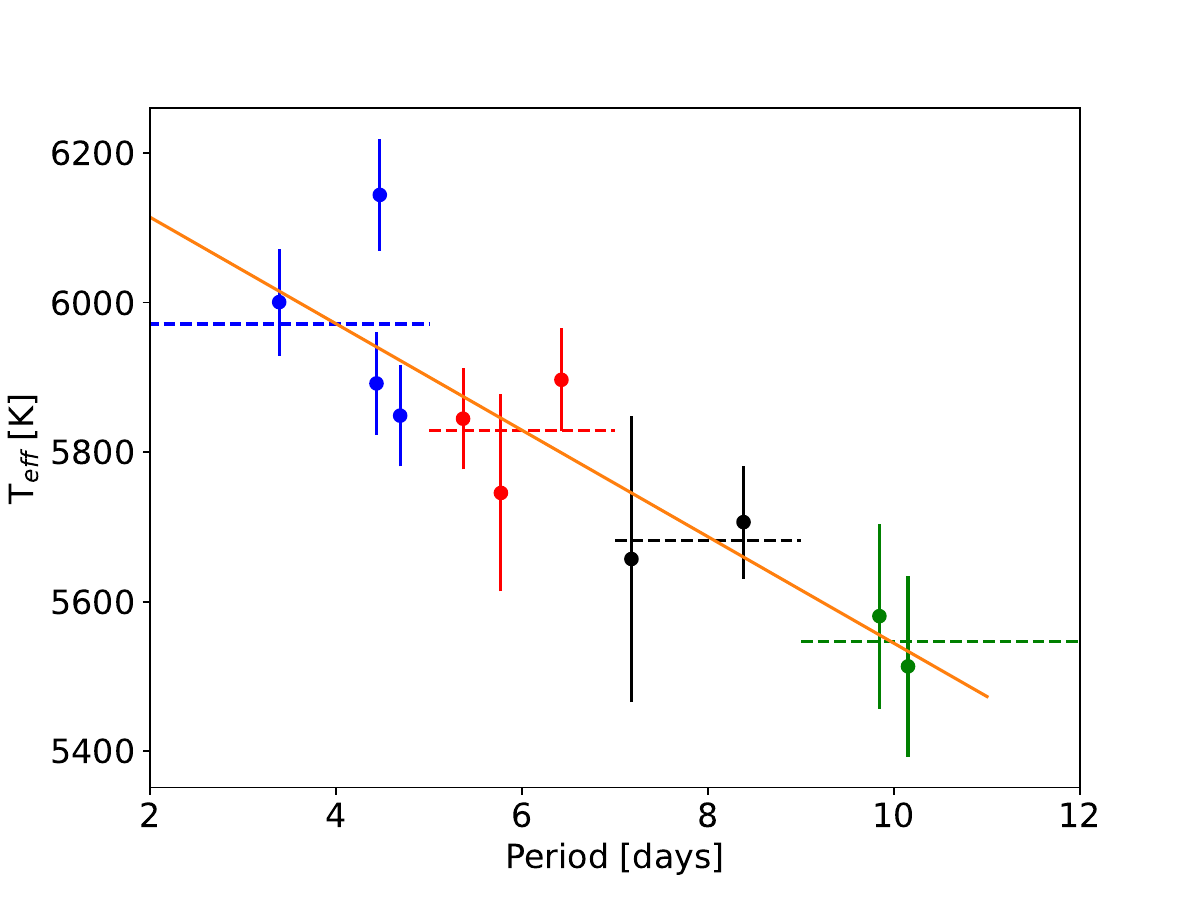}
\caption{Effective temperature versus period for CCs with full coverage 
of the pulsation cycle. The different colors indicate the different 
period bins of the temperature curve templates. The dashed lines 
indicate the average temperature for each bin in period. The orange 
line shows the linear fit to the data.}
\label{fig:teff_vs_period}
\end{center}
\end{figure}

However, the effective temperature curve templates are far from being an  
academic issue, since once a period--mean temperature relation is known, 
they can also provide the estimate of the effective temperature along the 
pulsation cycle. This is a far-reaching opportunity, since effective 
temperature estimates based on color-temperature relations are hampered 
by uncertainties affecting reddening estimates.   
This is a trivial effort in the optical regime with high-resolution 
and high signal-to-noise spectra, but it becomes more 
difficult either in the NIR, or at low spectral resolution, or in 
the very metal-poor regime due to the paucity of lines. 
Therefore, we performed a linear fit between pulsation 
period and mean effective temperature (see Fig.~\ref{fig:teff_vs_period}) 
and we found: 
\begin{equation}
T_{\rm eff} [K] = 6256.73 (\pm 89.45) - 71.23(\pm 13.29) \times P
\end{equation} 
with a standard deviation of 286 K. Once the mean $T_{\rm eff}$ is 
available, one can anchor the temperature template to the phase 
of the spectroscopic measurement and provide an estimate of the 
$T_{\rm eff}$ at any phase. 

Optical and NIR light curves of 
the calibrating Cepheids, together with effective temperature and 
radial velocity curves, can be adopted to perform a detailed 
comparison with nonlinear, convective hydrodynamical models 
of classical Cepheids. Dating back to \citet{Nataleetal2008}, 
it has been found that the simultaneous fit of both luminosity 
and radial-velocity variations provides solid constraints on the 
physical assumptions adopted to build pulsation models \citep{marconi13a}. 
Moreover, the use of the effective temperature curves (shapes and amplitudes), 
covering a broad range of pulsation periods, brings forward the opportunity 
to constrain, on a quantitative basis, the efficiency of the convective 
transport over the entire pulsation cycle.

% editor 20 
%%%%%%%%%%%%%%%%%%%%%%%%%%%%%%%%%%%%%%%%%%%%%%%%%%%%%%%%%%%%%%%%%%%%%%%%%%%%%%%%%%%%%%%%%%%%%%%%%%%%%%%
\subsection{Metallicity distribution in the Galactic thin disk} 
%===============================================================================

In discussing the metallicity distribution of Cepheids, we are following 
a series of papers 
\citep{romaniello08, pedicelli10, Lemasleetal2013,Lemasleetal2017, Innoetal2019,
genovali2013,genovali2014,daSilvaetal2016,proxauf2018, dasilva2022}
focused on the metallicity distribution of CCs. We are following these 
papers because they rely on the same strategy for the estimate of the atmospheric parameters and of the abundances. 

% editor 19 bis 
Spectroscopic abundances discussed in the following rely on 1D LTE
analysis. These investigations are based on different sets of
1D LTE atmosphere model provided by 
\citet{CastelliKurucz2004}, and by \citet{gustafsson08}. 
Theory and observations suggest that this approach
is prone to systematics when compared with 1D and/or 3D NLTE analysis. This
effect becomes larger when moving from dwarfs to giants and from metal-rich to
more metal-poor stellar populations 
\citep{thevenin1999, idiart2000, bergemann2012, hansen2013, fabrizio2021}. 
Different lines of the same element have different NLTE corrections 
\citep{collet2005, merle2011} 
and neutral lines are more affected than ionized lines. Theory and 
observations indicate that \ion{Fe}{II} lines are minimally affected 
by NLTE effects \citep{lind2012,hansen2013}. 
Moreover, the forbidden [\ion{O}{I}] line at 6300.3 \AA\ is not subject 
to any NLTE effect in cool stars 
\citep{kiselman2001, tautvaivsiene2015} 
and the same outcome applies to the \ion{S}{I} triplet at 6757 \AA\ as thoroughly 
discussed by \citet{duffau2017}. Fortunately, CCs are GK-type giants and they display 
in high resolution and high signal-to-noise ratio spectra from several to 
a few dozen of FeII lines. Moreover, the \ion{S}{I} triplet is unblended and relatively strong 
and the quoted forbidden line can be measured over a significant fraction of Galactic CCs 
\citep{dasilva2023}. The Galactic metallicity gradients based on other elements should 
be cautiously treated until the 1D and the 3D NLTE analysis will be performed.

%%%%%%%%%%%%%%%%%%%%%%%%%%%%%%%%%%%%%%%%%%%%%%%%%%%%%%%%%%%%%%%%%%%%%%%%%%%%%%%%%%%%%
% 			fig 25
%%%%%%%%%%%%%%%%%%%%%%%%%%%%%%%%%%%%%%%%%%%%%%%%%%%%%%%%%%%%%%%%%%%%%%%%%%%%%%%%%%%%%
\begin{figure}[htbp]
\centering
\begin{minipage}[t]{\textwidth}
\centering
\resizebox{\hsize}{!}{\includegraphics{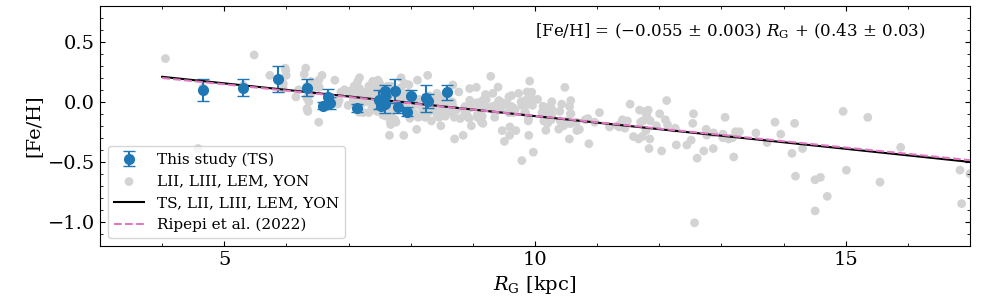}}
\end{minipage} \\
\begin{minipage}[t]{\textwidth}
\centering
\resizebox{\hsize}{!}{\includegraphics{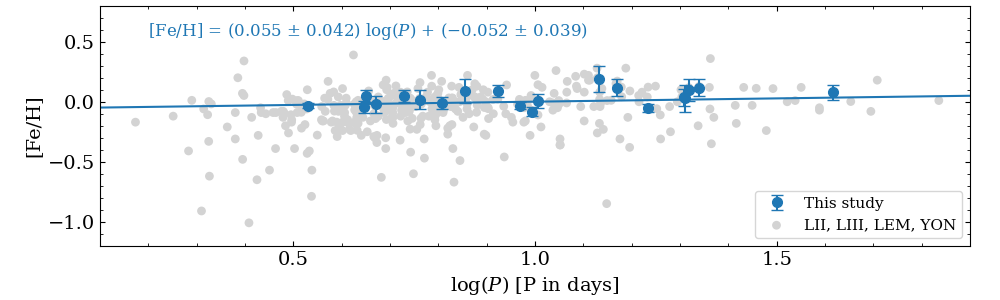}}
\end{minipage}
\caption{Top: Iron abundances as a function of the Galactocentric distance for a sample of 20 calibrating Cepheids (blue circles) compared with results from the literature (gray circles): \citet[][LII]{luck2011b}; \citet[][LIII]{luck2011a}; \citet[][LEM]{Lemasleetal2013}; \citet[][YON]{Yongetal2006}. A linear regression (solid black line plus equation) fitted to the entire sample is compared with the radial gradient provided by \citet{Ripepietal2022} (dashed magenta line). The latter was artificially shifted in order to coincide with the current radial gradient at ${\rm R}_G$ = 10~kpc. The ${\rm R}_G$ values are from 
\citet{genovali2014}. Bottom: same as the top, but as a function of the logarithmic pulsation period. 
The solid blue line plus the equation show the linear regression fitted to the current sample. Image reproduced with permission from \citet[][Fig.~10]{dasilva2022}, copyright by ESO.}
\label{figure:feh_period}
\end{figure}

\subsection{The iron distribution}

During the last few years, several investigations have addressed the open
problem concerning the space and age dependence of the metallicity 
% editor 20 
gradient in the Galactic thin disk. This issue has been studied not only from the 
empirical \citep[see, e.g.,][]{Maciel2003,Nordstrom2004,Henry2010,yong2012}
but also from the theoretical point of view. In particular, the role that 
different stellar tracers can play in constraining the chemical tagging 
in spatial distribution and in age distribution 
\citep{freeman2002}.

In order to investigate the iron distribution of the calibrating Cepheids (blue circles) 
provided by \citet{dasilva2022}, the top panel of Fig.~\ref{figure:feh_period} 
shows the individual iron abundances  as a function of Galactocentric 
distance $R_G$. Note that  \citet{dasilva2022} defined calibrating 
CCs only variables whose entire pulsation cycle is covered by high-resolution spectra. The gray circles display similar Cepheid abundances, 
but only for CCs available in the literature. A linear fit over 
the entire sample gives the following relation:
\begin{equation}
{\rm [Fe/H]} = (-0.055 \pm 0.003)\, R_G + (0.43 \pm 0.03)
\end{equation}
The current slope agrees quite well with similar estimates available in the literature and, 
in particular, with the recent estimate of the iron radial gradient (dashed line) 
provided by \citet{Ripepietal2022}. To overcome variations in the zero-point, 
mainly introduced by the sampling of innermost and outermost Cepheids, the gradient from 
\citet{Ripepietal2022} was artificially shifted, so as to coincide with the current radial 
gradient at $R_G = 10$~kpc.

The bottom panel of Fig.~\ref{figure:feh_period} shows the iron abundance as a function of the 
logarithmic pulsation period. Data plotted in this panel agree quite well with similar estimates 
available in the literature. Moreover, they do not show any significant variation in iron abundance 
when moving from young (long-period) to less young (short-period) classical Cepheids.

%_______________________________________________________________________
\subsection{Light metals and $\alpha$-element distribution}
%%%%%%%%%%%%%%%%%%%%%%%%%%%%%%%%%%%%%%%%%%%%%%%%%%%%%%%%%%%%%%%%%%%%%%%%%%%%%%%%%%%%%
% 			fig 26
%%%%%%%%%%%%%%%%%%%%%%%%%%%%%%%%%%%%%%%%%%%%%%%%%%%%%%%%%%%%%%%%%%%%%%%%%%%%%%%%%%%%%
\begin{figure}[htbp]
\centering
\includegraphics[width=0.66\columnwidth]{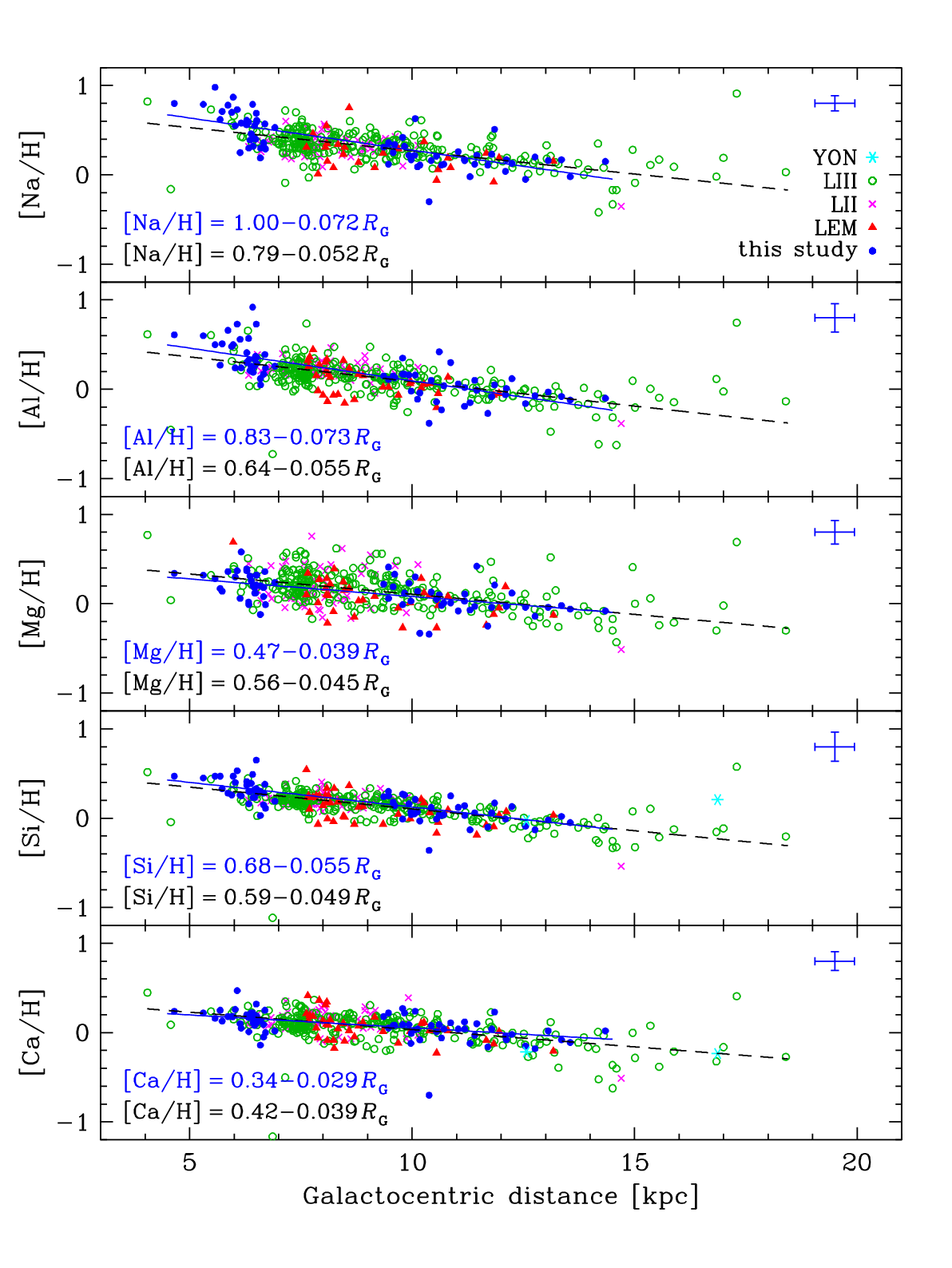}
% editor 21 
\caption{From top to bottom: Abundances for CCs of Na, Al, and $\alpha$ elements 
(Mg, Si, Ca) as a function of Galactocentric distance. The abundances by G15 
(filled blue circles) are compared with those of
\citet[][YON, cyan asterisks]{Yongetal2006}, \citet[][LII, magenta crosses]{luck2011b},
\citet[][LIII, open green circles]{luck2011a} and \citet[][LEM, 
red triangles)]{Lemasleetal2013}. The solid and the dashed lines display the 
linear regressions to only G15 data and to the entire sample. 
Image reproduced with permission from \citet[][Fig.~2]{genovali2015}, 
copyright by ESO.}
\label{xh_Gdist}
\end{figure}
%___________________________________________________________________________________

This section deals with  light metals and $\alpha$-element gradients 
traced by CCs across the Galactic thin disk.  Fig.~\ref{xh_Gdist} shows 
the abundances of Na, Al and of three $\alpha$-elements (Mg, Si, Ca) 
as a function of the Galactocentric distance. 
The five investigated elements display well-defined radial gradients. 
The difference in standard deviations is mainly correlated with the number 
of lines adopted to estimate the abundances. The dispersion of the 
Mg gradient is almost a factor of two larger than the Si gradient, because
the former is based on a single line, while the latter on more than a dozen.  
The radial gradients of the five investigated elements attain, within the errors, 
quite similar slopes as originally found by \citet{lemasle2007,Lemasleetal2013} and 
by LII+LIII. The above result becomes even more compelling if we consider the 
similarity with the iron radial gradient ($-0.060 \pm 0.002\,\mathrm{dex\ kpc^{-1}}$)  
found by G14 using the same CCs. The current radial gradients seem to show 
a ﬂattening for distances greater than $\sim$13 kpc. This supports a 
similar trend, but based on open clusters \citep{Carraro2007b,yong2012,magrini2023}. 
However, firm conclusions are hampered by the increased spread in abundance 
and by the scarcity of outer-disk CCs.

One of the most relevant issue in dealing with the chemical enrichment 
history of the thin disk is the age dependence. Indeed, chemo-dynamical 
models suggest a significant flattening of the metallicity gradients 
for ages older than 1-3 Gyr \citep{minchev2014,kubryk2015}. 
This means a steady decrease in the metallicity with increasing age. 
The same models also predict a strong dependence on stellar migrations.
To constrain the age dependence of the metallicity gradients, 
Fig.~2 in \citet{genovali2015} shows the same elemental abundances of Fig.~\ref{xh_Gdist}, 
but as a function of the logarithmic period. Data plotted in their figure 
show a well-defined positive gradient for increasing pulsation period. 
The slopes of the $\alpha$-elements attain similar values, while for 
Na and Al they are systematically larger (see labelled values). This 
evidence further supports the hydrostatic nature 
\citep{arnett1979,limongi2018}  
of both Na and Al, owing to their steady increase 
correlated with the pulsation period (stellar mass).

Data plotted in Fig.~5 of \citet{genovali2015} show that the ratio  
[element/Fe] is, on average, quite ﬂat across the entire thin disk. 
They performed a bi-weight linear least squares fit over the entire sample, 
and they found evidence of a positive slope for Ca. This finding brings 
forward a few interesting consequences. 

{\em i)}--Chemical enrichment history-- Iron and the other 
five elements have had quite similar chemical enrichment histories 
across the Galactic thin disk. 
Moreover, the quoted ratios have small standard deviations suggesting that 
it is dominated by measurement errors.  The [$\alpha$/Fe] ratios are typically 
considered as tracers of the star-formation activity and the lack of 
a clear negative gradient between the high (inner) and the low (outer) 
density regions of the thin disk suggests that either the iron and the 
quoted elements are produced by the same polluters, or the [$\alpha$/Fe] 
radial gradients are affected by other parameters, such as radial migration 
of stars or radial gas ﬂows.

{\em ii)}--Radial gradients-- The occurrence of a mild positive gradient in [Ca/H] as a function 
of period, and of a negative gradient of [Ca/H] as a function of Galactocentric 
distance are not correlated, since CCs located in the outer disk 
have a canonical period distribution (mostly between 2 and 20 days).
The negative trend shown by Ca when moving from older to younger 
CCs remains unclear.

In order to further investigate the ratio between hydrostatic and explosive elements,
the left panel of Fig.~11 in \citet{genovali2015} shows [Mg/Ca] as a function of the 
iron abundance. The CCs display typical solar abundances for iron abundances 
that are more metal-poor than the Sun, but the spread increases to 0.5--0.7 dex
for iron abundances more metal-rich than the Sun. On the other hand, the abundance
ratio of thin- and thick-disk dwarfs (B05, M14) display the typical decrease 
when moving from metal-poor to more metal-rich objects. The metal-rich regime
shows a similar trend, and indeed the metal-rich dwarfs (green diamonds, B05) 
display a mild increase in the [Mg/Ca] abundance ratio for [Fe/H]$\ge$0.15.

The reasons for the above difference are not clear, therefore G15 searched for 
a possible radial dependence. Data plotted in the middle panel of their Fig.~11
display that a significant fraction of the spread in [Mg/Ca] abundance ($\sim$0.7 dex) 
showed by metal-rich CCs is evident among objects placed either 
across or inside the solar circle ($\sim$8 kpc). 
However, the abundance ratio is quite constant across the disk regions covered 
by CCs and dwarfs. This trend is slightly counter intuitive, since the increase 
in iron, typical of the innermost disk regions, should also be followed by a steady 
increase in Ca, since they are explosive elements. Namely, there is a steady decrease 
in the [Mg/Ca] abundance ratio when approaching the innermost disk regions. 

The right panel of their Fig.~11 shows the [Mg/Ca] abundance ratio of CCs 
as a function of the logarithmic period. Data plotted in this figure show no clear trend. 
The spread in the [Mg/Ca] abundance ratio is almost constant over the entire period 
range, thus suggesting that there is no solid evidence of a dependence on stellar mass.

The scientific impact of the $\alpha$-element abundances on the 
chemical enrichment of the Galactic thin disk, and the difference between 
explosive and hydrostatic elements have been recently addressed by 
\citet{trentin2023}.

%_______________________________________________________________________
\subsection{Neutron capture distribution}

%%%%%%%%%%%%%%%%%%%%%%%%%%%%%%%%%%%%%%%%%%%%%%%%%%%%%%%%%%%%%%%%%%%%%%%%%%%%%%%%%%%%%
% 			fig 27
%%%%%%%%%%%%%%%%%%%%%%%%%%%%%%%%%%%%%%%%%%%%%%%%%%%%%%%%%%%%%%%%%%%%%%%%%%%%%%%%%%%%%
\begin{figure}[htbp]
\centering
\includegraphics[width=0.66\columnwidth]{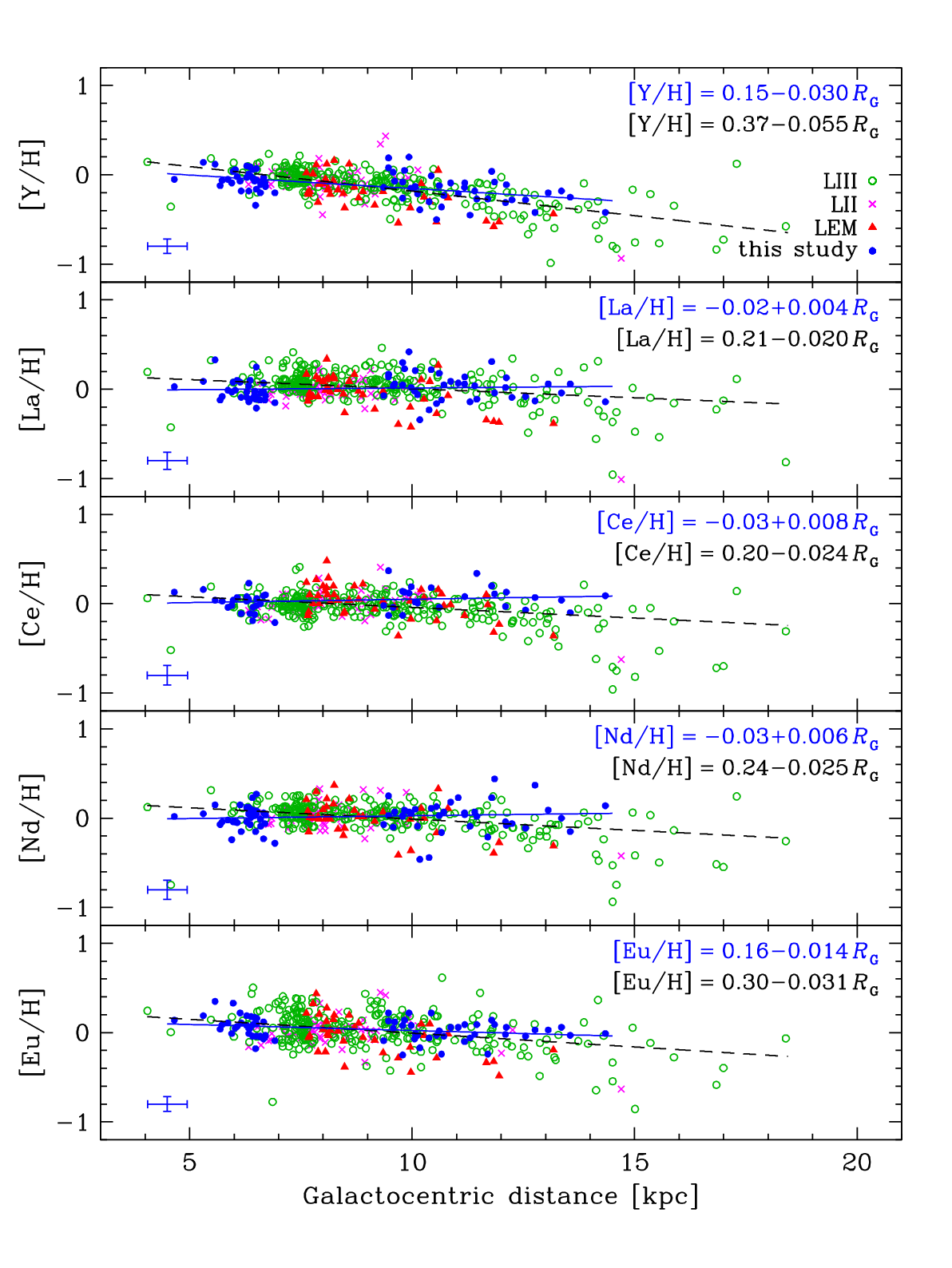}
\caption{From top to bottom: Abundances for CCs of neutron capture elements 
(Y, La, Ce, Nd, Eu) as a function of the Galactocentric distance. 
Abundances provided by \citet[][filled blue circles]{daSilvaetal2016}  
are compared with those of LII \citep[magenta crosses,][]{luck2011b}, 
LIII \citep[open green circles,][]{luck2011a}, and LEM 
\citep[red triangles,][]{Lemasleetal2013}. 
The blue solid line shows the linear regression of our Cepheid sample, while 
the black dashed line the linear regression of the entire Cepheid sample. The blue 
error bars display the mean spectroscopic error of the current sample. 
The abundances available in the literature have similar errors. 
Adapted from \citet[][Fig.~2]{daSilvaetal2016}. Image credit: R. da Silva.}
\label{xh_Gdist_rs}
\end{figure}
%_______________________________________________________________________________

In the following, we discuss the radial gradients of four s-process 
(Y, La, Ce, Nd) elements and a single r-process element (Eu). The 
n-capture elements can be split according to solar system abundances 
in pure s-process, pure r-process, and mixed-percentage isotopes. 
Among them, Eu is a pure r-process element, since the r-fraction 
abundance is 97\% \citep{burris2000,simmerer2004}, while for 
the s-process elements the s-fraction ranges from roughly 50\% 
\citep[Nd, 58\%, ][]{sneden2008} to more than 70\% 
(Y, 72\%; La, 75\%; Ce, 81\%). The quoted s- and r-fraction 
abundances should be treated cautiously since \citet{bisterzo2011}, 
by using a different approach, found similar fractions 
for Eu (94\%), Nd (52\%), La (71\%), and Ce (81\%), but 
a significantly larger s-fraction for Y (92\%).

Figure~\ref{xh_Gdist_rs} shows the abundance of the five selected 
neutron capture elements as a function of the Galactocentric 
distance. Data plotted in this figure bring out several interesting 
results. 

{\em i)}--Neutron-capture radial gradients-- The five investigated neutron-capture elements 
display well-defined radial gradients. This evidence, combined with 
similar results concerning iron peak and $\alpha$-elements  (LII, LIII) 
further indicates that young stellar tracers show radial gradients 
across the Galactic thin disk. There is only one exception,  
\citet{andrievsky2014} and \citet{martin2015} found that Ba does 
not display a radial gradient. However, Ba abundances in classical 
Cepheids are affected by severe limits. In particular, \citet{luck2014} 
noted that strong BaII lines are affected by line-formation effects, 
while \citet{andrievsky2013} discussed isotopic shift effects. 
The reason for the lack of a Ba gradient remains unclear.

{\em ii)}--Slopes of the radial gradients-- The slopes are quite similar, 
on average $-0.025 \pm 0.004\,\mathrm{dex\ kpc^{-1}}$ 
for La, Ce, Nd, and Eu. The only exception is Y, for which the slope is more 
than a factor of two steeper ($-0.053 \pm 0.003\,\mathrm{dex\ kpc^{-1}}$). The current slopes 
agree quite well, within the errors, with similar estimates available in the 
literature. 

{\em iii)}--Spread of the radial gradients-- The spread of the individual 
abundances attains similar values 
across the thin disk. The outermost disk regions are an exception, since the 
spread increases for $R_G$  larger than 13 kpc. The neutron-capture elements 
display the same trend of iron and $\alpha$-element abundances. Among the 
investigated elements, Y once again seems to be an exception, since the spread 
is homogeneous across the disk regions covered by the current CC sample. 

{\em iv)}--Theory and observations-- Radial gradients for 
 La ($\sim0.020 \pm 0.003\,\mathrm{dex\ kpc^{-1}}$) and 
 for Eu ($\sim0.030 \pm 0.004\,\mathrm{dex\ kpc^{-1}}$) agree quite well with theoretical 
 predictions by \citet{Cescuttietal2007} for Galactocentric distances covering 
 the entire thin disk (4$\le R_G\le$22 kpc). They found a slope of 
 $\sim0.021\,\mathrm{dex\ kpc^{-1}}$ for La and of $\sim0.030\,\mathrm{dex\ kpc^{-1}}$ for Eu. 
 The predicted slopes become steeper for Galactocentric distances shorter 
 than 14 kpc, and shallower for distances larger than 16 kpc (see their Table 5).

%_______________________________________________________________________
\subsection{CNO distribution}

CNO abundances are a solid diagnostic to constrain evolutionary 
predictions, based on different physical assumptions, concerning the size 
of the helium core and the efficiency of the first dredge-up. The key issue is 
that evolutionary models accounting for core rotation display an enhancement 
in N that is significantly larger when compared with models neglecting rotation
and only accounting for semi-convection and/or convective core overshooting during 
central H-burning phases. This means that CNO abundances are a solid observable 
to constrain the input physics of intermediate-mass models \citep{anderson2014}. 
Even though the abundances of these three elements can play a key role to constrain 
the physics of intermediate-mass stars, we still lack accurate and homogeneous 
measurements of short and long-period Galactic and Magellanic CCs 
\citep{Andrievskyetal2005,Kovtyukhetal2005,luck2011a,romaniello2022,trentin2023},
to constrain the impact that mixing due to the first dredge-up 
(see section~C.3 in Appendix) have on the surface 
abundances of CCs. 

%_______________________________________________________________________
\subsection{A new spin on iron and $\alpha$-element abundance gradients}
In a recent investigation \citet{dasilva2023} provided 
a new homogeneous and accurate investigation of the radial gradients (iron, 
$\alpha$-elements [O, S]) across the Galactic thin disk by using CCs. 
This paper offers several new insights concerning the MW chemical 
enrichment history. The main novelties when compared with similar 
investigations in the literature are the following. 

{\em i)}--Sample selections-- 
High-resolution, high SNR spectra for CCs were secured with four different 
spectrographs, the sample includes more than 1100 spectra for 356 variables. 
For 77 CCs accurate abundances are provided for the first time. 
The heliocentric distances are mainly based on Gaia EDR3 distances
\citep{BailerJonesetal2021} complemented with NIR and MIR diagnostics 
available in the literature.
Their Galactocentric distances range from the inner disk ($R_G$=5 kpc)
to the outskirts of the disk ($R_G$=30 kpc) with a handful (eight) of 
Cepheids located at distances larger than 18 kpc. Moreover, they also 
investigated the orbital properties of their spectroscopic sample and 
they found that the bulk of them have orbits typical of thin disk stars.
The sample of open 
clusters is from \citet{randich2022} and includes 62 stellar systems 
observed by the Gaia-ESO Survey (GES) plus 18 OCs retrieved from the ESO 
archive \citep{viscasillas22}. The age of the OCs ranges from a few Myrs 
to $\sim$7~Gyr, while their Galactocentric distances range from $\sim$6 
to 21~kpc.

{\em ii)}--Homogeneous abundances-- 
Atmospheric parameters and 
% editor 22
elemental abundances over the entire spectroscopic sample were based  
on individual spectra. The abundances for each spectrum are the mean value 
computed from individual lines. The final abundances for variables with 
multiple spectra are the mean over the individual spectra.  

{\em iii)}--Spectroscopic diagnostics-- 
The O abundances are based on the 
forbidden [\ion{O}{I}] line at 6300.3~\AA, in the LTE approximation, 
while S abundances are based on the \ion{S}{I} triplet at 6757~\AA. 

\begin{figure}[htbp]
\centering
\begin{minipage}[t]{0.8\textwidth}
\centering
\resizebox{\hsize}{!}{\includegraphics{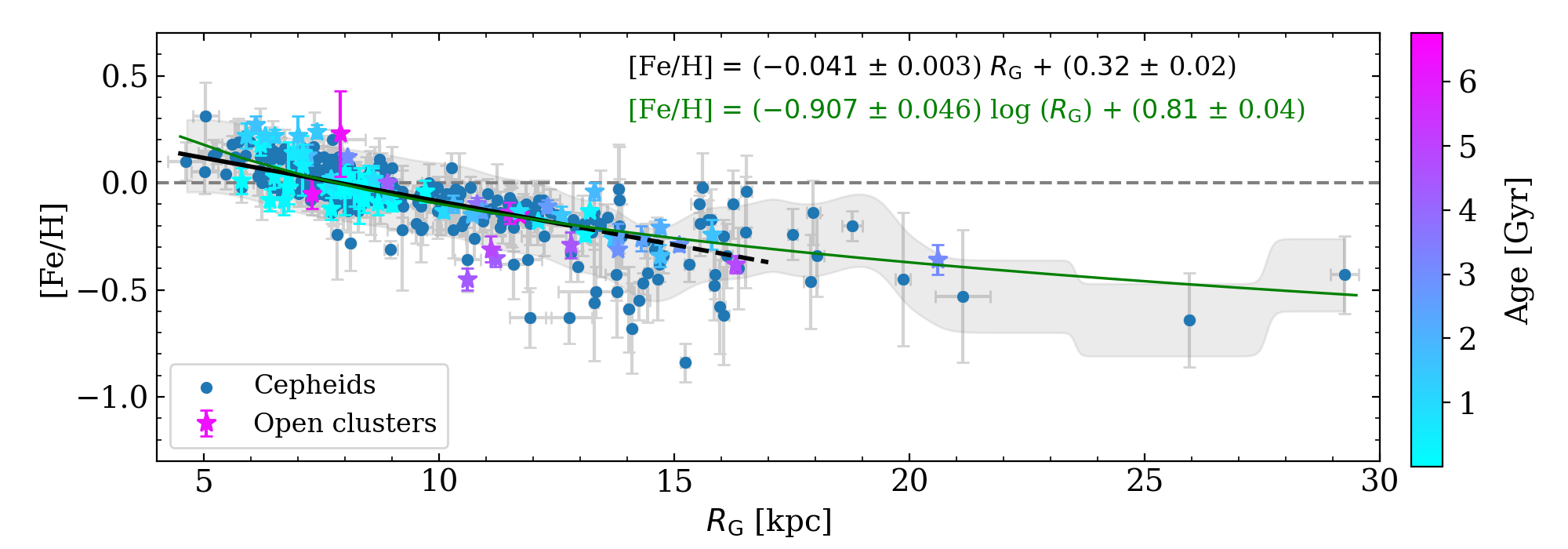}}
\end{minipage} \\
\begin{minipage}[t]{0.8\textwidth}
\centering
\resizebox{\hsize}{!}{\includegraphics{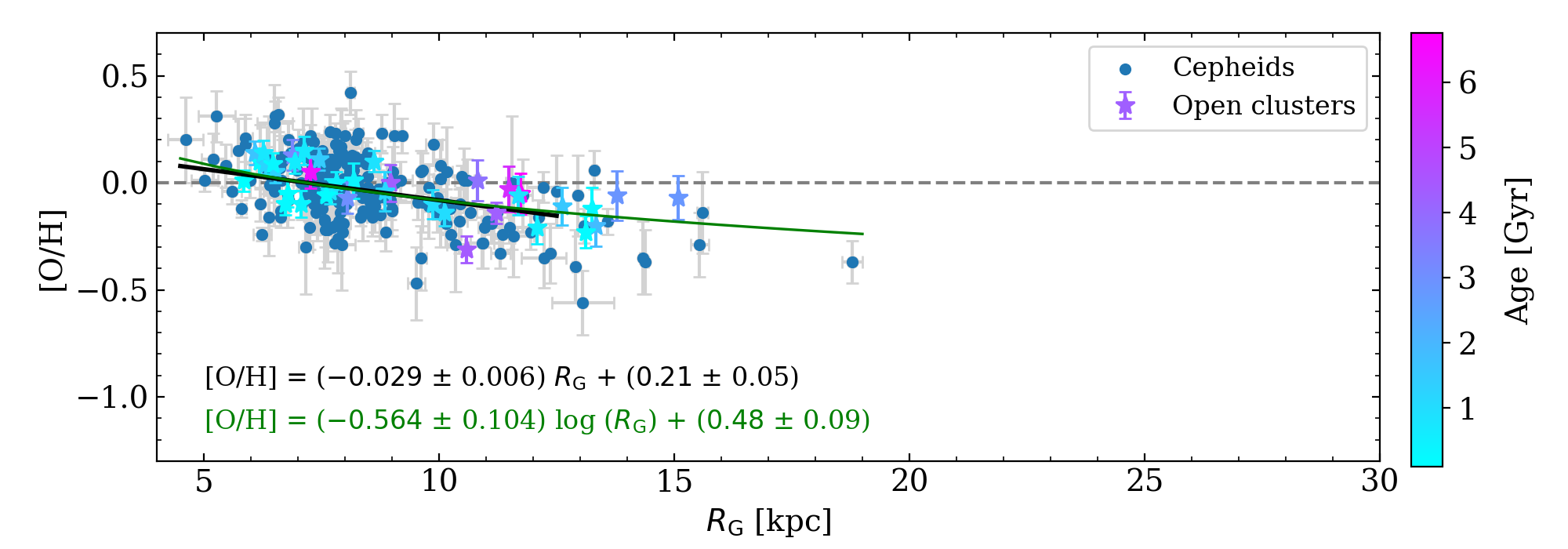}}
\end{minipage} \\
\begin{minipage}[t]{0.8\textwidth}
\centering
\resizebox{\hsize}{!}{\includegraphics{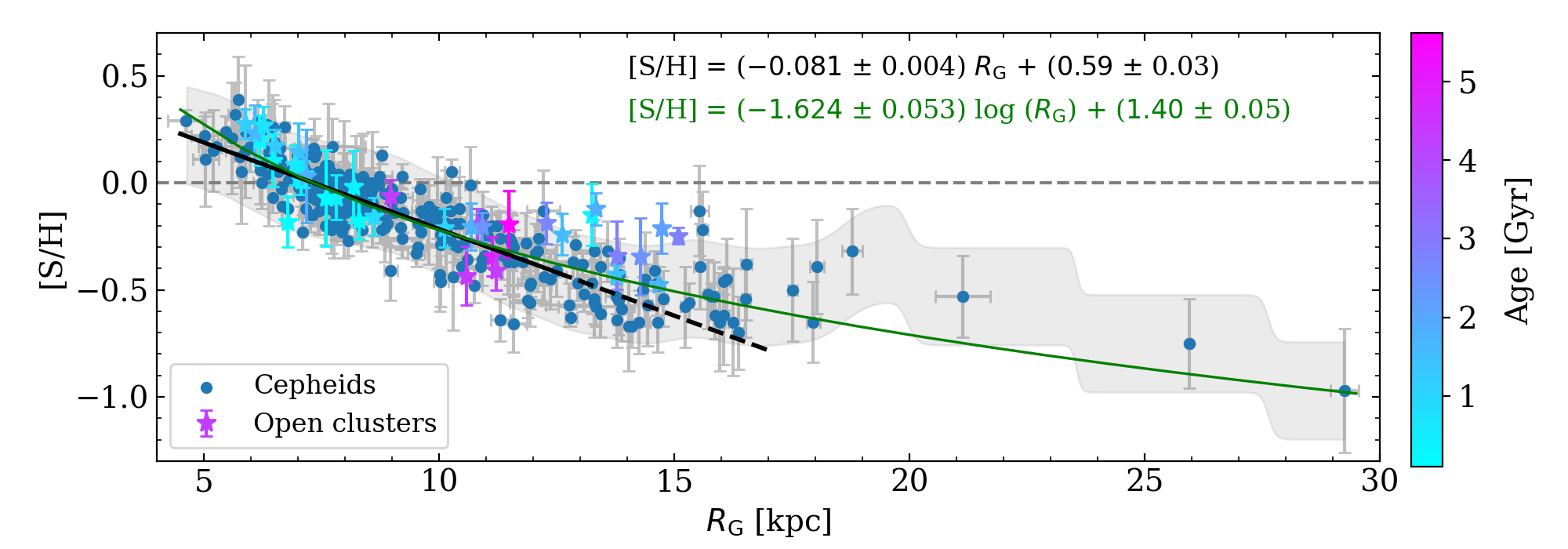}}
\end{minipage}
\caption{Top: Iron abundances for CCs as a function of the Galactocentric 
distance (kpc). CCs are marked with blue circles while OCs with colored 
symbols. Iron abundances for OCs are from \citet{magrini2023}. 
The solid black line shows the linear regression for CCs with Galactocentric 
distances ranging from 4 to 12.5~kpc, while the dashed black line is the 
extrapolation up to 17~kpc. The green line displays the logarithmic fit 
over the entire sample of CCs. The coefficients of the analytical relations 
are labelled. The shaded area shows the running average, and its standard 
deviation, weighted with a Gaussian function taking into account the errors.
Middle: same as the top, but for O abundances. 
Bottom: Same as the top, but for S abundances. 
Adapted from \citet[][Fig.~2]{dasilva2023}. Image credit: R. da Silva.
}
\label{figure:xh_rgal_open_cluster}
\end{figure}

The radial gradient for iron, oxygen and sulphur are plotted as a 
function of the Galactocentric distance in Fig.~\ref{figure:xh_rgal_open_cluster}. 
Simple inspection to the data plotted in this figure disclose several 
new findings.

{\em i)}--Linearity of the radial gradient--  
The iron radial gradient is linear for Galactocentric distances
smaller than $\sim$12 kpc, but it becomes flatter at larger distances 
($R_G\ge$15 kpc). They also found that a logarithmic fit takes into
account the linear gradient of the inner disk and for the 
flattening of the outer disk.  Their estimates of both linear and 
logarithmic fits agrees quite well with similar estimates available 
in the literature 
\citep{Carraro2007b,yong2012,donor2020,magrini2023}.
However, the iron radial gradient provided by 
\citet{dasilva2023} shows 
a well defined departure from linearity for Galactocentric distances 
larger than 12--14 kpc. A more quantitative analysis requires a larger
sample to assess whether the radial gradient shows either a change in the 
slope, or a steady variation with distance, as suggested by the logarithmic 
fit.

{\em ii)}--Comparison between CCs and OCs-- 
Data plotted in the top panel of Fig.~\ref{figure:xh_rgal_open_cluster}. 
show that CCs and OCs show similar iron radial gradients. 
Moreover, the ages of the GES OC sample cover more than one order 
of magnitude, whereas their radial distributions are similar. However, 
the flattening of the radial gradient in the outer disk is
only partially supported by OCs, since they typically have 
Galactocentric distance smaller than $\sim$15 kpc with a 
single cluster located at $\sim$20 kpc.

{\em iii)}--Oxygen radial gradient-- 
A glance at the oxygen abundances plotted in the middle panel of 
Fig.~\ref{figure:xh_rgal_open_cluster} shows that the oxygen
radial gradient is 25\% shallower than the iron gradient. CCs 
and OCs display very similar radial trends and a vanishing, if any, 
dependence on the cluster age. To overcome possible 
systematics, \citet{dasilva2023} used the same data provided 
by \citet{magrini2023} and performed an independent fit following 
the same approach adopted for CCs. They found that the two 
gradients, within the errors, agree quite well with each other. 
The comparison with similar estimates available in the literature 
based on CCs indicates that the current slope is shallower 
\citep{luck2011a, trentin2023}. 
However, the comparison should be used carefully, since the 
difference could be due to the different lines adopted for 
measuring oxygen abundances, and to the spectroscopic analysis.

{\em iv)}--Slope of the sulfur radial gradient-- 
Data plotted in the
bottom panel of  Fig.~\ref{figure:xh_rgal_open_cluster} show that 
the linear fit of the sulfur radial gradient is twice steeper than 
the slope of the iron gradient, while the difference in the logarithmic 
fit is $\approx$0.7~dex\,($\log{\rm kpc})^{-1}$. It is noteworthy that 
the sulfur radial gradients for both CCs and OCs agree quite well. 
Indeed, the zero-points and the slopes of the linear fits attain very 
similar values (see labels). This circumstantial evidence brings forward 
an $\alpha$-element with a radial gradient steeper than the iron gradient, 
and points toward a more significant role played by massive stars in
iron and sulfur chemical enrichment. 

%%%%%%%%%%%%%%%%%%%%%%%%%%%%%%%%%%%%%%%%%%%%%%%%%%%%%%%%%%%%%%%%%%%%%%%%%%%%%%%%%%%%%
% 			fig 28
%%%%%%%%%%%%%%%%%%%%%%%%%%%%%%%%%%%%%%%%%%%%%%%%%%%%%%%%%%%%%%%%%%%%%%%%%%%%%%%%%%%%%
%-----------------------------------------------------------------
%-----------------------------------------------------------------

The same authors also performed a detailed comparison between observations 
and chemical evolution models of the Galactic thin disk. They adopted the 
best-fit model recently provided by \citet{palla2020} and they found that
the use of a constant star formation efficiency for $R_G\ge12$~kpc takes into 
account the flattening observed in iron and in $\alpha$-element radial gradients.
Moreover, they found that inside the solar circle the current [S/H] and [S/Fe] 
gradients are well reproduced by canonical yield prescriptions by 
\citet{kobayashi2006} and by \citet{kobayashi2011}. The same comparison 
indicates that a decrease of a factor of four in the current S predictions 
is required to explain the flattening in the outermost regions. This means 
a relevant decrease in the S yields in the more metal-poor regime. 

Finally, \citet{matsunaga2023} took advantage of high resolution, 
near infrared spectra, covering the $Y$,$J$-bands, collected with WINERED 
at the Magellan telescope to investigate the iron abundance of 16 CCs 
located in the inner disk ($R_G$=3--5.6 kpc). They found that the iron 
radial gradient in this region is either shallower or even flat (see 
their Fig.~2 and also \citealt{kovtyukh2022}).
 
%_______________________________________________________________________________
\section{Omne ignoto pro magnifico} 

The next two subsections are focussed on open problems concerning Cepheids 
with possible/challenging near-term solutions.
%_______________________________________________________________________________
\subsection{Open problems with possible near future solutions}

The most interesting open problems at the crossroad between theory and 
observations appear to be the following. 

{\em i)}--Nonlinear, convective models for long-period TIIC Cepheids--
The current generation of nonlinear, convective hydrodynamical models 
of variable stars shacked off their speculative nature and are 
providing solid predictions of the current properties of radial 
variables. However, we are still missing detailed pulsation 
predictions concerning long-period W Virginis and RV Tauri.

{\em ii)}--Metallicity dependence of both PL, PLC and PW relations-- 
This is a sort of message in the bottle for future spectroscopic 
surveys. Gaia and long-term ground-based variability surveys 
(ASAS-SN\footnote{\url{https://asas-sn.osu.edu/}}, 
OGLE-IV\footnote{\url{http://ogle.astrouw.edu.pl/}}, 
ZTF\footnote{\url{https://www.ztf.caltech.edu/}}, 
PTF\footnote{\url{https://www.ptf.caltech.edu/page/ptf_about}}, 
KISOGP \citep[][]{takase1984} will provide solid identification 
of thousands of Galactic Cepheids. In a few years, Gaia will 
also provide accurate geometrical distances for a good fraction of them.
%ref_2_11
Moreover, the use of high resolution images collected with NIRCam on board 
James Webb Space Telescope\footnote{https://webb.nasa.gov} is providing 
the unique opportunity to test the crowding affecting extragalactic Cepheids 
and to validate earlier results based on HST photometry \citep{riess2023,riess2024} 
and in the near future to improve the accuracy of the extragalactic 
Population I and Population II distance scale \citep[][and references therein]{madore2023}.

Moreover, the unique opportunity of using 
the panoramic imaging capabilities of NIRCam on board James Webb Space 
Telescope\footnote{\url{https://webb.nasa.gov}}
will provide accurate NIR and MIR photometry to significantly improve 
the accuracy of the extragalactic Population I and Population II 
distance scale \citep[][and references therein]{madore2023}.

The near-future challenge for ongoing and forthcoming 
optical (SDSS~V\footnote{\url{https://www.sdss.org/}}, 
GALAH\footnote{\url{https://www.galah-survey.org/}}, 
WEAVE\footnote{\url{https://ingconfluence.ing.iac.es/confluence/display/WEAV/The+WEAVE+Project}}, 
4MOST\footnote{\url{https://www.4most.eu/cms/home/}}, 
MAVIS@VLT \citep{monty2021}\footnote{\url{https://mavis-ao.org/mavis/}}) and NIR 
(MOONS@VLT\footnote{\url{https://vltmoons.org/}}, 
APOGEE\footnote{\url{https://www.sdss.org/instruments/apogee-spectrographs/}}, 
ERIS@VLT\footnote{\url{https://www.eso.org/public/teles-instr/paranal-observatory/vlt/vlt-instr/eris/}}, 
WINERED@Magellan\footnote{\url{http://lihweb.kyoto-su.ac.jp/WINERED/overview_winered.html}})
spectroscopic surveys is to provide 
accurate and homogeneous abundances for iron peak, $\alpha$ and 
neutron capture elements. All of this will bring forward reliable 
predictionss for the impact of metallicity on pulsation properties 
and on the different diagnostics to estimate individual distances. 
In his seminal review on the absolute magnitude of CCs Bob Kraft 
suggested, on the basis of preliminary evidence, that MC and MW 
open clusters hosting CCs should cover a range in chemical 
compositions \citep{kraft1963}. In particular, more 
than half century ago, he argued 

{\em Would we then be justified in assuming universality of the 
period-luminosity law even among so-called classical Cepheids?}

This {\it vexata questio} has not been solved yet, but the 
solution, sixty years later, seems just around the corner.  

{\em iii)}--Hertzsprung et similia-- 
Homogeneous and accurate photometric 
data collected by microlensing (MACHO, EROS, OGLE) experiments
paved the way for a new golden age for the physics  of stellar 
oscillations and for the use of variable stars as stellar tracers. 
These surveys collected data mainly in a single band, plus an 
ancillary secondary band. The same area is going to experience a 
quantum jump in a couple of years thanks to the Vera C.\ Rubin Observatory 
and the LSST survey. This survey is going to observe the entire 
southern sky every three nights in six different photometric bands
($u,g,r,i,z,y$). In a time interval of ten years, the VRO will 
collect on average more than eighty phase point per objects in each band
down to limiting magnitudes of 24--27. This will provide a complete census 
of helium-burning variables across the Local Group, and of 
CCs and Miras across the Local Volume. There will be the unique 
opportunity to study long-term secondary modulations 
(mixed-mode, Blazhko), and in particular, pulsation phenomena 
showing amplitude modulations over a broad range in wavelengths.
In this context, the spectroscopic follow-up 
will play a crucial role not only to provide radial velocity 
measurements (low spectral resolution) and individual elemental 
abundances (medium/high spectral resolution), but also to further 
constrain the physics of variable stars. The most recent investigation 
concerning the phase lag between light and radial velocity curves in 
CCs dates back to \citet{szabo2007} and is based on a few dozen 
Galactic Cepheids. More accurate data can shed new light on the role 
that the different ionization regions play in shaping the light 
curves.

{\em iv)}--Homogeneous spectroscopic abundances -- 
Radial gradients appear to show quite similar trends when 
moving from the inner to the outer disk. Typically, they 
agree quite well with predictions based on chemical 
and on chemo-dynamical evolution models. There are a couple of 
exceptions like Ca among the $\alpha$-elements and Y among 
the neutron capture elements. The current empirical and 
theoretical evidence do not allow us to understand whether this 
is a limit in the atomic data we are using for these elements, 
or whether they are suggesting different paths in the 
chemical enrichment of the thin disk. Homogeneous, high-resolution
spectra based on a sizable number of lines can shed new 
light on the nature or nurture of the quoted elements.

%editor 19 
It is worth mentioning that the current abundances analyses are mainly based
on 1D stellar atmosphere models in LTE approximations, but the transition to
spectroscopic abundances based on 3D stellar atmosphere models in NLTE
approximations and with time-dependent convective energy transport is a
scientific goal, achievable in the next few years 
\citep{magic2015,chiavassa2018,gerber2023}. 
This transition is more
difficult in dealing with Cepheids, since they experience along the
pulsation cycle variations in effective temperature of several hundred
degrees, in surface gravity of half dex and almost a factor two in 
micro-turbulent velocity (see section~\ref{teff_logg_vt}). Indeed, in a
recent investigation \cite{vasilyev2019}, by using radiation hydrodynamics
simulations of Cepheid atmospheres (2D NLTE line formation), found a
significant variation in the oxygen abundance (\ion{O}{I} triplet at 7774 \AA) along
the pulsation cycle.

Moreover, large and homogeneous open cluster parameters 
(distance, reddening, proper motion, chemical composition) 
are required to trace the age dependence of the radial 
gradients across the thin disk. This seems well within the 
reach of Gaia and near future-optical/NIR spectroscopic 
survey.

%_______________________________________________________________________________
\subsection{Open problems with challenging near future solutions}\label{sec:open_problems}

Thinking about significant near-term achievements in the field of radial variables, 
some of the most interesting open problems appear to be the following. 

{\em i)}--To loop or not to loop?-- 
We are still missing the micro and 
macro physics driving the morphology of the blue loops. The same 
drawback applies to the efficiency of the mass loss during the 
red giant branch phase. Therefore, the comparison between 
predicted and observed period distributions does depend on the 
physical assumptions adopted to deal with these phenomena. 

{\em ii)}--What is the physical mechanism driving the occurrence of bump Cepheids?--
In spite of the paramount theoretical and 
observational effort,  we still lack {\it ab initio} calculations 
to explain the occurrence of secondary features (bumps, dips) 
along the pulsation cycle.

{\em iii)}--What is the physical mechanism driving the occurrence of mixed-mode Cepheids? Is there any connection with Blazhko RR Lyrae?-- 
The same outcome applies to the physical 
mechanism(s) driving the occurrence of secondary modulations. 
In particular, it is not clear whether the secondary modulations 
present in RR Lyrae and in Cepheids have a common origin. 

{\em iv)}--Reddening estimates-- 
Accurate estimates of the extinction in low and high reddened 
regions has been one the toughest problems in quantitative 
astrophysics. Even in the era of Gaia, as well as of large 
spectroscopic surveys, we are dealing with this long-standing 
problem. A significant and quantitative improvement has been 
provided by diffuse interstellar bands 
\citep{munari2008,zhao2021a,saydjari2023}, 
but this approach requires at least medium resolution spectra. 
The photometric diagnostics based either on variables stars, 
\citep{sturch1978} or on narrow-band photometry (Str\"{o}mgren) 
require large amount of telescope time.  A diagnostic that is 
solid and easy to use is still lacking.

%;_________________________________________________________________________________________
\section{Summary and future developments}

This review deals with different aspects of Cepheid variable stars. 
the content can be split into five different broad topics. 

{\em Phenomenology of variable stars}-- Ongoing space (Gaia, WISE) 
and ground-based long-term variability surveys (OGLE~IV) provided 
the opportunity to investigate the phenomenology of variable 
stars. In particular, the occurrence of single and mixed-mode 
variables in the MW and in the Magellanic Clouds. This means 
a detailed investigation of the diagnostics adopted for 
the identification (periods, pulsation amplitudes, shape of the 
light curves) of the pulsation mode. Moreover, the opportunity 
to use homogeneous data for the three different stellar systems, 
characterized by different chemical enrichment histories, allowed us 
to investigate the dependence of the pulsation parameters on the 
metal content. The circumstantial evidence concerning the 
pulsation parameters are the following. 

The period distribution of CCs steadily moves towards 
shorter periods when moving from more metal-rich (MW) to more 
metal-poor (SMC) stellar systems. Moreover, the mean colors 
of CCs become systematically bluer for a decrease in the 
metal content. 

The luminosity amplitude for FU and FO CCs display 
well-defined primary and secondary peaks moving towards 
shorter periods for a decrease in the metal content. 

The use of different populations ratios among 
single-mode variables and/or among mixed-mode variables 
provide an unique opportunity to analyze, on a quantitative 
basis,  pulsation (topology of the instability strip)  
and evolutionary (the stars formation episodes) properties, 
together with the chemical enrichment histories of the 
different stellar systems. 

{\em i)}--The population ratio 
between second overtone and fundamental 
CCs decreases for an increase in metallicity (0.03 [SMC] vs 0.01 [LMC]) 
and it is vanishing for the MW.
The fraction of mixed-mode variables, instead, 
does not seem to depend on metallicity (0.11$\pm$0.01, [SMC]; 
0.17$\pm$0.01, [LMC]; 0.14$\pm$0.01, [MW]).

{\em ii)}--The population ratio between first overtone and fundamental ACs 
is clearly higher in the MW (0.76$\pm$0.11) than in the SMC 
(0.54$\pm$0.10) and in the LMC (0.44$\pm$0.08). However, 
the MC sample is limited and affected by large Poissonian 
uncertainties.

{\em iii)}--The relative number of different subgroups of 
TIICs does display a positive trend with metallicity for BLHer
(0.28$\pm$0.01, [SMC]; 0.34$\pm$0.01, [LMC]; 0.38$\pm$0.02, [MW])
and a negative one for WVir 
(0.51$\pm$0.01, [SMC]; 0.46$\pm$0.01, [LMC]; 0.43$\pm$0.02, [MW]). 
On the other hand, the fraction of 
RVTau remains, within the errors, constant: 0.21$\pm$0.01 (SMC), 
0.19$\pm$0.01 (LMC) and 0.19$\pm$0.01 (MW). Note that only a few 
FOs have been currently identified among MW and MC TIICs.

{\em Hertzsprung progression}-- A detailed analysis of the pulsation 
amplitude and of the difference between the bump and the minimum in luminosity,
together with their difference in phase, allowed us to investigate 
on a quantitative basis the occurrence of the Hertzsprung progression 
in MW and Magellanic CCs. The pulsation amplitudes were adopted because, 
in a specific period range, the maximum of the light curve 
is given by the bump rather than by the pulsation maximum. We found that the 
center of the Hertzsprung progression is anti-correlated with the metal 
content. Indeed, the center moves from P(HP)=$9.0\pm0.2$ days in the 
MW to P(HP)=$9.4\pm0.2$ days in the LMC and to P(HP)=$10.0\pm0.2$ days
in the SMC. 

{\em Evolutionary constraints for Cepheids}-- The evolutionary properties 
of Cepheids were investigated by using optical, optical-NIR and NIR 
color--magnitude diagrams. In order to overcome possible systematics, we adopted 
homogeneous optical (OGLE~IV) and NIR (IRSF) photometry for both 
variable stars and common stars in the MW and in the MCs. The comparison 
between observations and theory was performed by using homogeneous sets 
of stellar isochrones and evolutionary models available in the BASTI-IAC data 
base. The key advantage of this database is that we adopted cluster 
isochrones, either considering or neglecting convective core overshooting 
during central hydrogen-burning phases. Moreover, evolutionary predictions are 
available for scaled-solar and $\alpha$-enhanced chemical mixtures. This is a 
crucial issue in dealing with young (CCs), intermediate (ACs) and 
old (TIICs) stellar tracers. The comparison provides ranges in 
age and in metallicity for the three different groups of variable stars, 
which agree quite well with similar estimates available in the literature.

We have studied in detail the transition between stellar structures that ignite helium in a core 
partially affected by electron degeneracy, and those that ignite helium 
quiescently. This difference marks 
the transition between low- and intermediate-mass stars. It goes without 
saying that the two groups obey to different mass-luminosity relations. 
The current circumstantial evidence indicates that the bright portion of 
the ACs is located across this boundary. This means that the brightest 
ACs are quite similar to CCs, whereas the others are the truly Anomalous 
Cepheids. 

It is worth mentioning that the central hydrogen-burning 
evolutionary time in the quoted transition is only a factor of 
four longer than the central helium-burning lifetime. The 
hydrogen-burning lifetime is typically two order of magnitude 
longer in the low-mass regime, and roughly 6/8 times longer 
in the intermediate-mass regime. This is why stellar 
structures in the transition between low- and intermediate-mass 
stars are the main contributors of central helium-burning stellar 
populations (red clump stars). 

We have also examined the impact of the first dredge-up among 
different evolutionary scenarios (classical, overshooting) to 
trace its efficiency. 
We further support the use of CNO abundances as a solid 
diagnostic to discriminate among the different physical assumptions 
driving the mixing during central hydrogen-burning phases 
in intermediate-mass stars. 

{\em Cepheids as primary distance indicators}-- The use of the 
different diagnostics have been investigated in detail. The 
current evidence based on MC Cepheids fully support the use 
of FOs as distance indicators. Moreover, pros and cons of 
different distance diagnostics PL, PW and PLC were addressed on a 
quantitative basis. We also discussed advantages and drawbacks 
in using optical, optical-NIR and NIR mean magnitudes. 

{\em Cepheids as stellar tracers}-- We studied the use of MW CCs 
as young stellar tracers of the Galactic thin disk. The current evidence 
indicates well-defined negative radial gradients in iron peak, $\alpha$- 
and neutron capture elements. The only exception is barium, and we still 
lack a quantitative explanation. Plain physical arguments suggest that 
the chemical enrichment has been more efficient in the high-density 
regions of the inner disk, and then moved into the outermost disk regions
\citep[][and references therein]{spitoni2019,matteucci2021}. 
Preliminary evidence based on both CCs 
and open clusters indicates a flattening of the gradient in the outskirts 
of the thin disk. This global trend is only minimally affected by the 
flaring of the disk \citep{lemasle2022}. 

Spectroscopic measurements based on high-resolution spectra  indicate that 
the slopes of several $\alpha$-elements are quite similar to the slope 
of iron. Indeed, the trend of the [$\alpha$/Fe] abundance ratio is 
mildly positive across the disk. This suggests a similar chemical enrichment 
history. This evidence is further supported by the radial gradients of 
s- (La, Ce, Nd) and r- (Eu) process elements. The slopes of these 
elements are steeper than those of the $\alpha$-elements, and the 
[n-capture/Fe] abundance ratios are slightly more positive across 
the entire disk. There is only one exception: Y shows a slope quite 
similar to iron, and in turn, the [Y/Fe] abundance ratio is flat. 

Interestingly enough, the same abundance ratios display a flat 
trend as a function of the age (logarithmic period). The pulsation 
period is a solid analog of the individual age of CCs, since the longer 
the period, the younger is the Cepheid. Therefore, the chemical 
enrichment history of CCs has been quite homogeneous during the 
last 200/300 Myrs. There is one exception: Ca, showing a mild 
positive gradient as a function of the period. The 
positive gradient in [Ca/H] versus period and the negative gradient 
of [Ca/H] vs Galactocentric distance are not related, since 
outer disk CCs have a canonical period distribution.

\bmhead{Acknowledgments}
We are grateful to two anonymous referees for their positive words concerning the
content and the cut of an earlier version of the present paper, and for their very 
relevant suggestions, which improved its readability. They also provided speciﬁc and 
appropriate hints on a couple of sections which improved the content of the paper. 
Special thanks to G. Giobbi for her valuable editing of the manuscript
and V. D’Orazi, G. Fiorentino, M. Marconi, N. Matsunaga and M. Salaris, for their 
detailed suggestions concerning the content and the layout of an early version of 
this manuscript. We are also happy to acknowledge informative interchanges and many 
interesting conversations with D. Lennon, W. Freedman, B. Madore, A. Udalski and 
their collaborators over more than 20 years. 

It is a real pleasure to thank the many colleagues that supported me with suggestions 
and advice along my research activity (V. Castellani, F. Caputo, R. Buonanno, R.F. Stellingwerf, 
A.J. Cox, P.B. Stetson, H. Smith, A.R. Walker, C. Sneden, H.-R. Kudritzki). Special thanks to 
my long-standing collaborators---M. Marconi, M. Dall’Ora, M. Monelli, M. Fabrizio, R. da Silva, 
B. Lemasle, L. Inno, V. D’Orazi, M. Marengo, J. Mullen, F. Thevenin, J. Storm, B. Chaboyer, S. Cassisi, 
M. Salaris, G. Coppola, V. Ripepi, I. Musella, M. di Criscienzo, G. Fiorentino, 
C. E. Martinez Vazquez, N. Matsunaga, W. Gieren, G. Pietrzynski, B. Pilecki, V. Kovtyukh, M. Tantalo, 
E. Valenti, M. Zoccali---who made and are still making the research in this ﬁeld a fantastic adventure. 

I am immensely grateful to my original family and to my current family, and in particular to $F^2G^2$, 
they never pointed me out for the many absences for astrophysical {\it otium}. This paper is dedicated to Francesco, 
who made our recent life ``really something special''. Special thanks to the A\&A Review editorial board, 
and in particular, to Francesca Matteucci, Bengt Gustafsson and Frank Schulz for this unique opportunity, 
and for being so patient with us over many years. This investigation was partially supported by Project 
PRIN MUR 2022 (2022ARWP9C) ``Early Formation and Evolution of Bulge and HalO (EFEBHO)'', PI: M. Marconi, 
funded by European Union – Next Generation EU and by the Large grant INAF 2022 MOVIE (PI: M. Marconi). 
This research was also partially supported by the Munich Institute for Astro–Particle and BioPhysics 
(MIAPbP) which is funded by the Deutsche Forschungsgemeinschaft(DFG,GermanResearch Foundation) under 
Germany´s Excellence Strategy – EXC-2094 – 390783311.
\clearpage

% \appendix
\begin{appendices}

\section{Metallicity dependence of the Hertzsprung progression}\label{sec:appendix_hp}   

\begin{figure}[htbp]
\begin{center}
\includegraphics[width=0.46\textwidth]{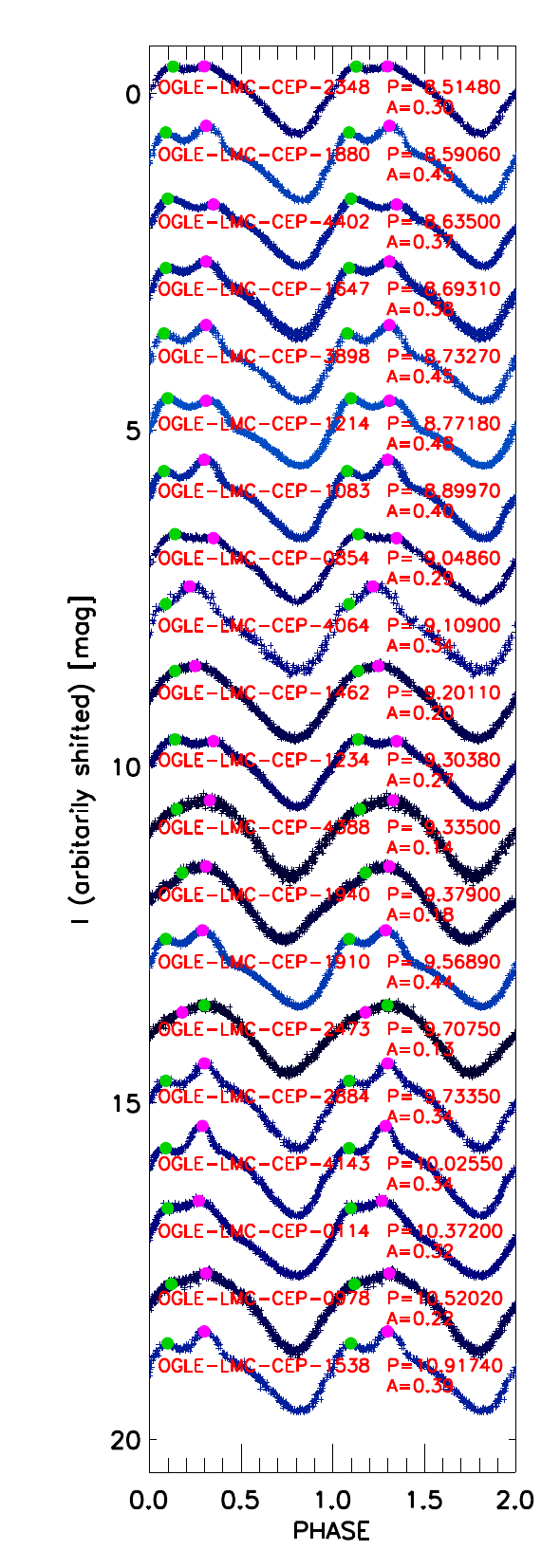}
\includegraphics[width=0.46\textwidth]{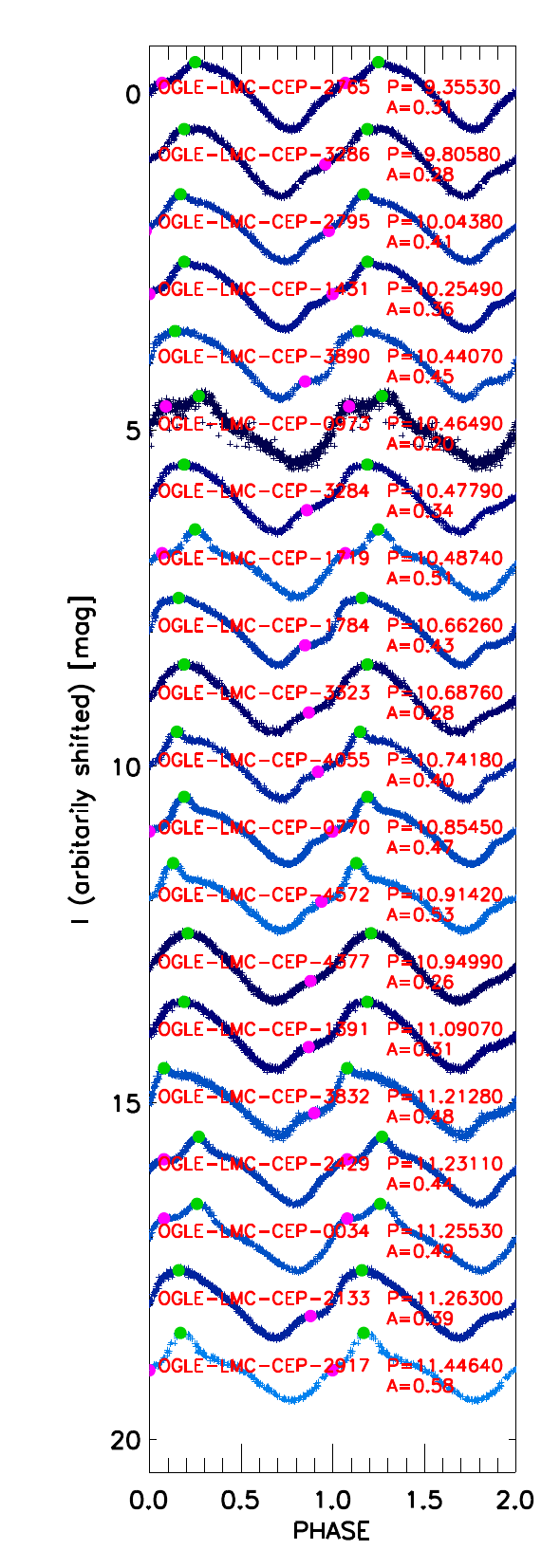}
%\captionsetup{labelformat=empty}
\caption{Phased $I$-band light curves of LMC bump Cepheids with a bump moving either
up along the decreasing branch (UPDB) or down along the rising branch (DORB).}
\label{fig_lmc_bump_light}
\end{center}
\end{figure}

\begin{figure}[htbp]
\begin{center}
\includegraphics[width=0.46\textwidth]{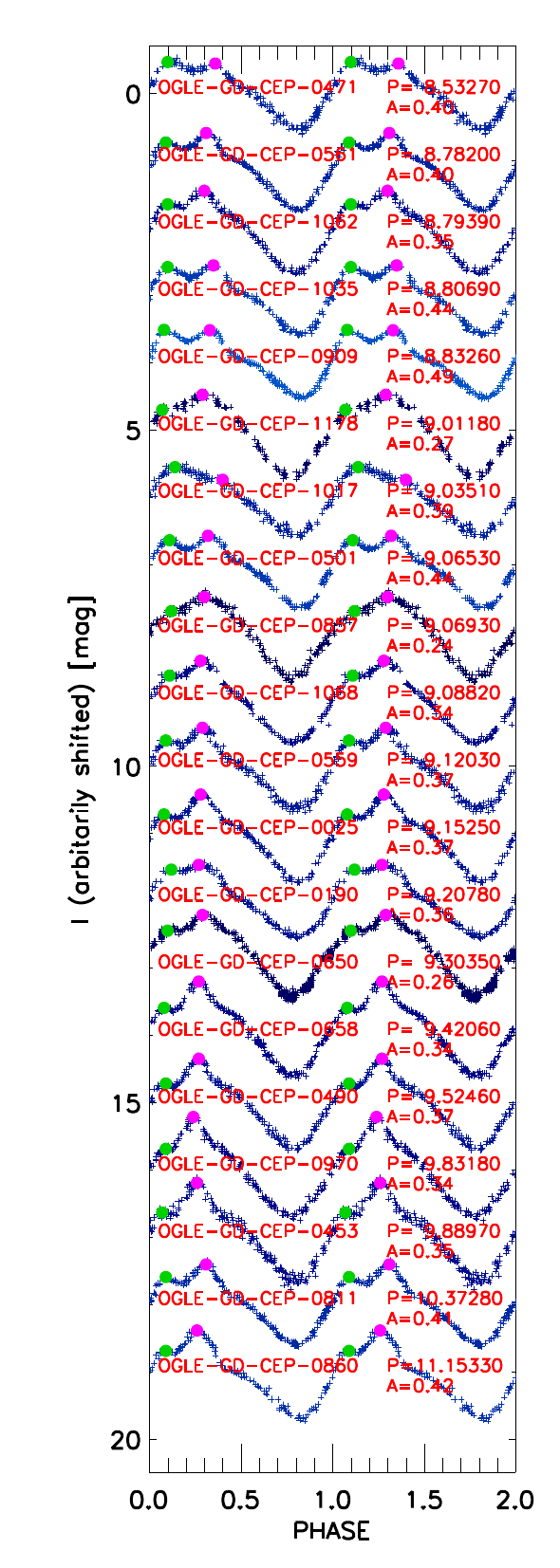}
\includegraphics[width=0.46\textwidth]{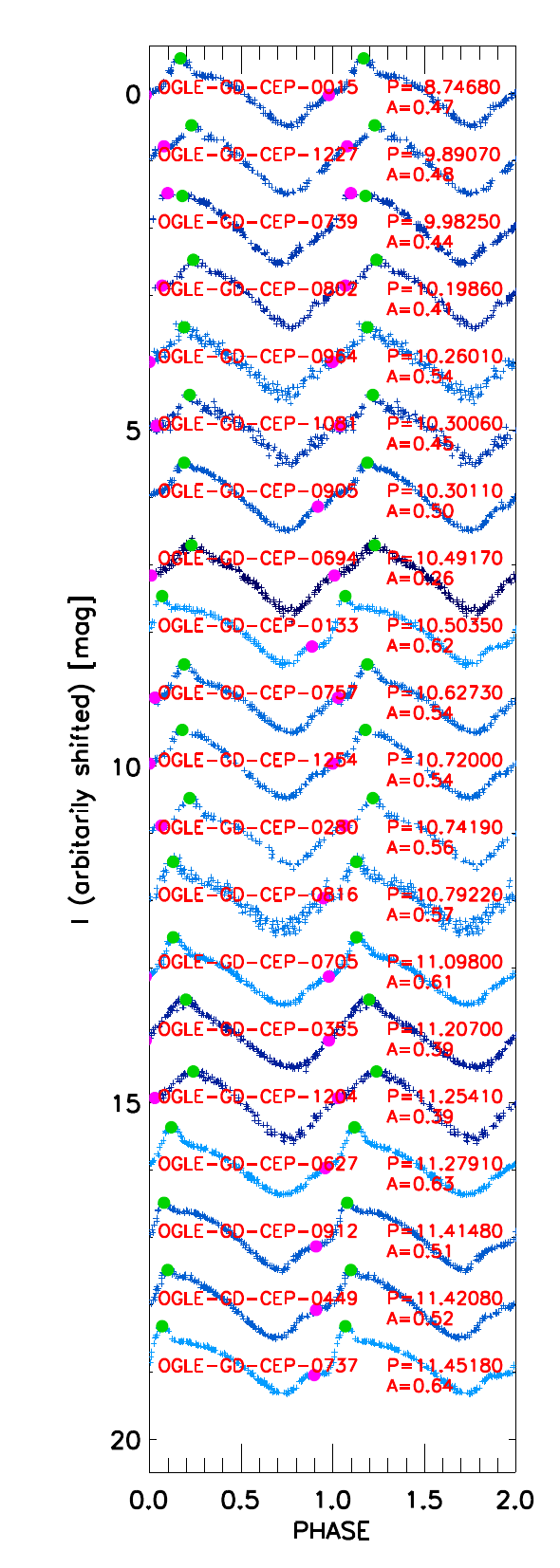}
\caption{Phased $I$-band light curves of LMC bump Cepheids with a bump moving either
up along the decreasing branch (UPDB) or down along the rising branch (DORB).}
\label{fig_gal_bump_light}
\end{center}
\end{figure}

In order to investigate on a more quantitative basis the variation of the bump among
UPDB and DORB Cepheids, Figs.~\ref{fig_lmc_bump_light} and \ref{fig_gal_bump_light}
display the $I$-band light curves  for LMC and Galactic bump Cepheids.

%%%%%%%%%%%%%%%%%%%%%%%%%%%%%%%%%%%%%%%%%%%%%%%%%%%%%%%%%%%%%%%%%%%%%%%%%%%%%%%%%%%%%
%                       fig  SF 1
%%%%%%%%%%%%%%%%%%%%%%%%%%%%%%%%%%%%%%%%%%%%%%%%%%%%%%%%%%%%%%%%%%%%%%%%%%%%%%%%%%%%%
%_______________________________________________________________________________
%_______________________________________________________________________________

%%%%%%%%%%%%%%%%%%%%%%%%%%%%%%%%%%%%%%%%%%%%%%%%%%%%%%%%%%%%%%%%%%%%%%%%%%%%%%%%%%%%%
%                        fig SF 2
%%%%%%%%%%%%%%%%%%%%%%%%%%%%%%%%%%%%%%%%%%%%%%%%%%%%%%%%%%%%%%%%%%%%%%%%%%%%%%%%%%%%%
%_______________________________________________________________________________
%_______________________________________________________________________________

\begin{figure}[htbp]
\begin{center}
\includegraphics[width=\textwidth]{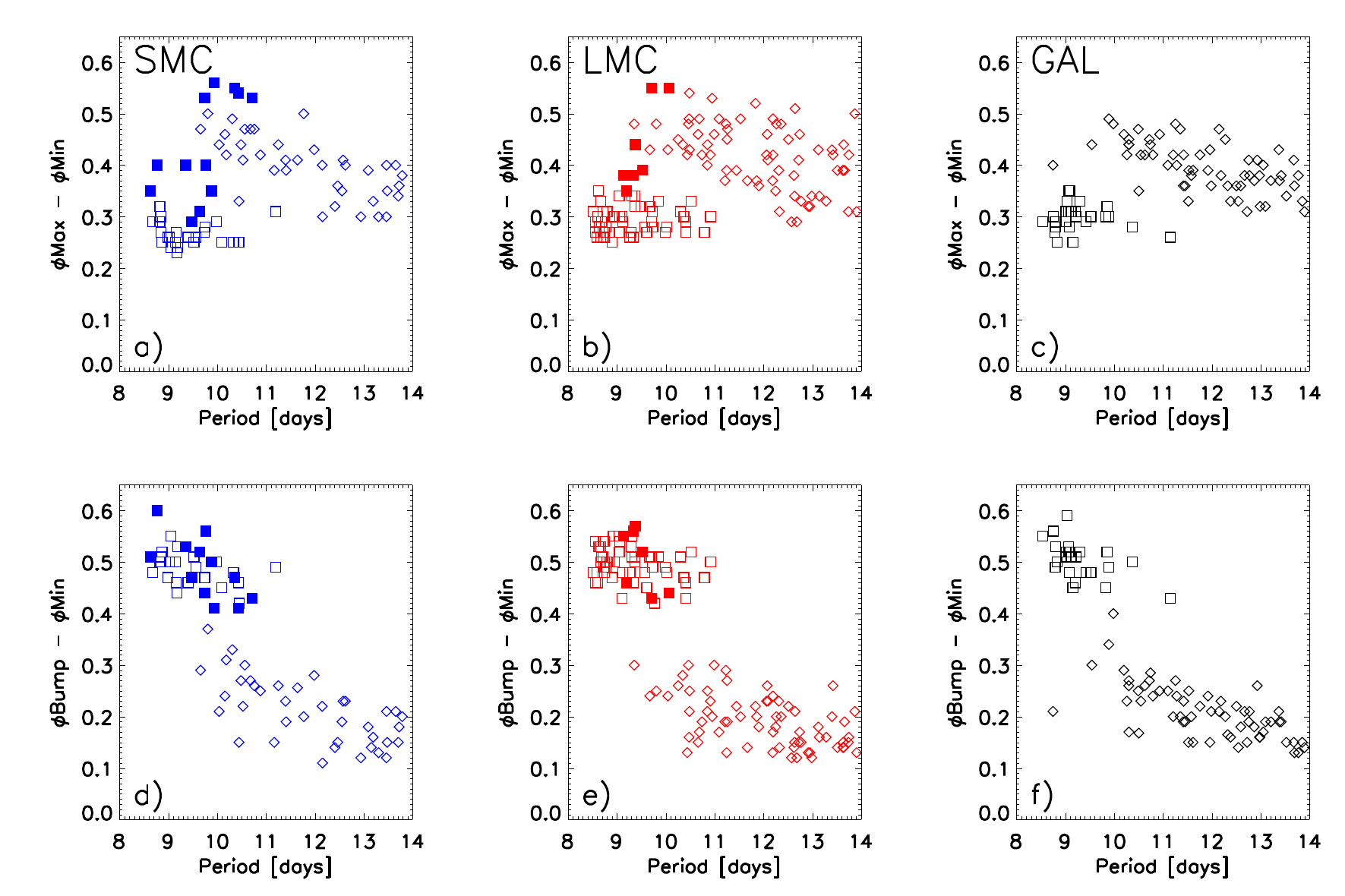}
\caption{
Top: from left to right difference between the phase of the maximum and the phase
of the minimum in $I$-band light curves as a function of the period for
SMC (panel a), LMC (panel b) and Galactic (panel c) bump Cepheids. Squares and
diamonds mark UPDB and DORB Cepheids.
Bottom: Same as the top, but the difference is between the phase of the bump and
the phase of the minimum along  $I$-band light curves (SMC, panel d; LMC, panel e;
Galactic, panel f).
}
\label{fig_phase_dif_bump}
\end{center}
\end{figure}

Moreover, we also estimated the difference in phase between
the pulsation maximum and the minimum in luminosity ($\Phi$Puls) and the
difference in phase between the bump and the minimum in luminosity ($\Phi$bump).
Data plotted in the top panels of Fig.~\ref{fig_phase_dif_bump} show that
UPDB Cepheids (squares) are characterized by an almost
constant phase difference between the pulsation maximum and the minimum.
The mean value are, within the errors, quite similar (0.26$\pm$0.02, SMC;
0.29$\pm$0.03, LMC; 0.30$\pm$0.03, MW) when moving from the more metal-poor
to the more metal-rich stellar systems. In the MCs there are a few outliers
which attain larger values, but they are typically located beyond the HP.

Conversely, the DORB Cepheids (diamonds) display a steady decrease from
$\approx$0.50 to $\approx$0.30 when moving from the short- to the long-period
range. There is a range in period between 9.5 and 11 days in which the two
subgroups overlap with each other. However, the separation
between UPDB and DORB Cepheids, with the exception of a few transitional
objects, appears quite clear.

%%%%%%%%%%%%%%%%%%%%%%%%%%%%%%%%%%%%%%%%%%%%%%%%%%%%%%%%%%%%%%%%%%%%%%%%%%%%%%%%%%%%%
%                        fig SF 3
%%%%%%%%%%%%%%%%%%%%%%%%%%%%%%%%%%%%%%%%%%%%%%%%%%%%%%%%%%%%%%%%%%%%%%%%%%%%%%%%%%%%%
%_______________________________________________________________________________
%_______________________________________________________________________________

The bottom panels of the same figure display the difference between the
phase of the bump and the phase of the minimum in the $I$-band light curve.
The separation between UPDB and DORB Cepheids becomes even clearer. Indeed,
the UPDBs are clustering around a constant mean value ranging from
0.49$\pm$0.02 (SMC) to 0.50$\pm$0.03 (LMC) and to 0.51$\pm$0.03 (MW).
The DORBs show a steady decrease when moving from the short- to the
long-period range. Once again, the two subgroups display
an overlap for periods ranging from 9.5 to 11 days, but they differ
on average by a factor of two.

\begin{figure}[htbp]
\begin{center}
\includegraphics[width=\textwidth]{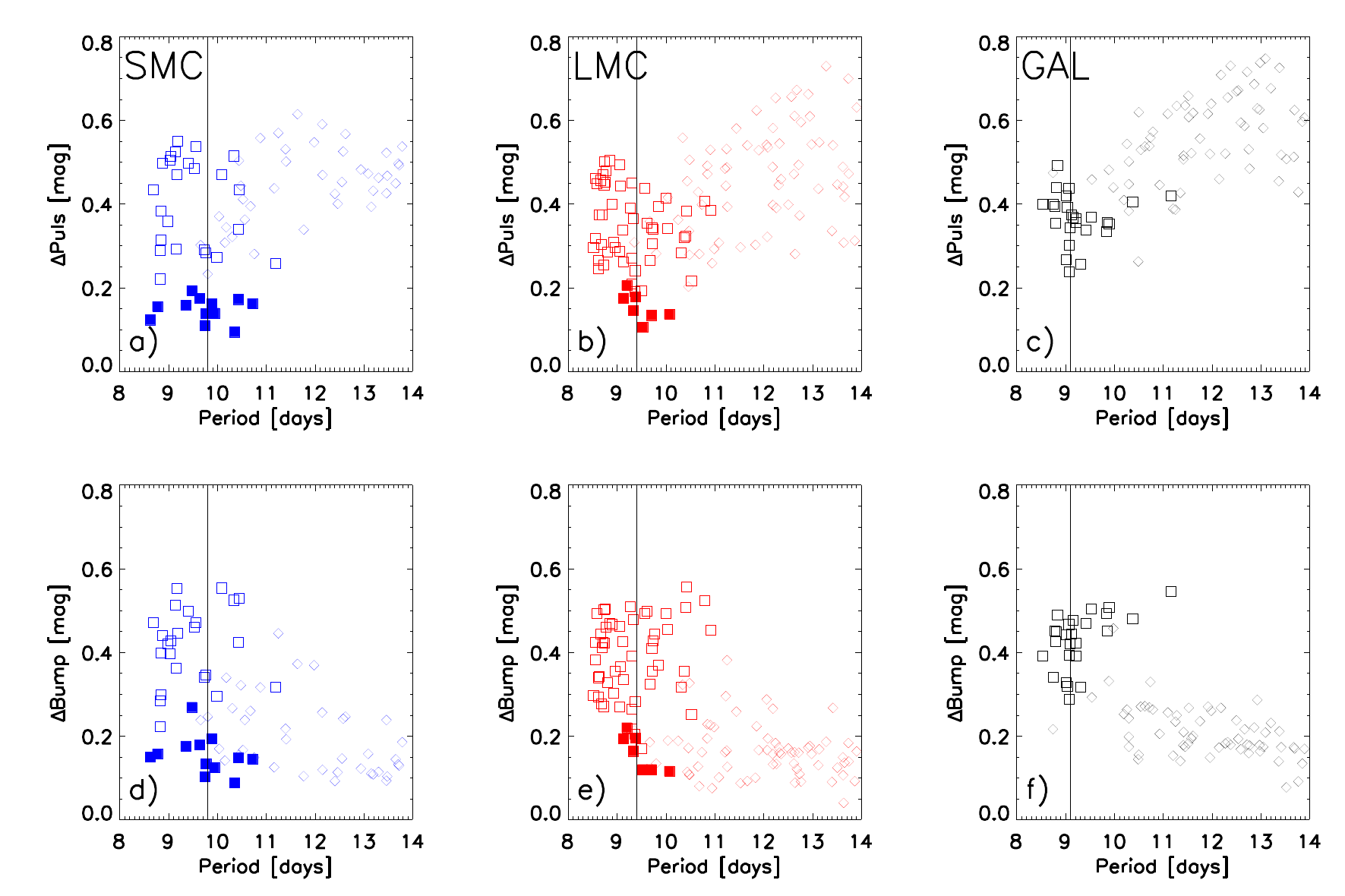}
\caption{
Top: from left to right difference between the pulsation maximum and the minimum
in the $I$-band light curves as a function of the period for SMC (panel a),
LMC (panel b) and Galactic (panel c) bump Cepheids. The black vertical line
shows the center of the HP. The symbols are the same as in
Fig.~\ref{fig_phase_dif_bump}.
Bottom: Same as the top, but the difference is between the magnitude of the bump
and the minimum along $I$-band light curves (SMC, panel d; LMC, panel e;
Galactic, panel f).
}
\label{fig_magn_dif_bump}
\end{center}
\end{figure}

In order to further investigate on a quantitative basis the difference between
the UPDB and DORB Cepheids, the top panels of Fig.~\ref{fig_magn_dif_bump}
display the difference in magnitude between the pulsation maximum and
the minimum in the $I$-band light curve (pulsation amplitude, $\Delta$Puls).
In this context, it is worth mentioning that $\Delta$Puls is not always the
luminosity amplitude, i.e. the difference in magnitude between the absolute
maximum and the minimum along the light curve. Indeed, there is a range in
period in which the bump is brighter than the pulsation maximum
(see 
%Figs.~\ref{fig_smc_bump_light}, 
Figs.~11, 
\ref{fig_lmc_bump_light} and
\ref{fig_gal_bump_light}). Data plotted in these panels display quite
clearly that the pulsation amplitudes attain a minimum that steadily
decreases from 9.5--10 days for the SMC to 9.0--9.5 days for the LMC
and to $\approx$9.0 days for the Galactic UPDB Cepheids (squares).
The DORB Cepheids (diamonds) display a steady increase when moving
from the short to the long-period regime, and the global trend is
steeper when moving from metal-poor to more metal-rich stellar systems.
They are characterized by a large spread at fixed pulsation period,
but the spread in magnitude is intrinsic, because it is well beyond
the typical photometric error along the light curve.

The center of the HP in the MW and in the MCs were marked with
vertical black lines in Fig.~\ref{fig_magn_dif_bump}. Data plotted
in this figure display some outliers across the minimum.
This variation across the center of the HP is intrinsic and expected.
Indeed, MC and MW CCs are characterized by a well defined metallicity
distribution, with a standard deviation of the order of 0.5 dex
\citep{romaniello08,hocde2023}. In a recent investigation, \citet{romaniello2022}
by using high resolution spectra of 89 LMC CCs suggested that the spread
in iron abundance is quite modest. However, spectroscopic measurements
based on B-type stars  \citep{hunter2009} and on red giants \citep{dobbie2014} 
indicate a spread in chemical composition larger than 0.5 dex.

%%%%%%%%%%%%%%%%%%%%%%%%%%%%%%%%%%%%%%%%%%%%%%%%%%%%%%%%%%%%%%%%%%%%%%%%%%%%%%%%%%%%%
%                        fig SF 4
%%%%%%%%%%%%%%%%%%%%%%%%%%%%%%%%%%%%%%%%%%%%%%%%%%%%%%%%%%%%%%%%%%%%%%%%%%%%%%%%%%%%%
%_______________________________________________________________________________
%_______________________________________________________________________________

The difference in magnitude  between the bump and the minimum versus the
pulsation period (bottom panels of Fig.~\ref{fig_magn_dif_bump}) display
an opposite trend. The UPDB Cepheids attain, as expected, the largest
amplitudes, while DORB Cepheids show once again a steady decrease when
moving from the short- to the long-period range. The
bump Cepheids that trace the center of the HP, i.e. those with the
smallest pulsation amplitudes, attain also the smallest bump amplitudes.
However, the minimum is less clear, since UPDBs and DORBs partially overlap
across the HP.

%%%%%%%%%%%%%%%%%%%%%%%%%%%%%%%%%%%%%%%%%%%%%%%%%%%%%%%%%%%%%%%%%%%%%%%%%%%%%%%%%%%%%
%                        fig SF 5
%%%%%%%%%%%%%%%%%%%%%%%%%%%%%%%%%%%%%%%%%%%%%%%%%%%%%%%%%%%%%%%%%%%%%%%%%%%%%%%%%%%%%
%_______________________________________________________________________________
\begin{figure}[htbp]
\begin{center}
\includegraphics[width=0.86\textwidth]{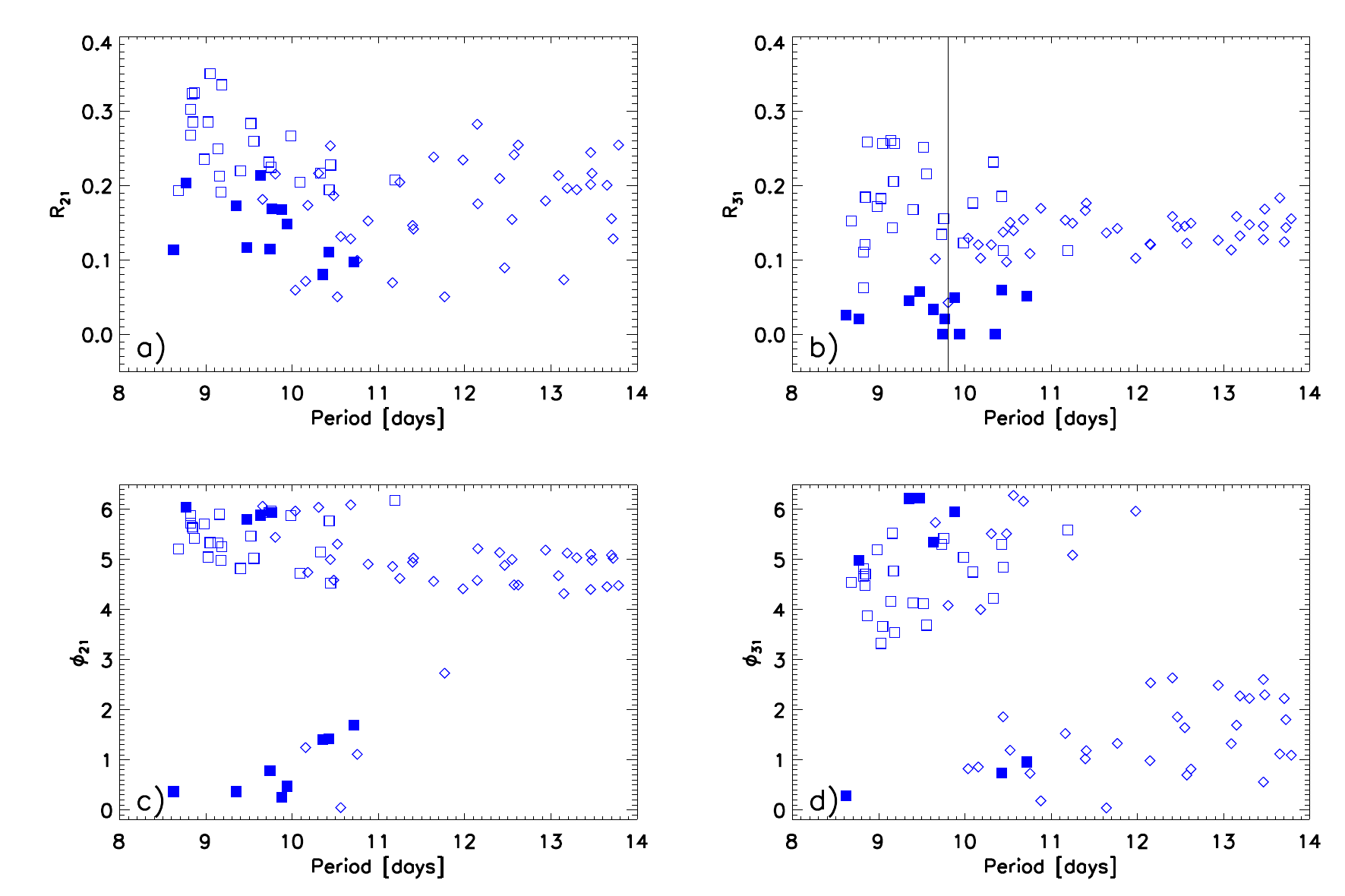}
\caption{
Top: $R_{21}$ (panel a) and $R_{31}$ (panel b) Fourier parameters as a function of
the pulsation period for $I$-band light curves of SMC bump Cepheids provided
by OGLE~IV. The symbols are the same as in Fig.~\ref{fig_phase_dif_bump}.
Bottom: Same as the top, but for $\phi_{21}$ (panel c)  and $\phi_{31}$ (panel d)
Fourier parameters .\label{fig:phi_r_smc}
}
\end{center}
\end{figure}
%_______________________________________________________________________________

\begin{figure}[htbp]
\begin{center}
\includegraphics[width=0.86\textwidth]{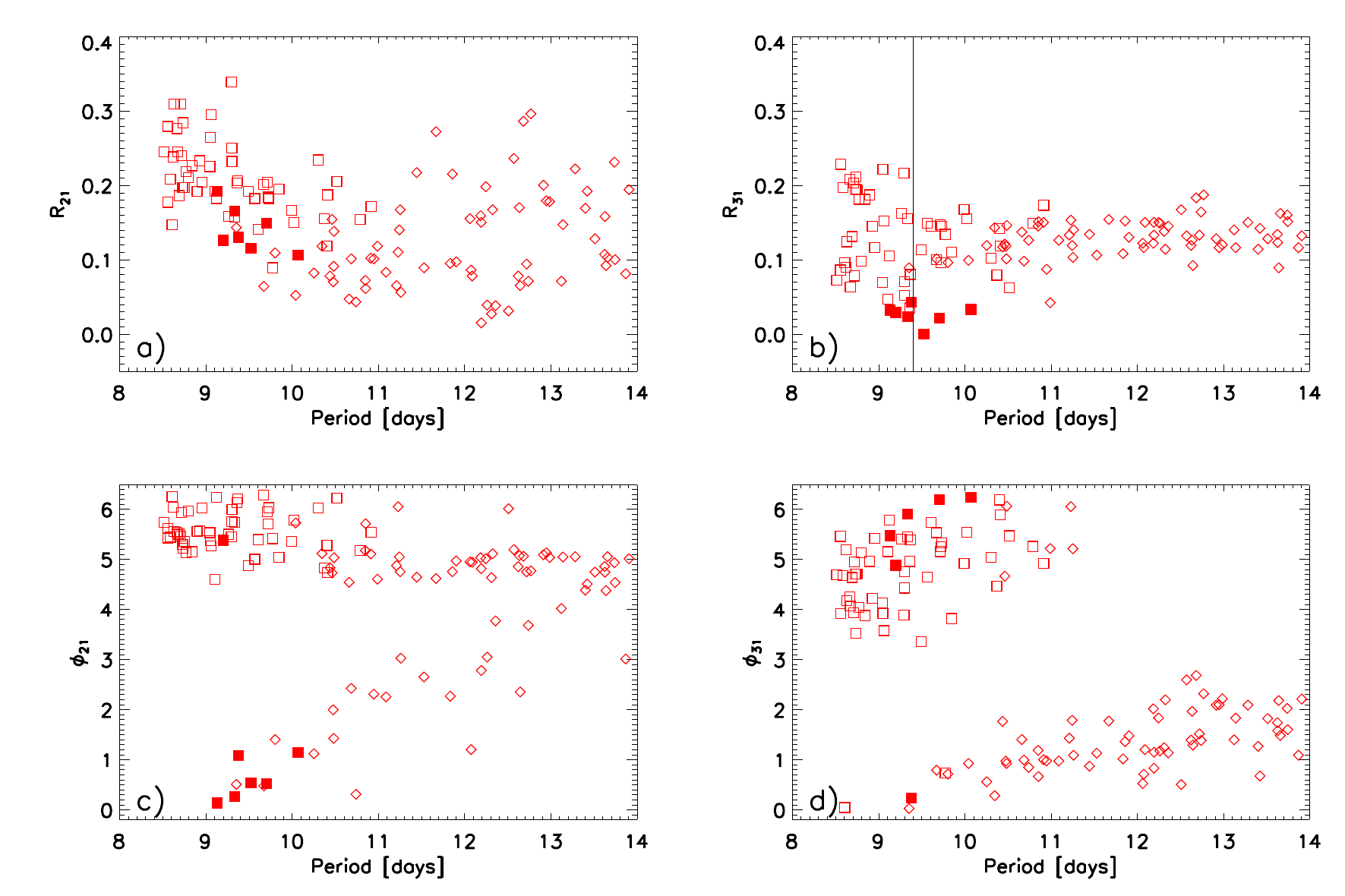}
\caption{
Top: $R_{21}$ (panel a) and $R_{31}$ (panel b) Fourier parameters as a function
of the pulsation period for $I$-band light curves of LMC bump Cepheids
provided by OGLE~IV ($R_{21}$, panel a; and $R_{31}$, panel b).
Symbols are the same as in Fig.~\ref{fig_phase_dif_bump}.
Bottom: Same as the top, but for $\phi_{21}$ (panel c)  and $\phi_{31}$ (panel d)
Fourier parameters.\label{fig:phi_r_lmc}
}
\end{center}
\end{figure}

\begin{figure}[htbp]
\begin{center}
\includegraphics[width=0.86\textwidth]{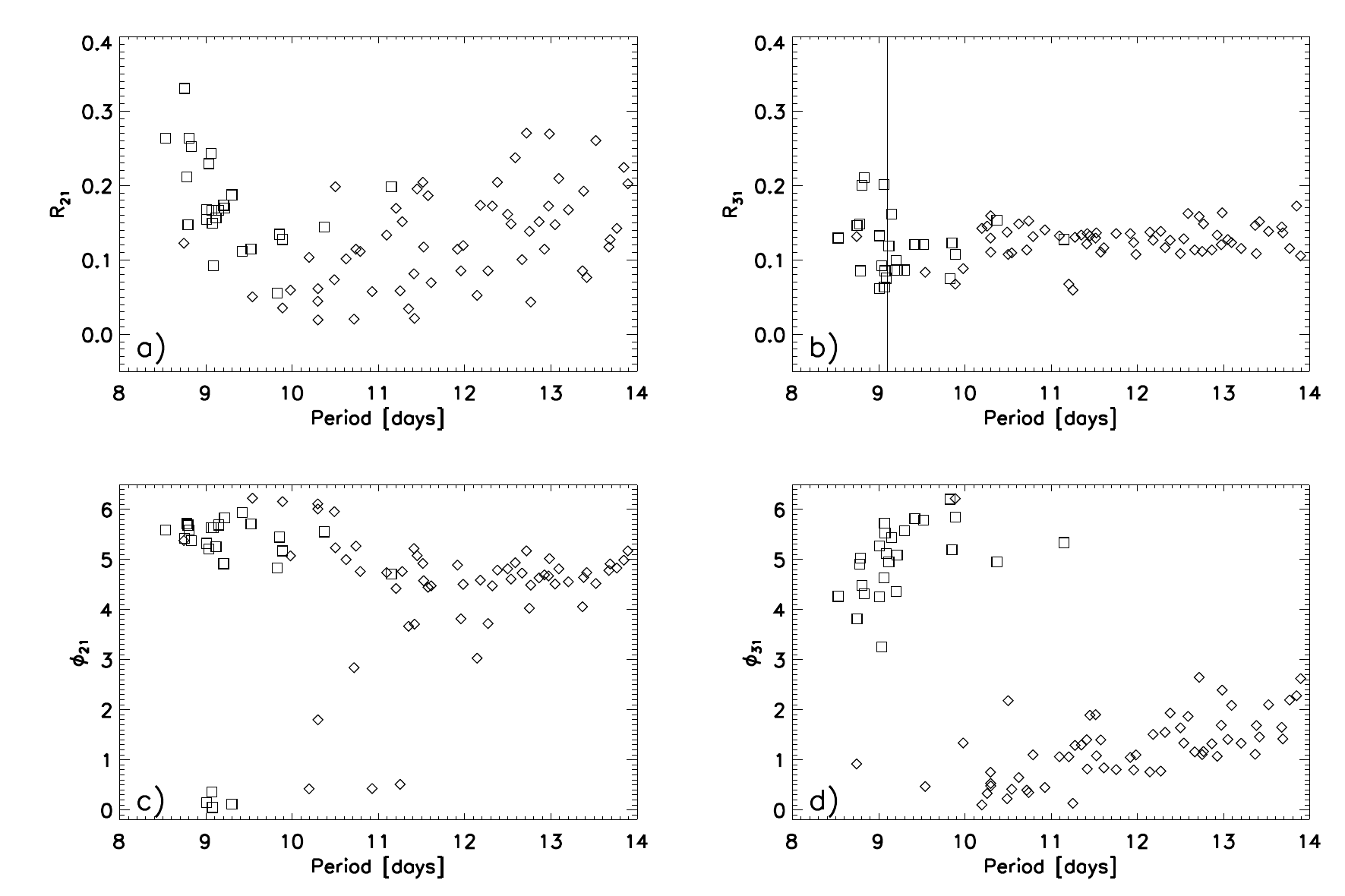}
\caption{
Top: $R_{21}$ (panel a) and $R_{31}$ (panel b) Fourier parameters as a function
of the pulsation period for $I$-band light curves of Galactic bump Cepheids
provided by OGLE~IV ($R_{21}$, panel a; R$_{31}$, panel b).
Symbols are the same as in Fig.~\ref{fig_phase_dif_bump}.
Bottom: Same as the top, but for $\phi_{21}$ (panel c)  and $\phi_{31}$ (panel d)
Fourier parameters.\label{fig:phi_r_gal}
}
\end{center}
\end{figure}

The current findings concerning the use of the Bailey (pulsation amplitude)
and pseudo-Bailey (difference in magnitude between the bump and the minimum)
diagrams together with the phase difference diagrams allow us to further
constrain the use of the Fourier parameters to trace the variation of the
shape of the light curves across the HP. In order to analyze this issue in depth by using
homogeneous and accurate estimates of the Fourier parameters, we took advantage
of the Fourier analysis provided by OGLE~IV
\citep{soszynski2015a,soszynski2015b,soszynski2017,soszynski2018}.
Data plotted in the top panels of Fig.~\ref{fig:phi_r_smc}, \ref{fig:phi_r_lmc}
and \ref{fig:phi_r_gal} display that the
$R_{21}$ (left) and the $R_{31}$ (right) Fourier parameters associated with
the amplitudes of the first three harmonics display either a mild
quadratic trend ($R_{21}$), or an almost constant value ($R_{31}$) across
the entire period range of the HP. The top right panels of these figures
show that the $R_{31}$ parameter shows minima similar to the
minima we identified by using the difference in magnitude between
pulsation maximum and minimum and the difference between bump
and minimum. In order to help the eye identify the minima,
we plotted the same black vertical lines plotted in
Fig.~\ref{fig_magn_dif_bump}. This suggests that the $R_{31}$
parameter is a solid diagnostic to constrain the center of the HP.

The trend of the phase difference among the same harmonics, plotted in the
bottom panels of the same figure, shows several interesting features.
The distribution of the $\phi_{21}$ (left) has a wedge shape with two
well-defined sequences. The horizontal sequence, which attains the largest values
and it is almost constant over the entire period range, while the slant
sequence moves from the center of the HP and approaches the horizontal
sequence in the long-period range. The UPDB Cepheids partially overlap with
DORB Cepheids. As a whole, the $\phi_{21}$ parameter does not seem to be
a solid diagnostic to constrain the center of the HP.
The separation between UPDBs and DORBs showed by the $\phi_{31}$ Fourier
parameter plotted in the bottom right panel is a very solid diagnostic
to separate bump Cepheids.

%%%%%%%%%%%%%%%%%%%%%%%%%%%%%%%%%%%%%%%%%%%%%%%%%%%%%%%%%%%%%%%%%%%%%%%%%%%%%%%%%%%%%
%                        fig SF 6
%%%%%%%%%%%%%%%%%%%%%%%%%%%%%%%%%%%%%%%%%%%%%%%%%%%%%%%%%%%%%%%%%%%%%%%%%%%%%%%%%%%%%
%_______________________________________________________________________________
%_______________________________________________________________________________

The key difference in the comparison between the Bailey and the pseudo-Bailey
diagram with the $R_{21}$ (left) and the $R_{31}$ Fourier parameters is that
the latter attain very similar values when moving from MC to Galactic Cepheids,
whereas the former ones display a clearer trend when moving from more metal-poor
to more metal-rich stellar systems. A similar outcome applies to the comparison
between the phase difference of pulsation maximum and of the bump with the
$\phi_{21}$  and $\phi_{31}$ Fourier parameters. The former parameters display
well separated linear trends for Cepheids, with a bump located either along the
decreasing or the rising branch of the light curve.

%%%%%%%%%%%%%%%%%%%%%%%%%%%%%%%%%%%%%%%%%%%%%%%%%%%%%%%%%%%%%%%%%%%%%%%%%%%%%%%%%%%%%
%                        fig SF 7
%%%%%%%%%%%%%%%%%%%%%%%%%%%%%%%%%%%%%%%%%%%%%%%%%%%%%%%%%%%%%%%%%%%%%%%%%%%%%%%%%%%%%
%_______________________________________________________________________________
%_______________________________________________________________________________

In closing this section it is worth mentioning that the HP can be considered as
the fingerprint of CCs. To our knowledge there is no other group of variable stars
showing a tight correlation of the phase of a secondary feature along the
light and the radial velocity curves with a specific range of pulsation periods.
This outcome applies to MW, MC and to extragalactic CCs.\footnote{This
paragraph was triggered by a chat we have had with Adam Riess
concerning the evolutionary and pulsation properties of extragalactic Cepheids.}

\clearpage

%%%%%%%%%%%%%%%%%%%%%%%%%%%%%%%%%%%%%%%%%%%%%%%%%%%%%%%%%%%%%%%%%%%%%%%%%%%%%%%%%%%%%
%                        fig SF 8
%%%%%%%%%%%%%%%%%%%%%%%%%%%%%%%%%%%%%%%%%%%%%%%%%%%%%%%%%%%%%%%%%%%%%%%%%%%%%%%%%%%%%
%_______________________________________________________________________________
\begin{figure}[hb!]
\begin{center}
\includegraphics[width=0.49\textwidth]{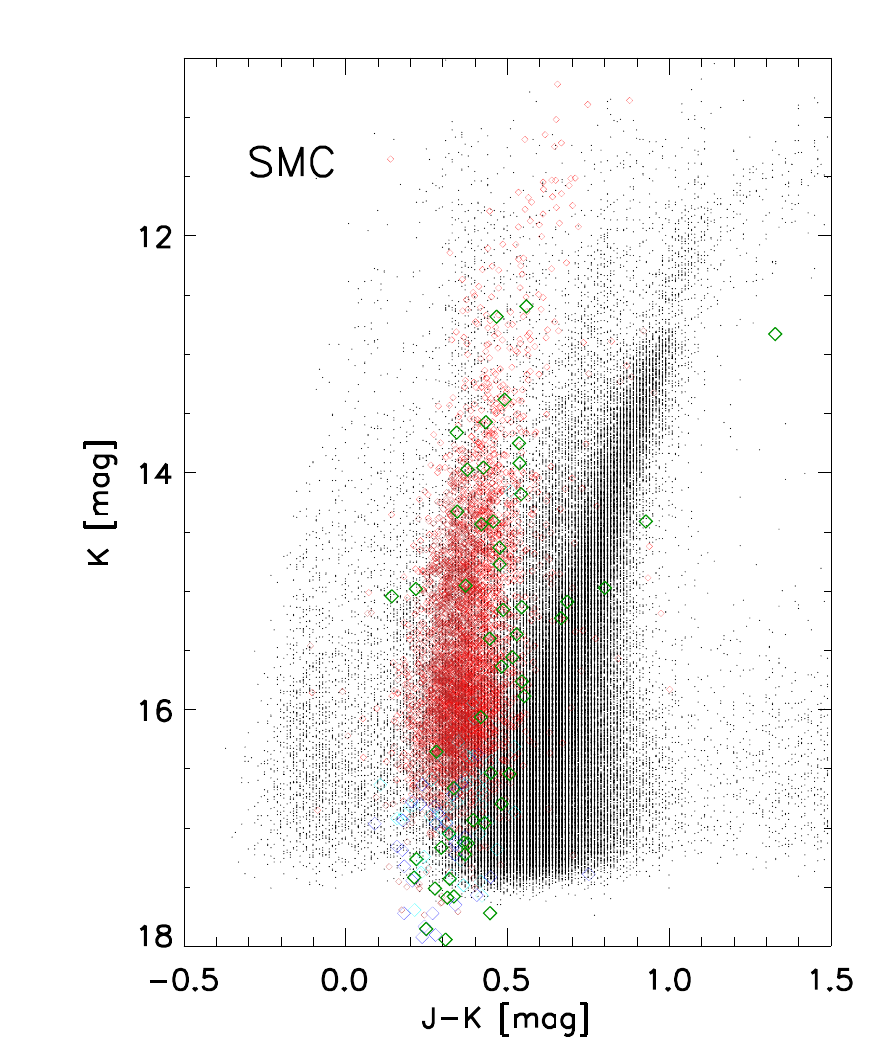}
\includegraphics[width=0.49\textwidth]{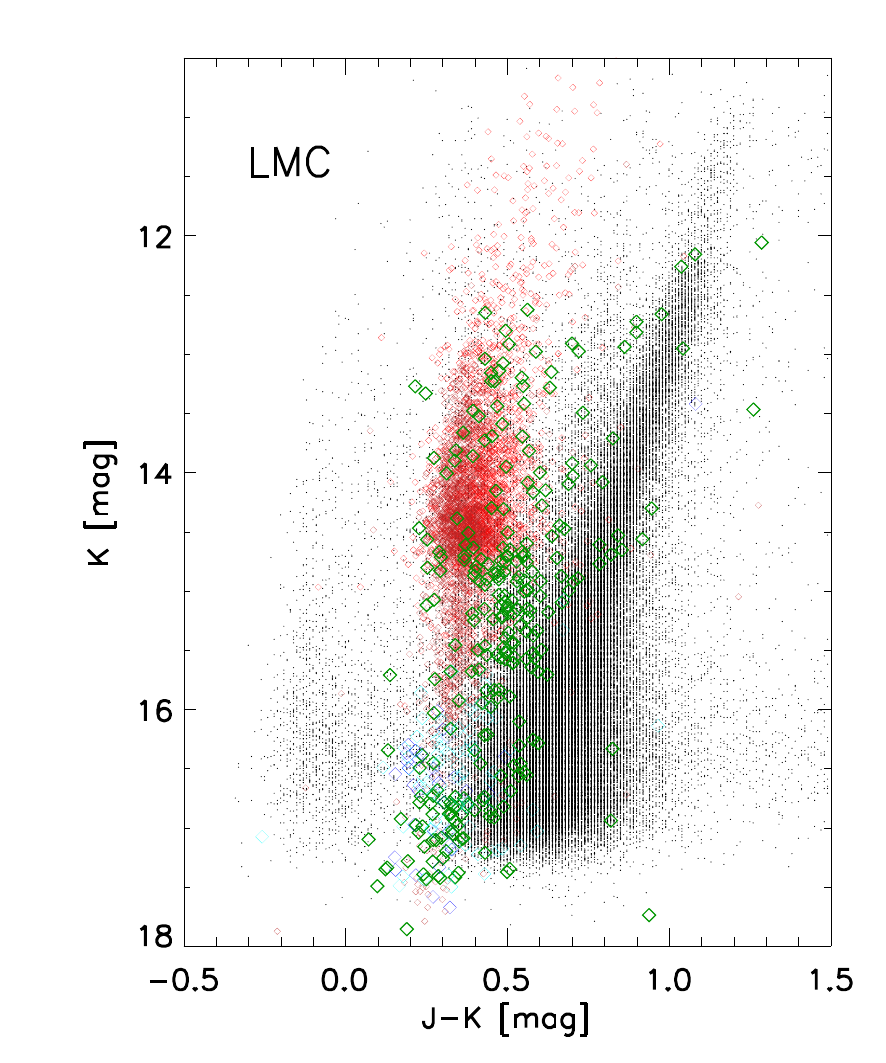}
\caption{Left:
distribution of fundamental (light red) and first overtone
(dark red) CCs, fundamental (light blue) and first overtone
(dark blue) ACs and fundamental (green) TIICs  in the
NIR $K$,$J-K$ CMD of the SMC. The overlap among
the different groups of Cepheids is even more severe, due to the reduced
sensitivity of the $J-K$ color to the effective temperature. Data plotted
in this panel come from the IRSF survey of the Magellanic Clouds
\citep{ita2004,ita2018}. The mean magnitudes of variable stars were
estimated as the mean of the measurements in magnitude.
Right: Same as the left, but for LMC Cepheids.} \label{fig_nir_cmd_MC}
\end{center}
\end{figure}

%_______________________________________________________________________________
\section{Optical/near-infrared and near-infrared diagnostics}\label{sec:appendix_nir}   
%_______________________________________________________________________________

%_______________________________________________________________________________
\subsection{Color magnitude diagrams}   
%_______________________________________________________________________________
This section deals with optical/NIR and NIR diagnostics, namely the
$K$,$V-K$ and the $K$,$J-K$ CMDs of MC stellar populations discussed in Sect.~4.
Moreover, the $K$-band PL relations and the $K$,$J-K$
PW relations are further investigated (see Sect.~6.2).
Figure~\ref{fig_nir_cmd_MC} shows the same variables discussed in section˜4,
but in the NIR $K$,$J-K$, CMD.
Data for both static and variables stars come from the NIR photometric survey
of the MCs performed by IRSF \citep{ita2004,ita2018}. The mean
magnitude of variable stars is the mean of the measurements.  Data plotted
in this figure display quite clearly that the degeneracy among the
different groups of variable stars and common stars is even more severe,
since they overlap over a significant fraction of the CMD.
The difference with optical CMDs is mainly due to the reduced sensitivity
of NIR colors to the effective temperature of F,G and early K-type stars.
Moreover, limiting magnitudes in NIR photometry are shallower than in the
optical regime, and evolutionary features such as the RC can hardly be
recognized.

%%%%%%%%%%%%%%%%%%%%%%%%%%%%%%%%%%%%%%%%%%%%%%%%%%%%%%%%%%%%%%%%%%%%%%%%%%%%%%%%%%%%%
%                        fig SF 9
%%%%%%%%%%%%%%%%%%%%%%%%%%%%%%%%%%%%%%%%%%%%%%%%%%%%%%%%%%%%%%%%%%%%%%%%%%%%%%%%%%%%%
%_______________________________________________________________________________
\begin{figure}[ht!]
\begin{center}
\includegraphics[width=0.43\textwidth]{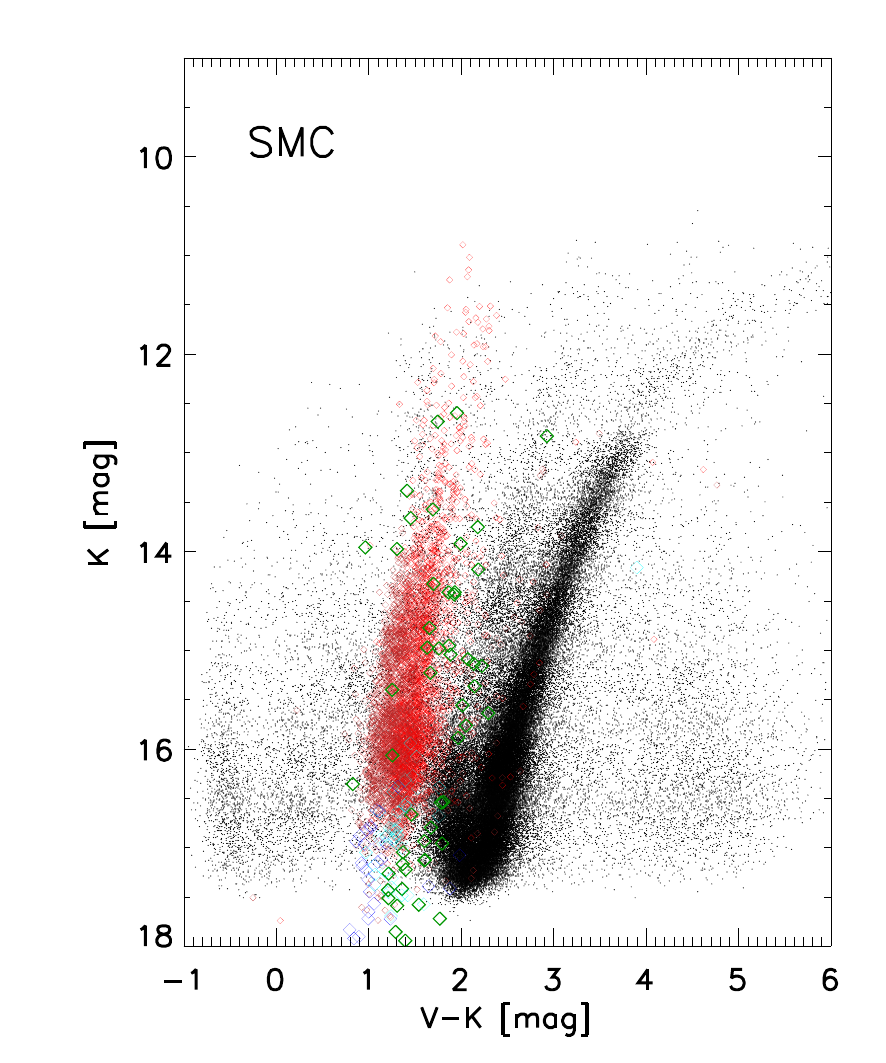}
\includegraphics[width=0.43\textwidth]{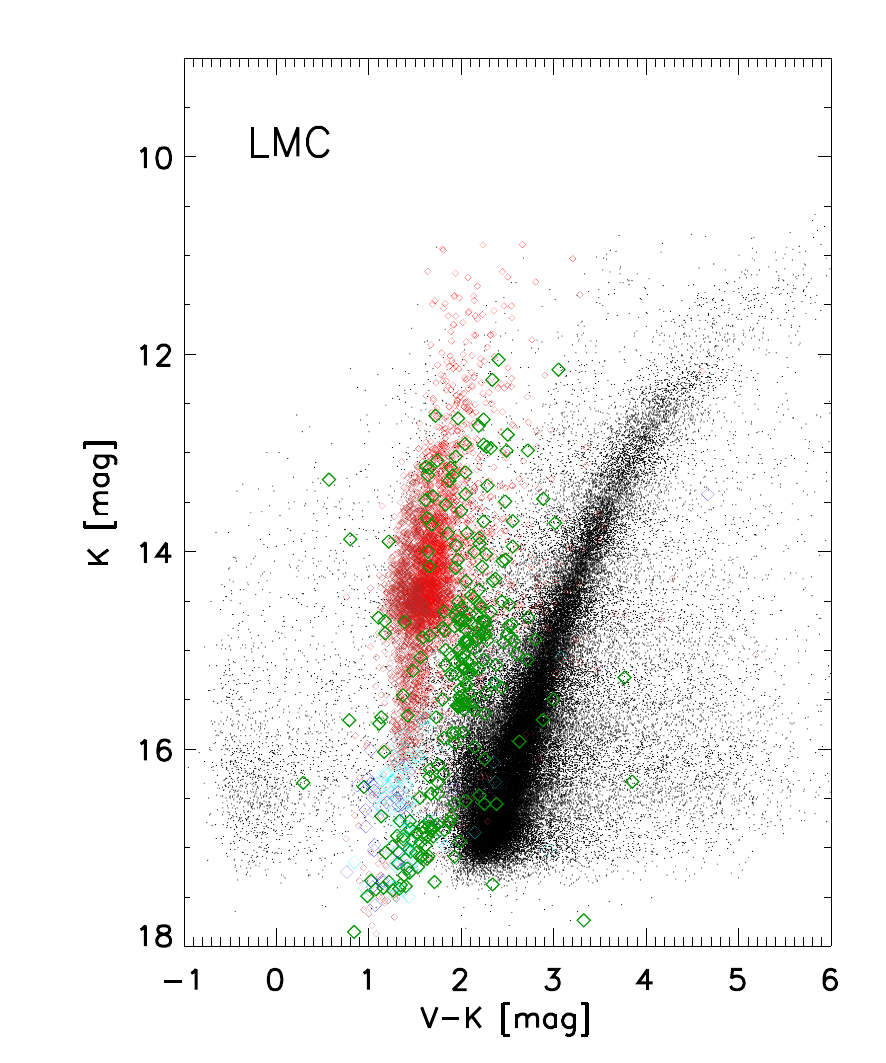}
\caption{Left:
distribution of fundamental (light red) and first overtone
(dark red) CCs, fundamental (light blue) and first overtone
(dark blue) ACs and fundamental (green) TIICs  in the
Optical-NIR $K$,$V-K$ CMD of the SMC. The different groups
of Cepheids are more separated, due to the increased sensitivity
of the $V-K$ color to the effective temperature.
Right: Same as the left, but for LMC Cepheids.}
\label{fig_optical_nir_cmd_MC}
\end{center}
\end{figure}
%_______________________________________________________________________________

To partially ease this difficulty,
Fig.~\ref{fig_optical_nir_cmd_MC} shows the same variables, but plotted
in the optical-NIR  $K$,$V-K$ CMD. There is still some overlap between
static and variable stars, but the increased sensitivity to the effective
temperature provides a better separation of the different groups of
variable stars. Indeed, they cover a range in color which is at least
a factor of four larger when compared with the optical CMD. Moreover,
the distribution  in magnitude and in color is significantly more
linear, since they become intrinsically brighter when moving from
hotter (shorter periods) to cooler (longer periods) variables.

%_______________________________________________________________________________
\subsection{Period-Luminosity and Period-Wesenheit relations}   
%_______________________________________________________________________________

\begin{figure}[htb]
\begin{center}
\includegraphics[width=0.76\textwidth]{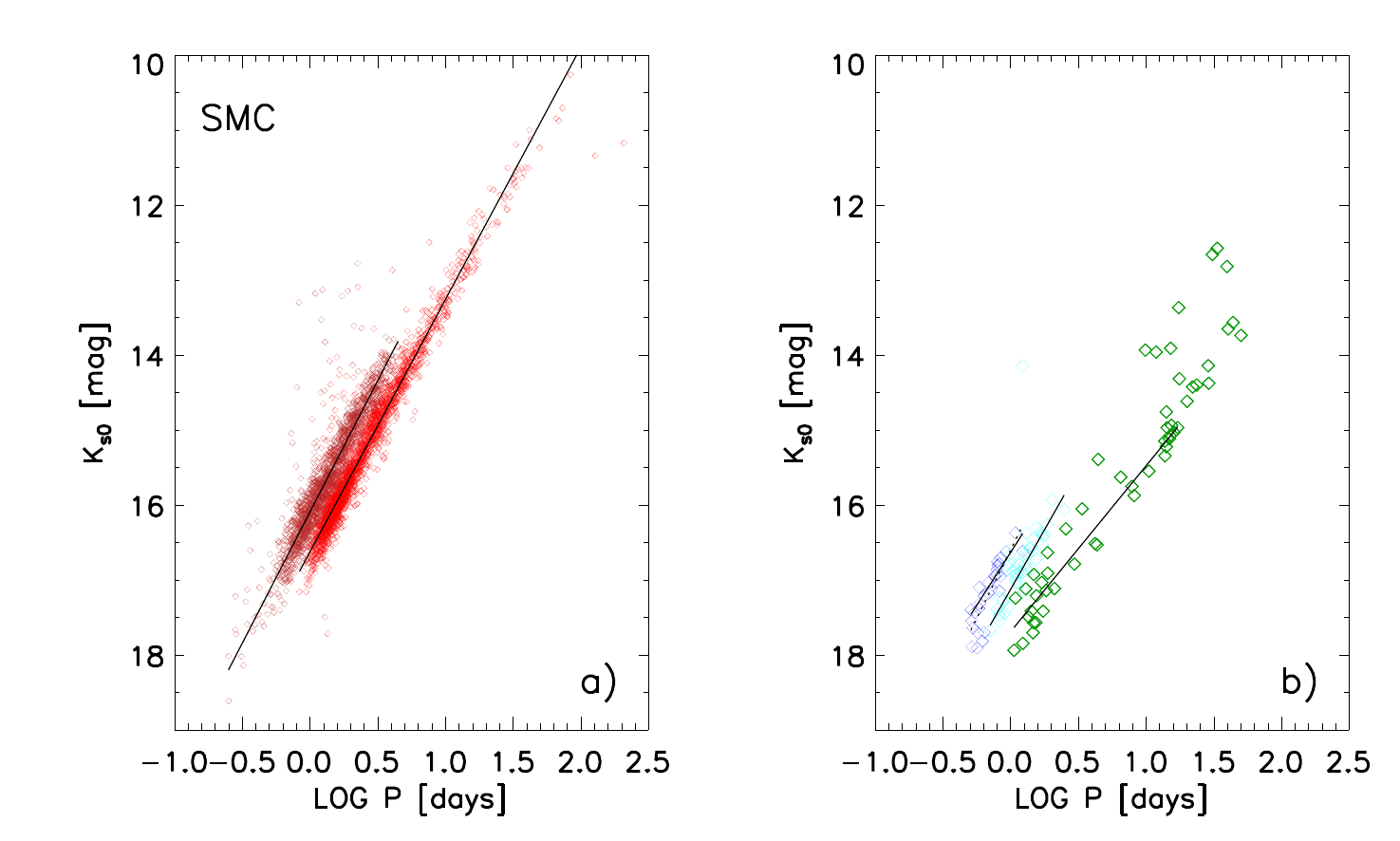}
\includegraphics[width=0.76\textwidth]{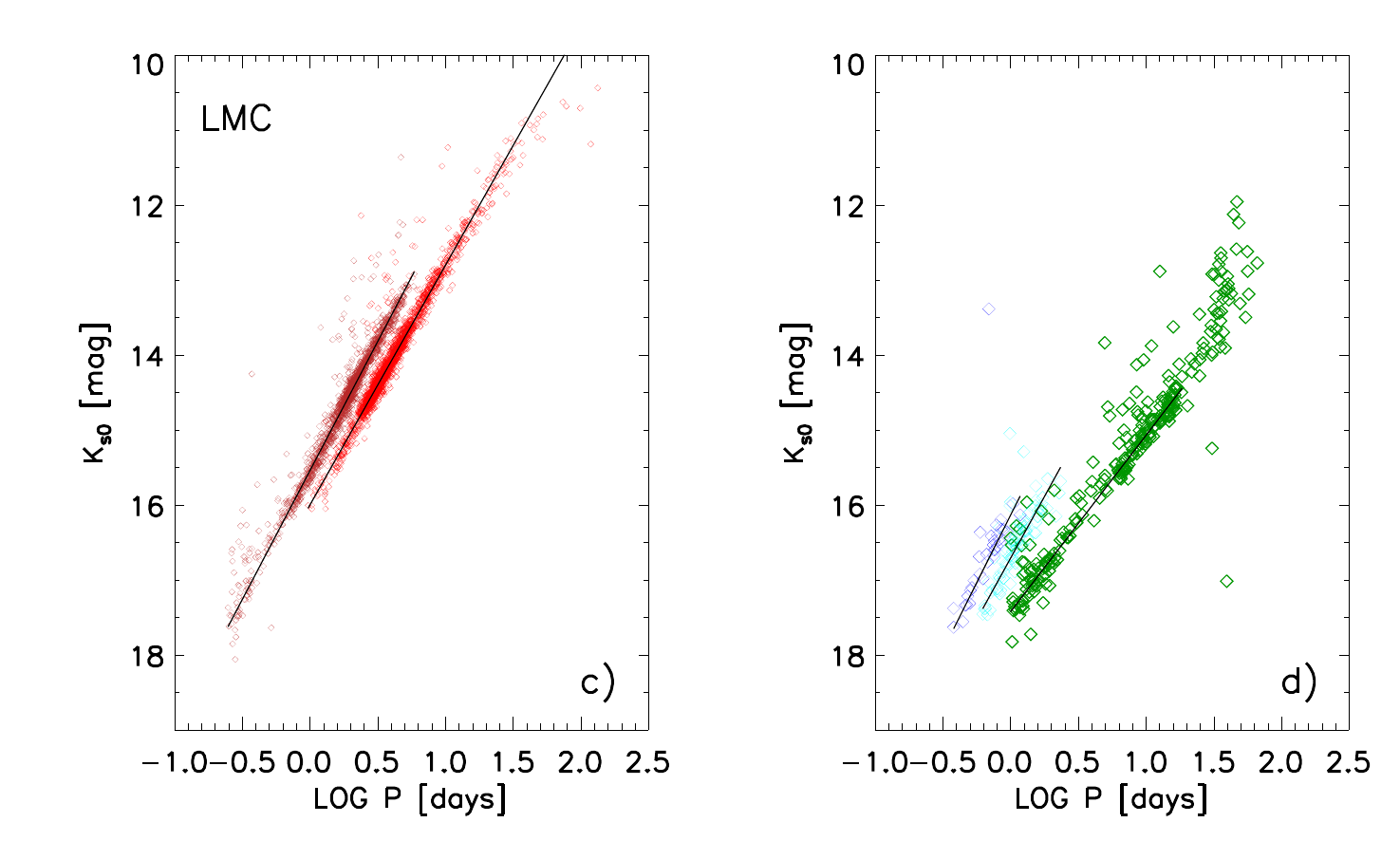}
\caption{Top: Dereddened $K$-band PL relations (see Table~\ref{tbl:cepheid_PL})
of SMC CCs (left) and ACs plus TIICs (right).  From shorter to longer periods, 
the dark red and the light red symbols display FO and FU CCs, 
while the dark blue and the light blue symbols display FO and FU  
ACs and the green symbols the FU TIICs.  The solid black lines display 
the linear fits to the PL relations. 
Bottom: Same as the top, but for LMC Cepheids}
\label{fig_PLK_MC}
\end{center}
\end{figure}

This subsection deals with NIR and optical-NIR diagnostics to estimate individual 
Cepheid distances. Data plotted in Fig.~\ref{fig_PLK_MC} clearly show the advantages 
in using NIR mean magnitudes. The spread for both FO and FU Cepheids is, at fixed 
pulsation period, systematically smaller and the slope becomes systematically 
steeper (see Table~\ref{tbl:cepheid_PL}). The same trend also applies to optical, 
optical-NIR and NIR PW relations (see Table~\ref{tbl:cepheid_PW}).

%%%%%%%%%%%%%%%%%%%%%%%%%%%%%%%%%%%%%%%%%%%%%%%%%%%%%%%%%%%%%%%%%%%%%%%%%%%%%%%%%%%%%
%                        fig  SF 10
%%%%%%%%%%%%%%%%%%%%%%%%%%%%%%%%%%%%%%%%%%%%%%%%%%%%%%%%%%%%%%%%%%%%%%%%%%%%%%%%%%%%%
%_______________________________________________________________________________
%_______________________________________________________________________________

\clearpage

%%%%%%%%%%%%%%%%%%%%%%%%%%%%%%%%%%%%%%%%%%%%%%%%%%%%%%%%%%%%%%%%
%			Table  ST 1 
%%%%%%%%%%%%%%%%%%%%%%%%%%%%%%%%%%%%%%%%%%%%%%%%%%%%%%%%%%%%%%%%
\begin{table}[htbp]
\centering
%\tiny
%\scriptsize
\footnotesize
\caption{\small Optical and NIR PL relations ($M_X\,=\,a\,+\,b \times \log P$) 
for Magellanic Cloud Cepheids.}\label{tbl:cepheid_PL}
\begin{tabular}{ccccccc}
\hline
\hline
band  & mode & a\footnotemark[1]       & b    & $\epsilon_a$ & $\epsilon_b$ & $\sigma$\\ 
\hline
\multicolumn{7}{c}{---SMC CCs---} \\
$V$     & FO &  17.155 &  --3.143 &   0.008 &   0.033 &   0.282 \\
$I$     & FO &  16.723 &  --3.299 &   0.006 &   0.026 &   0.231 \\
$J$     & FO &  16.388 &  --3.398 &   0.006 &   0.024 &   0.205 \\
$H$     & FO &  16.162 &  --3.537 &   0.006 &   0.023 &   0.179 \\
$K$     & FO &  16.101 &  --3.501 &   0.005 &   0.021 &   0.184 \\
$V$     & FU &  17.783 &  --2.907 &   0.009 &   0.018 &   0.267 \\
$I$     & FU &  17.315 &  --3.128 &   0.007 &   0.014 &   0.214 \\
$J$     & FU &  16.935 &  --3.211 &   0.006 &   0.013 &   0.202 \\
$H$     & FU &  16.701 &  --3.376 &   0.006 &   0.013 &   0.186 \\
$K$     & FU &  16.633 &  --3.365 &   0.005 &   0.011 &   0.172 \\
\multicolumn{7}{c}{---SMC ACs---} \\
$V$     & FO &  17.601 &  --2.370 &   0.048 &   0.344 &   0.171 \\
$I$     & FO &  17.216 &  --2.585 &   0.033 &   0.237 &   0.120 \\
$J$     & FO &  16.825 &  --2.859 &   0.027 &   0.214 &   0.055 \\
$H$     & FO &  16.540 &  --3.836 &   0.044 &   0.258 &   0.074 \\
$K$     & FO &  16.638 &  --2.820 &   0.044 &   0.272 &   0.124 \\
$V$     & FU &  18.173 &  --2.256 &   0.031 &   0.208 &   0.233 \\
$I$     & FU &  17.674 &  --2.233 &   0.018 &   0.113 &   0.105 \\
$J$     & FU &  17.392 &  --2.887 &   0.023 &   0.148 &   0.111 \\
$H$     & FU &  17.095 &  --2.649 &   0.069 &   0.393 &   0.230 \\
$K$     & FU &  17.132 &  --3.175 &   0.024 &   0.153 &   0.121 \\
\multicolumn{7}{c}{---SMC TIICs---} \\
$V$     & FU &  18.885 &  --1.677 &   0.092 &   0.127 &   0.325 \\
$I$     & FU &  18.401 &  --1.913 &   0.078 &   0.108 &   0.276 \\
$J$     & FU &  17.973 &  --2.027 &   0.080 &   0.112 &   0.286 \\
$H$     & FU &  17.705 &  --2.113 &   0.122 &   0.170 &   0.355 \\
$K$     & FU &  17.694 &  --2.211 &   0.070 &   0.097 &   0.248 \\
\multicolumn{7}{c}{---LMC CCs---} \\
$V$     & FO &  16.618 &  --3.241 &   0.007 &   0.018 &   0.187 \\
$I$     & FO &  16.207 &  --3.317 &   0.005 &   0.013 &   0.135 \\
$J$     & FO &  15.829 &  --3.364 &   0.004 &   0.011 &   0.115 \\
$H$     & FO &  15.614 &  --3.475 &   0.004 &   0.011 &   0.094 \\
$K$     & FO &  15.549 &  --3.441 &   0.003 &   0.009 &   0.089 \\
$V$     & FU &  17.056 &  --2.690 &   0.011 &   0.016 &   0.180 \\
$I$     & FU &  16.660 &  --2.924 &   0.008 &   0.012 &   0.123 \\
$J$     & FU &  16.266 &  --3.013 &   0.007 &   0.011 &   0.124 \\
$H$     & FU &  16.094 &  --3.232 &   0.007 &   0.011 &   0.117 \\
$K$     & FU &  16.006 &  --3.195 &   0.005 &   0.008 &   0.086 \\
\multicolumn{7}{c}{---LMC ACs---} \\
$V$     & FO &  16.988 &  --3.416 &   0.065 &   0.316 &   0.236 \\
$I$     & FO &  16.679 &  --3.362 &   0.046 &   0.214 &   0.168 \\
$J$     & FO &  16.354 &  --3.390 &   0.047 &   0.224 &   0.152 \\
$H$     & FO &  16.219 &  --3.504 &   0.064 &   0.264 &   0.158 \\
$K$     & FO &  16.143 &  --3.603 &   0.029 &   0.136 &   0.079 \\
$V$     & FU &  17.719 &  --2.582 &   0.028 &   0.186 &   0.252 \\
$I$     & FU &  17.353 &  --2.921 &   0.018 &   0.122 &   0.162 \\
$J$     & FU &  16.980 &  --3.136 &   0.020 &   0.139 &   0.162 \\
$H$     & FU &  16.754 &  --3.311 &   0.015 &   0.102 &   0.080 \\
$K$     & FU &  16.715 &  --3.284 &   0.015 &   0.112 &   0.122 \\
\multicolumn{7}{c}{---LMC TIICs---} \\
$V$     & FU &  18.414 &  --1.602 &   0.036 &   0.046 &   0.262 \\
$I$     & FU &  18.041 &  --1.954 &   0.022 &   0.028 &   0.170 \\
$J$     & FU &  17.601 &  --2.063 &   0.027 &   0.034 &   0.192 \\
$H$     & FU &  17.497 &  --2.326 &   0.035 &   0.043 &   0.206 \\
$K$     & FU &  17.428 &  --2.352 &   0.017 &   0.021 &   0.109 \\
\hline
\end{tabular}
\footnotetext{From left to right the columns list the photometric band, the mode, 
the zero-point, the slope and their uncertainties and the standard deviation 
(mag).}
\footnotemark[1]{To obtain the absolute zero points, one should subtract the geometrical distance modulus
of the LMC and SMC, based on eclipsing binaries (18.477$\pm$0.026 mag and 18.977$\pm$0.028 mag
for the LMC and SMC, respectively, \citep{pietrzynski2019,graczyk2020})}
\end{table}

%
%%%%%%%%%%%%%%%%%%%%%%%%%%%%%%%%%%%%%%%%%%%%%%%%%%%%%%%%%%%%%%%%
%			Table ST 2 
%%%%%%%%%%%%%%%%%%%%%%%%%%%%%%%%%%%%%%%%%%%%%%%%%%%%%%%%%%%%%%%%
\begin{table}[htbp]
\centering
\footnotesize
\caption{\small Optical, NIR and optical--NIR PW relations 
($M_X\,=\,a\,+\,b \times \log P\,+\,c \times CI$) 
for Magellanic Cloud Cepheids.}\label{tbl:cepheid_PW}
\begin{tabular}{lccccccc}
\hline
\hline
band/CI  & mode & a\footnotemark[1]       & b    & $\epsilon_a$ & $\epsilon_b$ & c & $\sigma$\\ 
\hline
\multicolumn{8}{c}{---SMC CCs---} \\
$I,V-I$ & FO &  15.966 &  --3.570 &   0.005 &   0.020 &  1.550 &   0.173 \\
$J,V-J$ & FO &  16.064 &  --3.554 &   0.005 &   0.022 &  0.406 &   0.186 \\
$H,V-H$ & FO &  15.940 &  --3.642 &   0.005 &   0.021 &  0.217 &   0.165 \\
$K,V-K$ & FO &  15.956 &  --3.569 &   0.005 &   0.021 &  0.134 &   0.174 \\
$K,I-K$ & FO &  15.949 &  --3.570 &   0.005 &   0.020 &  0.250 &   0.175 \\
$K,J-K$ & FO &  15.904 &  --3.576 &   0.005 &   0.021 &  0.690 &   0.178 \\
$K,V-I$ & FO &  16.047 &  --3.528 &   0.005 &   0.021 &  0.288 &   0.179 \\
$I,V-I$ & FU &  16.491 &  --3.457 &   0.005 &   0.011 &  1.550 &   0.159 \\
$J,V-J$ & FU &  16.580 &  --3.359 &   0.006 &   0.013 &  0.406 &   0.187 \\
$H,V-H$ & FU &  16.450 &  --3.471 &   0.006 &   0.013 &  0.217 &   0.184 \\
$K,V-K$ & FU &  16.473 &  --3.429 &   0.005 &   0.011 &  0.134 &   0.163 \\
$K,I-K$ & FU &  16.468 &  --3.432 &   0.005 &   0.011 &  0.250 &   0.164 \\
$K,J-K$ & FU &  16.422 &  --3.466 &   0.005 &   0.011 &  0.690 &   0.172 \\
$K,V-I$ & FU &  16.573 &  --3.396 &   0.005 &   0.011 &  0.288 &   0.166 \\
\multicolumn{8}{c}{---SMC ACs---} \\
$I,V-I$ & FO &  16.517 &  --3.470 &   0.026 &   0.150 &  1.550 &  0.083 \\
$J,V-J$ & FO &  16.519 &  --3.217 &   0.062 &   0.381 &  0.406 &  0.143 \\
$H,V-H$ & FO &  16.299 &  --4.136 &   0.044 &   0.236 &  0.217 &  0.075 \\
$K,V-K$ & FO &  16.502 &  --2.763 &   0.045 &   0.267 &  0.134 &  0.119 \\
$K,I-K$ & FO &  16.343 &  --3.674 &   0.018 &   0.110 &  0.250 &  0.037 \\
$K,J-K$ & FO &  16.424 &  --2.776 &   0.048 &   0.286 &  0.690 &  0.128 \\
$K,V-I$ & FO &  16.627 &  --2.527 &   0.043 &   0.261 &  0.288 &  0.100 \\
$I,V-I$ & FU &  16.936 &  --2.918 &   0.017 &   0.114 &  1.550 &  0.121 \\
$J,V-J$ & FU &  17.053 &  --3.042 &   0.027 &   0.175 &  0.406 &  0.134 \\
$H,V-H$ & FU &  16.842 &  --2.680 &   0.078 &   0.442 &  0.217 &  0.259 \\
$K,V-K$ & FU &  16.981 &  --3.234 &   0.025 &   0.159 &  0.134 &  0.126 \\
$K,I-K$ & FU &  16.976 &  --3.252 &   0.024 &   0.159 &  0.250 &  0.126 \\
$K,J-K$ & FU &  16.912 &  --3.158 &   0.022 &   0.145 &  0.690 &  0.108 \\
$K,V-I$ & FU &  17.073 &  --3.195 &   0.024 &   0.156 &  0.288 &  0.124 \\
\multicolumn{8}{c}{---SMC TIICs---} \\
$I,V-I$ & FU &  17.554 &  --2.309 &   0.056 &   0.079 &  1.550 & 0.201 \\
$J,V-J$ & FU &  17.604 &  --2.193 &   0.079 &   0.111 &  0.406 & 0.283 \\
$H,V-H$ & FU &  17.470 &  --2.244 &   0.124 &   0.173 &  0.217 & 0.361 \\
$K,V-K$ & FU &  17.536 &  --2.290 &   0.067 &   0.094 &  0.134 & 0.239 \\
$K,I-K$ & FU &  17.527 &  --2.291 &   0.069 &   0.097 &  0.250 & 0.247 \\
$K,J-K$ & FU &  17.502 &  --2.338 &   0.072 &   0.101 &  0.690 & 0.257 \\
$K,V-I$ & FU &  17.635 &  --2.249 &   0.067 &   0.093 &  0.288 & 0.238 \\
\multicolumn{8}{c}{---LMC CCs---} \\
$I,V-I$ & FO &  15.396 &  --3.437 &   0.003 &   0.008 &  1.550 &  0.079 \\
$J,V-J$ & FO &  15.479 &  --3.448 &   0.004 &   0.010 &  0.406 &  0.099 \\
$H,V-H$ & FO &  15.377 &  --3.538 &   0.004 &   0.010 &  0.217 &  0.083 \\
$K,V-K$ & FO &  15.398 &  --3.490 &   0.003 &   0.009 &  0.134 &  0.079 \\
$K,I-K$ & FO &  15.389 &  --3.493 &   0.003 &   0.009 &  0.250 &  0.082 \\
$K,J-K$ & FO &  15.355 &  --3.502 &   0.003 &   0.009 &  0.690 &  0.082 \\
$K,V-I$ & FO &  15.503 &  --3.465 &   0.003 &   0.009 &  0.288 &  0.086 \\
$I,V-I$ & FU &  15.889 &  --3.315 &   0.005 &   0.008 &  1.550 &  0.077 \\
$J,V-J$ & FU &  15.922 &  --3.183 &   0.007 &   0.011 &  0.406 &  0.120 \\
$H,V-H$ & FU &  15.858 &  --3.352 &   0.008 &   0.012 &  0.217 &  0.120 \\
$K,V-K$ & FU &  15.850 &  --3.263 &   0.005 &   0.007 &  0.134 &  0.078 \\
$K,I-K$ & FU &  15.844 &  --3.269 &   0.005 &   0.008 &  0.250 &  0.079 \\
$K,J-K$ & FU &  15.812 &  --3.299 &   0.005 &   0.008 &  0.690 &  0.091 \\
$K,V-I$ & FU &  15.961 &  --3.233 &   0.005 &   0.008 &  0.288 &  0.082 \\
\hline\noalign{\smallskip}
\multicolumn{4}{r}{\it {\footnotesize continued on next page}} \\
\end{tabular}
\end{table}
\addtocounter{table}{-1}
\begin{table}[htbp]
\centering
\footnotesize
\caption[]{continued.}
\begin{tabular}{lccccccc}
\hline
\hline
band/CI  & mode & a\footnotemark[1]       & b    & $\epsilon_a$ & $\epsilon_b$ & c & $\sigma$\\ 
\hline
\multicolumn{8}{c}{---LMC ACs---} \\
$I,V-I$ & FO &  16.012 &  --3.475 &   0.038 &   0.183 &  1.550 &  0.127 \\
$J,V-J$ & FO &  16.088 &  --3.380 &   0.045 &   0.225 &  0.406 &  0.142 \\
$H,V-H$ & FO &  16.003 &  --3.690 &   0.076 &   0.326 &  0.217 &  0.181 \\
$K,V-K$ & FO &  16.060 &  --3.462 &   0.030 &   0.152 &  0.134 &  0.089 \\
$K,I-K$ & FO &  16.022 &  --3.535 &   0.028 &   0.133 &  0.250 &  0.074 \\
$K,J-K$ & FO &  16.025 &  --3.355 &   0.037 &   0.172 &  0.690 &  0.115 \\
$K,V-I$ & FO &  16.112 &  --3.610 &   0.030 &   0.148 &  0.288 &  0.079 \\
$I,V-I$ & FU &  16.578 &  --2.825 &   0.014 &   0.089 &  1.550 &  0.108 \\
$K,J-K$ & FU &  16.514 &  --3.327 &   0.016 &   0.120 &  0.406 &  0.131 \\
$K,V-K$ & FU &  16.573 &  --3.396 &   0.016 &   0.115 &  0.217 &  0.123 \\
$K,V-I$ & FU &  16.669 &  --3.332 &   0.016 &   0.113 &  0.134 &  0.123 \\
$K,I-K$ & FU &  16.557 &  --3.233 &   0.017 &   0.119 &  0.250 &  0.136 \\
$J,V-J$ & FU &  16.642 &  --3.160 &   0.020 &   0.137 &  0.690 &  0.156 \\
$H,V-H$ & FU &  16.570 &  --3.406 &   0.023 &   0.142 &  0.288 &  0.127 \\
\multicolumn{8}{c}{---LMC TIICs---} \\
$I,V-I$ & FU &  17.326 &  --2.496 &   0.014 &   0.018 &  1.550 & 0.101 \\
$J,V-J$ & FU &  17.280 &  --2.307 &   0.031 &   0.038 &  0.406 & 0.202 \\
$H,V-H$ & FU &  17.260 &  --2.470 &   0.042 &   0.052 &  0.217 & 0.237 \\
$K,V-K$ & FU &  17.263 &  --2.433 &   0.017 &   0.021 &  0.134 & 0.107 \\
$K,I-K$ & FU &  17.264 &  --2.448 &   0.017 &   0.021 &  0.250 & 0.114 \\
$K,J-K$ & FU &  17.275 &  --2.523 &   0.020 &   0.024 &  0.690 & 0.129 \\
$K,V-I$ & FU &  17.368 &  --2.375 &   0.016 &   0.020 &  0.288 & 0.101 \\
\hline
\end{tabular}
\footnotetext{From left to right the columns list the photometric band and the color 
adopted to define the Wesenheit pseudo magnitude, the mode, the zero-point, 
the slope of the PW relation, the uncertainties on zero-point and slope, 
the color index and the standard deviation (mag).}
\footnotemark[1]{To obtain the absolute zero points, one should subtract the geometrical distance modulus
of the LMC and SMC, based on eclipsing binaries (18.477$\pm$0.026 mag and 18.977$\pm$0.028 mag
for the LMC and SMC, respectively \citep{pietrzynski2019,graczyk2020}).}
\end{table}

\clearpage
%_________________________________________________________________________
\section{Evolutionary constraints for Cepheids}\label{evolution}

Cepheid pulsation properties are the convolution between evolutionary and 
pulsation properties of low- and intermediate-mass stars. Panel a) of 
%Fig.~\ref{fig:hrd_three} shows the comparison in the HRD between 
Fig.~4 shows the comparison in the HRD between 
BASTI-IAC canonical scaled solar 
evolutionary models at solar chemical composition, covering a broad range in 
stellar masses and the CC instability strip \citep{fiorentino02}. 
The dashed lines display hydrogen-burning phases, while solid purple lines 
display helium-burning phases, and the solid cyan lines the double shell 
burning phases (AGB). The blue and the red almost vertical lines show the 
first overtone blue edge and the fundamental red edge for the same chemical 
composition. These boundaries bracket the region of the HRD, in which CCs 
are expected to be pulsationally stable. Evolutionary and pulsation 
predictions plotted in this panel indicate that CCs at solar iron abundance 
are expected to cover a broad range in pulsation period. The minimum 
mass crossing the Cepheid instability strip is of the order of $M=3.0$--$3.5\,M_\odot$,
while the most massive ones are still crossing the instability strip
during helium-burning phases.

%Panels b) and c)  of Fig.~\ref{fig:hrd_three} display the same predictions, 
Panels b) and c)  of Fig.~4 display the same predictions, 
but for stellar structures constructed by assuming chemical compositions that are 
half solar Z=0.01 and Z=0.001, respectively. A glance at the predictions plotted 
in these panels shows quite clearly that the minimum mass crossing the CC instability 
strip steadily decreases 
when moving into the more metal-poor regimes. This explains why the period 
distribution shifts toward shorter periods moving from the MW to the SMC.
Moreover, evolutionary tracks plotted in the bottom panel show that a 
significant fraction of more massive stellar structures cross the instability 
strip during the AGB (cyan color). This means a decrease in the evolutionary 
time spent inside the instability strip, and in turn, it explains the lack of 
long period Cepheids in metal-poor stellar systems like 
IC~1613 \citep{pietrzynski2006b} and WLM \citep{pietrzynski2007}.

In order to investigate this issue on a more quantitative basis, 
Table~\ref{tbl:cc_crossing_time} gives for four different chemical 
compositions the difference in luminosity and in evolutionary lifetime 
for the three different crossings based on evolutionary models taking 
into account convective core overshooting during central hydrogen 
burning phases.
%%%%%%%%%%%%%%%%%%%%%%%%%%%%%%%%%%%%%%%%%%%%%%%%%%%%%%%%%%%%%%%%%%%%%%%%%%%%%%%%%%%%%%%%%%%%%%%%%%%%
%			Table  ST 3 
%%%%%%%%%%%%%%%%%%%%%%%%%%%%%%%%%%%%%%%%%%%%%%%%%%%%%%%%%%%%%%%%%%%%%%%%%%%%%%%%%%%%%%%%%%%%%%%%%%%%
\begin{table}[htbp!]
\centering
\footnotesize
\caption{\small Mean luminosity and evolutionary lifetime for the three different crossings of 
intermediate-mass evolutionary models constructed by taking into account convective core 
overshooting during central hydrogen burning phases.}
\label{tbl:cc_crossing_time}
\begin{tabular}{rrrrrrr}
\hline
\hline
%M/M$_\odot$&log(L/L$_\odot$)$_{1st}$& t(1st) &log(L/L$_\odot$)$_{2nd}$& t(2nd) &log(L/L$_\odot$)$_{3rd}$& t(3rd) \\ 
$M/M_\odot$ &  $\langle \log(L/L_\odot) \rangle^a$          & t(Kyr) & $\langle \log(L/L_\odot) \rangle^a$ & t(Kyr)         & $\langle \log(L/L_\odot) \rangle^a$ & t(Kyr) \\
           &  \multicolumn{2}{c}{1st crossing} & \multicolumn{2}{c}{2nd crossing} & \multicolumn{2}{c}{3rd crossing} \\
\hline
%          &            &             &            &               &            &             \\
                         \multicolumn{7}{c}{[Fe/H]=+0.06} \\ 
    2.8   &   2.099  &     251.81    &    \ldots    &     \ldots   & \ldots     &      \ldots \\ 
    3.0   &   2.205  &     170.61    &    \ldots    &     \ldots   & \ldots     &      \ldots \\ 
    3.5   &   2.436  &      73.56    &    \ldots    &     \ldots   & \ldots     &      \ldots \\
    4.0   &   2.636  &      39.95    &    \ldots    &     \ldots   & \ldots     &      \ldots \\
    4.5   &   2.812  &      23.00    &    \ldots    &     \ldots   & \ldots     &      \ldots \\
    5.0   &   2.970  &      14.50    &    3.163     &  2880.21$^b$ &  \dots     &      \dots  \\
    6.0   &   3.242  &       7.00    &    3.459     &     138.81   &  3.501     &      191.63 \\
    7.0   &   3.472  &       4.00    &    3.704     &       53.40  &  3.739     &       59.52 \\
    8.0   &   3.649  &       2.50    &    3.922     &       16.96  &  3.933     &        27.52\\
    9.0   &   3.818  &       1.56    &    4.109     &        9.45  &  4.098     &        15.28\\
   10.0   &   3.937  &       1.20    &    \ldots    &     \ldots   & \ldots     &      \ldots \\
   11.0   &   3.998  &       0.79    &    4.402     &        5.22  &  4.339     &         6.46\\
   12.0   &   4.109  &       0.70    &    4.490     &        5.22  &  4.443     &         5.40\\
   13.0   &   4.303  &       1.12    &    \ldots    &     \ldots   & \ldots     &      \ldots \\ 
   14.0   &   4.622  &       9.09    &    \ldots    &     \ldots   & \ldots     &      \ldots \\ 
 %         &          &               &              &              &            &             \\
                             \multicolumn{7}{c}{[Fe/H]=-0.30} \\
    2.8  &   2.210   &   159.59    &    \ldots &     \ldots   & \ldots    &      \ldots \\
    3.0  &   2.324   &   111.45    &   \ldots  &     \ldots   & \ldots    &      \ldots \\
    3.5  &   2.539   &    52.57    &   \ldots  &     \ldots   & \ldots    &      \ldots \\ 
    4.0  &   2.736   &    29.07    &   2.907   &      45.04   &   2.982   &     2225.45 \\
    4.5  &   2.910   &    17.79    &   3.113   &     364.28   &   3.179   &     365.68 \\
    5.0  &   3.068   &    11.90    &   3.289   &    307.13    &   3.348   &     192.83 \\ 
    6.0  &   3.340   &     6.09    &   3.570   &    140.94    &   3.612   &      36.92 \\ 
    7.0  &   3.565   &     3.70    &   3.807   &     41.16    &   3.821   &       16.35 \\ 
    8.0  &   3.749   &     2.58    &   4.009   &      17.75   &   3.998   &      10.73 \\ 
    9.0  &   3.896   &     1.90    &   4.157   &     12.57    &   4.103   &       5.28 \\ 
   10.0  &   3.991   &     1.29    &   4.314   &     12.29    &   4.273   &       9.08 \\ 
   11.0  &   4.099   &     1.16    &   4.406   &       9.48   &   4.291   &       2.25 \\ 
   12.0  &   4.239   &     1.31    &   4.498   &     10.30    &   4.380   &       2.24 \\
   13.0  &   4.421   &     2.23    &   4.610   &     15.56    &   4.601   &      11.20 \\ 
%           &            &           &              &              &           &             \\
                             \multicolumn{7}{c}{[Fe/H]=-0.60} \\
    2.8   &   2.296    &   114.17    &  \ldots    &     \ldots    & \ldots     &      \ldots  \\
    3.0   &   2.397    &    82.87    &  \ldots    &     \ldots    & \ldots     &      \ldots  \\
    3.5   &   2.623    &    41.54    &    2.814   &     737.93    &   2.898    &     933.95    \\
    4.0   &   2.819    &    23.61    &    3.035   &     645.04    &   3.101    &     248.89   \\
    4.5   &   2.992    &    15.17    &    3.232   &     682.77    &   3.273    &     186.44   \\
    5.0   &   3.148    &    10.35    &    3.403   &     551.16    &   3.428    &     176.14   \\
    6.0   &   3.417    &    5.70     &   3.659    &    232.49     &   3.694    &      99.88   \\
    7.0   &   3.628    &    4.00     &   3.882    &     55.88     &   3.912    &      55.91   \\
    8.0   &   3.807    &    3.00     &   4.070    &     25.61     &   4.085    &      33.60   \\
    9.0   &   3.944    &    1.95     &   4.221    &    19.35      &   4.216    &      17.50   \\
   10.0   &   4.066    &    1.38     &   4.334    &     16.28     &   4.300    &       6.24   \\
   11.0   &   4.197    &    1.66     &   4.435    &     19.41     &   4.396    &       5.24   \\
   12.0   &   4.378    &    2.64     &   4.556    &     36.83     &   4.591    &      15.55   \\
   13.0   &   4.724    &   105.75    & \ldots     &    \ldots     &  \ldots    &      \ldots  \\
%             &            &           &              &              &           &             \\
                             \multicolumn{7}{c}{[Fe/H]=-0.90} \\
    2.6   &   2.262    &   129.10    &  \ldots    &     \ldots    & \ldots     &      \ldots  \\
    2.8   &   2.371    &    91.97    &   2.615    &     3449.14   &   2.646    &    4483.62   \\
    3.0   &   2.472    &    66.21    &   2.737    & 6159.92$^b$   &  \ldots    &      \ldots  \\
    3.5   &   2.698    &    35.14    &   2.964    & 5414.59$^b$   &  \ldots    &      \ldots  \\
    4.0   &   2.893    &    20.54    &   3.132    &     1129.51   &   3.186    &     315.84   \\
    4.5   &   3.066    &    13.50    &   3.291    &      595.31   &   3,359    &     134.27   \\
    5.0   &   3.221    &     9.50    &   3.435    &      370.46   &   3.509    &      66.84   \\
    6.0   &   3.481    &     6.00    &   3.686    &      107.64   &   3.754    &      24.09   \\
    7.0   &   3.692    &     3.71    &   3.904    &       31.16   &   3.947    &      13.18   \\
    8.0   &   3.868    &     2.71    &   4.087    &       22.71   &   4.106    &       9.37   \\
    9.0   &   4.013    &     2.41    &   4.237    &       25.18   &   4.212    &       4.92   \\
   10.0   &   4.148    &     2.32    &   4.358    &       34.52   &   4.341    &       4.40   \\
   11.0   &   4.317    &     3.14    &   4.474    &       74.24   &   4.534    &      10.85   \\
   12.0   &   4.652    &    28.52    &   \ldots   &    \ldots     &  \ldots    &      \ldots  \\
   13.0   &   4.738    &    18.84    &   \ldots   &    \ldots     &  \ldots    &      \ldots  \\
\hline
\hline
\end{tabular}
\footnotetext{$^a$ The mean luminosity was estimated as the mean along the crossing (see text for more details).
$^b$ The blue loop is entirely located inside the instability strip.}
\end{table}
%%%%%%%%%%%%%%%%%%%%%%%%%%%%%%%%%%%%%%%%%%%%%%%%%%%%%%%%%%%%%%%%%%%%%%%%%%%%%%%%%%%%%%%%%%%%%%%%%%%%

%Figure~\ref{fig:hrd_four} shows the same comparison as
%Fig.~\ref{fig:hrd_three}, but for intermediate-age 
Figure~7  shows the comparison in the HRD between 
BASTI-IAC canonical evolutionary models for intermediate-age 
stellar structures and the AC instability strip \citep{monelli2022}. 
The blue tips (the hottest point during helium 
burning phases) when moving from more metal-rich to more metal-poor stellar 
structures become systematically hotter and cross the Anomalous Cepheid 
instability strip (blue and red almost vertical lines). This trend is 
even more evident for stellar structures taking into account convective 
core overshooting.  

%_________________________________________________________________________
%                           Type II 
%_________________________________________________________________________

%Evolutionary and pulsation predictions for TIICs plotted in Fig.~\ref{theo_tracks}
Evolutionary and pulsation predictions for TIICs plotted in Fig.~10
% editor 16 
cover two dex in metal/hydrogen abundance ratio. However, TIICs evolutionary 
status is minimally 
affected by the metal content, since they are either AGB or post AGB stars during 
their lifetime. This property is independent of the pulsation period, i.e. 
from BL Herculis to RV Tauri variables. Moreover, the current evidence supports their 
presence both in metal-poor and in metal-rich stellar systems. The key point is 
the presence of hot and extreme HB stars, i.e. the low mass-tail of HB stars.

%%%%%%%%%%%%%%%%%%%%%%%%%%%%%%%%%%%%%%%%%%%%%%%%%%%%%%%%%%%%%%%%%%%%%%%%%%%%%%%%%%%%%%
\subsection{Helium burning lifetimes}\label{sec:He_burn}
%%%%%%%%%%%%%%%%%%%%%%%%%%%%%%%%%%%%%%%%%%%%%%%%%%%%%%%%%%%%%%%%%%%%%%%%%%%%%%%%%%%%%%%%%%%%%%%%%%%%

\begin{figure}[htbp!]
\centering
\includegraphics[width=0.49\textwidth]{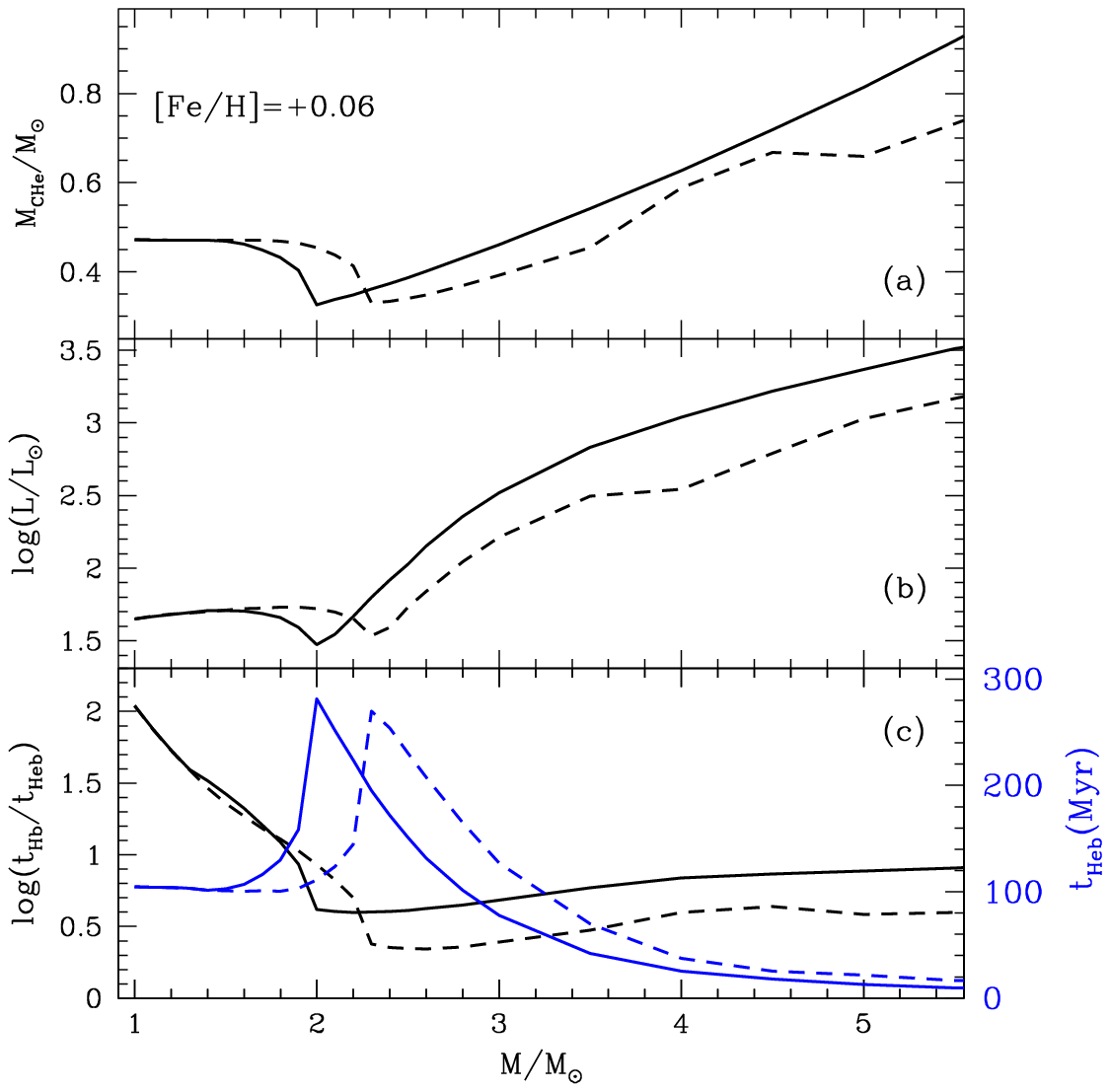}
\includegraphics[width=0.49\textwidth]{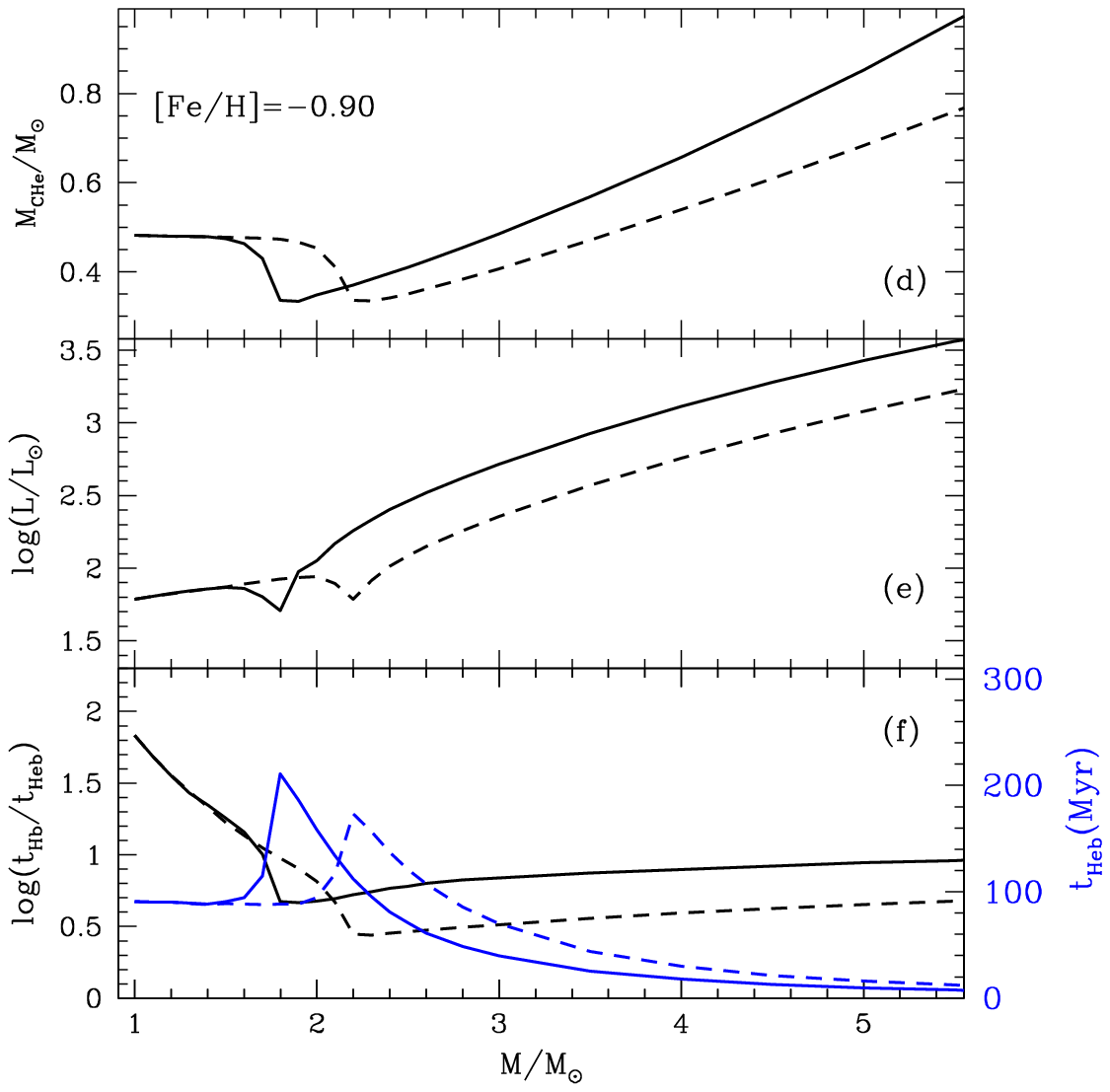}
\includegraphics[width=0.49\textwidth]{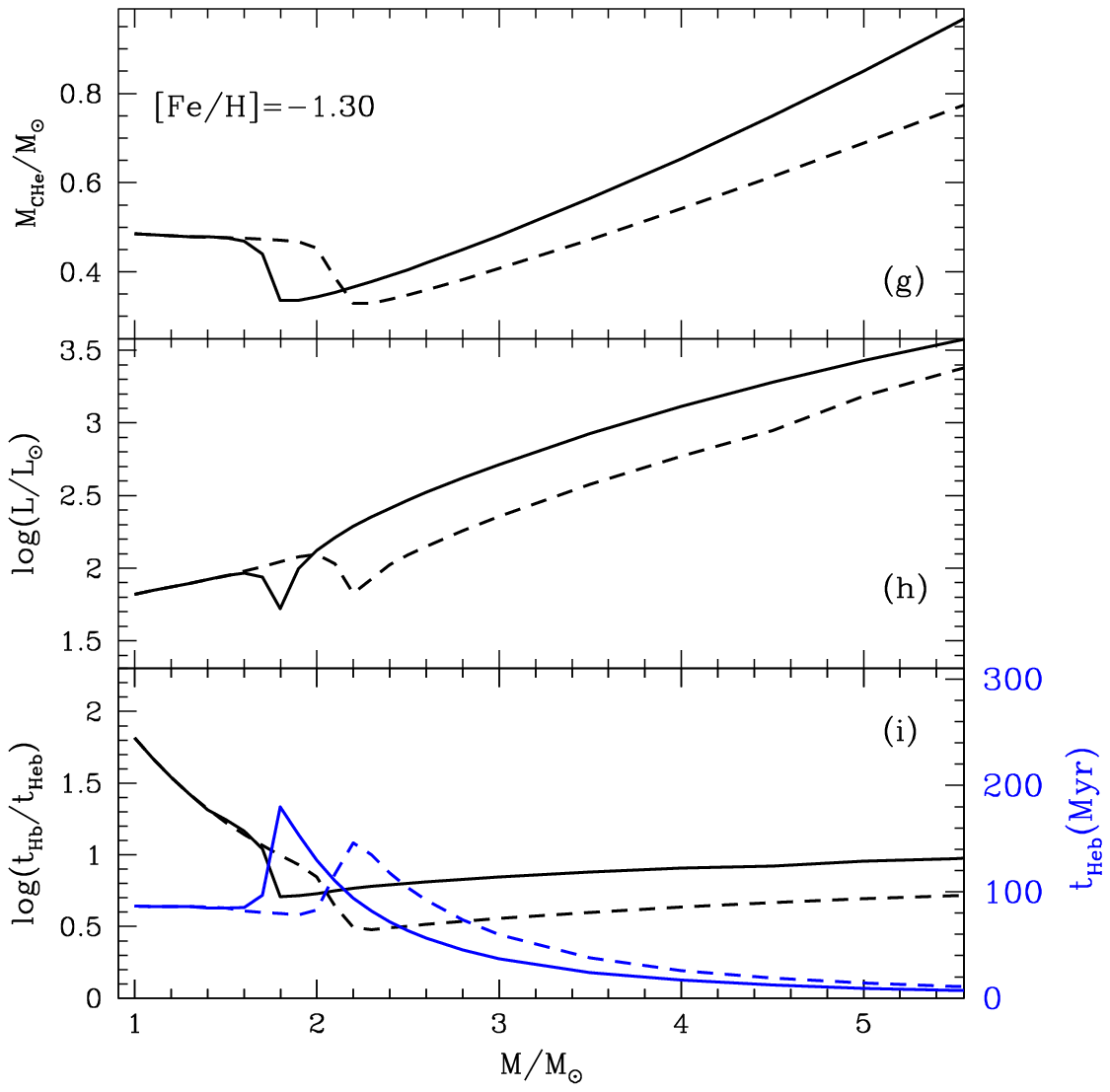}
\includegraphics[width=0.49\textwidth]{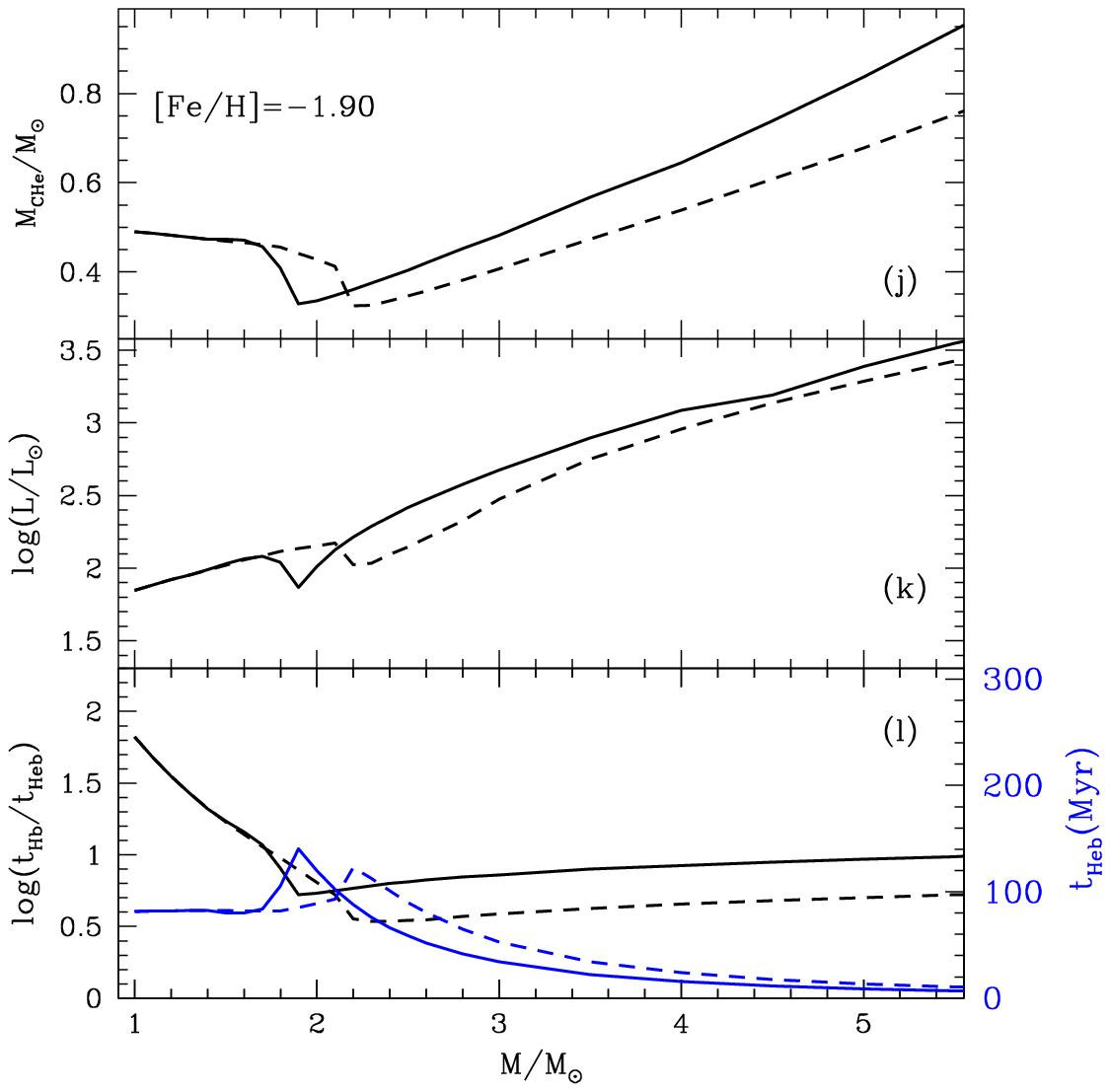}
\caption{
Top left: Panel a)-- Core helium mass (solar units) just before central 
helium burning phases as a function of the stellar mass (solar units) 
at solar metal content. The solid lines display predictions based on 
evolutionary models taking account of convective core overshooting, 
while the dashed lines on evolutionary models that neglect convective 
core overshooting. 
Panel b)-- Logarithmic luminosity (solar units) at the onset of central 
helium-burning phases as a function of the stellar mass (solar units).  
Panel c)-- Logarithmic ratio between central hydrogen and central 
helium-burning evolutionary times (black line). The blue line shows the 
central helium-burning evolutionary time, the scale is showed on the right 
side of the panel. 
Top right: Panels d,e,f) same as the top left, but for a more metal-poor chemical composition [Fe/H]$\,=-0.90$. 
Bottom left: Panels g,h,i) same as the top left, but for a more metal-poor chemical composition [Fe/H]$\,=-1.30$.
Bottom right: Panels j,k,l) same as the top left, but for a more metal-poor chemical composition [Fe/H]$\,=-1.60$.
}
\label{fig:times_h_he}
\end{figure}

The top left panels of Fig.~\ref{fig:times_h_he} display three key 
parameters for investigating central helium burning phases of both 
low- and intermediate-mass stellar structures at solar metal content. 
The solid lines plotted in panel a) show the helium-core mass (solar units) 
as a function of the stellar mass just before the onset of central helium 
burning. The helium-core mass is, as expected, almost constant 
($M_{\rm CHe}/M_\odot\sim 0.5$) for stellar masses ranging from 1 to $1.6\,M_\odot$, 
attains a well-defined minimum ($M_{\rm CHe}/M_\odot\sim 0.32$) for $M\sim1.8\,M_\odot$ 
and then steadily increases. This minimum marks the transition between low-mass stellar structures that 
ignite helium in electron-degenerate core through a core-helium flash and intermediate-mass stars that 
ignite helium quiescently.  Hydrogen burning for stellar structures at solar iron abundance 
and more massive than $1.6\,M_\odot$ is almost entirely driven by the CNO cycle. The strong 
dependence of the CNO cycle on the temperature causes an increase in the core temperature 
gradient, and in turn, the transition into the convective regime.  
The onset of convective cores causes the steady increase in the helium core mass soon 
after the minimum. This transition has been defined by \citet{sweigart1989} as the 
{\em RGB transition}. Solid and dashed line display predictions for evolutionary 
models constructed either taking account of or neglecting convective core overshooting. 
Canonical models display, as expected, the same transition, but at larger 
stellar masses ($M\sim 2.2\,M_\odot$). 

Panel b) shows the logarithmic luminosity (solar units) at the onset of central helium burning 
as a function of the stellar mass. Predictions plotted in this panel display a well 
known result. The luminosity of evolutionary models taking account 
for the convective core overshooting is, at fixed stellar mass, systematically 
brighter than canonical models. Indeed, the former models have, at fixed stellar mass, 
larger helium-core masses.

%%%%%%%%%%%%%%%%%%%%%%%%%%%%%%%%%%%%%%%%%%%%%%%%%%%%%%%%%%%%%%%%%%%%%%%%%%%%%%%%%%%%%
% 			fig  SF 11 
%%%%%%%%%%%%%%%%%%%%%%%%%%%%%%%%%%%%%%%%%%%%%%%%%%%%%%%%%%%%%%%%%%%%%%%%%%%%%%%%%%%%%
%________________________________________________________________________________

Panel c) shows the logarithm of the ratio between central hydrogen and helium 
burning time. Predictions indicate that this ratio decreases by 1.5 dex when moving 
from the low- to the intermediate-mass regime. This decrease is caused by the fact that, 
across the transition, the luminosity of central helium-burning structures decreases and 
the central helium-burning lifetime increases by more than a factor of three. Indeed, the 
central helium-burning time (solid blue line) increases from $\approx$100 Myr to almost 
 $\approx$300 Myr. The main consequence of this increase is that central hydrogen 
burning lifetime across the transition is  only a factor of four 
longer than central helium-burning lifetime. The hydrogen burning lifetime is 
two order of magnitude longer in the low-mass regime, and roughly one order of magnitude 
longer in the intermediate-mass regime. The stellar structures at the 
transition between low- and intermediate-mass stellar structures are strong 
contributors of central helium-burning stellar populations \citep[red clump,][]{girardi2016}. 
Predictions based on canonical evolutionary models display the same trend. 
The main difference being that central hydrogen burning lifetime is, at the transition, 
only a factor of two longer than central helium lifetime. Thus further increases the 
fraction of red clump stars compared with main-sequence stars. 

Evolutionary predictions plotted in panels d), e), and f) and in the bottom 
panels of Fig.~\ref{fig:times_h_he} display the same predictions, but for more 
metal-poor chemical compositions (see labelled values). The decrease in 
the metal content causes, as expected, a mild decrease in the stellar mass 
at which the transition between low- and intermediate-mass stars takes place.  
Indeed, it decreases from $M\sim2.3\,M_\odot$ at solar iron abundance to 
$M\sim2.2\,M_\odot$ at [Fe/H]$=-1.90$. The decrease is caused by the fact that 
more metal-poor stellar structure are, at fixed stellar mass, brighter and 
hotter than more metal-rich stellar structures. Therefore, the former stellar 
structures ignite quiescently helium at lower stellar masses  \citep{castellani95}.

Evolutionary models constructed by taking into account convective core overshooting 
during central hydrogen burning show a similar trend. The difference is also 
discussed in section~\ref{sec:ML_ACs}; here we simply mention that the transition 
decreases from  $M\sim2.0\,M_\odot$ to $M\sim1.90\,M_\odot$ when moving 
from solar iron abundance to [Fe/H]$=-1.90$.

%_____________________________________________________________________________________________
%%%%%%%%%%%%%%%%%%%%%%%%%%%%%%%%%%%%%%%%%%%%%%%%%%%%%%%%%%%%%%%%%%%%%%%%%%%%%%%%%%%%%%
\subsection{On the mass luminosity relation of ACs}\label{sec:ML_ACs}

\begin{figure}[htbp]
\begin{center}
\includegraphics[width=0.8\textwidth]{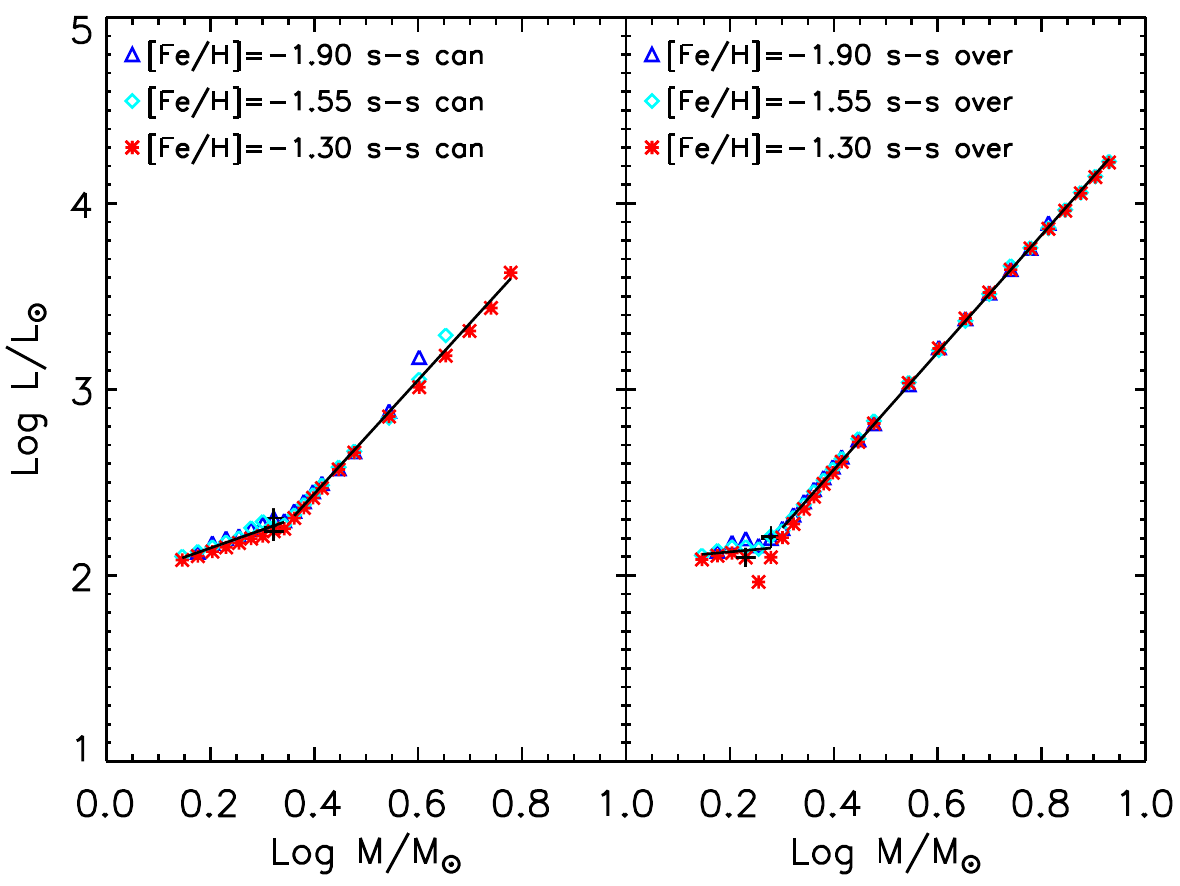}
\caption{Left: Logarithmic stellar mass versus logarithmic luminosity. 
Different symbols display canonical evolutionary models constructed by assuming 
different scaled-solar chemical compositions (see labelled values). The black 
crosses mark the transition between stellar structures, in which core 
helium burning is partially affected by electron degeneracy (less massive), 
and those with a quiescent ignition of core-helium burning (more massive). 
The dashed and the solid black lines show the linear fit for the two different 
groups. 
Right: Same as the left, but for evolutionary models constructed by assuming a 
mild convective core overshooting during central hydrogen burning phases.
%
%Right: 
}
\label{fig_ML_tra}
\end{center}
\end{figure}

Evolutionary predictions plotted in Fig.~\ref{fig_ML_tra} display the 
mass--luminosity relations in a log-log plane for stellar structures during 
central helium-burning phases. Intermediate-mass stars perform the so-called 
blue loop, during which they cross the Cepheid instability strip. The luminosity 
for these stellar structures was estimated at the center of the instability 
strip ($\log T_{\rm eff}=3.85$), as the mean between the blue-ward and the red-ward 
excursion. Stellar structures in which the helium ignition in the core takes 
place in partially electron degenerate conditions (low-mass regime) do not 
become hot enough to cross the Cepheid instability strip. The luminosity for 
these stellar structures was defined as the mean luminosity between the luminosity 
at the helium ignition and central helium exhaustion. The black crosses 
show the transition ($M/M_\odot$=2.3--2.2) between stellar structures experiencing 
a mild core-helium flash and stellar structures that ignite helium quiescently. 
In order to provide a more quantitative estimate of the transition between low and 
intermediate-mass stars, we also performed linear fits of the two groups.
We found that the ML relation for the intermediate-mass stars is: 
$$\log (M/M_{\odot}) = 1.21 \pm 0.03 + 3.06(\pm 0.05) \times \log (L/L_{\odot})$$
where the symbols have their usual meaning, and the standard deviation is $\sigma$= 0.03 dex, 
whereas for the low-mass stars it is
$$\log (M/M_{\odot}) = 1.95 \pm 0.02 + 0.98(\pm 0.08) \times \log (L/L_{\odot})$$
with $\sigma$= 0.02 dex. The current findings are suggesting that the coefficient of 
the luminosity term is three times larger when moving from the low- to the 
intermediate-mass regime. 

%%%%%%%%%%%%%%%%%%%%%%%%%%%%%%%%%%%%%%%%%%%%%%%%%%%%%%%%%%%%%%%%%%%%%%%%%%%%%%%%%%%%%
% 			fig  SF 12  
%%%%%%%%%%%%%%%%%%%%%%%%%%%%%%%%%%%%%%%%%%%%%%%%%%%%%%%%%%%%%%%%%%%%%%%%%%%%%%%%%%%%%
%_______________________________________________________________________________
%_____________________________________________________________________________________________

In order to provide a comprehensive analysis of the transition between low- and intermediate-mass stars 
we performed the same calculations, but for evolutionary models constructed by assuming a mild 
convective core overshooting during central hydrogen-burning phases. Predictions plotted in the 
right panel of Fig.~\ref{fig_ML_tra} clearly show that models taking account of core 
overshooting are at a fixed stellar mass, as expected, systematically brighter than canonical 
models. Moreover, the transition between low- and intermediate mass stars takes place 
at lower stellar masses when compared with canonical models, typically between 
$M/M_\odot$=2.0 and 1.9. The difference is due to the fact that evolutionary models 
taking into account core overshooting have, at a fixed stellar mass, larger core helium 
masses, and in turn hotter central temperatures. Therefore, they are less affected by 
electron degeneracy. We also performed the linear fits to the two different regimes 
and we found: 
$$\log (M/M_{\odot}) = 1.31 \pm 0.01 + 3.16(\pm 0.01) \times \log (L/L_{\odot})$$
with $\sigma$= 0.02 dex (intermediate-mass)
$$\log (M/M_{\odot}) = 2.08 \pm 0.07 + 0.25(\pm 0.33) \times \log (L/L_{\odot})$$
with $\sigma$= 0.06 dex (low-mass). The difference between low- and intermediate-mass 
regime is smaller than for canonical models, but the variation is still at 50\% level.
This suggests, once again, a relevant change between faint and bright ACs.

This direct evidence, together with the circumstantial evidence brought 
forward in Sect.~5 discloses a more complex evolutionary framework for ACs. 
Indeed, the current findings are suggesting, as originally found by 
\citet{caputo04} and \citet{fiorentino14a}, that ACs are a mix of stellar structures that are 
either partially affected by electron degeneracy or stellar structures that 
quiescently ignite helium. The latter group is from the evolutionary point 
of view identical to classical Cepheids. This means that more massive 
ACs are indeed, nothing else but the more metal-poor tail of classical Cepheids 
\citep{caputo04}. Therefore, the truly ACs appear to be the less massive 
ACs. This plain evidence also explains why the PL relations for 
ACs explicitly include a term for the stellar mass. The ACs identified 
in dwarf galaxies are a mix of objects which are either minimally or 
partially affected by electron degeneracy. This means that they are 
fundamental laboratories to constrain the impact of electron degeneracy 
on the pulsation properties of ACs.

%%%%%%%%%%%%%%%%%%%%%%%%%%%%%%%%%%%%%%%%%%%%%%%%%%%%%%%%%%%%%%%%%%%%%%%%%%%%%%%%%%%%%
% 			I dredge up canonical 
%%%%%%%%%%%%%%%%%%%%%%%%%%%%%%%%%%%%%%%%%%%%%%%%%%%%%%%%%%%%%%%%%%%%%%%%%%%%%%%%%%%%%
\subsection{First dredge up}\label{sec:first_DUP}

\begin{figure}[bhtp!]
\begin{center}
\includegraphics[width=0.8\textwidth]{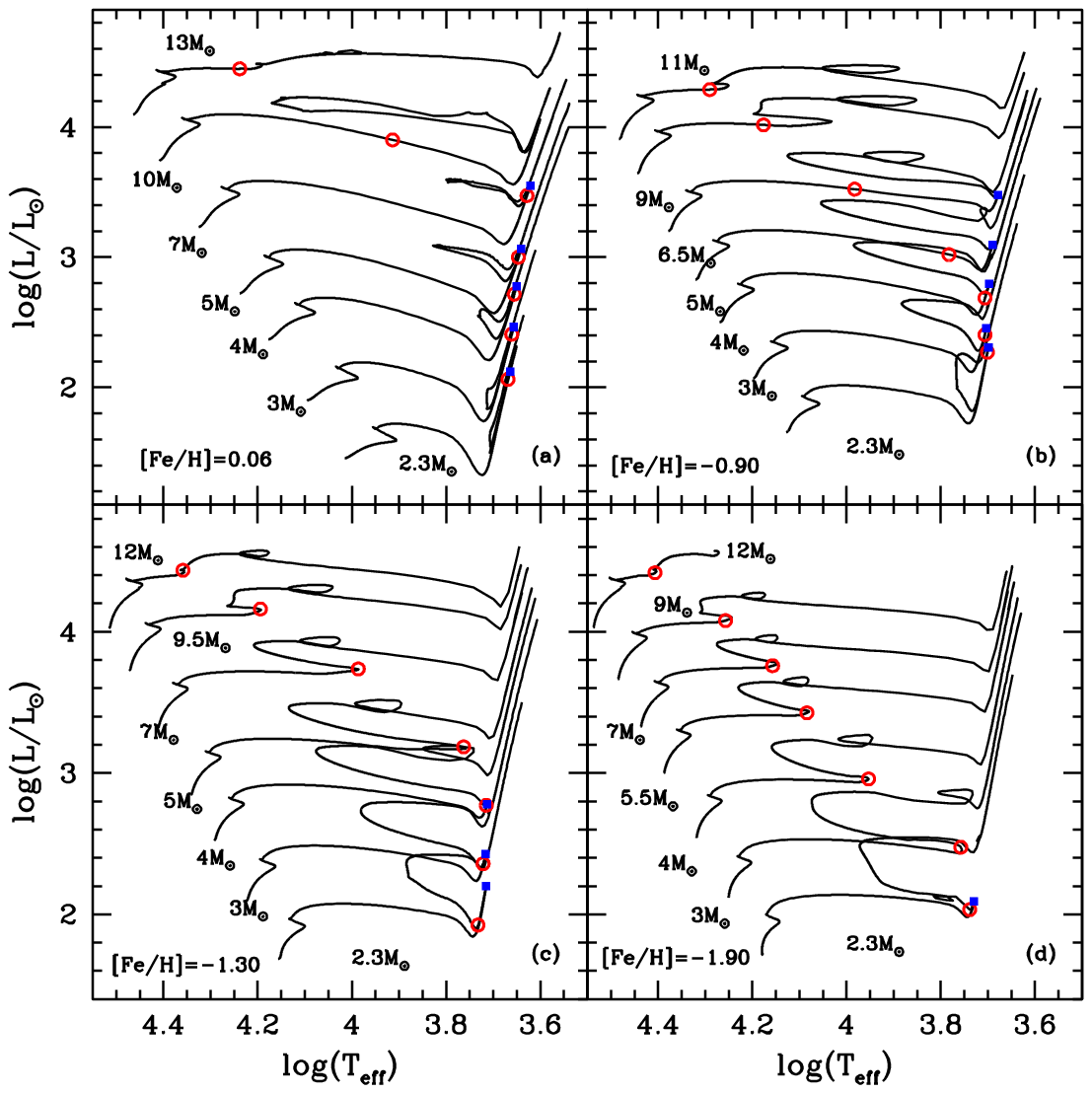}
\caption{
Panel a): Hertzsprung--Russell Diagram for young stellar structures neglecting 
convective overshooting during core hydrogen-burning phase with a scaled solar 
chemical mixture and solar iron abundance. The stellar masses range from 
2.8 to $13\,M_\odot$. The blue squares mark the position of the I dredge up, whereas 
the red circles the ignition of the 3$\alpha$ reactions (core He burning).  
Panel b): Same as the left, but for a more metal-poor ([Fe/H]$=-0.90$) chemical 
composition. 
Bottom: Same as the top, but for two more metal-poor chemical compositions:
[Fe/H]$=-1.30$ (panel c) and [Fe/H]$=-1.90$ (panel d). 
\label{fig:hrd_fdp_can}
}
\end{center}
\end{figure}

The transition between low- and intermediate-mass stars brings forward several 
interesting properties concerning the mixing and the variation of surface abundances. 
Panel a) of Fig.~\ref{fig:hrd_fdp_can} shows a set of evolutionary model at 
solar iron abundance and stellar masses ranging from 2.3 to $13\,M_\odot$. The red 
circles plotted along the individual tracks mark the central helium ignition. 
%%%%%%%%%%%%%%%%%%%%%%%%%%%%%%%%%%%%%%%%%%%%%%%%%%%%%%%%%%%%%%%%%%%%%%%%%%%%%%%%%%%%%
% 			fig SF 13 
%%%%%%%%%%%%%%%%%%%%%%%%%%%%%%%%%%%%%%%%%%%%%%%%%%%%%%%%%%%%%%%%%%%%%%%%%%%%%%%%%%%%%
%_____________________________________________________________________________________________
This key evolutionary phase moves from the tip of the red giant branch (TRGB) in the 
low- and in the intermediate-mass regime to earlier and earlier phases in the 
massive stellar regime. The helium ignition for $M\sim10\,M_\odot$ takes place 
in the so-called Hertzsprung gap, well before the stellar structure approached 
the TRGB, and soon after central hydrogen exhaustion for $M\sim13\,M_\odot$. 
The latter stellar structure does not perform the blue loop, therefore, 
its contribution to classical Cepheids becomes either minimal or vanishing. 

The blue squares plotted along the same tracks mark the phases in which the 
I dredge up (IDUP) takes place. This is a well-known mixing phenomenon, stellar 
structures moving towards their Hayashi track become systematically cooler. The 
increase in the envelope opacity causes the convective envelope to become thicker 
and thicker. During subsequent evolutionary phases the hydrogen shell moves 
outwards (the helium-core mass increases) and the bottom of the convective 
envelope recedes. In this specific phase, the IDUP occurs, during which 
nuclear processed material (central hydrogen burning phases) is dredged-up 
into the surface regions. Predictions plotted in this panel 
show that stellar structures in which central-helium ignition takes place 
before the stellar structures approach their Hayashi track either do not 
experience or experience a very mild IDUP.

Evolutionary tracks plotted in panel b) and in the panels c), d)  
display similar predictions, but for more metal-poor stellar structures
(see labelled values). The main consequence of the decrease in metal 
content is that central helium ignition in more metal-poor stellar 
structures takes place before the approach to the Hayashi tracks at 
lower stellar masses, indeed it ranges from $M=6.5\,M_\odot$ for [Fe/H]$=-0.90$ 
to $M=2.3\,M_\odot$ at [Fe/]$=-1.90$. This means that only a minor fraction, if any, 
of intermediate mass stars in the metal-poor regime experience the IDUP. This 
change has implications not only for the surface chemical abundances, 
but also on the evolutionary lifetime spent inside the instability strip.

% BG_short FS
The key evolutionary parameters at central He-burning ignition 
and at the  IDUP for canonical and convective core overshooting evolutionary 
models are listed in Table~\ref{He_burn} and in Table~\ref{He_burn_over} 
available in this Appendix.

%
%%%%%%%%%%%%%%%%%%%%%%%%%%%%%%%%%%%%%%%%%%%%%%%%%%%%%%%%%%%%%%%%%%%%%%%%%%%%%%%%%%%%%
% 			 I dredge up     
%%%%%%%%%%%%%%%%%%%%%%%%%%%%%%%%%%%%%%%%%%%%%%%%%%%%%%%%%%%%%%%%%%%%%%%%%%%%%%%%%%%%%

\begin{figure}[htbp!]
\centering
\includegraphics[width=0.49\textwidth]{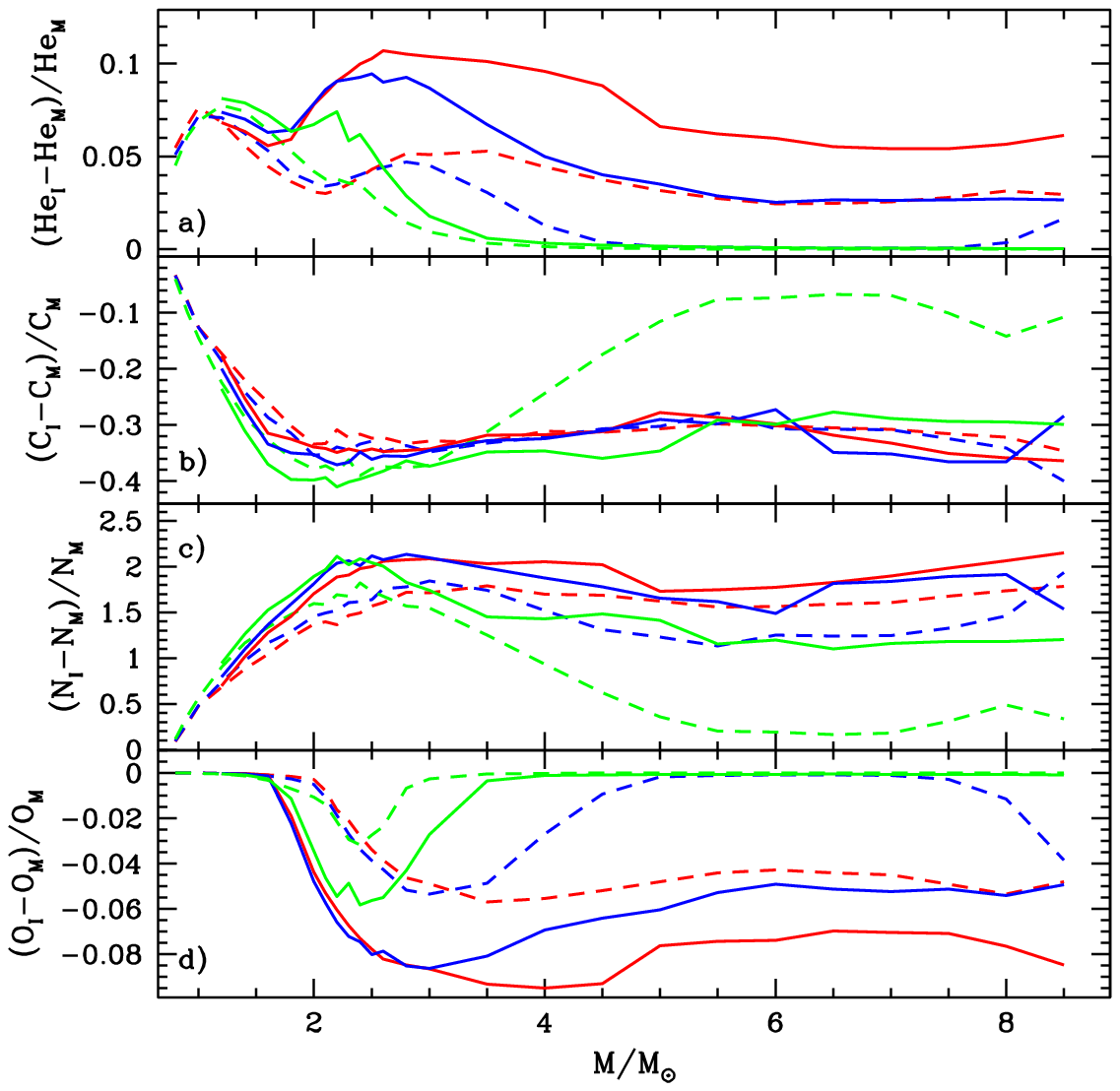}
\includegraphics[width=0.49\textwidth]{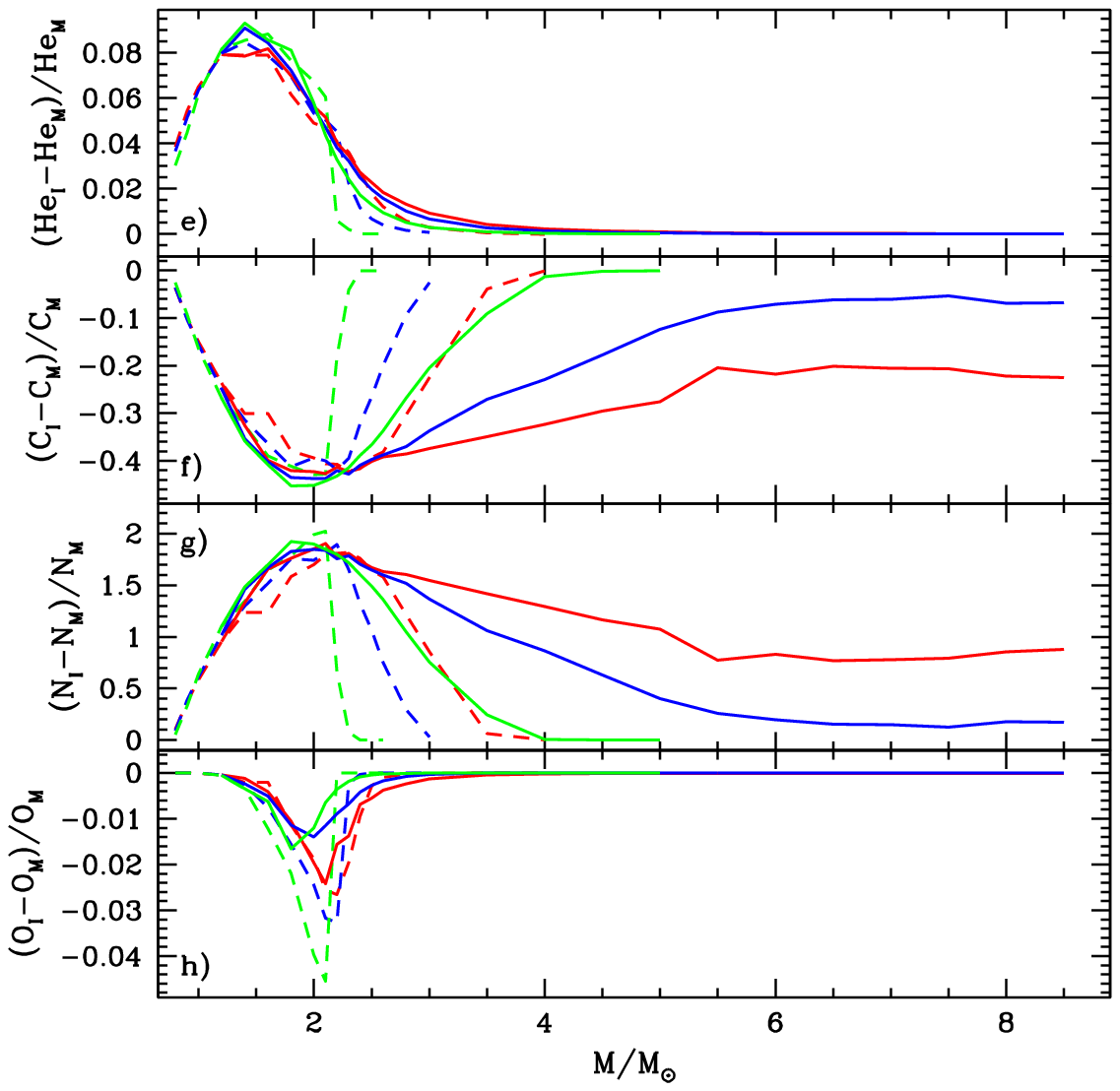}
\caption{
% editor 1 
Top: Relative difference between the surface chemical composition at the I dredge up and the 
surface chemical composition on the main sequence for He (panel a), C (panel b), N (panel c) and O (panel d)
normalized to the abundance on the main sequence. The iron abundance of the different evolutionary models is color-coded and changes from solar iron abundance ([Fe/H]=0.06, red) to half-solar ([Fe/H]$=-0.30$, blue) and to Z=0.001 ([Fe/H]$=-0.90$, green). Dashed lines display predictions based on evolutionary models neglecting convective core overshooting, while the solid ones are based on models taking into account convective core overshooting during central hydrogen-burning phases. Note that only evolutionary models that underwent the I dredge up have been plotted. 
Bottom: Same as the top panels, but for the three more metal-poor chemical compositions: [Fe/H]$=-1.30$, red; [Fe/H]$=-1.55$, blue; [Fe/H]$=-1.90$, green.}
\label{fig:abunance_ratio}
\end{figure}

To further constrain the abundance patterns of different groups of variable stars, 
we decided to analyze the impact of the I dredge up on their surface abundance. 
Figure~\ref{fig:abunance_ratio} shows the fractional variation in the abundances 
of the elements affected by the I dredge up.  The top panels display the relative difference between 
the surface chemical composition at the I dredge up and the surface chemical composition 
on the main sequence for He (panel a), C (panel b), N (panel c) and O (panel d)
normalized to the abundance on the main sequence.  These panels only 
includes stellar structures which undergo  the I dredge up and lines of different 
colors show predictions for different chemical compositions. The bottom panels 
of Fig.~\ref{fig:abunance_ratio} display the predicted fractional variation in the 
abundance of He (panel e), C (panel f), N (panel g) and O (panel h), but for 
more metal-poor chemical compositions. 

Panel a) of Fig.~\ref{fig:abunance_ratio} shows that the impact on 
the He abundance in the metal-rich regime is, as expected, modest and 
typically smaller than 10\% with a local minimum for stellar masses 
of 1.5--2.0$\,M_\odot$. The variation in helium abundance is vanishing 
for stellar masses larger than $\sim2.5\,M_\odot$ over the entire 
metallicity range. The impact is also minimal in the metal-poor regime. 
Indeed, canonical (dashed lines) and non-canonical (solid lines) 
stellar structures display  variations of the order of a few percents.

%%%%%%%%%%%%%%%%%%%%%%%%%%%%%%%%%%%%%%%%%%%%%%%%%%%%%%%%%%%%%%%%%%%%%%%%%%%%%%%%%%%%%
% 			fig SF 14   
%%%%%%%%%%%%%%%%%%%%%%%%%%%%%%%%%%%%%%%%%%%%%%%%%%%%%%%%%%%%%%%%%%%%%%%%%%%%%%%%%%%%%
%%%%%%%%%%%%%%%%%%%%%%%%%%%%%%%%%%%%%%%%%%%%%%%%%%%%%%%%%%%%%%%%%%%%%%%%%%%%%%%%
 
The variation concerning the C abundance is more complex. Indeed, the surface 
C abundance steadily decreases in more metal-rich stellar structures (left panels) 
that undergo a core helium flash. The variation in C (panel b) ranges from a few 
percent for old ($M/M_\odot=0.8$) stellar structures to 30\%--40\% for 
stellar structures at the transition between low- and intermediate-mass 
($M/M_\odot$=2.0--2.2) regime. The metallicity has a minimal impact, since they 
remain under-abundant over the entire range in stellar mass. More metal-poor stellar 
structures (panel f) display a steady increase soon after 
the maximum, thus suggesting that these structures experience a very mild IDUP. 
The difference is mainly caused by the fact that 
metal-poor stellar structures are systematically brighter and hotter than metal-rich ones. 
This means that along the RGB, convective transport is less efficient and the 
bottom of the convective envelope is not deep enough to approach the regions 
in which the CNO cycle has been efficient. The difference between canonical (dashed lines) 
and non-canonical (solid) is minimal with the former ones experiencing the IDUP 
in a very limited range in mass. 

The variation of the surface N abundance is, as expected for 
the CN cycle, the mirror image of the C abundance. 
The N abundance (panel c) steadily increases by more than a factor of two 
when moving from old  ($M/M_\odot=0.8$) to intermediate-age  
($M/M_\odot$=2.0--2.2) stellar structures. The current  
predictions indicate that more metal-rich (solar, half solar) stellar 
structures are overabundant in N over the entire mass range. On the other 
hand, metal-intermediate and metal-poor stellar structures (panel g) are 
less affected by the IDUP for stellar masses larger than $4\,M_\odot$. 
Canonical evolutionary models indicate that only stellar structures less 
massive than $4\,M_\odot$ experience the IDUP. 

The variation of the surface O abundance for metal-rich (panel d) 
stellar structures is similar to C, but there are some key differences. 
In the low-mass regime ($M/M_\odot \le 1.5$) O abundance is not affected 
by the IDUP. This trend is expected because in these structures the NO cycle 
has a very low efficiency. The variation becomes more relevant and 
approaches 5\%--6\% for stellar structures with masses ranging from 2 to 4\,$M_\odot$. 
The trend in the metal-poor regime (panel h) is the 
mirror image of He abundance, indeed, for stellar masses of 
1.8--2.5\,$M_\odot$ decreases by roughly 3\%--4\%, while it is vanishing for 
more massive stellar structures. 

The difference between metal-rich and metal-poor stellar structures is twofold: 
the metal-poor ones are brighter and hotter; this means that with a fixed stellar 
mass the central temperature is higher and the NO cycle becomes more efficient. 
The convective envelopes are less extended, since convection is less efficient 
along the RGB, but they deepen enough to dredge up material processed by both 
CN and NO cycles.  

The current predictions concerning the efficiency of the IDUP as a function 
of stellar mass and chemical composition bring forward two key features 
worth being discussed in more detail.

a) Stellar structures at the transition between low- and intermediate-mass stars after the IDUP are predicted to show an enhancement in N abundance that is two times the abundance on the main sequence. The decrease in C abundance for the same stellar structures is 30\%--40\% when compared with the MS abundance. Note that this metallicity range does not produce ACs, since their helium-burning locus is systematically cooler (redder) than the instability strip. 
 
b) The current predictions indicate that Galactic CCs should show larger abundance variations due to the IDUP when compared with LMC CCs, and in particular, with SMC CCs. In this context, it is worth mentioning that canonical evolutionary models for intermediate-mass stars in the metal-poor regime do not experience any IDUP. Therefore, CCs in very metal-poor stellar systems (IC~1613, SagDig, NGC~6822) should show N and C surface abundances similar to their main sequence progenitors. 

\clearpage
%%%%%%%%%%%%%%%%%%%%%%%%%%%%%%%%%%%%%%%%%%%%%%%%%%%%%%%%%%%%%%%%%%%%%%%%%%%%%%%%%%%%%%%%%%%%%%%%%%%%
%			Table  ST 4 
%%%%%%%%%%%%%%%%%%%%%%%%%%%%%%%%%%%%%%%%%%%%%%%%%%%%%%%%%%%%%%%%%%%%%%%%%%%%%%%%%%%%%%%%%%%%%%%%%%%%
\begin{table}[htbp!]
\centering
\footnotesize
\caption{
Evolutionary parameters at central He-burning ignition and at I dredge-up (IDUP) from BaSTI2 scaled solar \citep{caffauetal2011} canonical evolutionary models.
}
\label{He_burn}
{%\small
\begin{tabular}{ccccccc}
\noalign{\smallskip}\hline\hline\noalign{\smallskip}
    & \multicolumn{3}{c}{He burning}    & \multicolumn{3}{c}{IDUP}     \\ 
   $M/M_\odot$  & $\log\,t$ & $\log\,L/L_\odot$  &  $\log\,T_{\rm eff}$ & $\log\,t$ & $\log\,L/L_\odot$  &  $\log\,T_{\rm eff}$ \\
\noalign{\smallskip}\hline\noalign{\smallskip}
\multicolumn{7}{c}{[Fe/H]=+0.06, Y=0.2695} \\
%    \multicolumn{7}{c}{}   \\ 
    2.5  &    8.726  &  2.353  &  3.653 &    2.268  &  3.659   &  8.724 \\ 
    2.6  &    8.676  &  2.342  &  3.656 &    2.347  &  3.656   &  8.676 \\ 
    2.8  &    8.587  &  2.365  &  3.659 &    2.400  &  3.657   &  8.587 \\ 
    3.0  &    8.505  &  2.409  &  3.661 &    2.436  &  3.659   &  8.506 \\ 
    3.5  &    8.328  &  2.562  &  3.659 &    2.534  &  3.661   &  8.330 \\ 
    4.0  &    8.177  &  2.716  &  3.655 &    2.621  &  3.663   &  8.177 \\ 
    4.5  &    8.048  &  2.863  &  3.651 &    2.691  &  3.666   &  8.047 \\ 
    5.0  &    7.931  &  3.000  &  3.646 &    3.064  &  3.641   &  7.931 \\ 
    5.5  &    7.833  &  3.129  &  3.642 &    3.194  &  3.636   &  7.833 \\ 
    6.0  &    7.745  &  3.249  &  3.637 &    3.320  &  3.631   &  7.745 \\ 
    6.5  &    7.667  &  3.364  &  3.633 &    3.437  &  3.626   &  7.668 \\ 
    7.0  &    7.598  &  3.473  &  3.628 &    3.549  &  3.621   &  7.598 \\ 
    7.5  &    7.536  &  3.568  &  3.625 &    3.655  &  3.617   &  7.536 \\ 
    8.0  &    7.479  &  3.601  &  3.627 &    3.755  &  3.612   &  7.480 \\ 
    8.5  &    7.428  &  3.493  &  3.642 &    3.848  &  3.608   &  7.429 \\ 
    9.0  &    7.381  &  3.449  &  3.665 &    3.937  &  3.604   &  7.382 \\ 
    9.5  &    7.338  &  3.702  &  3.767 &    4.011  &  3.601   &  7.339 \\ 
   10.0  &    7.299  &  3.900  &  3.913 &    4.057  &  3.600   &  7.300 \\ 
   11.0  &    7.229  &  4.139  &  4.091 &    4.070  &  3.605   &  7.230 \\ 
   12.0  &    7.189  &  4.324  &  4.210 &    4.328  &  3.596   &  7.257 \\ 
   13.0  &    7.147  &  4.447  &  4.238 &    4.537  &  3.577   &  7.216 \\ 
   14.0  &    7.140  &  4.558  &  4.279 &    4.715  &  3.564   &  7.197 \\ 
%    \multicolumn{7}{c}{}   \\ 
\multicolumn{7}{c}{[Fe/H]=--0.30, Y=0.2571} \\
%    \multicolumn{7}{c}{}   \\ 
%====================================================================================
%   M/Msun      logt       logL      logTe              
%%====================================================================================
    2.5  &    8.647  &  2.330  &  3.676 &    2.358  &  3.674   &  8.648 \\
    2.6  &    8.602  &  2.332  &  3.678 &    2.370  &  3.676   &  8.602 \\
    2.8  &    8.517  &  2.367  &  3.680 &    2.412  &  3.677   &  8.518 \\
    3.0  &    8.440  &  2.419  &  3.681 &    2.468  &  3.678   &  8.440 \\
    3.5  &    8.271  &  2.579  &  3.679 &    2.632  &  3.675   &  8.271 \\
    4.0  &    8.130  &  2.731  &  3.676 &    2.735  &  3.676   &  8.130 \\
    4.5  &    8.008  &  2.880  &  3.671 &    2.826  &  3.676   &  8.008 \\
    5.0  &    7.903  &  3.017  &  3.668 &    3.078  &  3.663   &  7.902 \\
    5.5  &    7.806  &  3.133  &  3.665 &    3.211  &  3.658   &  7.806 \\
    6.0  &    7.724  &  3.193  &  3.666 &    3.338  &  3.653   &  7.725 \\
    6.5  &    7.651  &  3.139  &  3.681 &    3.456  &  3.648   &  7.652 \\
    7.0  &    7.586  &  3.236  &  3.698 &    3.570  &  3.644   &  7.587 \\
    7.5  &    7.528  &  3.431  &  3.729 &    3.673  &  3.639   &  7.528 \\
    8.0  &    7.475  &  3.592  &  3.801 &    3.773  &  3.633   &  7.476 \\
    8.5  &    7.427  &  3.728  &  3.879 &    3.862  &  3.631   &  7.427 \\
    9.0  &    7.383  &  3.850  &  3.950 &    3.940  &  3.628   &  7.383 \\
    9.5  &    7.343  &  3.981  &  4.040 &    3.974  &  3.629   &  7.343 \\
   10.0  &    7.306  &  4.068  &  4.104 &    3.976  &  3.633   &  7.307 \\
   11.0  &    7.239  &  4.243  &  4.208 &    4.364  &  3.608   &  7.307 \\
   12.0  &    7.200  &  4.362  &  4.268 &    4.445  &  3.607   &  7.266 \\
\hline\noalign{\smallskip}
\multicolumn{7}{r}{\it {\footnotesize continued on next page}} \\
\end{tabular}}
\end{table}
%---------------------------------------------------------------------------
%				new page 
%---------------------------------------------------------------------------
\addtocounter{table}{-1}
\begin{table}[htbp!]
\centering
\footnotesize
\caption[]{continued.}
{%\small
\begin{tabular}{ccccccc}
\noalign{\smallskip}\hline\hline\noalign{\smallskip}
    & \multicolumn{3}{c}{He burning}    & \multicolumn{3}{c}{IDUP}     \\ 
   $M/M_\odot$  & $\log\,t$ & $\log\,L/L_\odot$  &  $\log\,T_{\rm eff}$ & $\log\,t$ & $\log\,L/L_\odot$  &  $\log\,T_{\rm eff}$ \\
\noalign{\smallskip}\hline\noalign{\smallskip}
%
%    \multicolumn{7}{c}{}   \\ 
\multicolumn{7}{c}{[Fe/H]=--0.60, Y=0.2521} \\
%    \multicolumn{7}{c}{}   \\ 
%======================================================
%   M/Msun      logt       logL      logTe              
%======================================================
    2.5  &      8.598  &  2.296  &  3.692 &   2.341  & 3.690   & 8.599 \\
    2.6  &      8.554  &  2.309  &  3.694 &   2.330  & 3.693   & 8.554 \\
    2.8  &      8.472  &  2.355  &  3.695 &   2.404  & 3.692   & 8.473 \\
    3.0  &      8.398  &  2.418  &  3.695 &   2.466  & 3.692   & 8.398 \\
    3.5  &      8.236  &  2.581  &  3.693 &   2.633  & 3.689   & 8.236 \\
    4.0  &      8.099  &  2.737  &  3.690 &   2.795  & 3.686   & 8.100 \\
    4.5  &      7.982  &  2.878  &  3.687 &   2.944  & 3.682   & 7.983 \\
    5.0  &      7.880  &  2.917  &  3.692 &   3.057  & 3.681   & 7.885 \\
    5.5  &      7.788  &  2.960  &  3.714 &   3.225  & 3.673   & 7.789 \\
    6.0  &      7.709  &  3.196  &  3.756 &   3.351  & 3.668   & 7.710 \\
    6.5  &      7.640  &  3.381  &  3.837 &   3.468  & 3.664   & 7.640 \\
    7.0  &      7.577  &  3.530  &  3.894 &   3.582  & 3.660   & 7.578 \\
    7.5  &      7.521  &  3.656  &  3.939 &   3.688  & 3.656   & 7.522 \\
    8.0  &      7.470  &  3.764  &  3.986 &   3.782  & 3.652   & 7.470 \\
    8.5  &      7.423  &  3.866  &  4.031 &   3.866  & 3.648   & 7.424 \\
    9.0  &      7.381  &  3.964  &  4.088 &   3.913  & 3.648   & 7.382 \\
    9.5  &      7.342  &  4.056  &  4.149 &   4.183  & 3.634   & 7.409 \\
   10.0  &      7.306  &  4.130  &  4.186 &   4.256  & 3.631   & 7.373 \\
%    \multicolumn{7}{c}{}   \\ 
\multicolumn{7}{c}{[Fe/H]=--0.90, Y=0.2496} \\
%    \multicolumn{7}{c}{}   \\ 
    2.5  &      8.564  &  2.255  &  3.706 &   2.302 &3.703  &8.565 \\
    2.6  &      8.521  &  2.277  &  3.706 &   2.326 &3.704  &8.522 \\
    2.8  &      8.442  &  2.336  &  3.707 &   2.386 &3.704  &8.442 \\
    3.0  &      8.369  &  2.404  &  3.706 &   2.454 &3.703  &8.370 \\
    3.5  &      8.212  &  2.574  &  3.705 &   2.627 &3.701  &8.212 \\
    4.0  &      8.079  &  2.686  &  3.706 &   2.795 &3.698  &8.079 \\
    4.5  &      7.964  &  2.747  &  3.726 &   2.944 &3.694  &7.965 \\
    5.0  &      7.866  &  3.020  &  3.783 &   3.096 &3.690  &7.867 \\
    5.5  &      7.776  &  3.227  &  3.880 &   3.231 &3.686  &7.777 \\
    6.0  &      7.700  &  3.387  &  3.939 &   3.358 &3.682  &7.701 \\
    6.5  &      7.632  &  3.523  &  3.983 &   3.477 &3.678  &7.633 \\
    7.0  &      7.571  &  3.643  &  4.021 &   3.587 &3.675  &7.572 \\
    7.5  &      7.516  &  3.750  &  4.051 &   3.691 &3.671  &7.518 \\
    8.0  &      7.467  &  3.843  &  4.081 &   3.784 &3.667  &7.468 \\
    8.5  &      7.421  &  3.929  &  4.120 &   3.848 &3.666  &7.423 \\
    9.0  &      7.379  &  4.016  &  4.176 &   4.289 &3.635  &7.447 \\
    9.5  &      7.341  &  4.095  &  4.206 &   4.375 &3.631  &7.407 \\
\hline\noalign{\smallskip}
\end{tabular}}
\end{table}
%%%%%%%%%%%%%%%%%%%%%%%%%%%%%%%%%%%%%%%%%%%%%%%%%%%%%%%%%%%%%%%%%%%%%%%%%%%%%%%%%%%%%%%%%%%%%%%%%%%%

%%%%%%%%%%%%%%%%%%%%%%%%%%%%%%%%%%%%%%%%%%%%%%%%%%%%%%%%%%%%%%%%%%%%%%%%%%%%%%%%%%%%%%%%%%%%%%%%%%%%
%			Table  ST 5 
%%%%%%%%%%%%%%%%%%%%%%%%%%%%%%%%%%%%%%%%%%%%%%%%%%%%%%%%%%%%%%%%%%%%%%%%%%%%%%%%%%%%%%%%%%%%%%%%%%%%
\begin{table}[bhtp!]
\centering
\footnotesize
\caption{\small 
Evolutionary parameters at central He-burning ignition and at I dredge-up (IDUP) from BaSTI2 scaled-solar \citep{caffauetal2011} evolutionary models taking into account of convective core overshooting during central hydrogen burning phases.  
}
\label{He_burn_over}
{%\small
\begin{tabular}{ccccccc}
\noalign{\smallskip}\hline\hline\noalign{\smallskip}
    & \multicolumn{3}{c}{He burning}    & \multicolumn{3}{c}{IDUP}    \\ 
   $M/M_\odot$  & $\log\,t$ & $\log\,L/L_\odot$  &  $\log\,T_{\rm eff}$ & $\log\,t$ & $\log\,L/L_\odot$  &  $\log\,T_{\rm eff}$ \\
\noalign{\smallskip}\hline\noalign{\smallskip}
\multicolumn{7}{c}{[Fe/H]=+0.06, Y=0.2695} \\
%    \multicolumn{7}{c}{}   \\ 
%
    % 2.0   &   9.08007889 & 1.474264 & 3.701339 \\
    % 2.1   &   9.01507665 & 2.448737 & 3.637050 \\
    % 2.2   &   8.95512166 & 2.463532 & 3.639037 \\
    % 2.3   &   8.89910102 & 2.466122 & 3.641789 \\
    % 2.4   &   8.84662428 & 2.478242 & 3.643688 \\
    2.5   &   8.796 & 2.501 & 3.645 &   2.538 & 3.642  & 8.797\\
    2.6   &   8.749 & 2.523 & 3.645 &   2.567 & 3.642  & 8.749\\
    2.8   &   8.660 & 2.590 & 3.645 &   2.635 & 3.641  & 8.660\\
    3.0   &   8.578 & 2.661 & 3.644 &   2.712 & 3.640  & 8.579\\
    3.5   &   8.400 & 2.840 & 3.639 &   2.894 & 3.635  & 8.400\\
    4.0   &   8.249 & 3.013 & 3.634 &   3.071 & 3.629  & 8.250\\
    4.5   &   8.120 & 3.173 & 3.628 &   3.231 & 3.622  & 8.121\\
    5.0   &   8.004 & 3.312 & 3.622 &   3.380 & 3.616  & 8.004\\
    5.5   &   7.905 & 3.449 & 3.617 &   3.523 & 3.610  & 7.906\\
    6.0   &   7.818 & 3.576 & 3.612 &   3.649 & 3.604  & 7.818\\
    6.5   &   7.740 & 3.692 & 3.607 &   3.765 & 3.599  & 7.741\\
    7.0   &   7.670 & 3.798 & 3.602 &   3.872 & 3.595  & 7.671\\
    7.5   &   7.607 & 3.896 & 3.598 &   3.972 & 3.590  & 7.608\\
    8.0   &   7.550 & 3.987 & 3.594 &   4.064 & 3.586  & 7.551\\
    8.5   &   7.498 & 4.069 & 3.591 &   4.150 & 3.583  & 7.499\\
    9.0   &   7.451 & 4.153 & 3.587 &   4.229 & 3.579  & 7.451\\
    9.5   &   7.407 & 4.236 & 3.583 &   4.303 & 3.576  & 7.408\\
   10.0   &   7.367 & 4.312 & 3.580 &   4.372 & 3.574  & 7.367\\
   11.0   &   7.295 & 3.893 & 3.640 &   4.496 & 3.569  & 7.295\\
   12.0   &   7.234 & 4.329 & 3.942 &   4.583 & 3.568  & 7.234\\
   13.0   &   7.181 & 4.513 & 4.081 &   4.556 & 3.578  & 7.181\\
   14.0   &   7.135 & 4.653 & 4.173 &   4.787 & 3.561  & 7.177\\
   % 15.0   &   7.09023960 & 4.738234 & 4.211248 \\
%    \multicolumn{7}{c}{}   \\ 
\multicolumn{7}{c}{[Fe/H]=--0.30, Y=0.2571} \\
%    \multicolumn{7}{c}{}   \\ 
    % 2.0   &   8.98490728 & 2.470072 & 3.655262 \\
    % 2.1   &   8.92461751 & 2.463430 & 3.658792 \\
    % 2.2   &   8.86794334 & 2.462231 & 3.661809 \\
    % 2.3   &   8.81562425 & 2.474309 & 3.663726 \\
    % 2.4   &   8.76599189 & 2.494028 & 3.664848 \\
    2.5   &   8.718 & 2.520 & 3.665 &   2.460 & 3.670 &  8.718\\
    2.6   &   8.674 & 2.553 & 3.665 &   2.481 & 3.671 &  8.673\\
    2.8   &   8.589 & 2.621 & 3.665 &   2.540 & 3.671 &  8.589\\
    3.0   &   8.512 & 2.693 & 3.663 &   2.584 & 3.672 &  8.512\\
    3.5   &   8.344 & 2.877 & 3.659 &   2.685 & 3.674 &  8.344\\
    4.0   &   8.203 & 3.046 & 3.654 &   2.770 & 3.678 &  8.202\\
    4.5   &   8.081 & 3.206 & 3.648 &   2.878 & 3.680 &  8.081\\
    5.0   &   7.976 & 3.351 & 3.643 &   3.302 & 3.647 &  7.976\\
    5.5   &   7.879 & 3.477 & 3.639 &   3.548 & 3.632 &  7.880\\
    6.0   &   7.798 & 3.601 & 3.634 &   3.675 & 3.627 &  7.798\\
    6.5   &   7.725 & 3.716 & 3.629 &   3.788 & 3.622 &  7.725\\
    7.0   &   7.659 & 3.817 & 3.625 &   3.893 & 3.618 &  7.659\\
    7.5   &   7.600 & 3.912 & 3.621 &   3.992 & 3.614 &  7.600\\
    8.0   &   7.546 & 3.978 & 3.620 &   4.082 & 3.610 &  7.546\\
    8.5   &   7.497 & 4.082 & 3.615 &   4.168 & 3.607 &  7.498\\
    9.0   &   7.452 & 4.128 & 3.615 &   4.246 & 3.604 &  7.452\\
    9.5   &   7.410 & 3.949 & 3.638 &   4.317 & 3.601 &  7.411\\
   10.0   &   7.373 & 4.060 & 3.757 &   4.385 & 3.599 &  7.373\\
   11.0   &   7.304 & 4.281 & 3.954 &   4.496 & 3.595 &  7.304\\
   12.0   &   7.244 & 4.436 & 4.088 &   4.553 & 3.596 &  7.245\\
   13.0   &   7.192 & 4.552 & 4.175 &   4.579 & 3.601 &  7.192\\
   14.0   &   7.145 & 4.680 & 4.255 &   4.813 & 3.586 &  7.185\\
   % 15.0   &   7.10713008 & 4.768887 & 4.272048 \\
\hline\noalign{\smallskip}
\multicolumn{7}{r}{\it {\footnotesize continued on next page}} \\
\end{tabular}}
\end{table}
%---------------------------------------------------------------------------
%				new page 
%---------------------------------------------------------------------------
\addtocounter{table}{-1}
\begin{table}[bhtp!]
\centering
\footnotesize
\caption[]{continued.}
{%\small
\begin{tabular}{ccccccc}
\noalign{\smallskip}\hline\hline\noalign{\smallskip}
    & \multicolumn{3}{c}{He burning}    & \multicolumn{3}{c}{IDUP}    \\ 
   $M/M_\odot$  & $\log\,t$ & $\log\,L/L_\odot$  &  $\log\,T_{\rm eff}$ & $\log\,t$ & $\log\,L/L_\odot$  &  $\log\,T_{\rm eff}$ \\
\noalign{\smallskip}\hline\noalign{\smallskip}
%
%    \multicolumn{7}{c}{}   \\ 
\multicolumn{7}{c}{[Fe/H]=--0.60, Y=0.2521} \\
%    \multicolumn{7}{c}{}   \\ 
    % 2.0   &   8.92651738 & 2.446150 & 3.672684 \\
    % 2.1   &   8.86765588 & 2.439822 & 3.675864 \\
    % 2.2   &   8.81412121 & 2.449039 & 3.677873  \\
    % 2.3   &   8.76330705 & 2.464390 & 3.679437  \\
    % 2.4   &   8.71533205 & 2.491241 & 3.679864  \\
    2.5   &   8.670 & 2.527 & 3.680 &    2.573 & 3.677 &  8.670 \\
    2.6   &   8.626 & 2.562 & 3.679 &    2.610 & 3.676 &  8.627 \\
    2.8   &   8.545 & 2.636 & 3.679 &    2.685 & 3.675 &  8.546 \\
    3.0   &   8.471 & 2.713 & 3.677 &    2.718 & 3.677 &  8.471 \\
    3.5   &   8.309 & 2.895 & 3.673 &    2.824 & 3.678 &  8.308 \\
    4.0   &   8.173 & 3.066 & 3.668 &    2.931 & 3.680 &  8.172 \\
    4.5   &   8.056 & 3.223 & 3.663 &    3.244 & 3.662 &  8.056 \\
    5.0   &   7.954 & 3.369 & 3.658 &    3.426 & 3.654 &  7.955 \\
    5.5   &   7.862 & 3.493 & 3.655 &    3.565 & 3.649 &  7.862 \\
    6.0   &   7.783 & 3.614 & 3.650 &    3.687 & 3.644 &  7.784 \\
    6.5   &   7.713 & 3.720 & 3.647 &    3.801 & 3.640 &  7.714 \\
    7.0   &   7.650 & 3.759 & 3.649 &    3.905 & 3.635 &  7.650 \\
    7.5   &   7.593 & 3.670 & 3.666 &    4.002 & 3.632 &  7.593 \\
    8.0   &   7.541 & 3.733 & 3.681 &    4.091 & 3.628 &  7.541 \\
    8.5   &   7.493 & 3.896 & 3.722 &    4.174 & 3.625 &  7.493 \\
    9.0   &   7.450 & 4.019 & 3.797 &    4.252 & 3.622 &  7.450 \\
    9.5   &   7.409 & 4.122 & 3.878 &    4.323 & 3.620 &  7.410 \\
   10.0   &   7.372 & 4.209 & 3.954 &    4.386 & 3.618 &  7.373 \\
   % 11.0   &   7.30538749 & 4.357397 & 4.074689  \\
   % 12.0   &   7.24700827 & 4.483211 & 4.173480  \\
   % 13.0   &   7.19562903 & 4.587816 & 4.240733  \\
   % 14.0   &   7.15006586 & 4.691456 & 4.288631  \\
   % 15.0   &   7.10927467 & 4.768207 & 4.319907  \\
%    \multicolumn{7}{c}{}   \\ 
\multicolumn{7}{c}{[Fe/H]=--0.90, Y=0.2496} \\
%    \multicolumn{7}{c}{}   \\ 
    % 2.0   &   8.88591854 & 2.408472 & 3.687955 \\
    % 2.1   &   8.83017738 & 2.414715 & 3.689956 \\
    % 2.2   &   8.77726651 & 2.426683 & 3.691555 \\
    % 2.3   &   8.72786997 & 2.456485 & 3.691896 \\
    % 2.4   &   8.68106008 & 2.491549 & 3.691952 \\
    2.5   &   8.636 & 2.527 & 3.692 &   2.571 & 3.689 &  8.637\\
    2.6   &   8.594 & 2.563 & 3.692 &   2.611 & 3.689 &  8.594\\
    2.8   &   8.515 & 2.638 & 3.691 &   2.688 & 3.688 &  8.515\\
    3.0   &   8.443 & 2.719 & 3.689 &   2.766 & 3.686 &  8.443\\
    3.5   &   8.285 & 2.902 & 3.685 &   2.955 & 3.682 &  8.285\\
    4.0   &   8.152 & 3.075 & 3.681 &   3.126 & 3.677 &  8.153\\
    4.5   &   8.038 & 3.230 & 3.677 &   3.262 & 3.675 &  8.042\\
    5.0   &   7.940 & 3.373 & 3.673 &   3.404 & 3.670 &  7.940\\
    5.5   &   7.850 & 3.482 & 3.671 &   3.569 & 3.663 &  7.851\\
    6.0   &   7.774 & 3.482 & 3.678 &   3.690 & 3.659 &  7.774\\
    6.5   &   7.705 & 3.515 & 3.693 &   3.803 & 3.655 &  7.706\\
    7.0   &   7.644 & 3.683 & 3.717 &   3.907 & 3.651 &  7.644\\
    7.5   &   7.588 & 3.831 & 3.798 &   4.003 & 3.648 &  7.588\\
    8.0   &   7.537 & 3.943 & 3.865 &   4.090 & 3.645 &  7.537\\
    8.5   &   7.490 & 4.038 & 3.918 &   4.175 & 3.642 &  7.491\\
    9.0   &   7.447 & 4.124 & 3.965 &   4.252 & 3.640 &  7.448\\
    9.5   &   7.408 & 4.199 & 4.013 &   4.322 & 3.637 &  7.408\\
   % 10.0   &   7.37110227 & 4.268677 & 4.070103 \\
   % 11.0   &   7.30534985 & 4.397286 & 4.157821 \\
   % 12.0   &   7.24749617 & 4.506278 & 4.238860 \\
   % 13.0   &   7.19678660 & 4.605303 & 4.298623 \\
\hline\noalign{\smallskip}
\end{tabular}}
\end{table}

% BG_short FS 
%\include{tab_heburn+Idup_reviewstyle}
%\include{tab_heburn+Idup_over_reviewstyle}
%\include{table/can/tab_Idup_sss_basti2_reviewstyle} 
%\include{table/over/tab_Idup_sso_basti2_chim_reviewstyle} 
\end{appendices}
\clearpage

%%%%%%%%%%%%%%%%%%%%%%%%%%%%%%%%%%%%%%%%%%%%%%%%%%%%%%%%%%%%%%%%%%%%%%%%%%%%%
%\clearpage
\phantomsection
\addcontentsline{toc}{section}{References}
% \bibliography{ms_nomacro} % if your bibtex file is called example.bib
\bibliography{ms} % if your bibtex file is called example.bib

\end{document}